\documentclass[preprint,12pt,numberedappendix,appendixfloats]{emulateapj}

\slugcomment{submitted to ApJ}  % comment overwriting the draft version and date
\shorttitle{High-$z$ Sub-kpc Resolution Metallicity Maps from \hst}
\shortauthors{Wang et al. (2016)}

\usepackage{ccaption}
\usepackage[english]{babel}
\usepackage[utf8]{inputenc} % useful to type directly diacritic characters
\usepackage[T1]{fontenc}
\usefont{T1}{tnr}{m}{sl}
\usepackage{fancyhdr}       % fancy header
\usepackage{pgf,pgfarrows,pgfnodes,pgfautomata,pgfheaps}
\usepackage{graphicx}       % you can specify [driver] (e.g. [pdftex]) which is the dvi to postscript program
\usepackage{amsmath,amssymb,amsxtra,amsfonts}   % to use pmatrix, etc.
\usepackage{txfonts}       % <<160621>> seems contradicting w/ ams packages at first, but maybe not... delete previous .aux file!
\usepackage{natbib}         % enables the use of \citep and others; see http://merkel.zoneo.net/Latex/natbib.php
\usepackage{url}            % enables the use of url web links
\usepackage{mathtools}      % enables the use of various environments work, e.g., |align|, |dcases|, define symbols (see below), etc.
\usepackage{bbold}          % enables the use of \mathbb{1} as the identity matrix

\usepackage{bbm}            % enables the use of bold Greek symbols

\usepackage{multirow}       % enables the use of \multirow{num}{*}{text} in deluxetable
\usepackage{xspace}         % enables the use of \xspace in defining macros

 \usepackage{color}         % using color package; put color boxes around text: \fcolorbox{frame colour}{background colour}{text}
 \definecolor{gold}{rgb}{1,0.80,0}
 \definecolor{orange}{rgb}{1,0.5,0}
 \definecolor{midgray}{gray}{0.3}
 \definecolor{lblue}{rgb}{0,0.2,0.6}
 \definecolor{dgreen}{rgb}{0.1,0.6,0.3}
\usepackage[colorlinks=true,citecolor=lblue,linkcolor=dgreen]{hyperref}    % does not work when enabling draft, nor with beamer

\newcommand{\be}{\begin{equation}}
\newcommand{\ee}{\end{equation}}

\newcommand{\ba}{\begin{align}}
\newcommand{\ea}{\end{align}}

\newcommand{\avg}[1]{\left<#1\right>}
\newcommand{\defeq}{\vcentcolon=}

\newcommand{\Msun}{\ensuremath{M_\odot}\xspace}

\newcommand{\chisq}{\ensuremath{\chi^2}\xspace}

\newcommand{\Mstar}{\ensuremath{M_\ast}\xspace}
\newcommand{\Lstar}{\ensuremath{L_\ast}\xspace}
\newcommand{\oh}{\ensuremath{12+\log({\rm O/H})}\xspace}
\newcommand{\Av}{\ensuremath{A_{\rm V}}\xspace}
\newcommand{\Rv}{\ensuremath{R_{\rm V}}\xspace}

\newcommand{\kpc}{\ensuremath{\rm kpc}\xspace}

\newcommand{\kms}{\ensuremath{\rm km~s^{-1}}\xspace}
\newcommand{\Hunit}{\ensuremath{\rm km~s^{-1}~Mpc^{-1}}\xspace}
\newcommand{\Funit}{\ensuremath{\rm erg~s^{-1}~cm^{-2}}\xspace}

\def\micron{\ensuremath{\mu\textrm{m}}\xspace}  % better than the default \micron, which does not use \xspace
\newcommand{\Ha}{\textrm{H}\ensuremath{\alpha}\xspace}
\newcommand{\Hb}{\textrm{H}\ensuremath{\beta}\xspace}
\newcommand{\Hg}{\textrm{H}\ensuremath{\gamma}\xspace}
\newcommand{\HII}{\textrm{H}\textsc{ii}\xspace}

\newcommand{\OII}{[\textrm{O}~\textsc{ii}]\xspace}
\newcommand{\OIII}{[\textrm{O}~\textsc{iii}]\xspace}
\newcommand{\CIII}{\textrm{C}~\textsc{iii}]\xspace}
\newcommand{\NII}{[\textrm{N}~\textsc{ii}]\xspace}
\newcommand{\SII}{[\textrm{S}~\textsc{ii}]\xspace}

\def\B{\ensuremath{B_{435}}\xspace}

\def\I{\ensuremath{I_{814}}\xspace}

\def\H{\ensuremath{H_{160}}\xspace}

\newcommand{\clsan}{MACS1149.6+2223\xspace}

\newcommand{\sex}{\textsc{SExtractor}\xspace}
\newcommand{\emc}{\textsc{Emcee}\xspace}
\newcommand{\linmix}{\textsc{linmix}\xspace}
\newcommand{\adriz}{\textsc{AstroDrizzle}\xspace}
\newcommand{\dpac}{\textsc{DrizzlePac}\xspace}
\newcommand{\fast}{\textsc{FAST}\xspace}
\newcommand{\axe}{\textsc{aXe}\xspace}

\newcommand{\glafic}{\textsc{Glafic}\xspace}
\newcommand{\gasoline}{\textsc{Gasoline}\xspace}
\newcommand{\ramses}{\textsc{Ramses}\xspace}
\newcommand{\SJ}{\textsc{Sharon \& Johnson}\xspace}

\newcommand{\hst}{\textit{HST}\xspace}
\newcommand{\jwst}{\textit{JWST}\xspace}

\newcommand{\herschel}{\textit{Herschel}\xspace}

\newcommand{\glass}{\textit{GLASS}\xspace}

\newcommand{\hff}{\textit{HFF}\xspace}
\newcommand{\muse}{\textit{MUSE}\xspace}
\newcommand{\kmos}{\textit{KMOS}\xspace}
\newcommand{\deimos}{\textit{DEIMOS}\xspace}
\newcommand{\mosfire}{\textit{MOSFIRE}\xspace}

\newcommand{\kd}{\textit{KMOS}$^{3\rm D}$\xspace}
\newcommand{\candels}{\textit{CANDELS}\xspace}

\def\mosdef{\textit{MOSDEF}\xspace}
\newcommand{\vlt}{\textit{VLT}\xspace}
\newcommand{\osiris}{\textit{OSIRIS}\xspace}
\newcommand{\keck}{\textit{Keck}\xspace}
\newcommand{\sinf}{\textit{SINFONI}\xspace}

\def\ie{i.e.\xspace}
\def\eg{e.g.\xspace}

\def\vsv{vis-\'a-vis\xspace}
\renewcommand\({\left(}
\renewcommand\){\right)}
\newcommand\mg{metallicity gradient\xspace}
\newcommand\mgs{metallicity gradients\xspace}

\newcommand\mgm{metallicity gradient measurement\xspace}
\newcommand\mgms{metallicity gradient measurements\xspace}

\newcommand\sra{spatially resolved analysis\xspace}

\newcommand\gpm{gas-phase metallicity\xspace}
\newcommand\subr{surface brightness\xspace}        % <<160715>> NOTE: cannot re-DEF \sb
\def\sf{star-forming\xspace}
\newcommand\sfr{star-formation rate\xspace}
\newcommand\sfh{star-formation history\xspace}
\newcommand\sfms{star-formation main sequence\xspace}

\newcommand{\el}[1]{\ensuremath{\textrm{EL}_{#1}}}
\newcommand{\obs}{\textrm{o}}
\newcommand{\theo}{\textrm{t}}

\newcommand\refe{\textrm{ref}}
\newcommand\pa{\textrm{PA}}

\newcommand{\Om} {\ensuremath{\Omega_{\rm{m}}}\xspace}

\newcommand{\Ol} {\ensuremath{\Omega_{\Lambda}}\xspace}

\begin{document}

%% LaTeX will automatically break titles if they run longer than one line.
%% However, you may use \\ to force a line break if you desire.

\title{The Grism Lens-Amplified Survey from Space (GLASS) X. Sub-kpc resolution gas-phase
metallicity maps at cosmic noon behind the Hubble Frontier Fields cluster \clsan}

%% Use \author, \affil, and the \and command to format author and affiliation information.
\author{
% === Main authors ===
Xin~Wang$^{1}$,
Tucker~A.~Jones$^{2,3,18}$,
Tommaso~Treu$^{1}$,
Takahiro~Morishita$^{1,4,5}$,
Louis~E.~Abramson$^{1}$,
Gabriel~B.~Brammer$^{6}$,
Kuang-Han Huang$^{3}$,
Matthew~A.~Malkan$^{1}$,
Kasper~B.~Schmidt$^{7}$,
% === Alphabetical order ===
Adriano~Fontana$^8$,
Claudio~Grillo$^{9,10}$,
Alaina~L.~Henry$^{6,11}$,
Wouter~Karman$^{12}$,
Patrick~L.~Kelly$^{13}$,
Charlotte~A.~Mason$^{1}$,
Amata~Mercurio$^{14}$,
Piero~Rosati$^{15}$,
Keren~Sharon$^{16}$,
Michele~Trenti$^{17}$,
Benedetta~Vulcani$^{17}$
}
\affil{$^{1}$ Department of Physics and Astronomy, University of California, Los Angeles, CA, USA 90095-1547}
\affil{$^{2}$ Institute of Astronomy, University of Hawaii, 2680 Woodlawn Drive, Honolulu, HI 96822, USA}
\affil{$^{3}$ University of California Davis, 1 Shields Avenue, Davis, CA 95616, USA}
\affil{$^{4}$ Astronomical Institute, Tohoku University, Aramaki, Aoba, Sendai 980-8578, Japan}
\affil{$^{5}$ Institute for International Advanced Research and Education, Tohoku University, Aramaki, Aoba, Sendai 980-8578, Japan}
\affil{$^{6}$ Space Telescope Science Institute, 3700 San Martin Drive, Baltimore, MD, 21218, USA}
\affil{$^{7}$ Leibniz-Institut f\"ur Astrophysik Potsdam (AIP), An der Sternwarte 16, D-14482 Potsdam, Germany}
\affil{$^{8}$ INAF - Osservatorio Astronomico di Roma Via Frascati 33 - 00040 Monte Porzio Catone, Italy}
\affil{$^{9}$ Dipartimento di Fisica, Universit\`a  degli Studi di Milano, via Celoria 16, I-20133 Milano, Italy}
\affil{$^{10}$ Dark Cosmology Centre, Niels Bohr Institute, University of Copenhagen, Juliane Maries Vej 30, 2100 Copenhagen, Denmark}
\affil{$^{11}$ Astrophysics Science Division, Goddard Space Flight Center, Code 665, Greenbelt, MD 20771}
\affil{$^{12}$ Kapteyn Astronomical Institute, University of Groningen, Postbus 800, 9700 AV Groningen, The Netherlands}
\affil{$^{13}$ Department of Astronomy, University of California, Berkeley, CA 94720-3411, USA}
\affil{$^{14}$ INAF - Osservatorio Astronomico di Capodimonte, Via Moiariello 16, I-80131 Napoli, Italy}
\affil{$^{15}$ Dipartimento di Fisica e Scienze della Terra, Universit\'a degli Studi di Ferrara, via Saragat 1, 44122 Ferrara, Italy}
\affil{$^{16}$ Department of Astronomy, University of Michigan, 1085 S. University Avenue, Ann Arbor, MI 48109, USA}
\affil{$^{17}$ School of Physics, University of Melbourne, VIC 3010, Australia}
\affil{$^{18}$ Hubble Fellow}
\email{xwang@astro.ucla.edu}
% ======================================================================

\begin{abstract}
We combine deep Hubble Space Telescope grism spectroscopy with a new Bayesian method to derive maps of \gpm, nebular dust
extinction, and \sfr for 10 star-forming galaxies at high redshift ($1.2\lesssim z\lesssim2.3$).  Exploiting lensing magnification
by the foreground cluster \clsan, we reach sub-kpc spatial resolution and push the limit of stellar mass associated with such
high-$z$ spatially resolved measurements below $10^8\Msun$ for the first time.
Our maps exhibit diverse morphologies, indicative of various effects such as efficient radial mixing from tidal torques, rapid
accretion of low-metallicity gas, and other physical processes which can affect the gas and metallicity distributions in
individual galaxies.
Based upon an exhaustive sample of all existing sub-kpc resolution \mgms at high-$z$, we find that predictions given by
analytical chemical evolution models assuming a relatively extended star-formation profile in the early disk formation phase can
explain the majority of observed \mgs, without involving galactic feedback or radial outflows.
We observe a tentative correlation between stellar mass and \mg, consistent with the ``downsizing'' galaxy formation picture that
more massive galaxies are more evolved into a later phase of disk growth, where they experience more coherent mass assembly at all
radii and thus show shallower \mgs.
In addition to the spatially resolved analysis, we compile a sample of homogeneously cross-calibrated integrated metallicity
measurements spanning three orders of magnitude in stellar mass at $z$$\sim$1.8.
We use this sample to study the mass-metallicity relation and test whether star-formation rates drive the scatter in the
mass-metallicity relation (\ie the ``fundamental metallicity relation'').
The slope of the observed mass-metallicity relation can rule out the momentum-driven wind model at 3-$\sigma$ confidence level.
We find no significant offset with respect to the fundamental metallicity relation, taking into account the intrinsic scatter and
measurement uncertainties.
\end{abstract}


\section{Introduction}\label{sect:intro}

Galaxies are complex ecosystems, particularly at the peak epoch of cosmic star formation,
corresponding to the redshift range of $z$$\sim$1-3, also known as the ``cosmic noon''
\citep[see][for a recent review]{2014ARA&A..52..415M}.  During this approximately 4 Gyr, the
Hubble sequence gradually breaks down and the predominant morphology of galaxies transforms
from irregular systems at high redshifts to symmetric disks and bulges at low redshifts
\citep{Mortlock:2013dg}.  This complexity is to a large extent induced by the interplay
between the process of star formation, and the diverse aspects of baryonic cycling, \eg,
galactic feedback, gas inflows/outflows, and major/minor mergers
\citep{2011MNRAS.415...11D,Martin:2012dx}.  The effect of environment surrounding galaxies
can further complicate the spatial distribution of star formation, as recently revealed by
\citet{2015ApJ...814..161V,TheGrismlensampli:-R7Qg0z6}.
Measurements of gas-phase metallicity, \ie, the chemical abundances of elements heavier than
hydrogen and helium in the interstellar medium (ISM), are a powerful means to shed light on
this complexity\footnote{Hereafter throughout the paper, we refer to \gpm as metallicity for
simplicity.}, because the metal enrichment history is strongly tied to the mass assembly history in galaxy evolution
\citep{2011MNRAS.416.1354D,Lu:2015ic}.  Since detailed elemental abundances are not directly measurable at extragalactic
distances, the relative oxygen abundance in ionized gaseous nebulae, \ie, \oh, is often chosen as the observational proxy of
metallicity.

For over a decade, a tight correlation between metallicity and galaxy stellar mass (\Mstar),
\ie, the mass-metallicity relation (MZR), has been quantitatively established, from the vast
database of local galaxies observed by the Sloan Digital Sky Survey
\citep{Tremonti:2004ed,Zahid:2012fp,Andrews:2013dn}. This relation has been further extended to high redshifts, using
deep near infrared (IR) spectroscopy facilitated by large ground-based and space-based
telescopes \citep{Erb:2006kn,2008A&A...488..463M,Zahid:2011bb,Henry:2013gx,2014ApJ...795..165S,Sanders:2015gk,Guo:2016wk}.
The measurements of the MZR as a function of redshift can cast useful constraints on various galaxy evolution models, since the
slope of the MZR is sensitive to the properties of outflows, such as the mass loading factor and the outflow speed \citep[see,
\eg,][]{2012MNRAS.421...98D,Lu:2015kh}).
This slope can also be explained by variations of star-formation efficiency and gas mass fraction in galaxies with different
stellar masses \citep[see, \eg,][]{Baldry:2008hm,TheUniversalRelati:2014kx}.
The normalization of the MZR can shed light upon the stellar chemical yield across cosmic time \citep{2008MNRAS.385.2181F}.

It was first suggested by \citet{2010MNRAS.408.2115M} that there exists a so-called fundamental metallicity relation (FMR) in
the 3D parameter space spanned by \Mstar, \sfr (SFR), and metallicity such that the MZR is merely a 2D projection of this more
fundamental 3D manifold \citep[see also][]{Hunt:2016ui}.
This 3D scaling relation shows a tight scatter ($\sim$0.05 dex) in metallicity and is speculated to not evolve with $z$.

This proposed lack of evolution of the FMR is conjectured to be due to the interplay of two separate physical processes in control
at different cosmic epochs.\\
\indent\textbullet~At low $z$, outflows of metal-enriched material are the predominant driver
of the FMR. For local galaxies below $10^{10}\Msun$, metallicity is in strong
anti-correlation with SFR \citep{2010MNRAS.408.2115M}. This is attributed to the fact that
more violent star formation leads to more powerful galactic feedback\footnote{Galactic
feedback can suppress the star-formation efficiency by heating up the gas thereby preventing
it from forming stars. So the interplay between star formation and feedback is
self-regulating.} and gas outflows, which are capable of removing the stellar nucleosynthesis
yields from the galaxy, if the gravitational well is shallow. The different properties of the outflow (in particular the
mass-loading factor) therefore set the slope of the MZR \citep{2012MNRAS.421...98D}.
However more massive galaxies are less prone to the loss of heavy elements from metal-loaded outflows because of the steeper
gravitational potential wells that these outflows have to overcome. Thus more massive galaxies can more easily retain their metal
production. This naturally explains the existence of the MZR and the observation that no correlation between metallicity and SFR
is seen in galaxies with $\Mstar\gtrsim10^{10.9}\Msun$ \citep{2010MNRAS.408.2115M}.

\indent\textbullet~At high $z$, inflows of pristine/metal-poor gas are key factor that
governs the behavior of the FMR. The inverse correlation between metallicity and SFR is
ascribed to the fact that the infall of metal-poor gas lowers the global metallicity and
triggers active star formation activities, because the gas reservoir in the
galaxy is replenished \citep[as suggested by, e.g.,][]{Hunt:2016wz}.

In this context, the apparent redshift evolution of the MZR normalization originates primarily from sampling the FMR in terms of
galaxies with different SFR. This picture is in accord with the gas regulator model proposed by \citet{Lilly:2013ko}, even though
mergers can also play a subtle role in shaping the form of the FMR by increasing the scatter \citep{MichelDansac:2008gp}. However,
at high redshifts, the validity of the FMR is still under investigation \citep[see, \eg,][]{Sanders:2015gk,Wuyts:2014ed}.

Spatially resolved chemical information provides a more powerful diagnostic tool about galaxy baryonic assembly than
integrated metallicity measurements, especially at high redshifts. Because for non-interacting galaxies, their \mgs are found to
be highly sensitive to the properties of gas, \ie, the surface density, the existence of inflows/outflows, and the kinematic
structure \citep{Cresci:2010hr,2013ApJ...765...48J,2014A&A...563A..49S,Metallicityevolutio:2014kg}.
In the past few years, radial metallicity gradients, measured from spectroscopic data
acquired by ground-based instruments with natural seeing ($\gtrsim$$0\farcs6$), have been
reported at high redshifts
\citep{Queyrel:2012hw,Metallicityevolutio:2014kg,2014MNRAS.443.2695S}. In particular, a large
mass-selected sample of galaxies at $0.7\lesssim z\lesssim2.7$ were observed with spatially
resolved gas kinematics and star formation, using the K-band Multi Object Spectrograph
(\kmos) on the Very Large Telescope (\vlt) \citep[\ie, the \kd
survey,][]{2015ApJ...799..209W}.  As a result, \citet{Wuyts:2016th} measured \mgs for 180 \sf
galaxies in three redshift intervals (\ie [0.8, 1.0], [1.3, 1.7], and [2.0, 2.6]), and found
that the majority of the \mgs are flat and no statistical significant correlations between
\mgs, and stellar, kinematic, and structural properties of the galaxy.

However, seeing-limited measurements typically have insufficient resolution to resolve the inner structure of galaxies at
$z\gtrsim1$, and the inferred metallicity gradient can be potentially biased. For example, \citet{2013ApJ...767..106Y} showed that
coarse spatial sampling ($\geq$1 \kpc) can result in artificially flat \mgs, inferring that sufficiently high spatial resolution,
\ie, \emph{sub-kpc} scale (corresponding to an angular resolution of $<0\farcs2$-$0\farcs3$), is crucial in yielding precise
information of how metals are distributed spatially in extragalactic \HII regions.  Only a few \mgms meet this requirement,
including a sample of 9 \Ha-selected galaxies at $z\in[0.84,2.23]$ (the majority at $z\sim1.45$), observed with the adaptive
optics (AO) assisted integral field unit (IFU) spectrograph \sinf onboard the \vlt \citep{2012MNRAS.426..935S}. Even higher
resolution can be attained through the combination of diffraction-limited data and gravitational lensing as shown by
\citet{2010ApJ...725L.176J,2013ApJ...765...48J,Yuan:2011hj}, targeting strongly lensed galaxies with the laser guide star AO aided
IFU spectrograph \osiris at the \keck telescope. Following this approach, \citet{2015arXiv150901279L} recently analyzed a sample
of 11 lensed galaxies (3 at $z\sim1.45$ and the rest at $z\gtrsim2$), deriving maps of both metallicity and emission line (EL)
kinematics.  Although a great amount of effort has been invested in enlarging the sample size of high-$z$ \mgs obtained with
sub-kpc scale spatial resolution, the current sample consists of only $\lesssim30$ such gradients and is still statistically
insufficient to explore trends with stellar mass and redshift.

In order to enlarge the sample of sub-kpc resolution measurements, we recently demonstrated that such metallicity maps can be
derived using space-based data \citep{2015AJ....149..107J}. We measured a flat metallicity gradient in a multiply lensed
interacting system at $z=1.85$, using diffraction-limited \hst grism data, and confirmed that flat \mgs can be caused by
gravitational interactions in merging systems.

The main goal of the work presented here is to collect a uniformly analyzed large sample of high-$z$ metallicity maps obtained
with sub-kpc spatial resolution. To meet our goal, we improve upon the methodology proposed in our pilot paper
\citep{2015AJ....149..107J}, in particular via developing a novel Bayesian method to imply metallicity from multiple EL
diagnostics simultaneously. We apply our advanced analysis to ultra-deep grism data of the massive galaxy cluster \clsan, thus
exploiting the powerful synergy of \hst diffraction-limited spectroscopy and lensing magnification.

The outline of this paper is as follows. In Section~\ref{sect:spec}, we introduce the spectroscopic observations used in this work
and the selection of our \mg sample. The photometric data and galaxy cluster lens models are briefed in Section~\ref{sect:phot}.
We then present our entire analysis process in Section~\ref{sect:analysis}, and show our results in terms of global demographics
and spatially resolved analysis in Sections~\ref{sect:global} and \ref{sect:sra}, respectively.  Finally,
Section~\ref{sect:conclu} will conclude and discuss our study. The concordance cosmological model (\Om=0.3, \Ol=0.7,
$H_0=70\Hunit$) and the AB magnitude system \citep{1983ApJ...266..713O} are used throughout this paper.  Without specific number
showing the wavelength, the names of forbidden lines are simplified as $\OII~\lambda\lambda3726,3729\defeq\OII$,
$\OIII~\lambda5008\defeq\OIII$, $\NII~\lambda6585\defeq\NII$, $\SII~\lambda\lambda6718,6732\defeq\SII$ in this paper\footnote{The
wavelength values are taken from \url{http://classic.sdss.org/dr7/algorithms/linestable.html}}.

\section{Sample selection, observations, and data reduction}\label{sect:spec}

In Section~\ref{subsect:hstdata}, we first describe our \hst grism observations, data quality, and data reduction.
In Section~\ref{subsect:srcselect}, we then describe our sample selection criteria.
In Section~\ref{subsect:ifu}, we summarize the ground based kinematic data that we will use to supplement and interpret the
metallicity and star formation maps obtained from the \hst grism spectroscopy.

\subsection{Hubble Space Telescope Grism Data}\label{subsect:hstdata}

%= = = = = = = = = = = = = = = = = = = = = = = = = = = = = = = = = = = = = = = =
% = = = = = = = = = = = = = = = = = = = = = = = = = = = = = = = = = = = = = = = = = =
% Include this table with \input{filename.tex}
% To rotate in emulateapj do: \begin{turnpage}\input{filename.tex}\end{turnpage}
% To display it on multiple pages do: \LongTables\input{filename.tex}
% - - - - - - - - - - - - - - - - - - - - - - - - - - - - - - - - - - - - - - - - - -
{\setlength\tabcolsep{2pt}
\begin{deluxetable}{cccp{2.5cm}ccccccccc} \tablecolumns{13}
\tablewidth{0pt}
\tablecaption{Properties of the grism spectroscopic data used in this work}
% - - - - - - - - - - - - - - - - - - - - - - - - - - - - - - - - - - - - - - - - - -
\tablehead{
    \colhead{PA\tablenotemark{a}} &
    \colhead{Grism} &
    \colhead{Exposure Time} &
    \colhead{Program}  &
    \colhead{Time of completion}\\
    \colhead{(deg.)} & & 
    \colhead{(s)} & & &
}
%---------------------------------------------------------------
\startdata
\multirow{2}{*}{032}    & G102 & 8723 & \glass & \multirow{2}{*}{Feburary 2014} \\
                        & G141 & 4412 & \glass  \\
111                     & G141 & 36088 & SN Refsdal follow-up & December 2014 \\
119                     & G141 & 36088 & SN Refsdal follow-up & January 2015 \\
\multirow{2}{*}{125}    & G102 & 8623 & \glass & \multirow{2}{*}{November 2014} \\
                        & G141 & 4412 & \glass
\enddata
% - - - - - - - - - - - - - - - - - - - - - - - - - - - - - - - - - - - - - - - - - -
\tablecomments{Here we only include the grism observations targetted on the prime field of
\clsan.}
\tablenotetext{a}{The position angle shown here corresponds to the ``PA\_V3'' value
    reported in the WFC3/IR image headers. The position angle of the dispersion axis of the
grism spectra is given by $\mathrm{PA_{disp}} \approx \mathrm{PA\_V3} - 45$.}
\label{tab:obsdata}
\end{deluxetable}
}

%= = = = = = = = = = = = = = = = = = = = = = = = = = = = = = = = = = = = = = = =

\subsubsection{The Grism Lens-Amplified Survey from Space}\label{subsect:glass}

\begin{figure*}
    \centering
    \includegraphics[width=\textwidth]{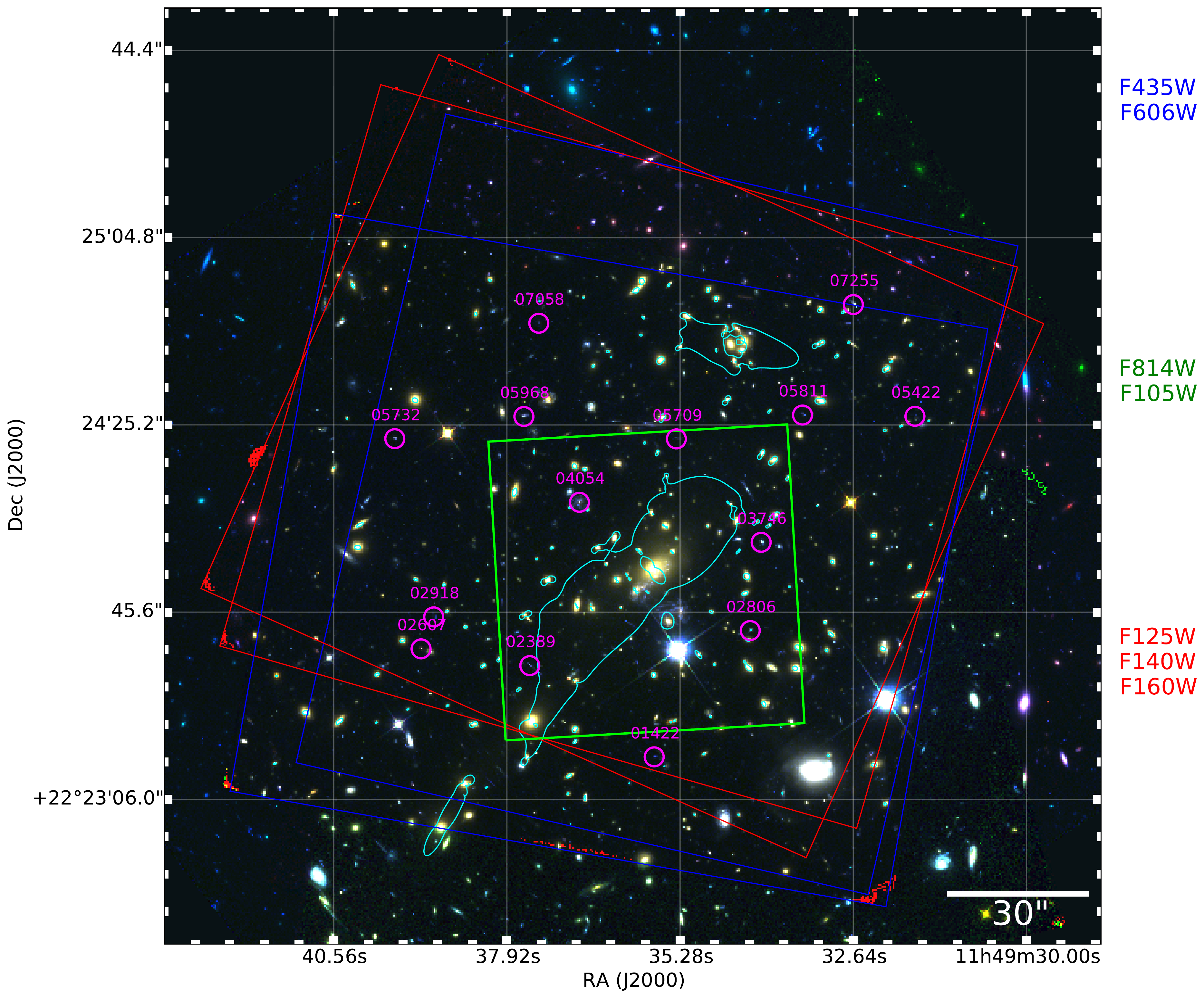}
    \caption{Color composite image of \clsan from the full-depth 7-filter \hff photometry.
    The blue, green, and red channels are comprised of the \hst broad-band filters shown on
    the right. The blue, red and green squares mark the footprints of \glass, the Refsdal
    follow-up, and \muse programs utilized in this work.  The cyan contours are the critical
    curves at $z=1.8$ predicted by the \glafic version 3 lens model.  The magenta circles
    show the positions of our \mg sample. 3$\times$3 arcsec$^2$ zoom-in postage stamps around
    these positions are shown in Figure~\ref{fig:RGBstamps}.}
    \label{fig:RGBfullFoV}
\end{figure*}

The Grism Lens-Amplified Survey from Space\footnote{\url{http://glass.astro.ucla.edu}} \citep[\glass; Proposal ID 13459;
P.I. Treu,][]{2014ApJ...782L..36S,2015ApJ...812..114T} is an \hst cycle 21 large general observing (GO) program.  \glass
observed 10 massive galaxy
clusters with the Wide Field Camera 3 Infrared (WFC3/IR) grisms (G102 and G141; 10 and 4 orbits per cluster,
respectively) targeted at their centers and the Advanced Camera for Survey (ACS) Optical grism (G800L) at their infall
regions.
The exposure on each cluster core is split into two nearly orthogonal
position angles (PAs), in order to disentangle contamination from neighboring objects.
Data acquisition was completed in January 2015. Here we focus on the grism data targeted on the center of
\clsan taken in February 2014 (PA=$032^{\circ}$) and November 2014 (PA=$125^{\circ}$), marked by blue squares in
Figure~\ref{fig:RGBfullFoV} (also see Table~\ref{tab:obsdata}). \glass provides $\sim$10\% of the G141
exposures used in this work, and 100\% of the G102 exposures.  Details on \glass data reduction can be found in
\citet{2014ApJ...782L..36S,2015ApJ...812..114T}.

Shallow images through filters F105W or F140W were taken to aid the
alignment and extraction of grism spectra. The imaging exposures are
combined with the exposures obtained by the Hubble Frontier
Fields (\hff) initiative and other programs to produce the deep
stacks, released as part of the \hff program (see Section~\ref{sect:phot}
for more details).

\subsubsection{SN Refsdal Follow-up Program with \protect\hst G141}\label{subsect:refsdal}

The majority of the G141 exposures used in this work (30 orbits) were taken as part of the follow-up \hst GO/DDT
campaign (Proposal ID 14041, P.I. Kelly, Brammer et al. in prep.) of the first multiply
imaged supernova, SN Refsdal \citep{2015Sci...347.1123K}. Two pointings
(shown by the red squares in Figure~\ref{fig:RGBfullFoV}) were exposed between December 2014 and January 2015 with 8 degrees apart
(\ie
at PA=$111^{\circ}$, and $119^{\circ}$ respectively, as shown in Tabel~\ref{tab:obsdata}), in order to optimize the spectroscopy
of SN Refsdal.  The analysis of the
spectra of SN Refsdal, which showed that it was SN 1987A-like, is described by \citet{2015arXiv151209093K}.

\subsubsection{Grism data reduction}

The combined grism dataset was reduced following the procedure of the
3D-\hst survey \citep{Brammer:2012bu,Momcheva:2015wa}.  An updated
3D-\hst pipeline \footnote{\url{http://code.google.com/p/threedhst/}}
was employed to reduce the images and spectra. The \adriz software
from the \dpac package and \verb+tweakreg+ were used to align and
combine individual grism exposures, subtracting the sky images
provided by \citet{Brammer:2012bu}. The time-varying sky background
due to helium glow (at 10,830 \AA) in the Earth exosphere was also
accounted for according to \citet{Brammer:2014wl}. After these initial
steps, the mosaics were co-added through interlacing onto a grid of
0\farcs065 resolution, Nyquist sampling the point spread function (PSF)
full width half maximum (FWHM).  The average spectral resolution is $R$$\sim$210
and $R$$\sim$130 for G102 and G141, respectively. The average pixel scale
after interlacing is 12 \AA{} and 22 \AA{} respectively for G102 and G141.

We used the co-added \H-band mosaics as the detection image, on which \sex
was run. An object catalog was generated and a
corresponding segmentation map was created for each object,
determining which spatial pixels (spaxels) belong to this object. Thus for a source
registered in the catalog and falling within the WFC3 grism field-of-view (FoV), we
extracted its spectra from grism mosaics through adding up the
dispersed flux for all spaxels within an area defined by the
segmentation image of this source, after flat-field correction and
background subtraction.  The ``fluxcube'' model from the \axe package
\citep{Kummel:2009dn} was employed to generate the 2D models of stellar
spectrum, based upon the spatial profile of the source, the color
information from direct image mosaics (if available), and the grism
calibration configurations.  For sources of interest, this model
serves as the continuum model which we re-scaled and subtracted from
the observed 2D spectra (see Section~\ref{subsect:combELmaps} for more
details) in order to obtain pure EL maps. At the same time, if the object is a bright contaminating neighbour,
this 2D model also functions as spectral contamination, which is subtracted before the continuum removal.

The data used in this work contain a total of 34 orbits of G141 and 10 orbits of G102,
reaching a 1-$\sigma$ flux limit (uncorrected for lensing magnification) of 3.5(1.2)$\times$$10^{-18}$ \Funit over the wavelength
range of G102(G141), calculated from \citet{Schmidt:2016ez}. Our data provide an uninterrupted wavelength coverage in the range of
0.8-1.7 \micron and make this particular field one of the deepest fields probed by \hst spectroscopy to date.

\subsection{Sample selection}\label{subsect:srcselect}

%= = = = = = = = = = = = = = = = = = = = = = = = = = = = = = = = = = = = = = = =
% = = = = = = = = = = = = = = = = = = = = = = = = = = = = = = = = = = = = = = = = = =
% Include this table with \input{filename.tex}
% To rotate in emulateapj do: \begin{turnpage}\input{filename.tex}\end{turnpage}
% To display it on multiple pages do: \LongTables\input{filename.tex}
% - - - - - - - - - - - - - - - - - - - - - - - - - - - - - - - - - - - - - - - - - -
\begin{deluxetable*}{ccccccccccccc} \tablecolumns{13}
\tablewidth{0pt}
\tablecaption{Global demographic properties of the \mg sample analyzed in this work}
% - - - - - - - - - - - - - - - - - - - - - - - - - - - - - - - - - - - - - - - - - -
\tablehead{
  \colhead{ID} &
  \colhead{R.A.} &
  \colhead{Dec.} &
  \colhead{$z_{\textrm{spec}}$} &
  \colhead{$\mu$\tablenotemark{a}} &
  \colhead{\H magnitude} &
  \multicolumn{2}{c}{SED fitting} &
  \multicolumn{3}{c}{EL diagnostics} \\
  & [deg.] & [deg.] &  &  & [ABmag] & \multicolumn{2}{c}{\hrulefill} & \multicolumn{3}{c}{\hrulefill} & \\
    & & & & & & log(\Mstar\tablenotemark{b}/\Msun) & $\Av^{\rm S}$ &  \oh  & SFR\tablenotemark{b} [\Msun/yr] & $\Av^{\rm N}$
}
%---------------------------------------------------------------
\startdata
01422  &  177.398643  &  22.387499  &  2.28  &  2.35 [2.09, 2.26]  &  24.46  &  $9.29_{-0.01}^{+0.07}$  &  $<0.01$  &  $8.26_{-0.13}^{+0.11}$  &  $15.10_{-7.11}^{+16.76}$  &  $<1.07$  \\
02389  &  177.406546  &  22.392860  &  1.89  &  43.25 [37.47, 50.28]  &  23.29  &  $8.43_{-0.07}^{+0.06}$  &  $<0.01$  &  $7.88_{-0.15}^{+0.16}$  &  $<6.37$  &  $1.36_{-0.82}^{+0.69}$  \\
02607\tablenotemark{c}  &  177.413454  &  22.393843  &  1.86  &  3.01 [2.79, 2.98]  &  22.63  &  $10.25_{-0.10}^{+0.06}$  &  $1.70_{-0.18}^{+0.42}$  &  $7.70_{-0.11}^{+0.13}$  &  $>36.51$  &  $2.26_{-0.58}^{+0.31}$  \\
02806  &  177.392541  &  22.394921  &  1.50  &  4.82 [3.11, 3.51]  &  23.35  &  $8.72_{-0.01}^{+0.19}$  &  $0.50_{-0.01}^{+0.01}$  &  $7.96_{-0.16}^{+0.15}$  &  $1.96_{-0.13}^{+1.37}$  &  $<0.18$  \\
02918  &  177.412652  &  22.395723  &  1.78  &  2.95 [2.72, 2.91]  &  25.22  &  $8.37_{-0.08}^{+0.15}$  &  $0.10_{-0.10}^{+0.34}$  &  $8.42_{-0.14}^{+0.11}$  &  $<70.32$  &  $1.25_{-0.80}^{+0.72}$  \\
03746  &  177.391848  &  22.400105  &  1.25  &  3.78 [3.56, 3.83]  &  22.48  &  $8.81_{-0.01}^{+0.03}$  &  $<0.01$  &  $8.11_{-0.11}^{+0.10}$  &  $10.95_{-0.58}^{+1.61}$  &  $<0.17$  \\
04054  &  177.403393  &  22.402456  &  1.49  &  3.39 [3.04, 3.28]  &  21.27  &  $9.64_{-0.01}^{+0.05}$  &  $1.10_{-0.01}^{+0.01}$  &  $8.70_{-0.11}^{+0.09}$  &  $16.99_{-2.80}^{+5.71}$  &  $0.72_{-0.23}^{+0.23}$  \\
05422  &  177.382077  &  22.407510  &  1.97  &  3.63 [3.11, 4.41]  &  25.05  &  $8.45_{-0.46}^{+0.15}$  &  $0.00_{-0.01}^{+0.56}$  &  $7.56_{-0.12}^{+0.16}$  &  $<38.18$  &  $<1.85$  \\
05709  &  177.397234  &  22.406181  &  1.68  &  7.10 [6.74, 7.53]  &  25.44  &  $7.90_{-0.03}^{+0.02}$  &  $<0.01$  &  $8.21_{-0.21}^{+0.13}$  &  $<17.97$  &  $<1.18$  \\
05732  &  177.415126  &  22.406195  &  1.68  &  1.56 [1.51, 1.61]  &  23.45  &  $9.09_{-0.01}^{+0.01}$  &  $<0.01$  &  $8.41_{-0.12}^{+0.10}$  &  $<84.70$  &  $<1.31$  \\
05811  &  177.389220  &  22.407583  &  2.31  &  7.50 [6.88, 9.34]  &  23.79  &  $8.85_{-0.10}^{+0.11}$  &  $<0.01$  &  $8.16_{-0.21}^{+0.14}$  &  $4.94_{-2.47}^{+3.13}$  &  $<0.74$  \\
05968  &  177.406922  &  22.407499  &  1.48  &  1.84 [1.78, 1.96]  &  22.24  &  $9.34_{-0.03}^{+0.02}$  &  $0.60_{-0.01}^{+0.20}$  &  $8.47_{-0.08}^{+0.07}$  &  $27.95_{-4.41}^{+4.70}$  &  $0.23_{-0.13}^{+0.15}$  \\
07058  &  177.405976  &  22.412977  &  1.79  &  2.13 [1.97, 3.01]  &  24.28  &  $8.74_{-0.15}^{+0.17}$  &  $0.00_{-0.01}^{+0.12}$  &  $8.38_{-0.19}^{+0.13}$  &  $<44.99$  &  $<1.85$  \\
07255  &  177.385990  &  22.414074  &  1.27  &  2.34 [2.31, 2.99]  &  22.47  &  $9.36_{-0.21}^{+0.01}$  &  $0.30_{-0.01}^{+0.21}$  &  $8.38_{-0.08}^{+0.08}$  &  $8.73_{-3.10}^{+3.53}$  &  $<0.79$
\enddata
% - - - - - - - - - - - - - - - - - - - - - - - - - - - - - - - - - - - - - - - - - -
\tablecomments{The error bars and upper/lower limits shown in the columns of SED fitting and EL diagnostics correspond
to 1-$\sigma$ confidence ranges.}
\tablenotetext{a}{Best-fit magnification values and 1-$\sigma$ confidence intervals. Except for galaxy ID 02389, the
magnification results are from the \glafic version 3 model, calculated by the \hff interactive online magnification
calculator available at \url{https://archive.stsci.edu/prepds/frontier/lensmodels/webtool/magnif.html}. 
For galaxy ID 02389, we use the \SJ version 3 model instead to compute the magnification results and correct for
lensing magnification}
\tablenotetext{b}{Values presented here are corrected for lensing magnification.}
\tablenotetext{c}{The EL diagnostic result on this source is not trustworthy, since it is classified as an AGN
    candidate (see Sect.~\ref{sect:global}).}
\label{tab:srcprop}
\end{deluxetable*}

The selection of our \mg sample is based upon the master redshift
catalog for \clsan\footnote{available at \url{https://archive.stsci.edu/prepds/glass/}}, published by the \glass collaboration.
As described by \citet{2015arXiv151005750T}, redshifts were determined by combining
spectroscopic information from \glass and the supernova (SN) Refsdal follow-up \hst
grism program, ground-based \muse observations and Keck \deimos data.

\begin{figure*}
    \centering
    \includegraphics[width=\textwidth]{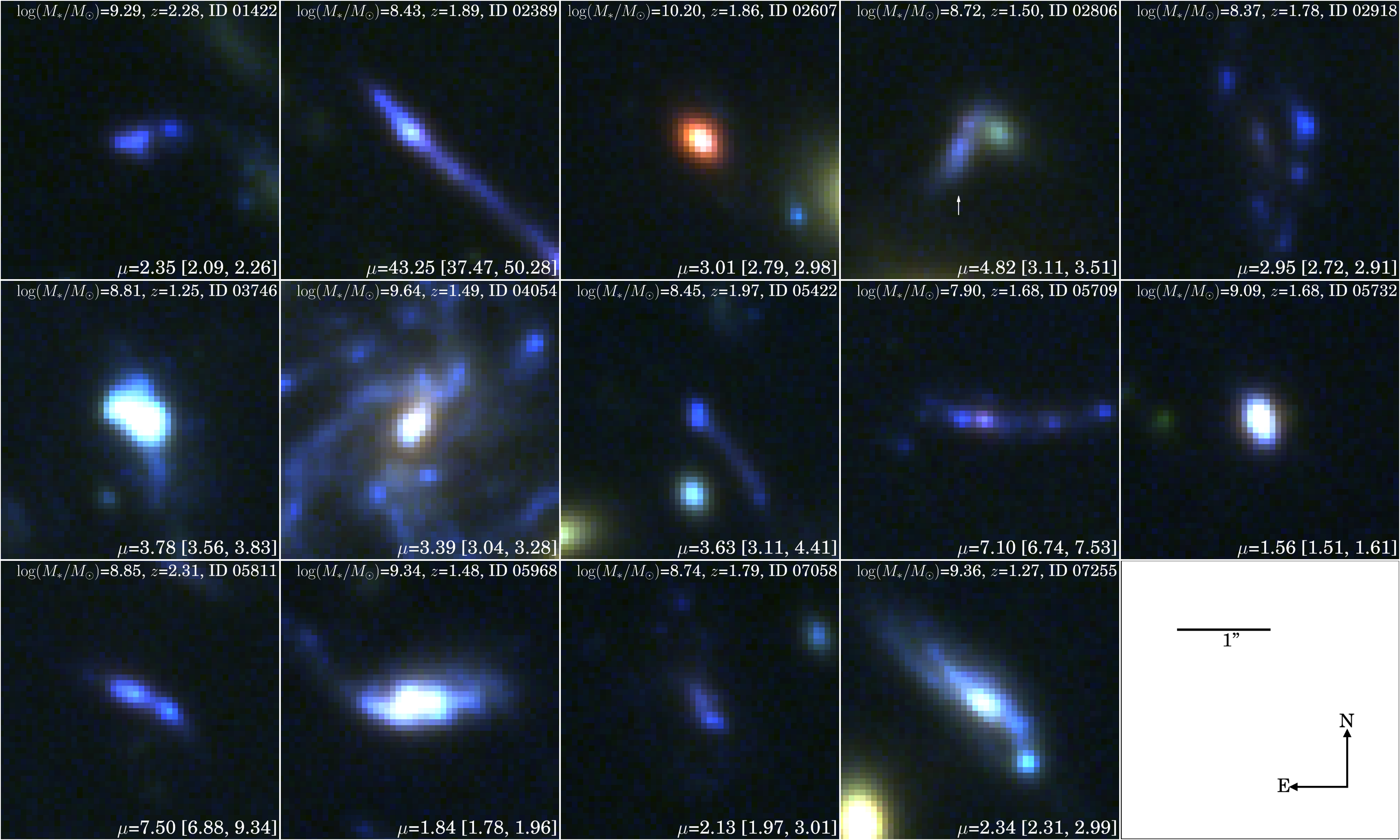}
    \caption{Zoom-in color composite stamps of our \mg sample, cut out from the full FoV
    image shown in Figure~\ref{fig:RGBfullFoV}. In each stamp, we also show the spectroscopic
    redshift, stellar mass and lensing magnification (best-fit value followed by 1-$\sigma$
    confidence interval) of the galaxy.
    In the panel of ID 02806, we use the white arrow to point to our galaxy of interest (ID
    02806). All stamps are on the same spatial scale. The 1\arcsec{} scale bar and north-east
    compass are shown in the lower right corner.}
    \label{fig:RGBstamps}
\end{figure*}

\begin{figure*}
    \includegraphics[width=.33\textwidth]{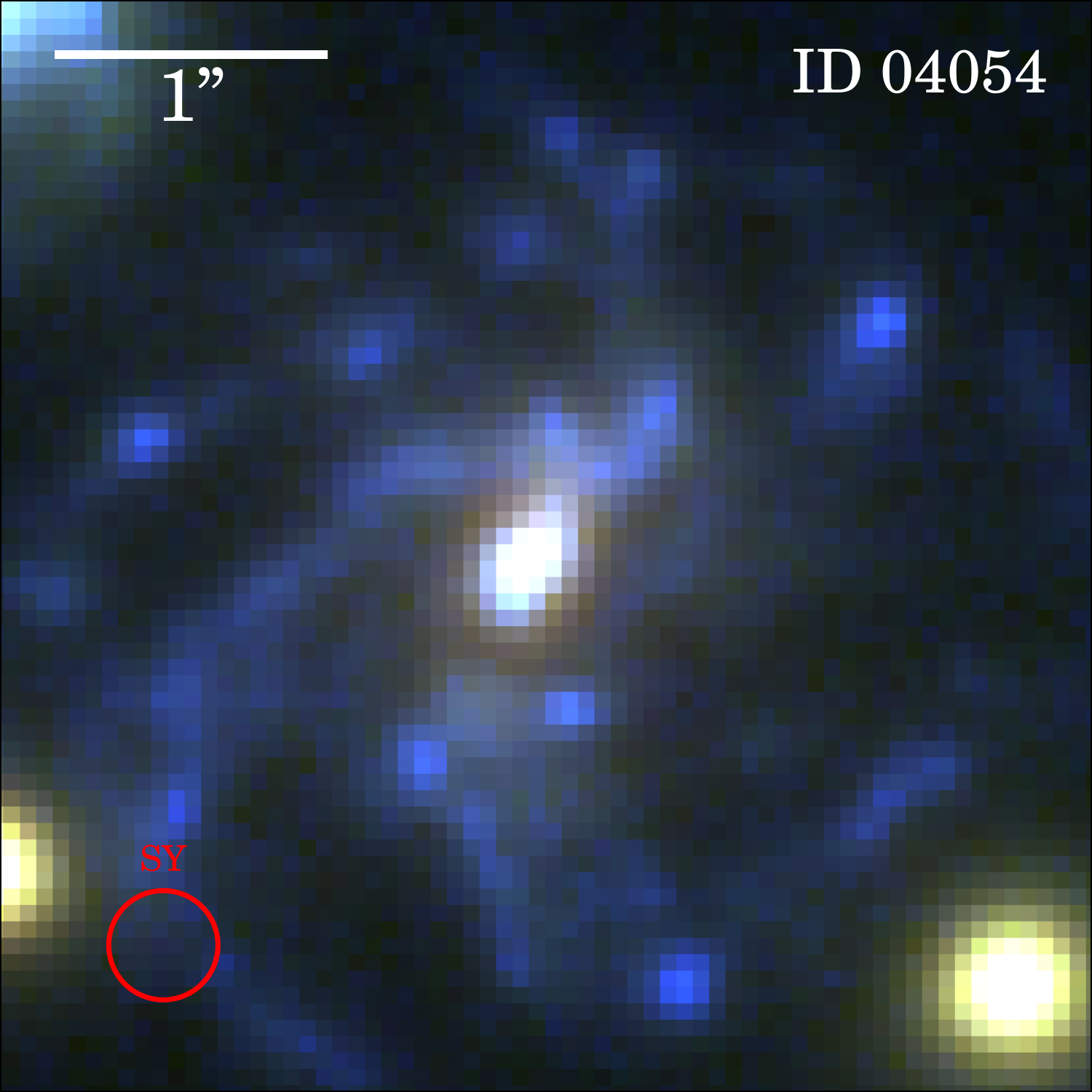}
    \includegraphics[width=.33\textwidth]{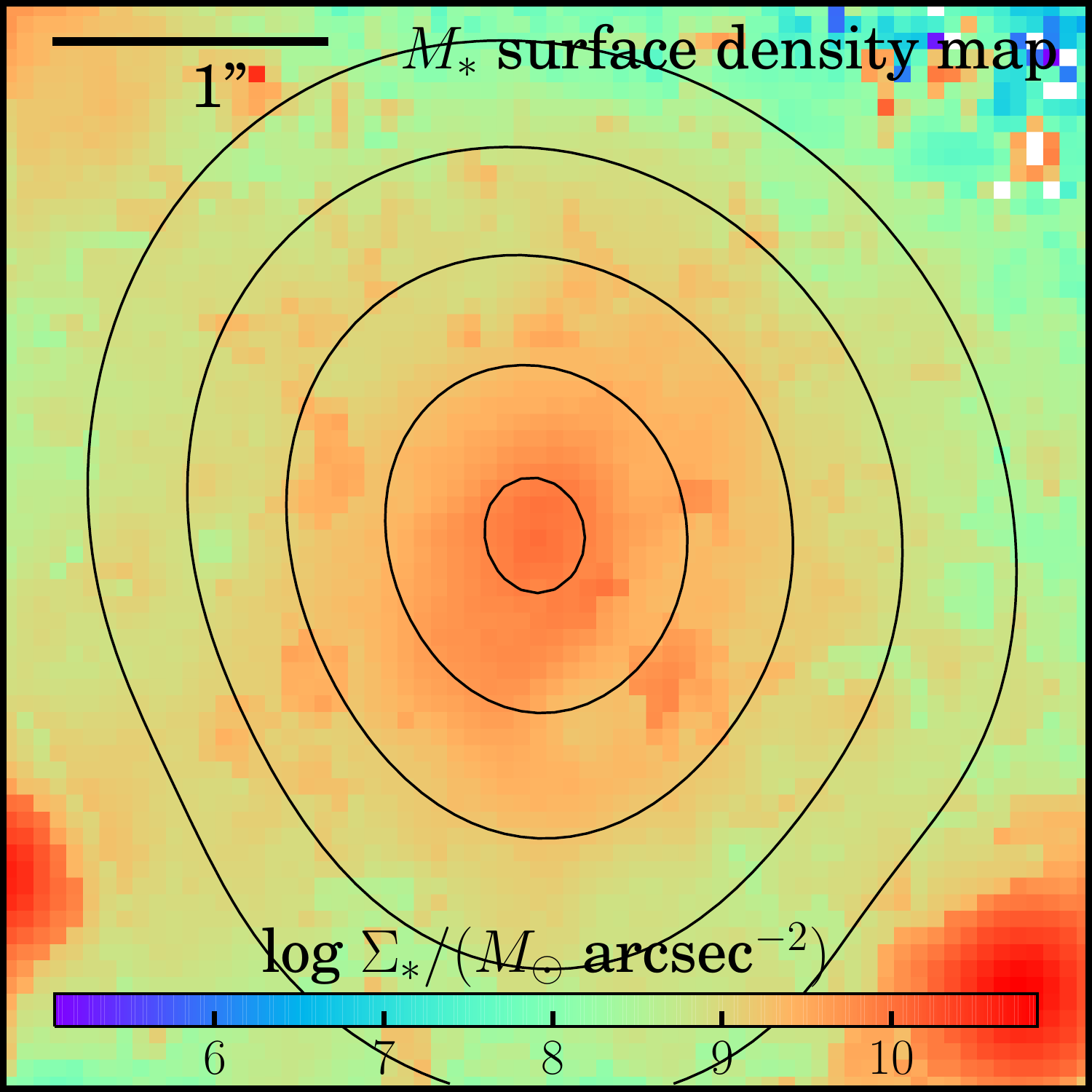}
	\includegraphics[width=.33\textwidth]{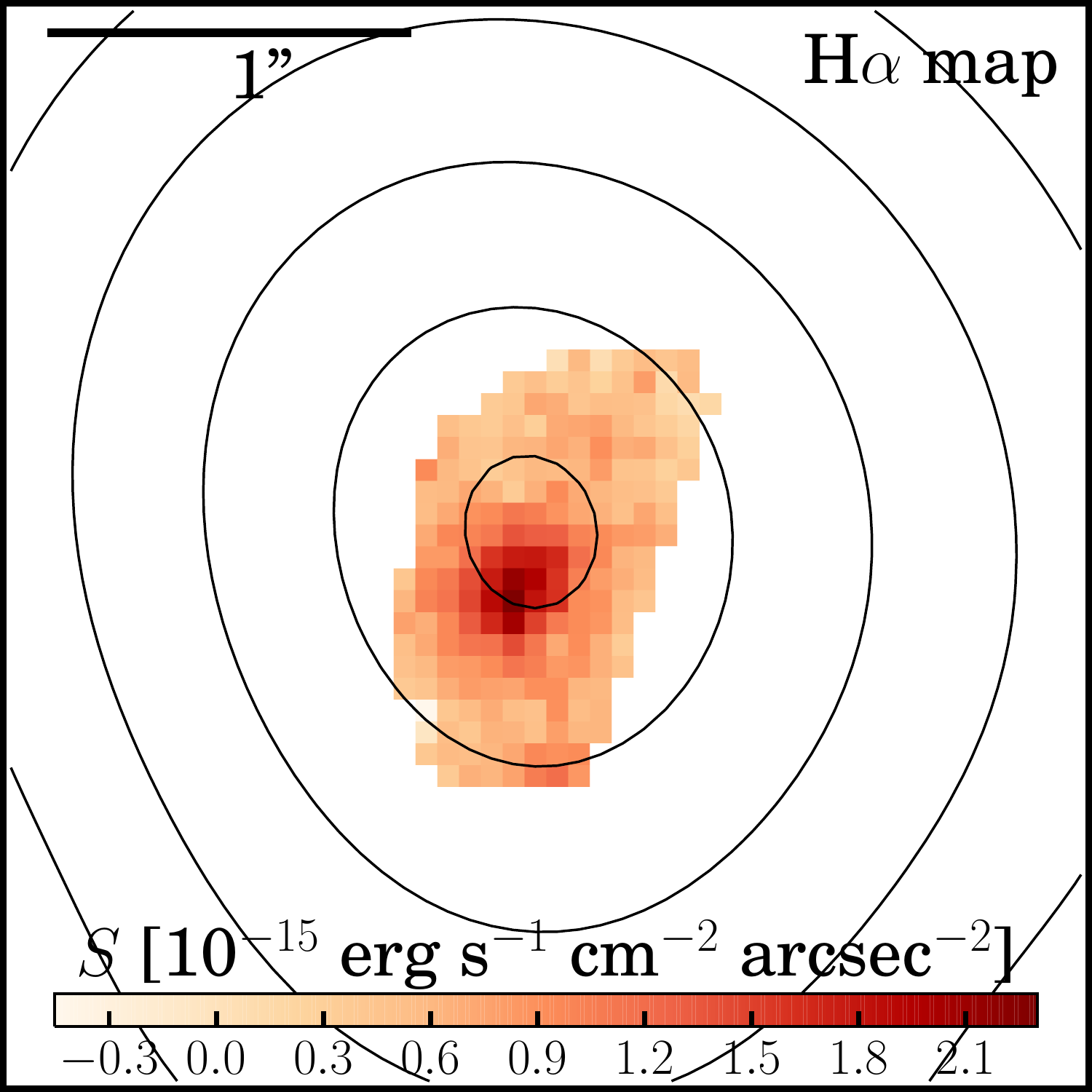}\\
	\includegraphics[width=.33\textwidth]{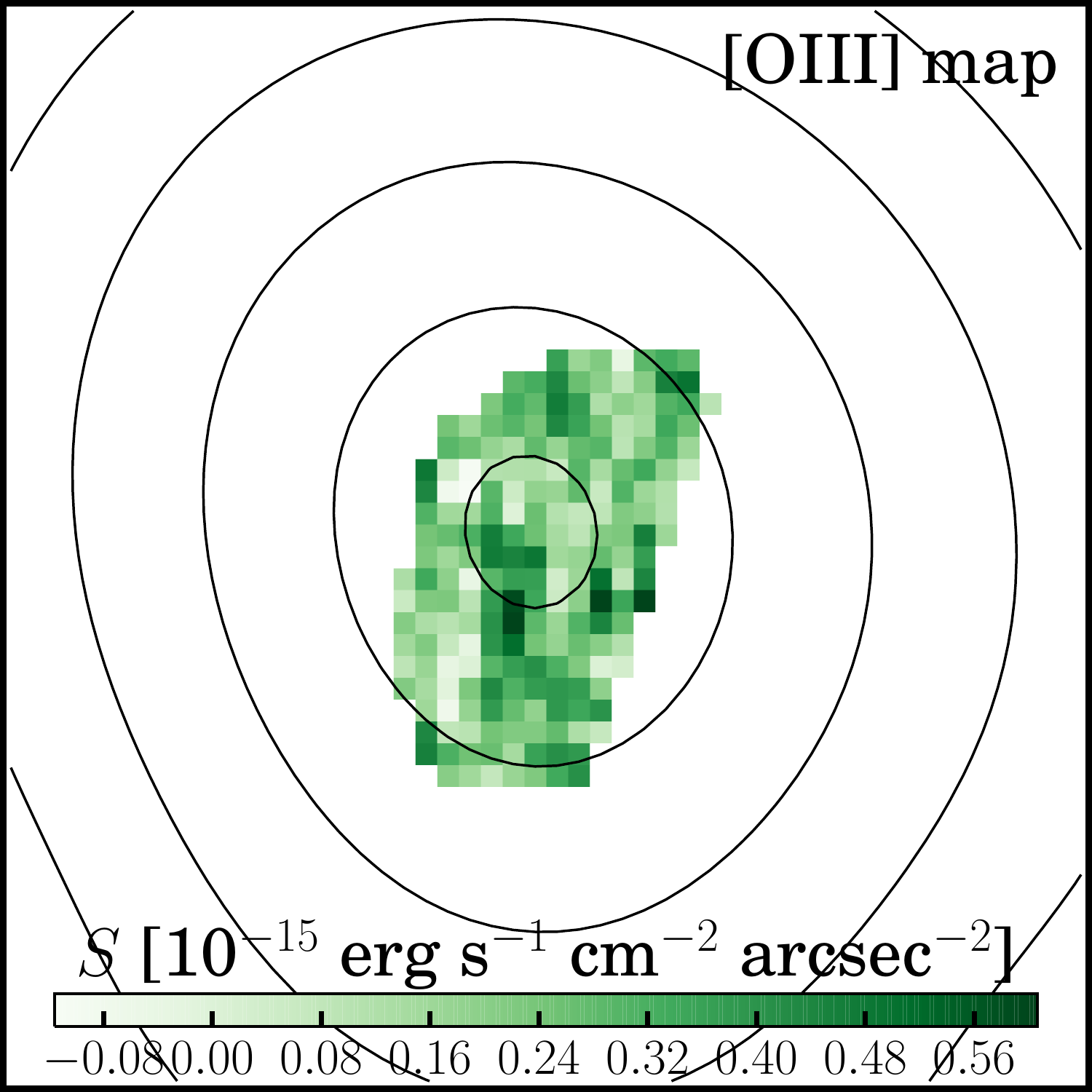}
	\includegraphics[width=.33\textwidth]{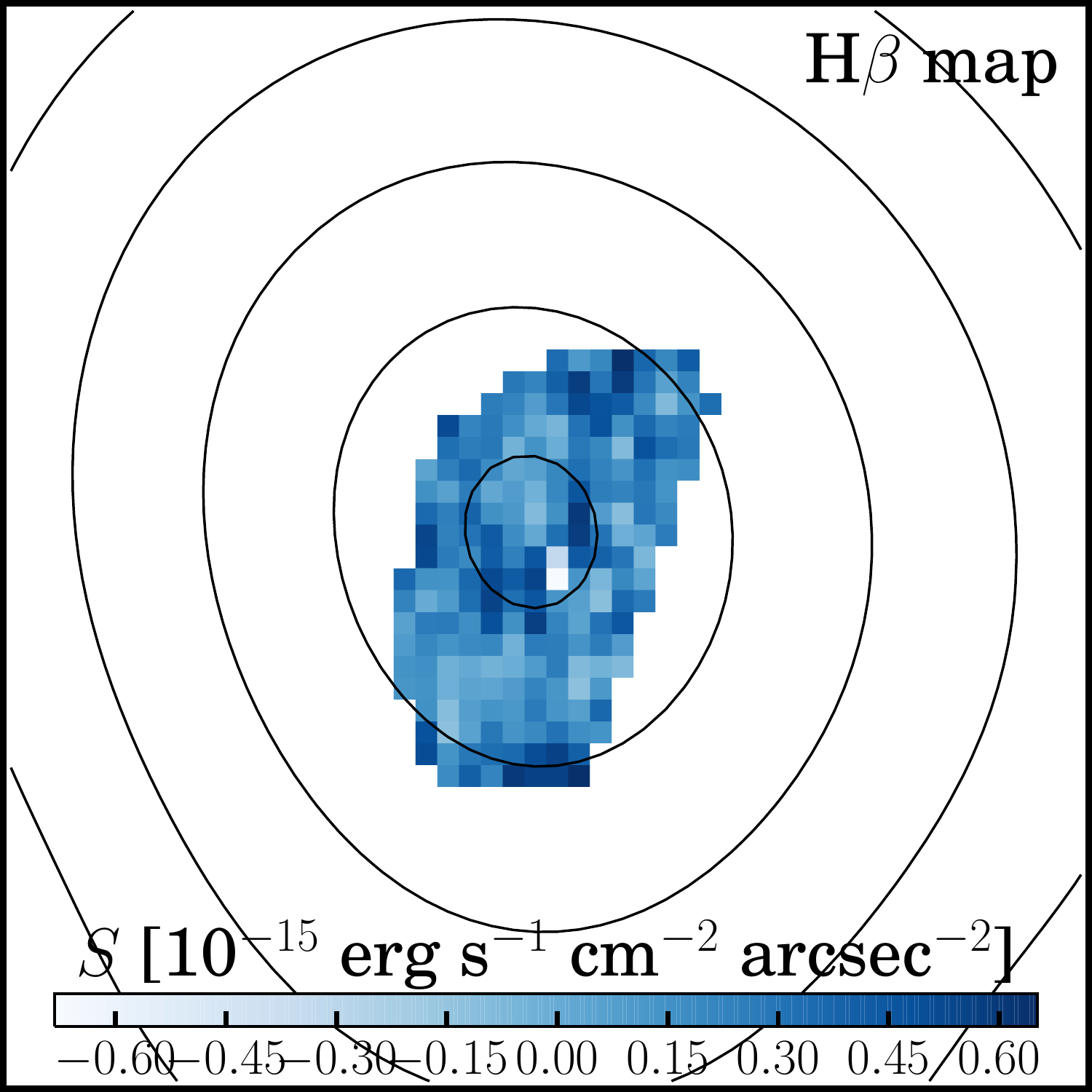}
	\includegraphics[width=.33\textwidth]{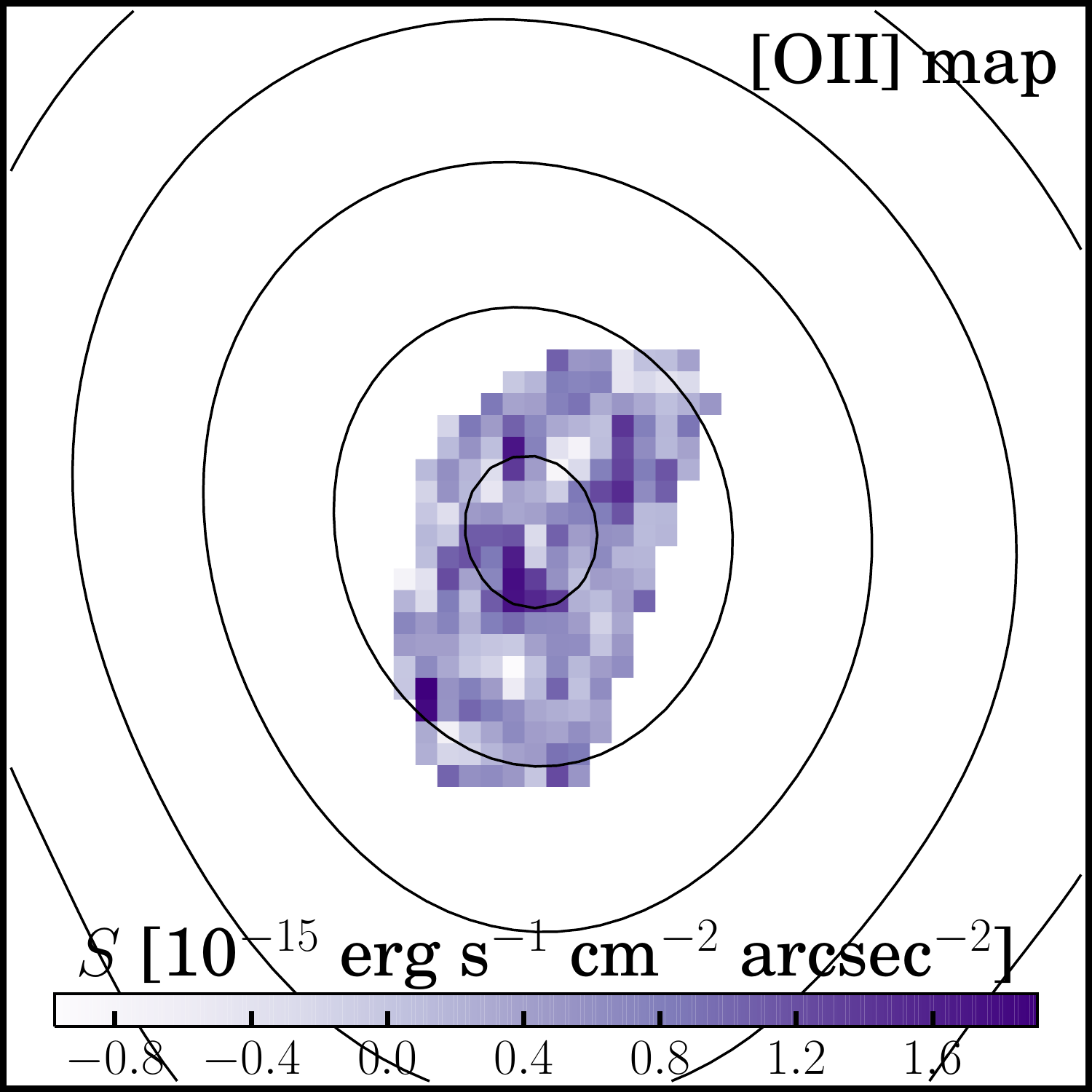}
    \caption{Multi-perspective view of the SN Refsdal host galaxy multiple image 1.3, \ie, ID
    04054 in our \mg sample.  \textbf{Upper left}: zoom-in color composite stamp cut out from
    Figure~\ref{fig:RGBfullFoV}, where the estimated site of the un-observed past appearance
    of SN Refsdal is denoted by the red circle. \textbf{Upper middle}: stellar mass surface
    density map for this galaxy, where the source plane de-projected galactocentric radius
    contours are overlaid, in 2 kpc intervals, starting from 1 kpc.  \textbf{Upper right}:
    \Ha surface brightness map combined from the \hst grism exposures at all PAs from the
    \glass and Refsdal follow-up programs.
    The black contours again show the de-projected galactocentric radii in the same fashion
    presented in the stellar mass surface density map in the upper middle panel.
    \textbf{Lower rows}: combined surface brightness maps of \OIII, \Hb, and \OII, presented
    in the same fashion as in the \Ha emission line map in the upper right panel.}
    \label{fig:multiP_4054}
\end{figure*}

For the purpose of this study, we compiled an exhaustive list of spatially extended sources with secure spectroscopic redshift
measurements in the range of $z\in[1.2, 2.3]$, showing strong ELs (primarily \OIII and \Hb) in the 2D grism spectra. This redshift
range avoids the region where grism sensitivity and throughput drop significantly.  This results in a sample of 14 galaxies. The
positions of these 14 galaxies relative to the cluster are denoted by the magenta circles in Figure~\ref{fig:RGBfullFoV}, while
the postage stamp images of these galaxy are shown in Figure~\ref{fig:RGBstamps}.
The full sample is also listed in Table~\ref{tab:srcprop}.
In particular, one source in our sample, \ie, galaxy ID 04054, is one of the multiple images of SN Refsdal's host
galaxy (see Figure~\ref{fig:multiP_4054} for its appearances from
various perspectives). A steep metallicity gradient ($-0.16\pm0.02$
dex/kpc) has been previously measured on another multiple image of
this grand-design spiral \citep{Yuan:2011hj}. Our work is the first
\mgm on this particular image, which is least distorted/contaminated
by foreground cluster members. We also take advantage of a more
precise lens model of both the entire cluster and the SN Refsdal host,
from the extensive lens modeling effort summarized by \citet{2015arXiv151005750T}.

\subsection{Ground-based IFU observations}\label{subsect:ifu}

When available, we use ground based spectroscopic integral field unit (IFU) data
to investigate the 2D kinematics of the sources for which we obtain
star formation and metallicity maps from the \hst grism data.  A first
set of ground-based data were obtained in 2015 with the instrument
\muse on the \vlt, taken as part of the Director's Discretionary Time
program 294.A-5032 (P.I. Grillo) aimed at assisting
with the modeling of SN Refsdal.  The \muse pointing is denoted
by the green square in Figure~\ref{fig:RGBfullFoV}.  The observations,
data reduction, first results, and applications to the host galaxy of
Refsdal and mass modeling of the cluster are described by \citet{2016ApJ...822...78G,Karman:2016cg}.
Five objects in our \mg sample fall within the \muse pointing, with one on the edge of the FoV.
Hence four maps showing complete kinematic structures can be extracted for sources IDs 02389, 02806, 03746, and 04054.
A second set of ground based observations were obtained with the instrument \kmos also on the
\vlt, as part of our \glass follow-up \kmos Large Program 196.A-0778 (P.I. Fontana). Two objects (IDs 05709 and 05968) in our
sample were targeted with the deployable IFUs, using the YJ grating.
The observations, data reduction, and first results from this program are described by \citet{Mason:2016ww}.

\section{Photometry and lens models from the \hff}\label{sect:phot}

The Hubble Frontier Fields
initiative\footnote{\url{http://www.stsci.edu/hst/campaigns/frontier-fields}} \citep[\hff;
P.I.  Lotz,][]{Lotz:2016ca} is an ongoing multi-cycle treasury program enabled by an \hst
director's discretionary time allocation. \hff targets the cores (prime fields) and infall regions (parallel fields)
of six massive galaxy clusters, reaching an ultra-faint intrinsic magnitude of 30-33, made possible by the synergy of diffraction
limited photometry and lensing magnification. The large wavelength coverage, uninterrupted from \B to \H passbands, is crucial for
photometric redshift determinations and stellar population synthesis.

Besides the deep image mosaics, the \hff collaboration also provides
the community with cluster lens models
\footnote{\url{http://archive.stsci.edu/prepds/frontier/lensmodels/}}.
Several groups of scientists were invited prior to the beginning of
the campaign to provide independent models depicting the total mass distribution
of the six \hff clusters, using a number of distinct techniques.
As data accumulate and more multiply imaged systems are identified, the lens models are continuously improved.
For our cluster of interest, \clsan, the most up-to-date publicly available model is the \glafic version 3 model,
constructed using a simply parametrized method proposed by \citet{2010PASJ...62.1017O}.
For an in-depth description of the various modeling techniques, their advantages and limitations, we
refer to \citet{2015arXiv151005750T,Meneghetti:2016wg}.

In this work, we need lens models to trace the observed metallicity
maps back to the source plane. Following our previous work \citep{2015AJ....149..107J},
we experiment with several publicly available models to find the ones that give the
reconstruction of our target sources the highest fidelity (judging by
how well the source plane morphologies of multiply imaged sources
match each other). Based upon the SN Refsdal test (see also \citet{2015arXiv151204654K}),
the \glafic version 3 model \citep{2015arXiv151006400K}, the \textsc{Grillo} model \citep{2016ApJ...822...78G}, and
the \SJ version 3 model were the most accurate ones, so we considered those three.
As an illustration, the \glafic version 3 critical curves at $z=1.8$, the
median redshift of our \mg sample, are overlaid in cyan in Figure~\ref{fig:RGBfullFoV}.

The most highly magnified source in our sample is galaxy ID 02389, one of
the folded arcs which straddle the critical curve. For that particular
source, we used as our default the \SJ version 3 model, updated from the earlier versions
presented in \citet{Johnson:2014cf,Sharon:2015hr}. The \SJ version 3 model leads
to a more precise reconstruction of the source plane morphology of
galaxy ID 02389.  We also tested that switching entirely from the
\glafic version 3 model to the \SJ version 3 model does not affect our
measurements significantly, as also pointed out by \citet{2015arXiv150901279L}.
The lensing magnification results from the considered lens models are given in
Table~\ref{tab:srcprop}.

\section{Analysis procedures}\label{sect:analysis}

Here we describe the stellar population synthesis analysis of our sample in
Section~\ref{subsect:sed}, and source plane morphology reconstruction in
Section~\ref{subsect:morph}.
The entire steps necessary for extracting EL maps from 2D grism spectra combined from
different PAs are detailed in Section~\ref{subsect:combELmaps}.
Our new Bayesian method to jointly infer metallicity, nebular dust extinction, and \sfr from a
simultaneous use of multiple strong ELs is presented in Section~\ref{subsect:bayes}.

\subsection{Spectral energy distribution fitting}\label{subsect:sed}

The full-depth 7-filter \hff photometry was fitted with the template spectra from
\citet{Bruzual:2003ck} using the stellar population synthesis code \fast
\citep{Kriek:2009cs}, in order to derive global estimates of stellar mass (\Mstar) and the
dust extinction of stellar continuum ($\Av^{\rm S}$) for each source in our sample. We take a
grid of stellar population parameters that include: exponentially declining star formation
histories with $e$-folding times ranging from 10~Myr to 10~Gyr, stellar ages ranging from 5~Myr to
the age of the universe at the observed redshift, and $\Av^{\rm S}$=0-4 magnitudes for a
\citet{Calzetti:2000iy} dust extinction law. We assume the \citet{Chabrier:2003ki}
initial mass function (IMF) and fix the stellar metallicity to solar.
Table~\ref{tab:srcprop} lists the best-fit \Mstar values, corrected for lensing
magnification.
Although the adopted solar metallicity is higher than the typical gas-phase abundances we
infer for the sample, this has little effect on the derived stellar mass. Fixing the stellar
metallicity to $1/5$ solar (i.e. $Z=0.004$ or $\oh=8.0$, corresponding to the sample median)
reduces the best-fit \Mstar by only $\sim0.03$ dex.

\subsection{Source plane morphology}\label{subsect:morph}

Measurements of metallicity gradients require knowledge of the galaxy center, major axis orientation, and inclination or
axis ratio. We derive these quantities from spatially resolved maps of the stellar mass surface density following the
methodology described in \citet{2015AJ....149..107J}. Briefly, we smooth the \hff photometric
images to a common point spread function of 0\farcs2 FWHM and fit the spectral energy
distribution (SED) of each spaxel using the same procedure described in
Section~\ref{subsect:sed}. The resulting \Mstar maps in the image plane are shown in Figures~\ref{fig:multiP_4054} and
\ref{fig:multiP_rest}. We reconstruct the \Mstar maps in the source plane using the adopted
lens models and determine the centroid, axis ratio, inclination and major axis orientation from a 2D
elliptical Gaussian fit to the \Mstar distribution. This allows us to measure the
galactocentric radius at each point, assuming that contours of \Mstar surface density
correspond to constant de-projected radius \citep[following][]{2015AJ....149..107J}.

Since the following analysis is mostly done with the robust image plane data, we reconstruct
the derived radius maps in the image plane. Contours of constant galactocentric radius are
shown in Figures~\ref{fig:multiP_4054} and \ref{fig:multiP_rest}.
For half of our sample we also have seeing-limited IFU data described in
Section~\ref{subsect:ifu}, and verify that for galaxies with significant rotation support (\eg
IDs 05709 and 05968), the morphological major axis agrees at 2-$\sigma$ with
the pseudo-slit that maximizes velocity shear according to our kinematic analysis (see Section~\ref{subsect:kinem}).

\begin{figure*}
    \includegraphics[width=.163\textwidth]{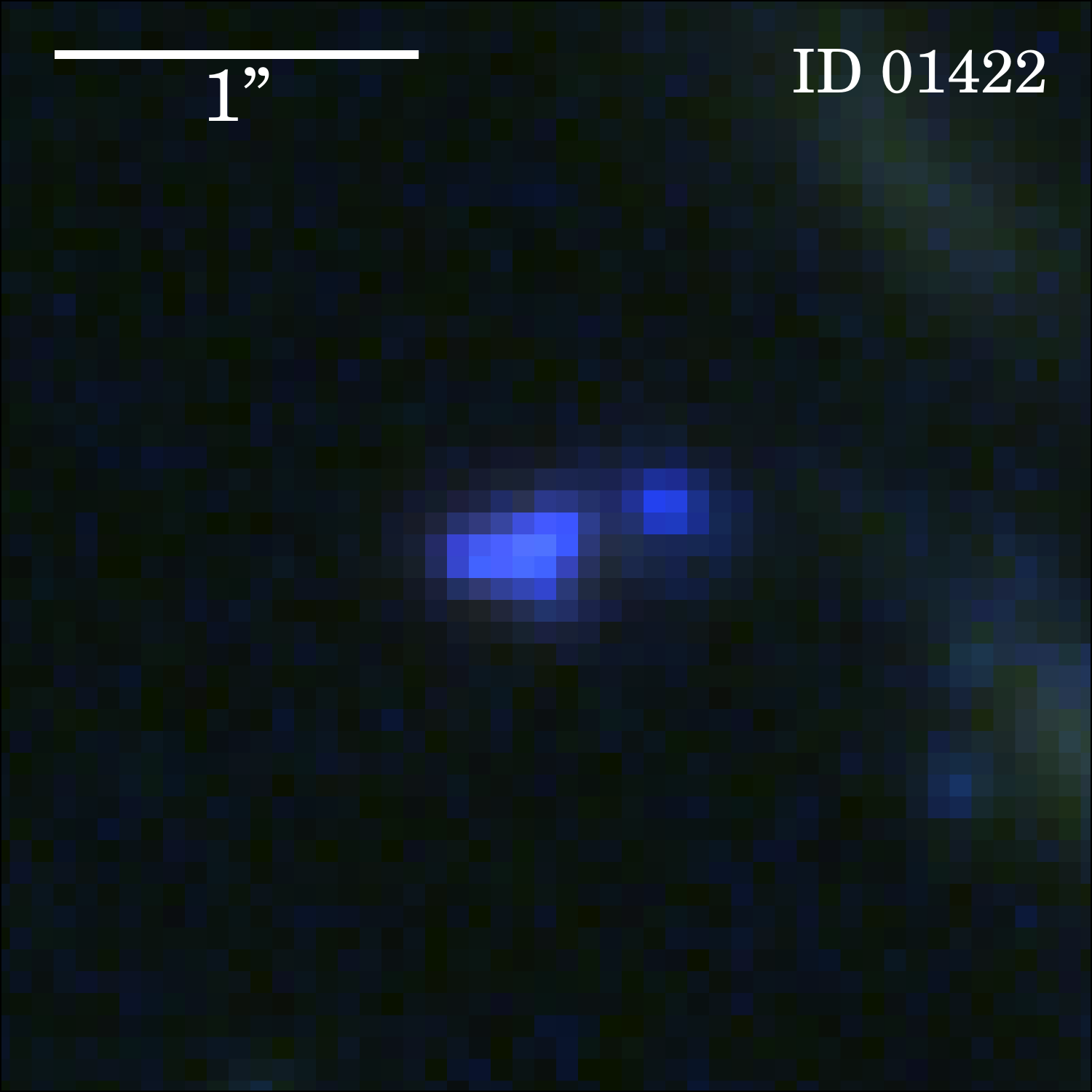}
    \includegraphics[width=.163\textwidth]{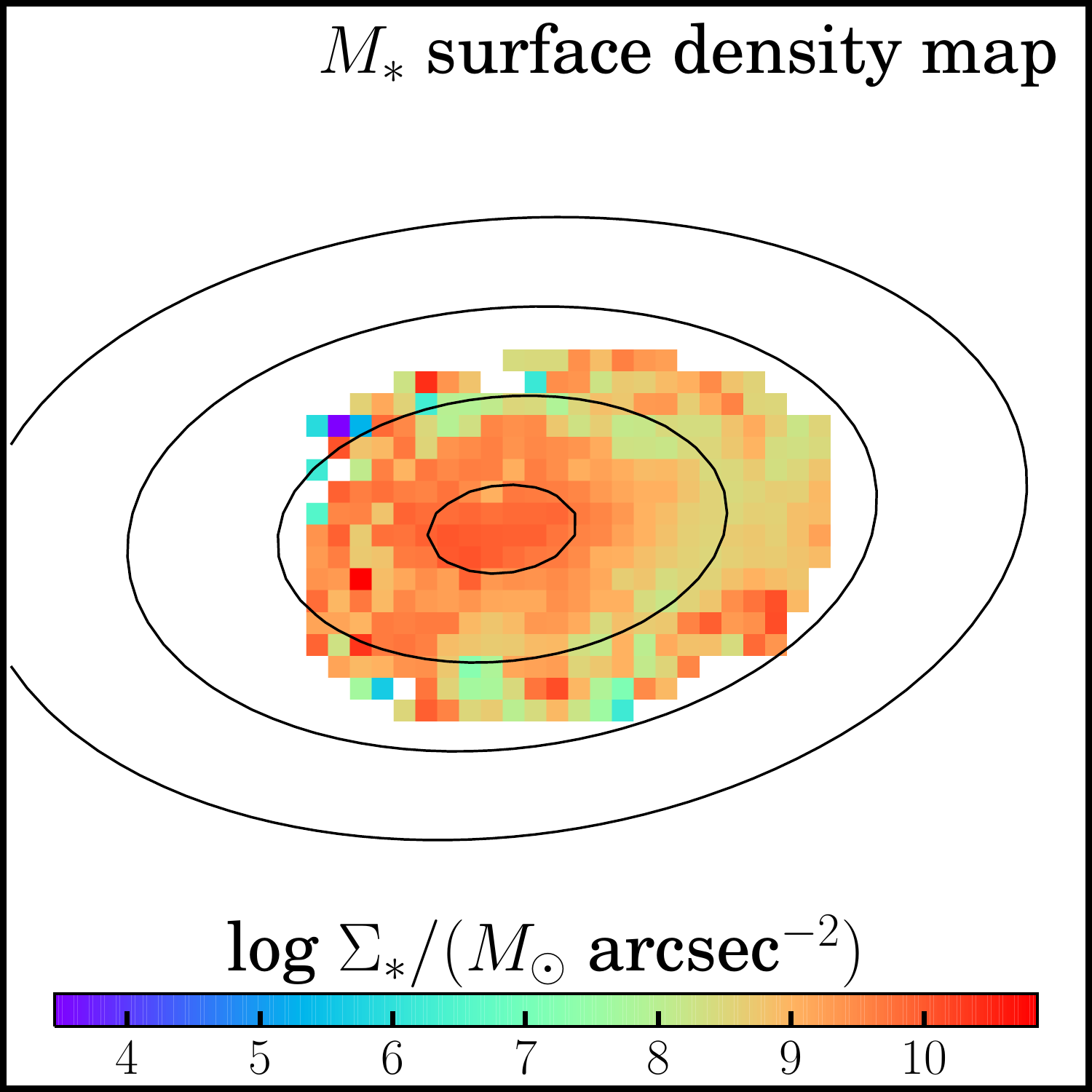}
    \includegraphics[width=.163\textwidth]{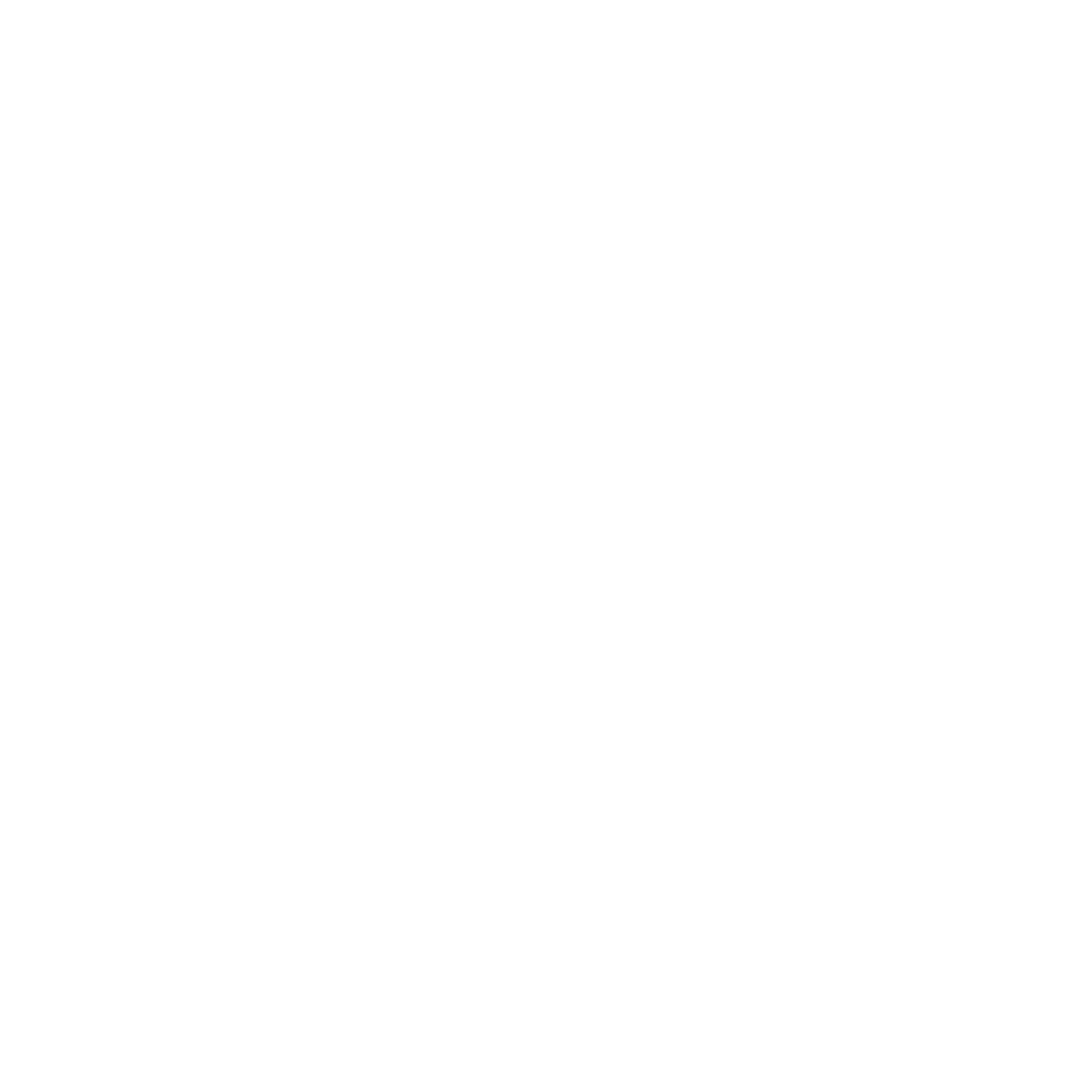}
    \includegraphics[width=.163\textwidth]{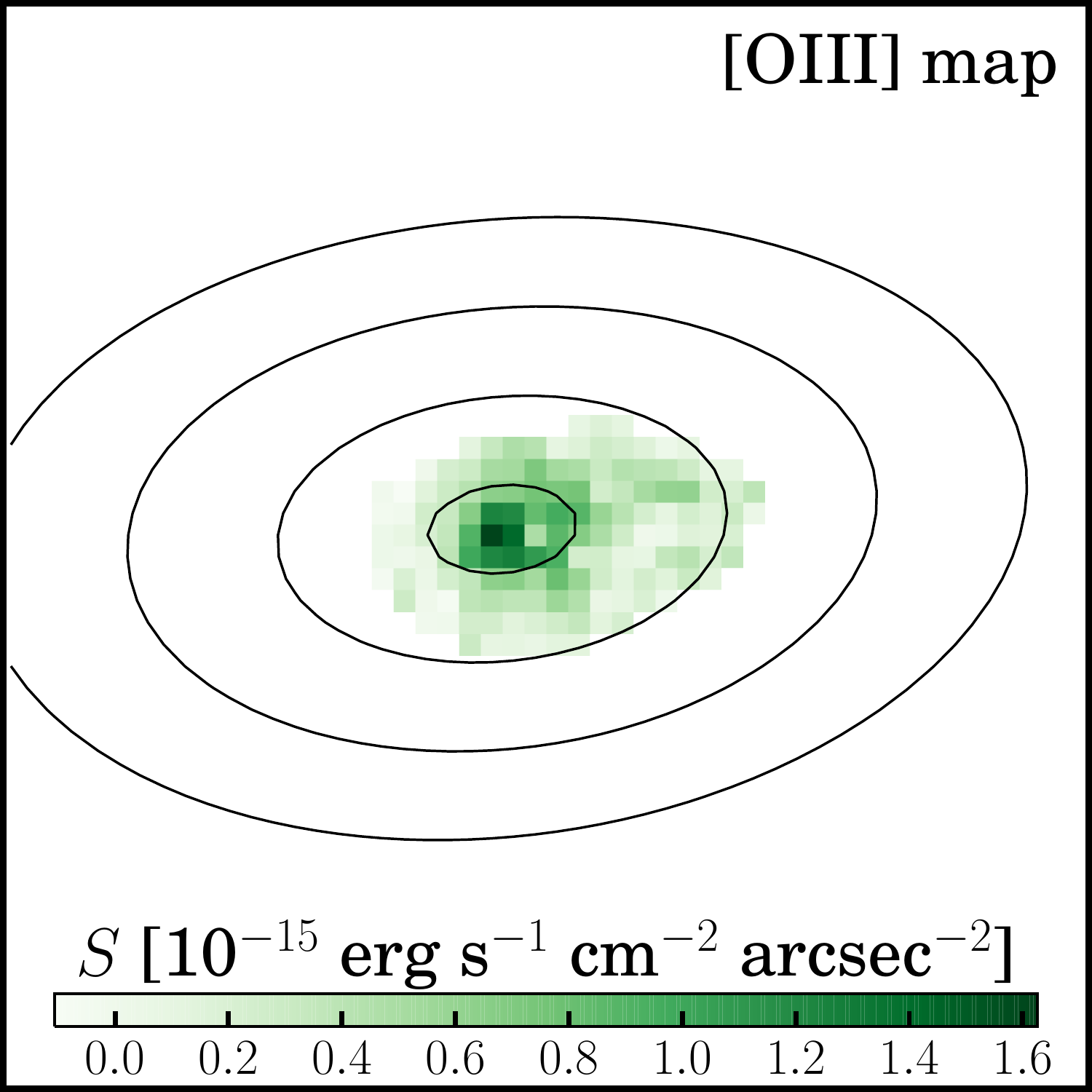}
    \includegraphics[width=.163\textwidth]{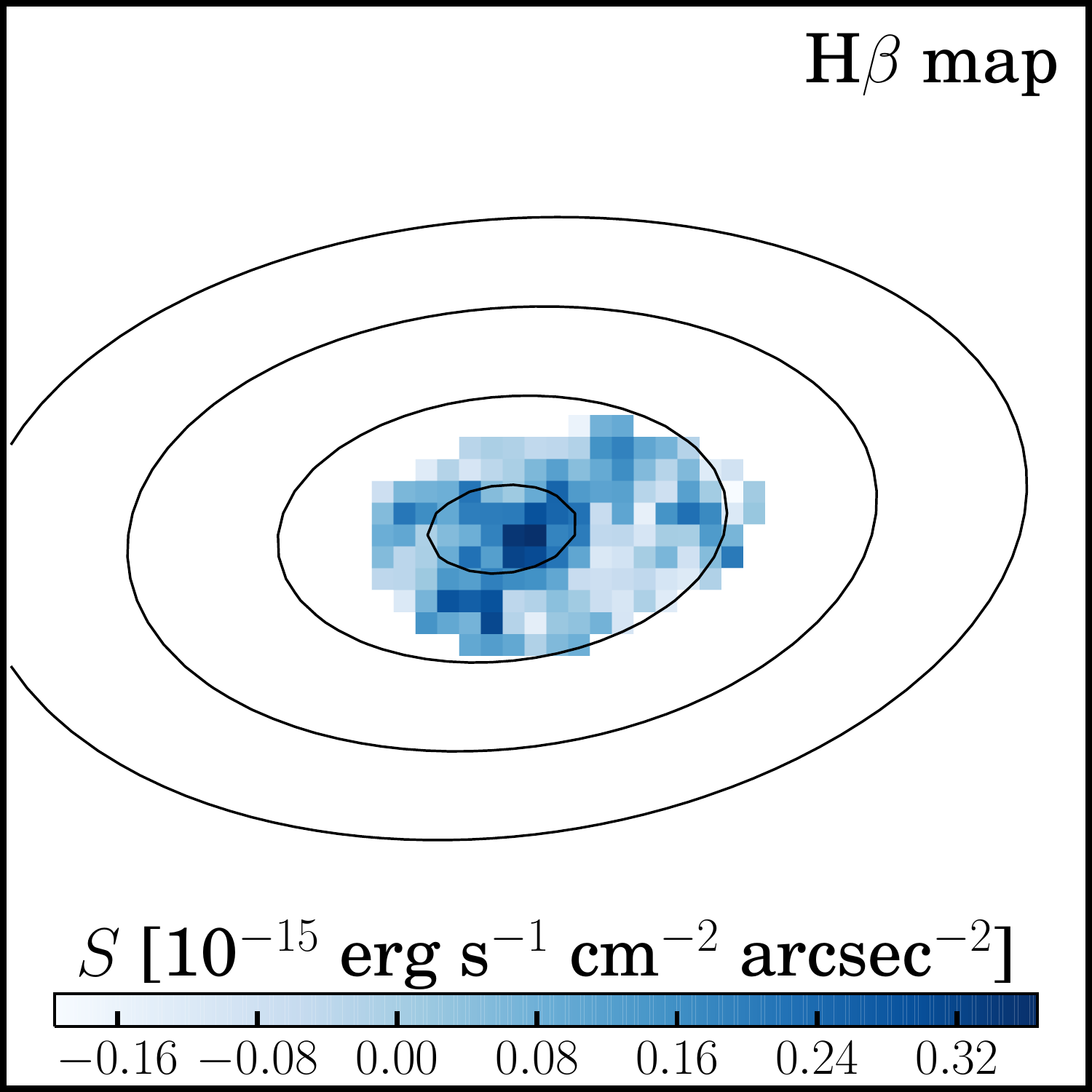}
    \includegraphics[width=.163\textwidth]{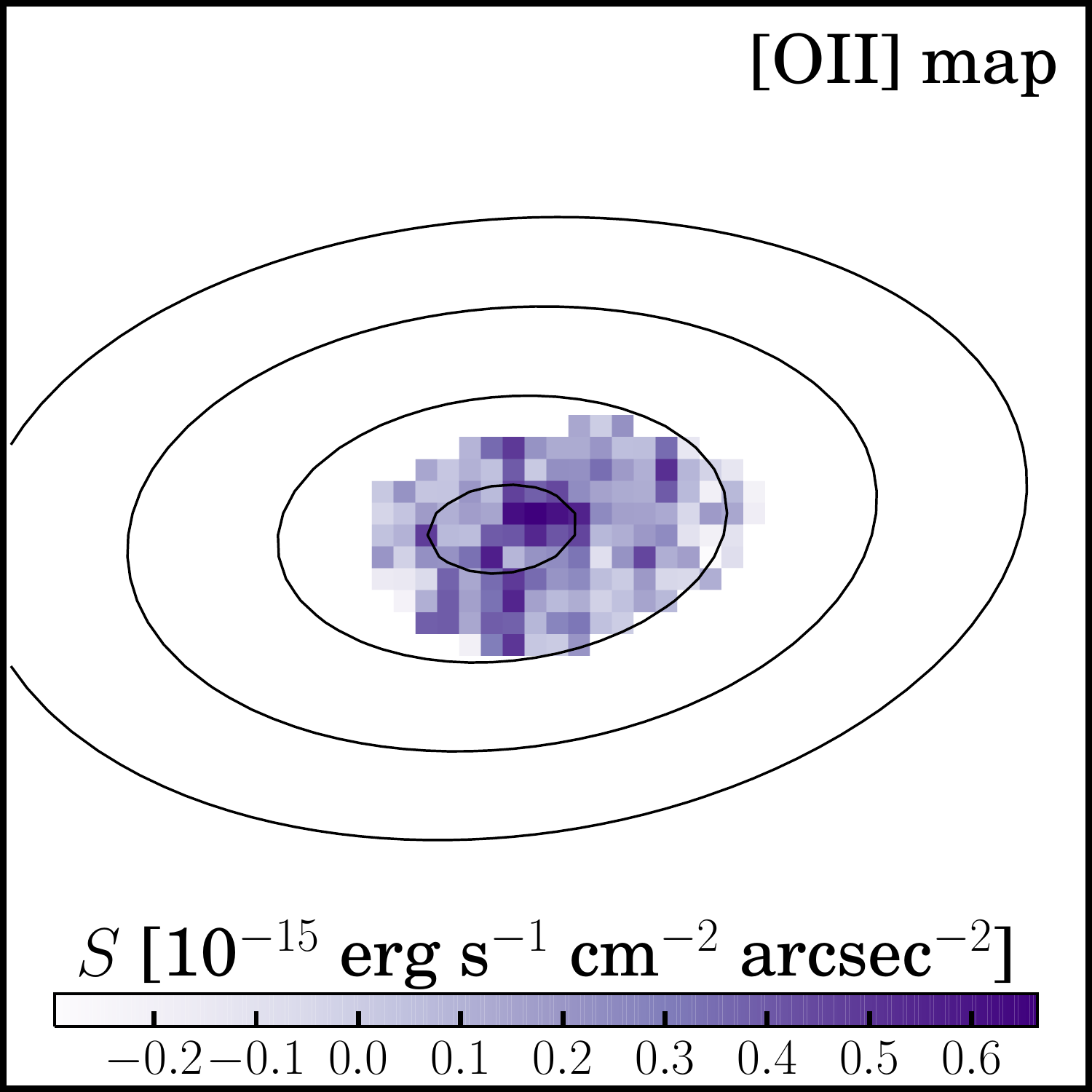}\\
    \includegraphics[width=.163\textwidth]{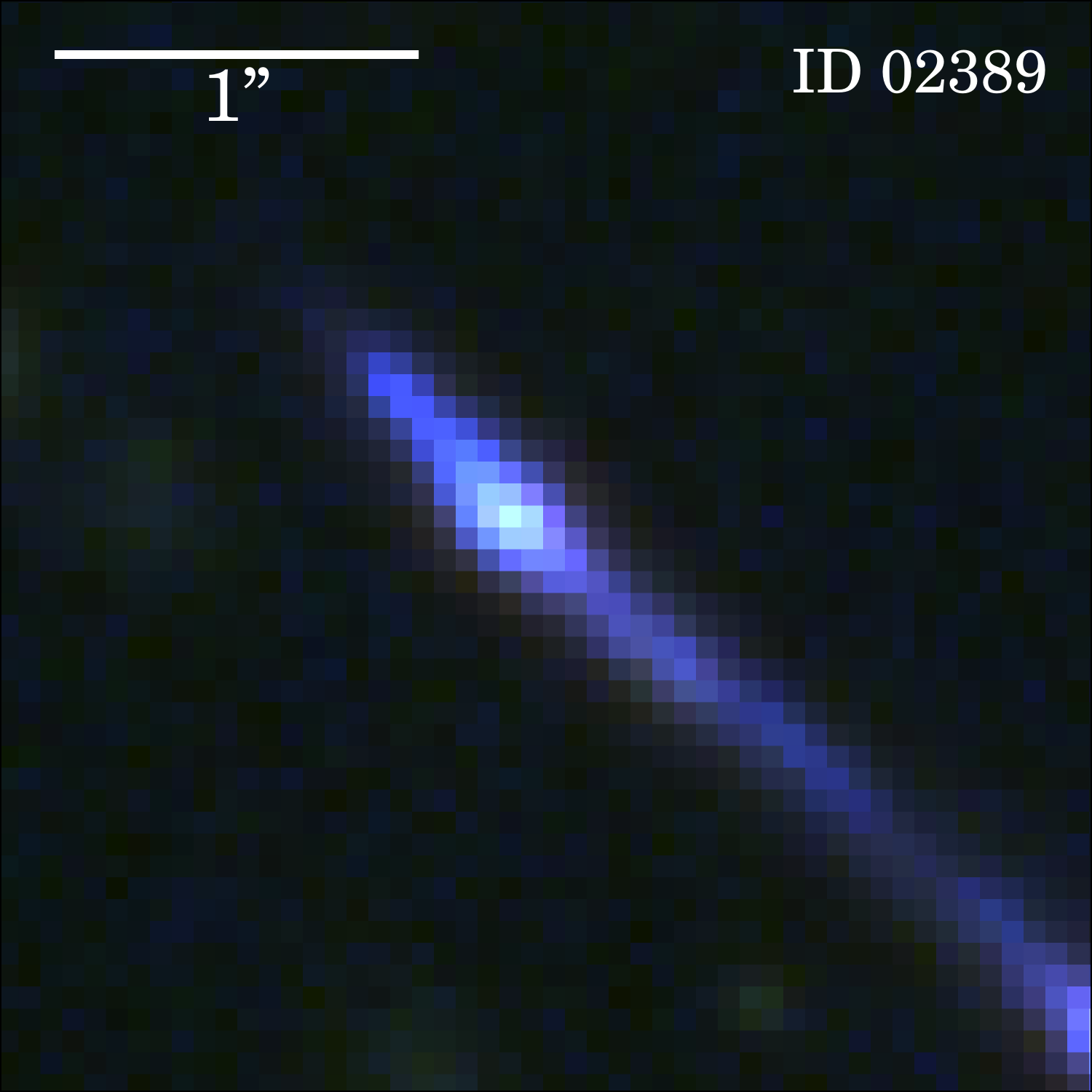}
    \includegraphics[width=.163\textwidth]{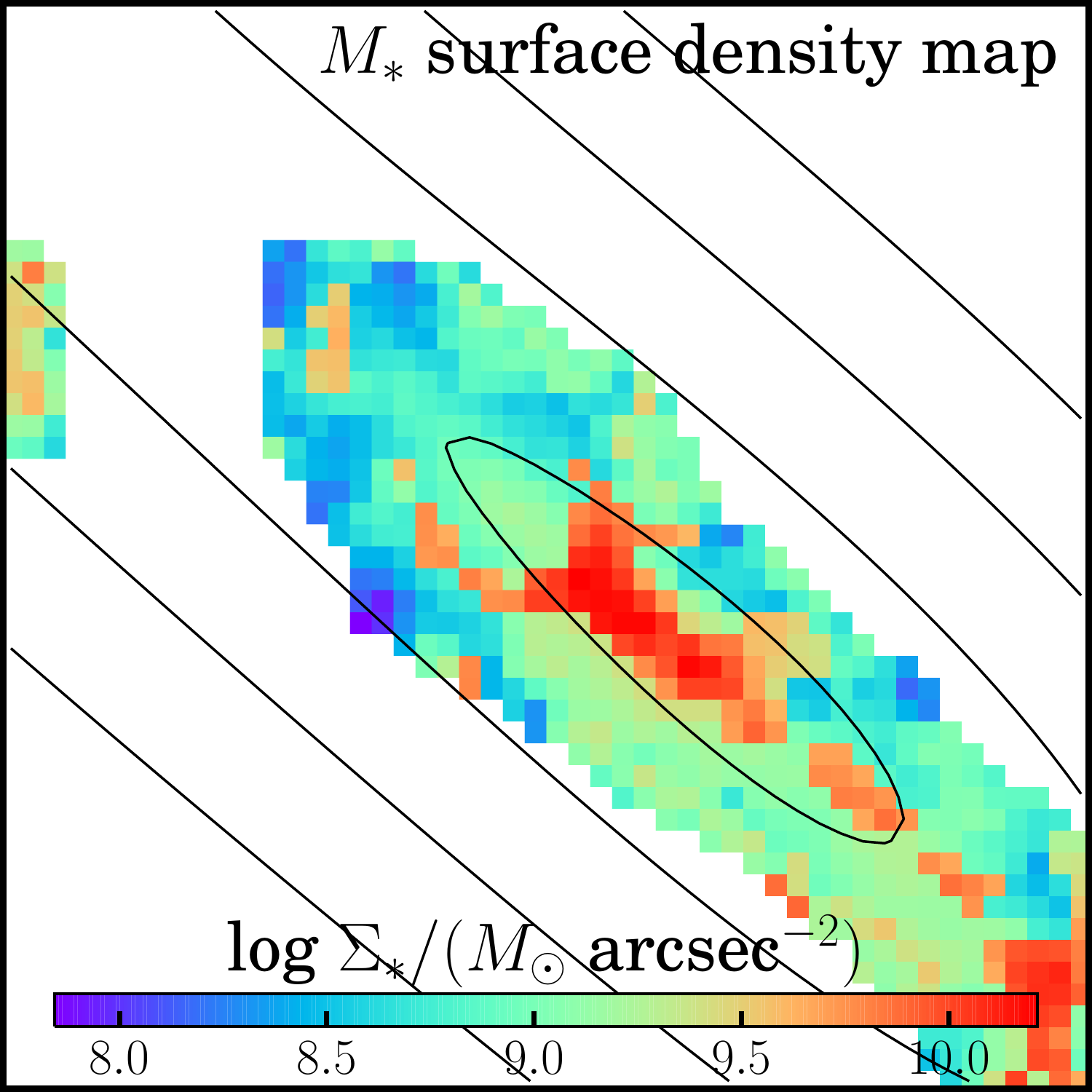}
    \includegraphics[width=.163\textwidth]{fig/baiban.png}
    \includegraphics[width=.163\textwidth]{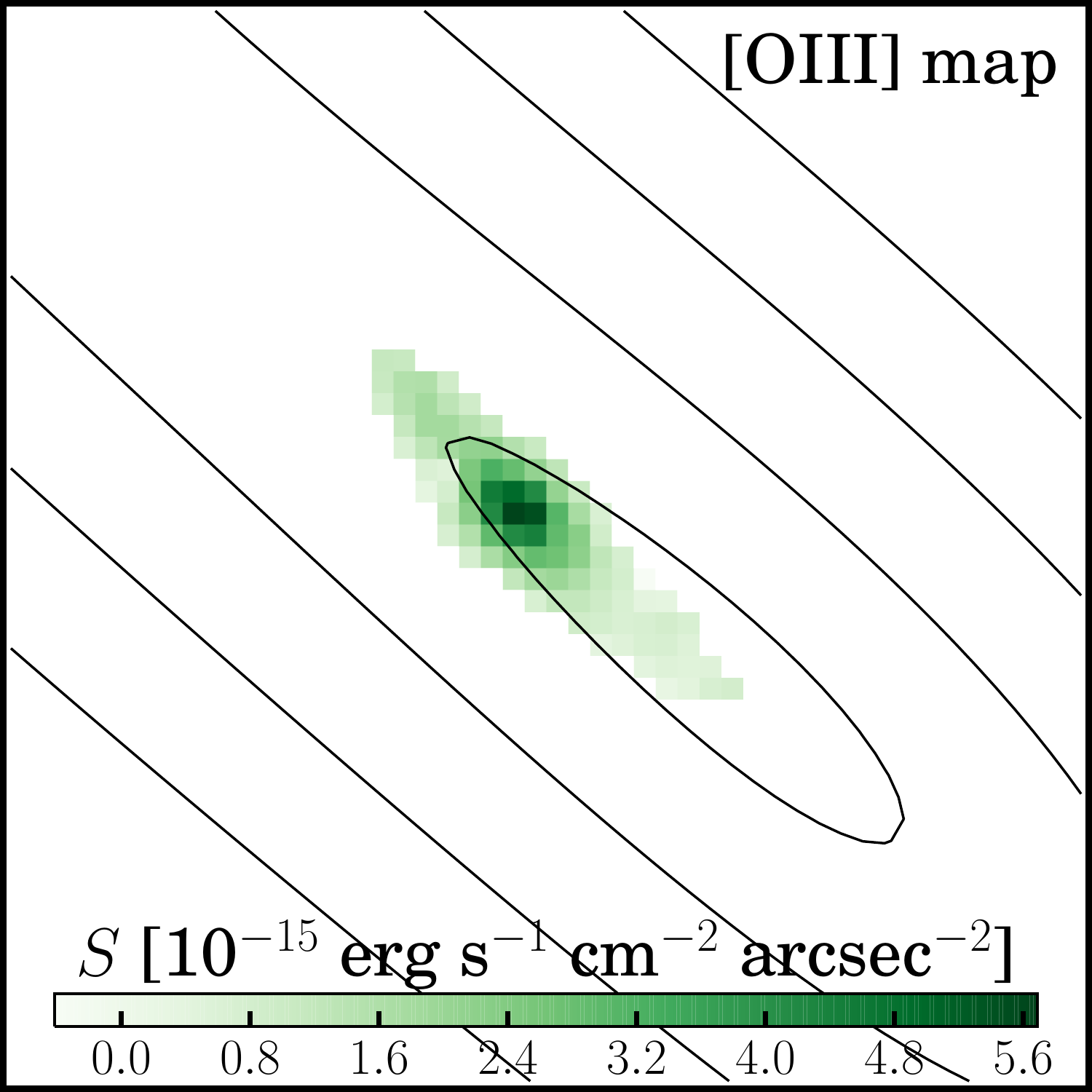}
    \includegraphics[width=.163\textwidth]{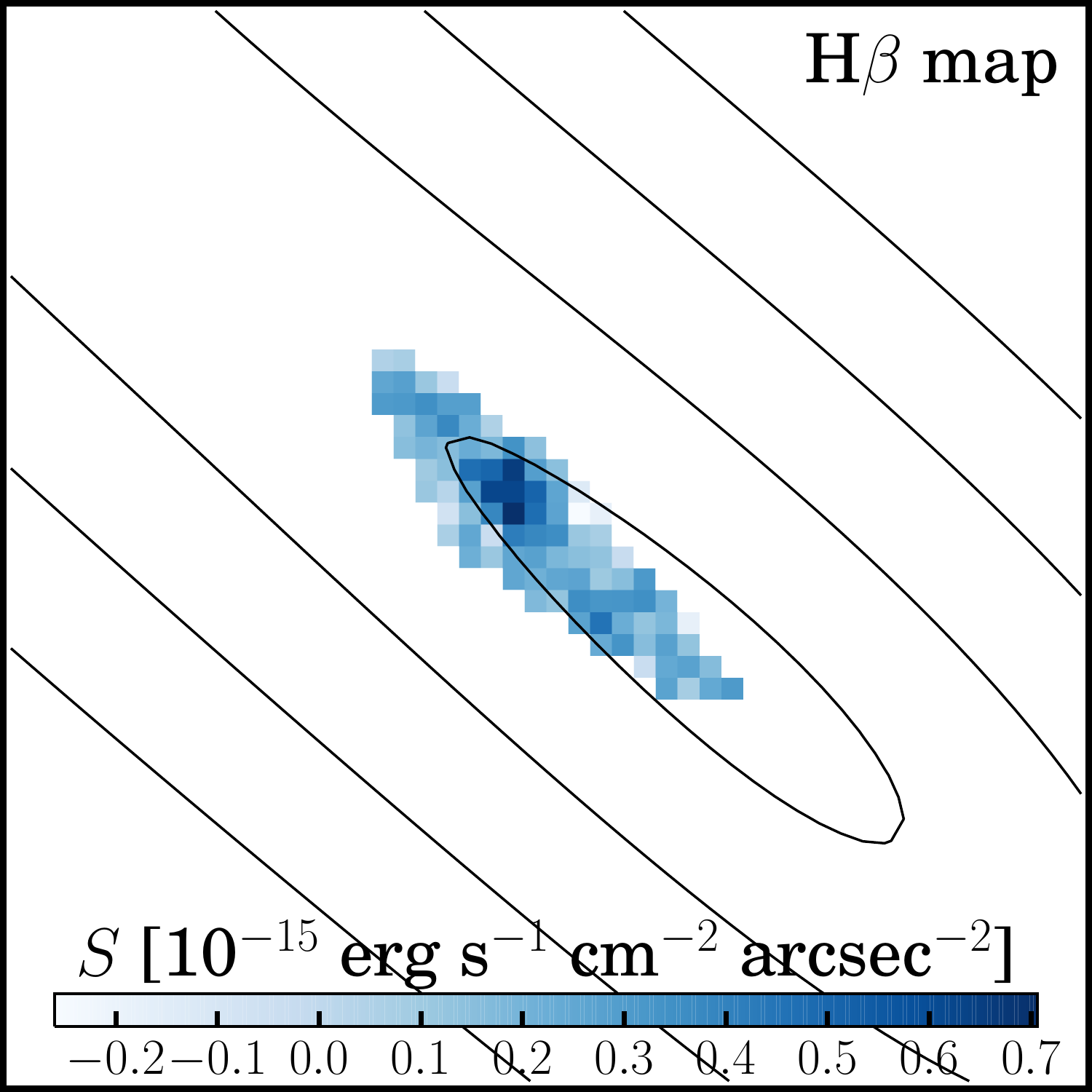}
    \includegraphics[width=.163\textwidth]{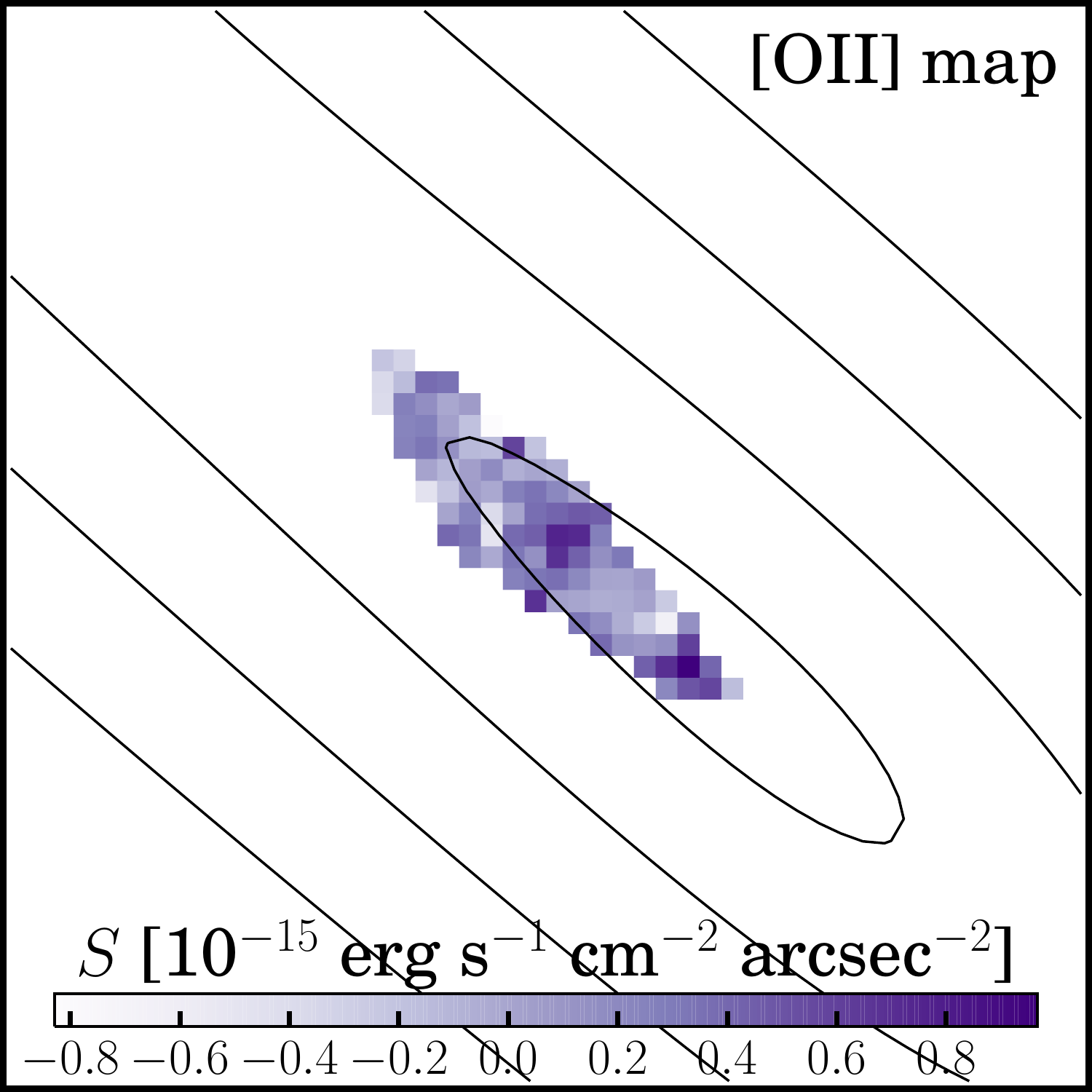}\\
    \includegraphics[width=.163\textwidth]{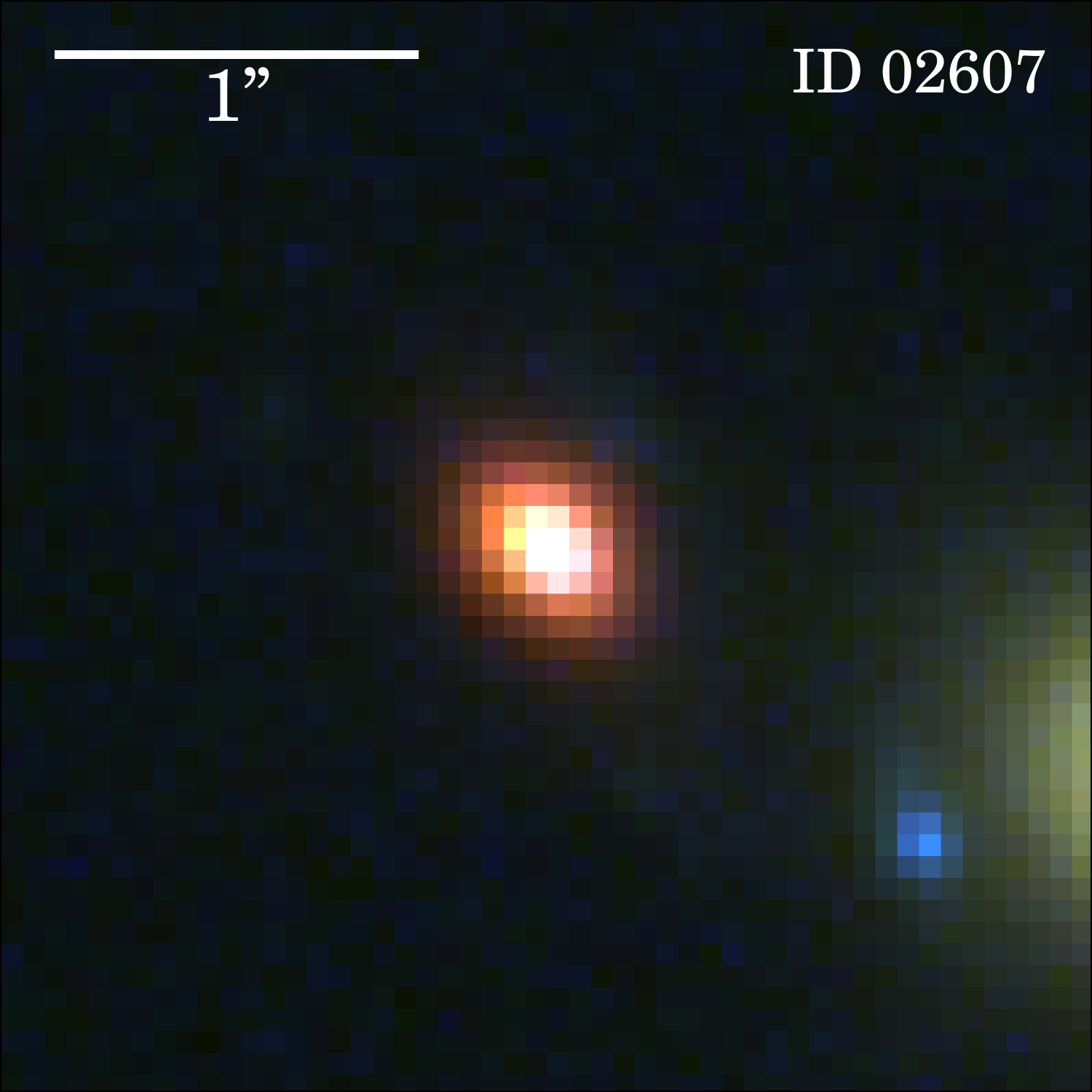}
    \includegraphics[width=.163\textwidth]{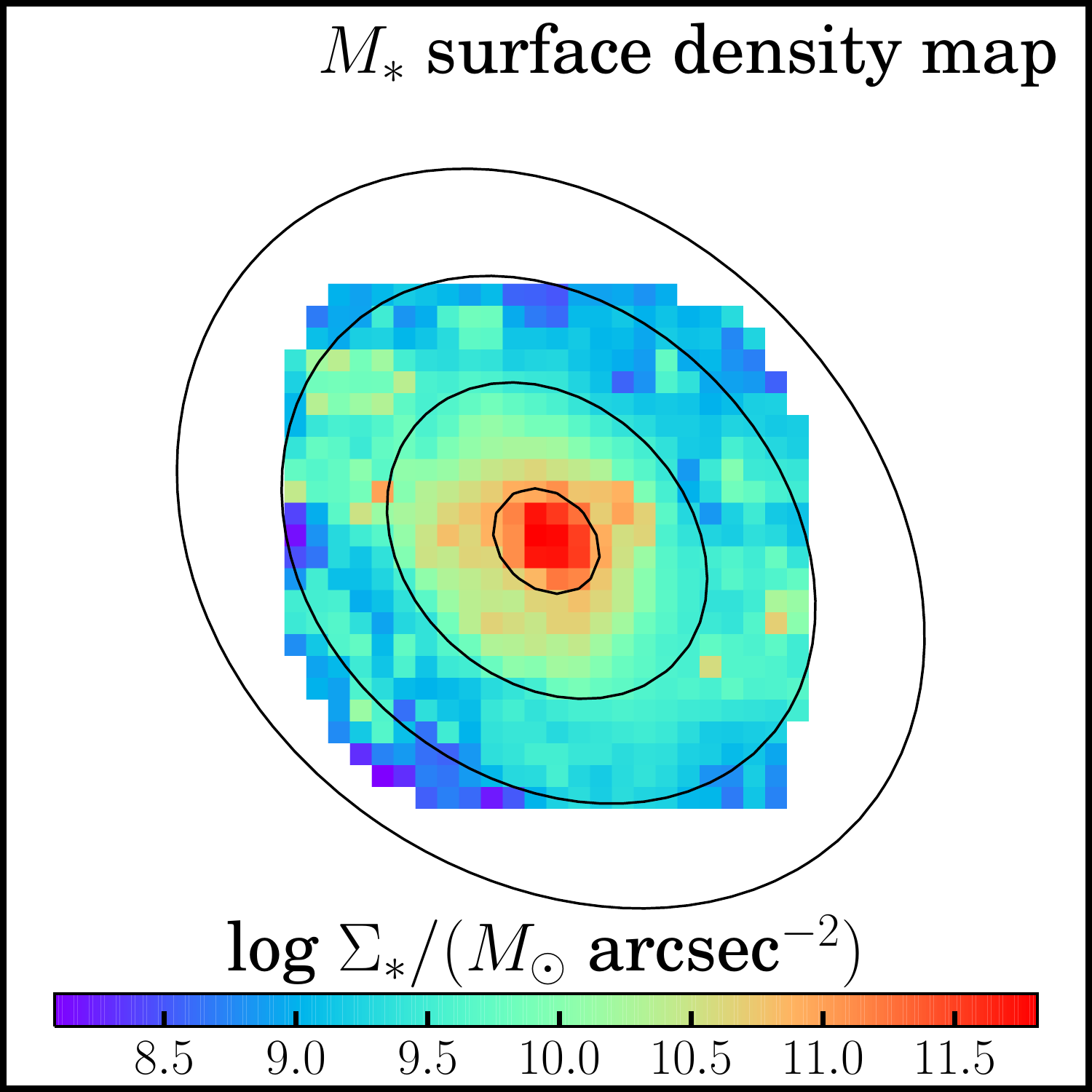}
    \includegraphics[width=.163\textwidth]{fig/baiban.png}
    \includegraphics[width=.163\textwidth]{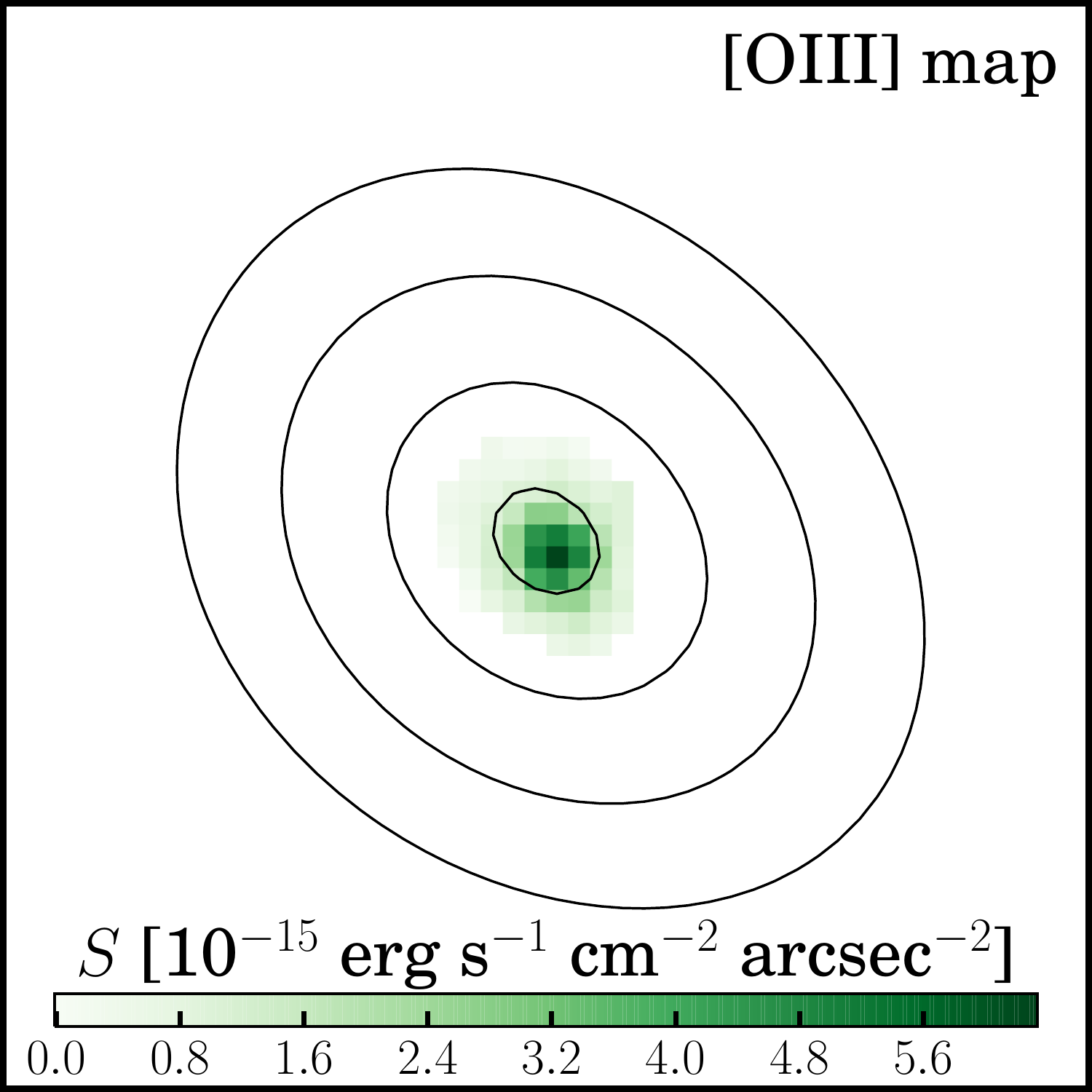}
    \includegraphics[width=.163\textwidth]{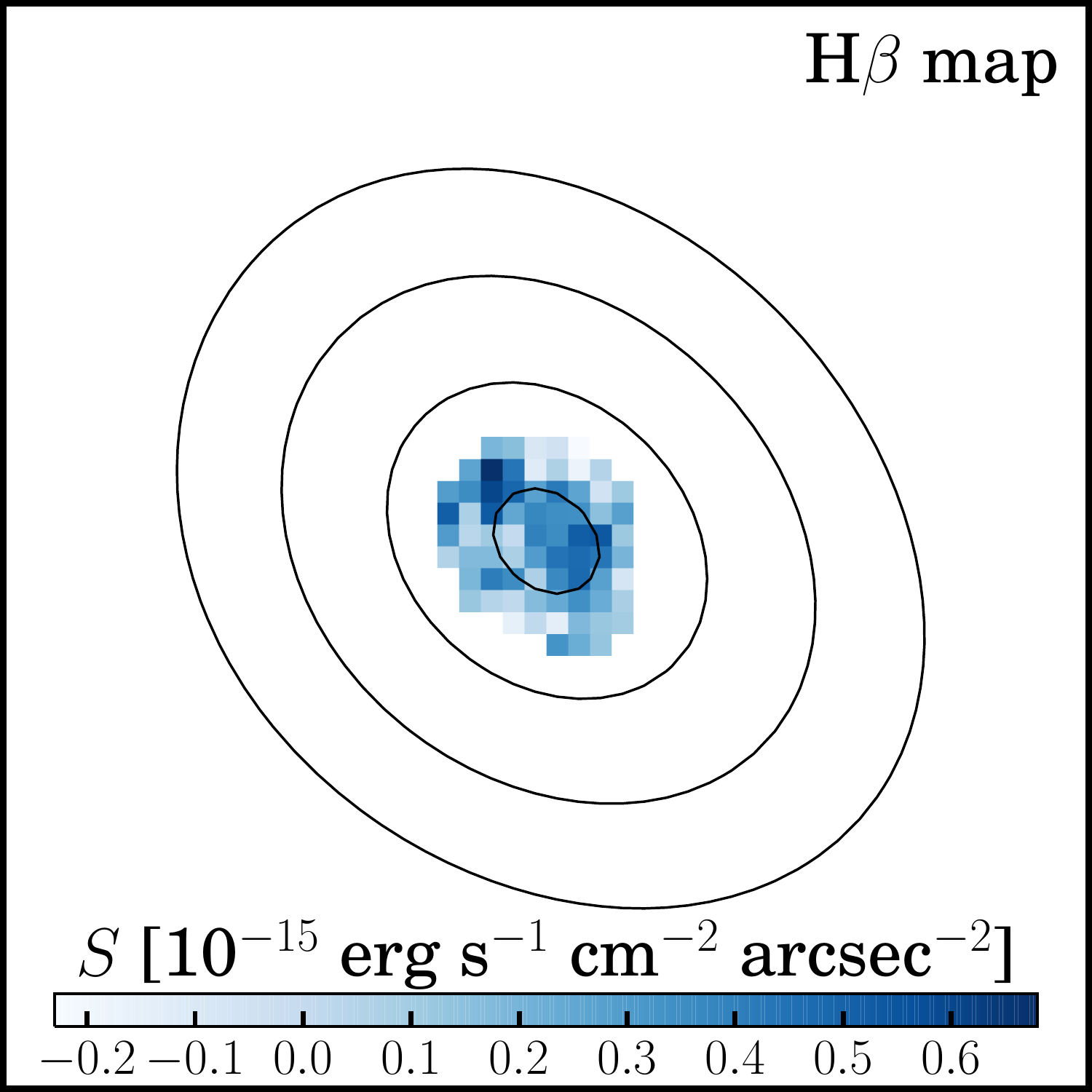}
    \includegraphics[width=.163\textwidth]{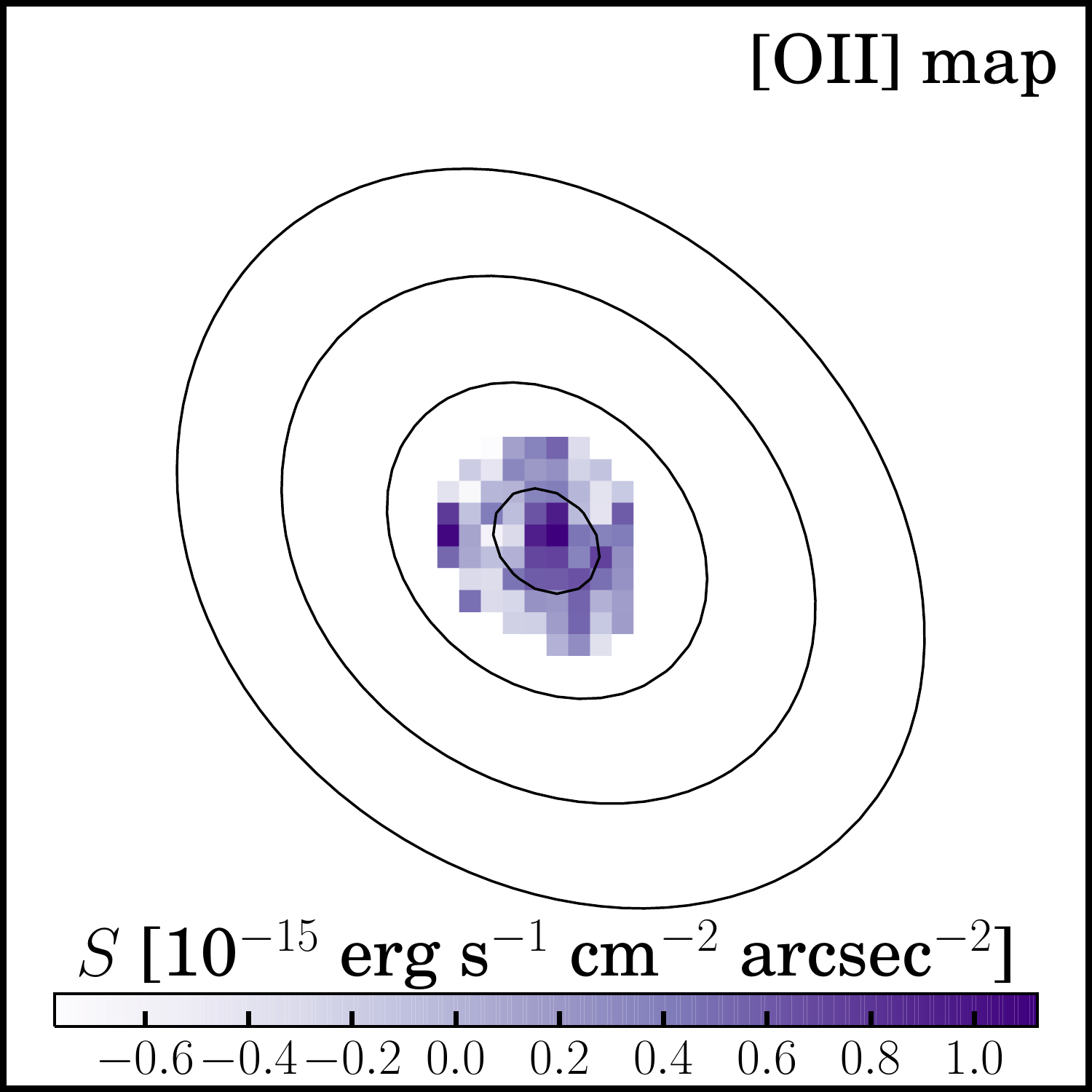}\\
    \includegraphics[width=.163\textwidth]{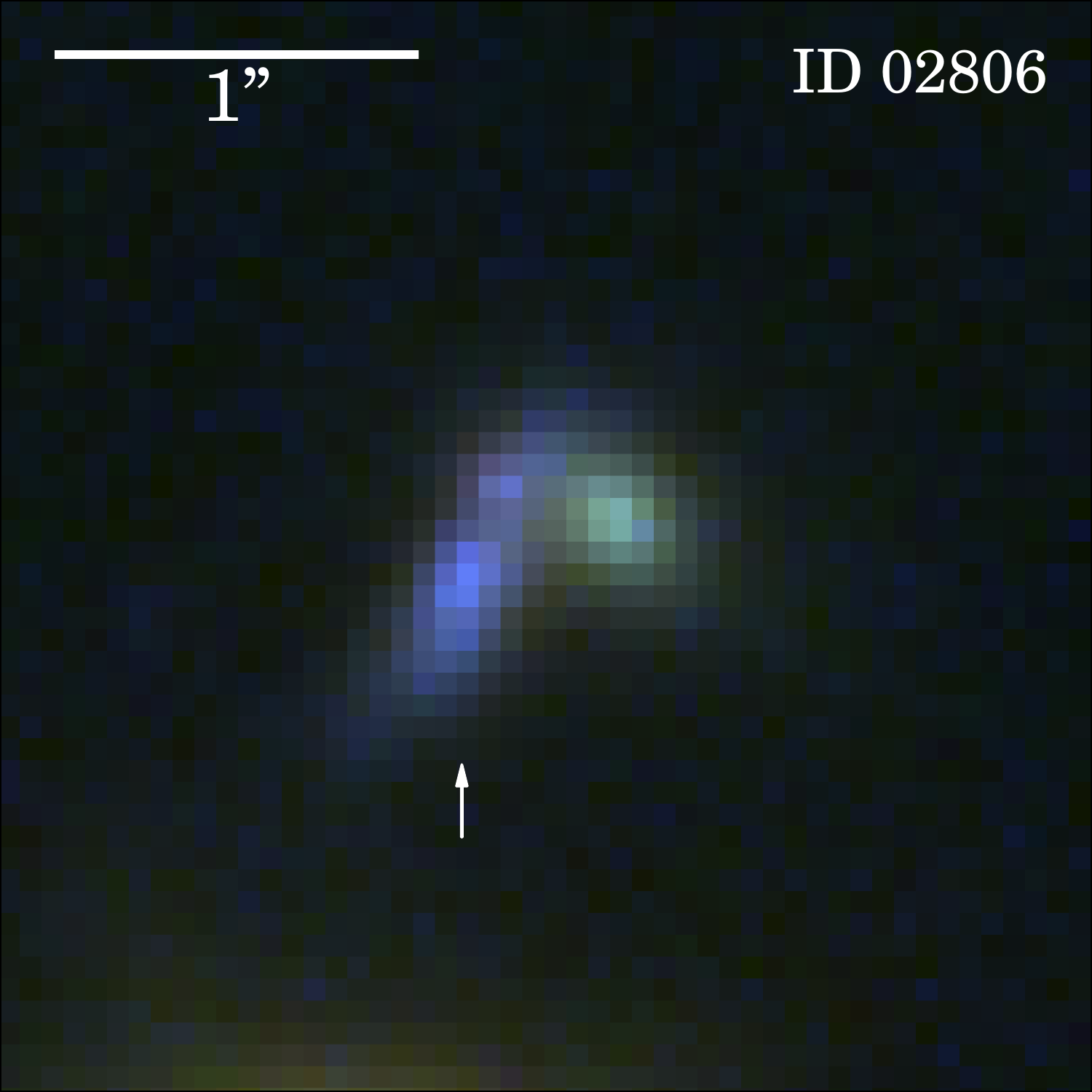}
    \includegraphics[width=.163\textwidth]{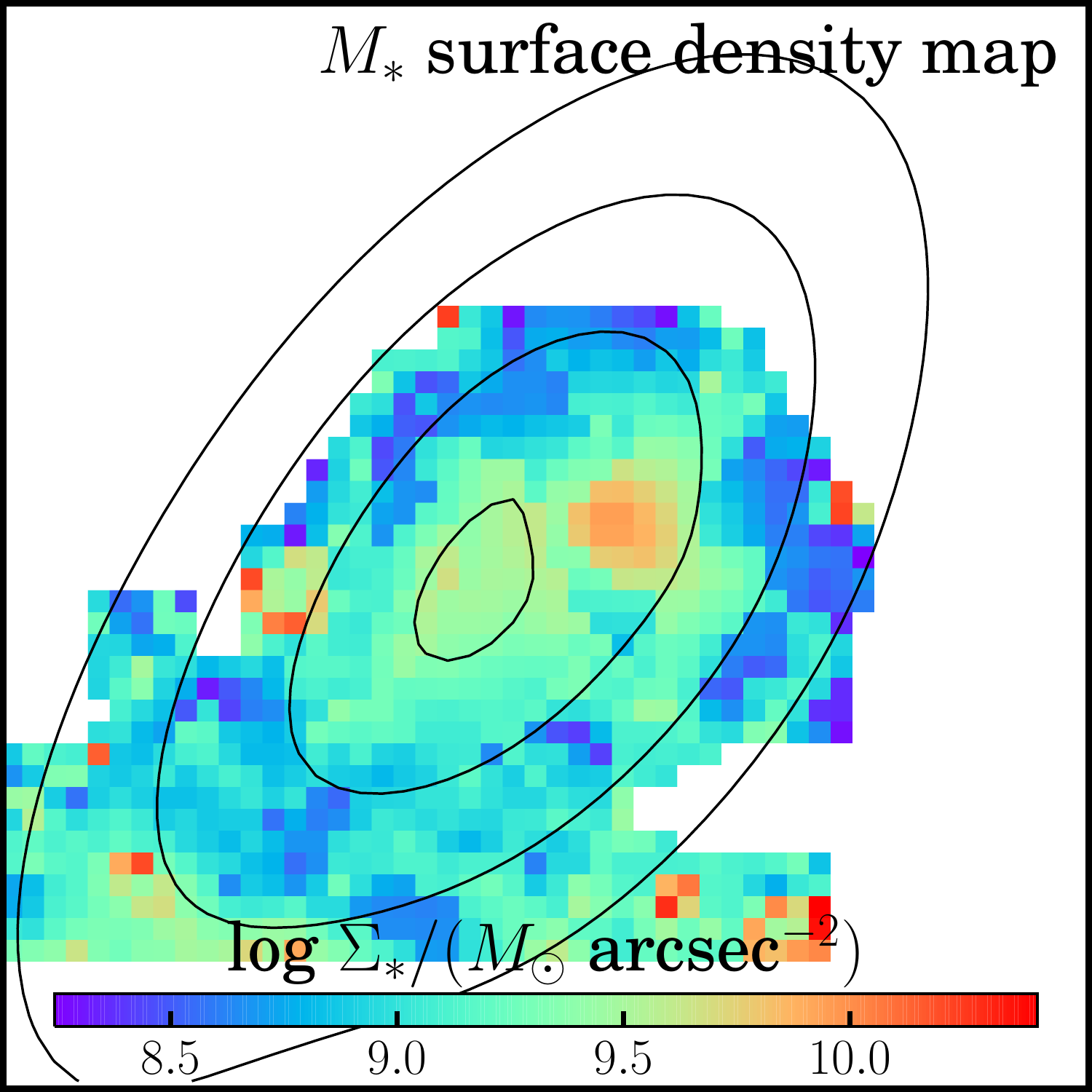}
    \includegraphics[width=.163\textwidth]{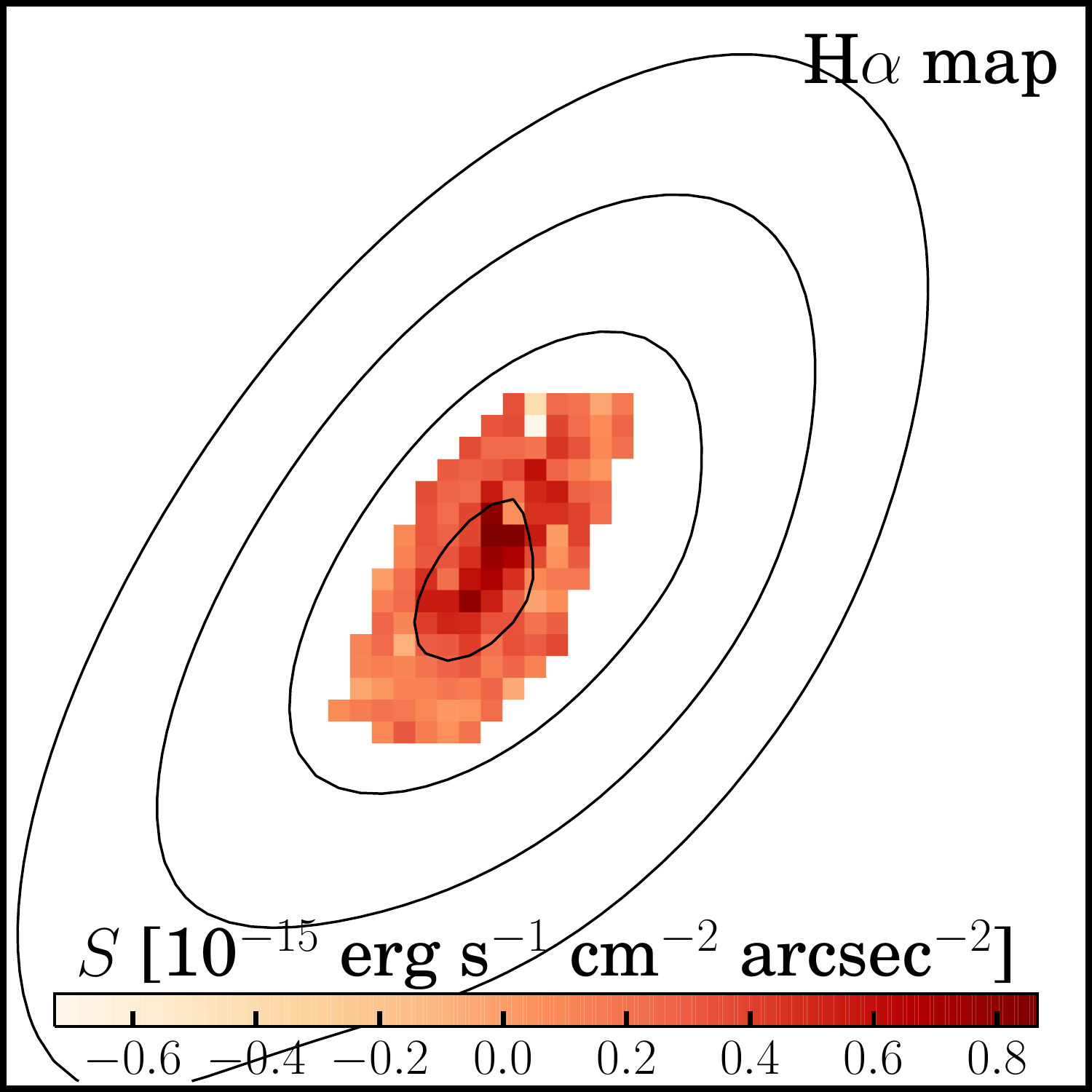}
    \includegraphics[width=.163\textwidth]{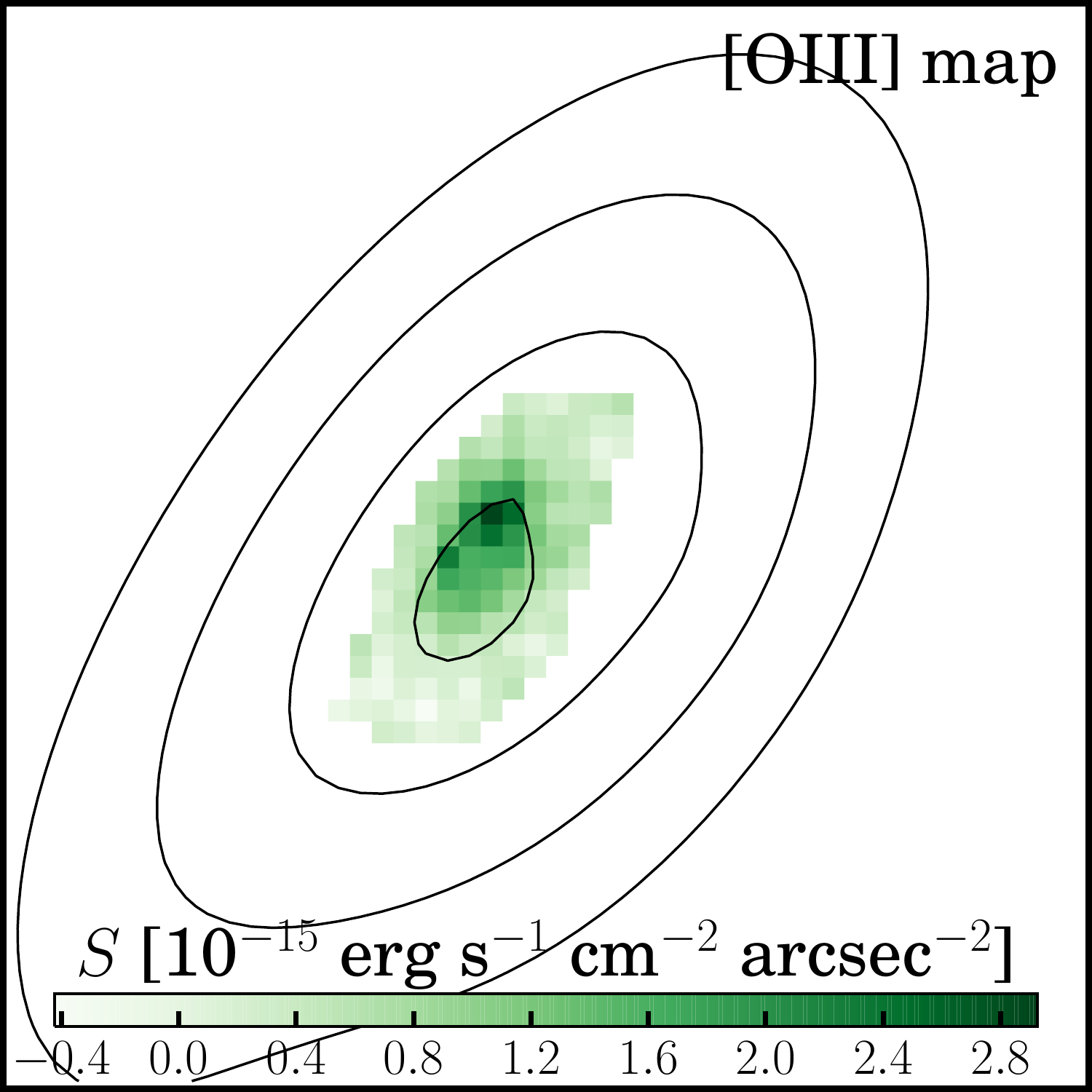}
    \includegraphics[width=.163\textwidth]{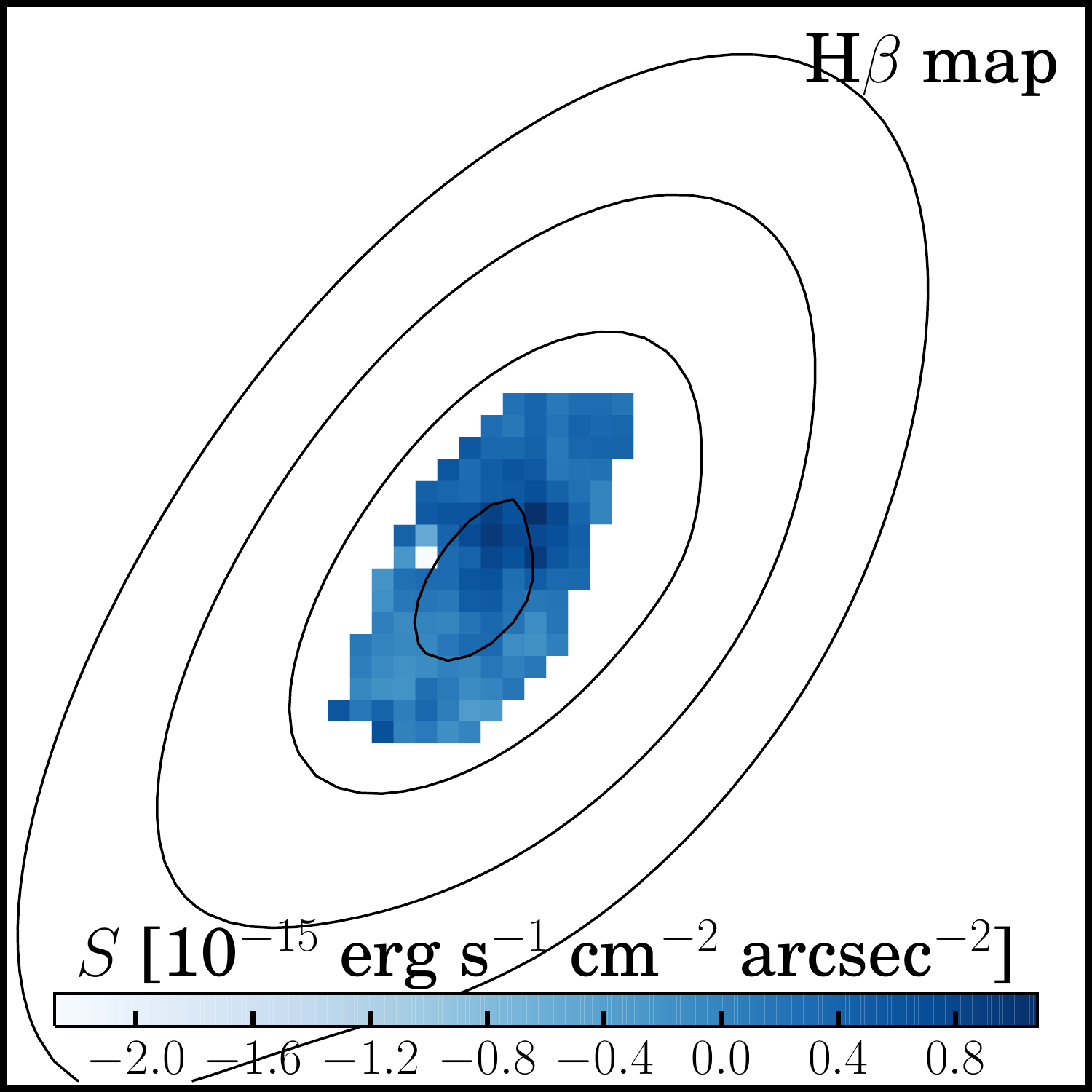}
    \includegraphics[width=.163\textwidth]{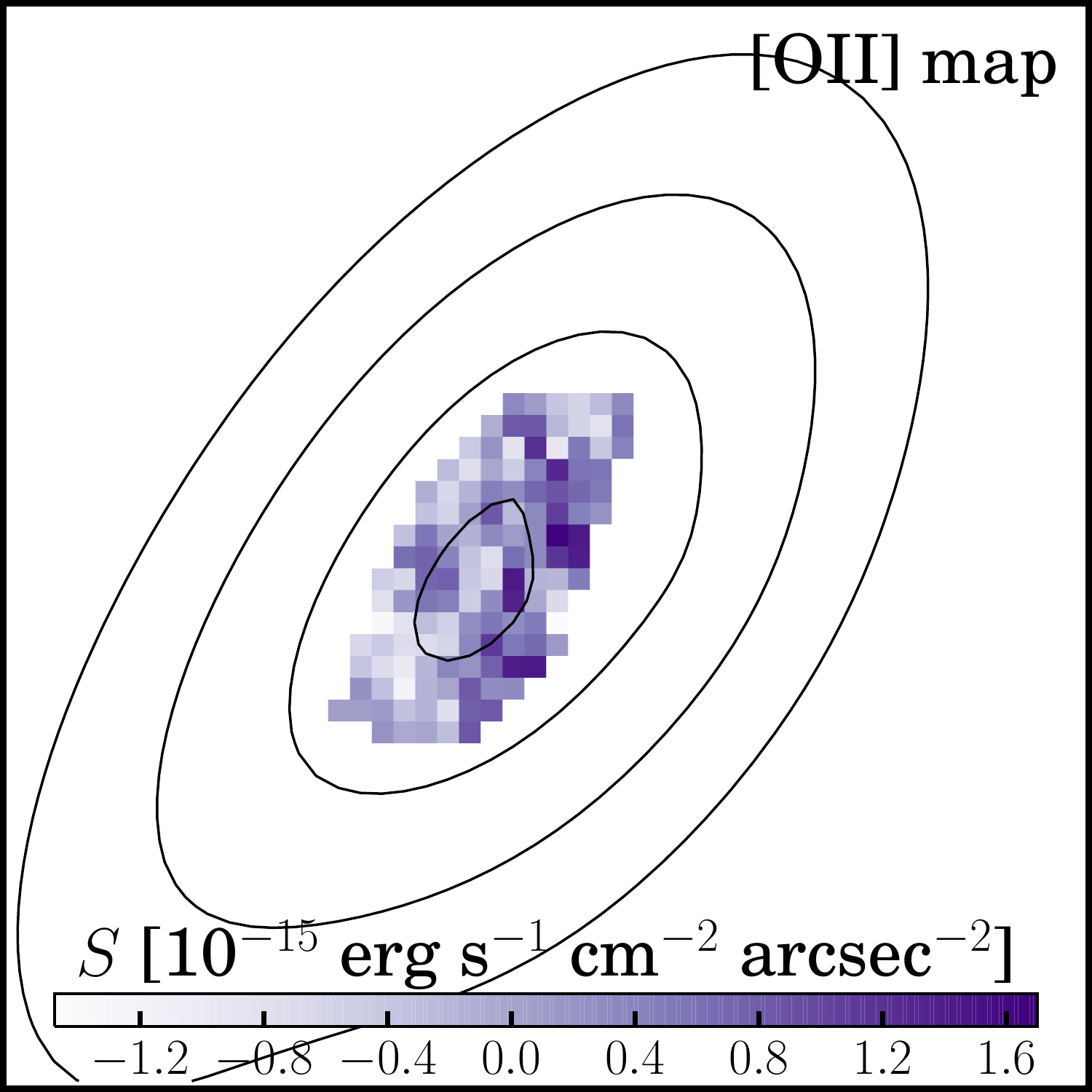}\\
    \includegraphics[width=.163\textwidth]{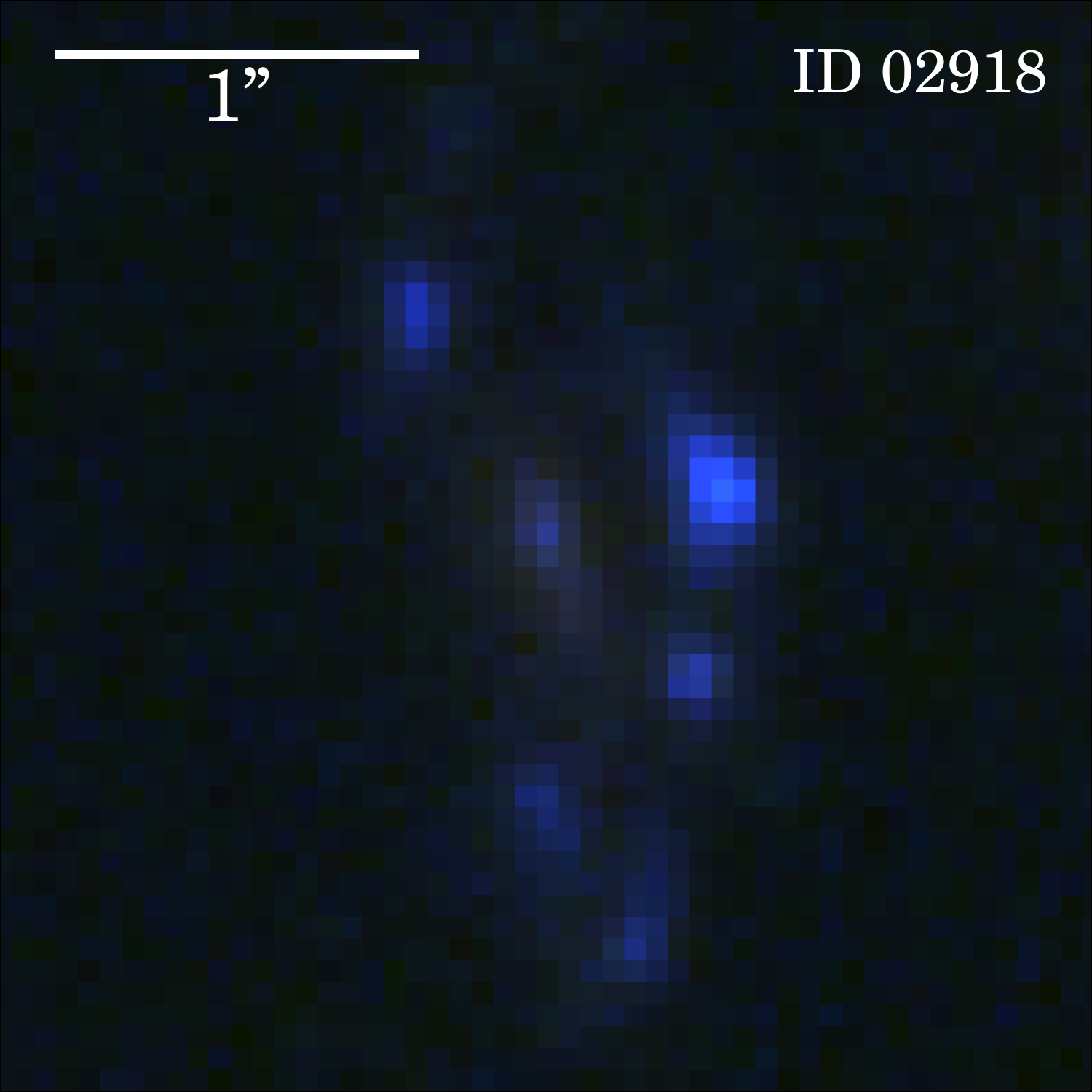}
    \includegraphics[width=.163\textwidth]{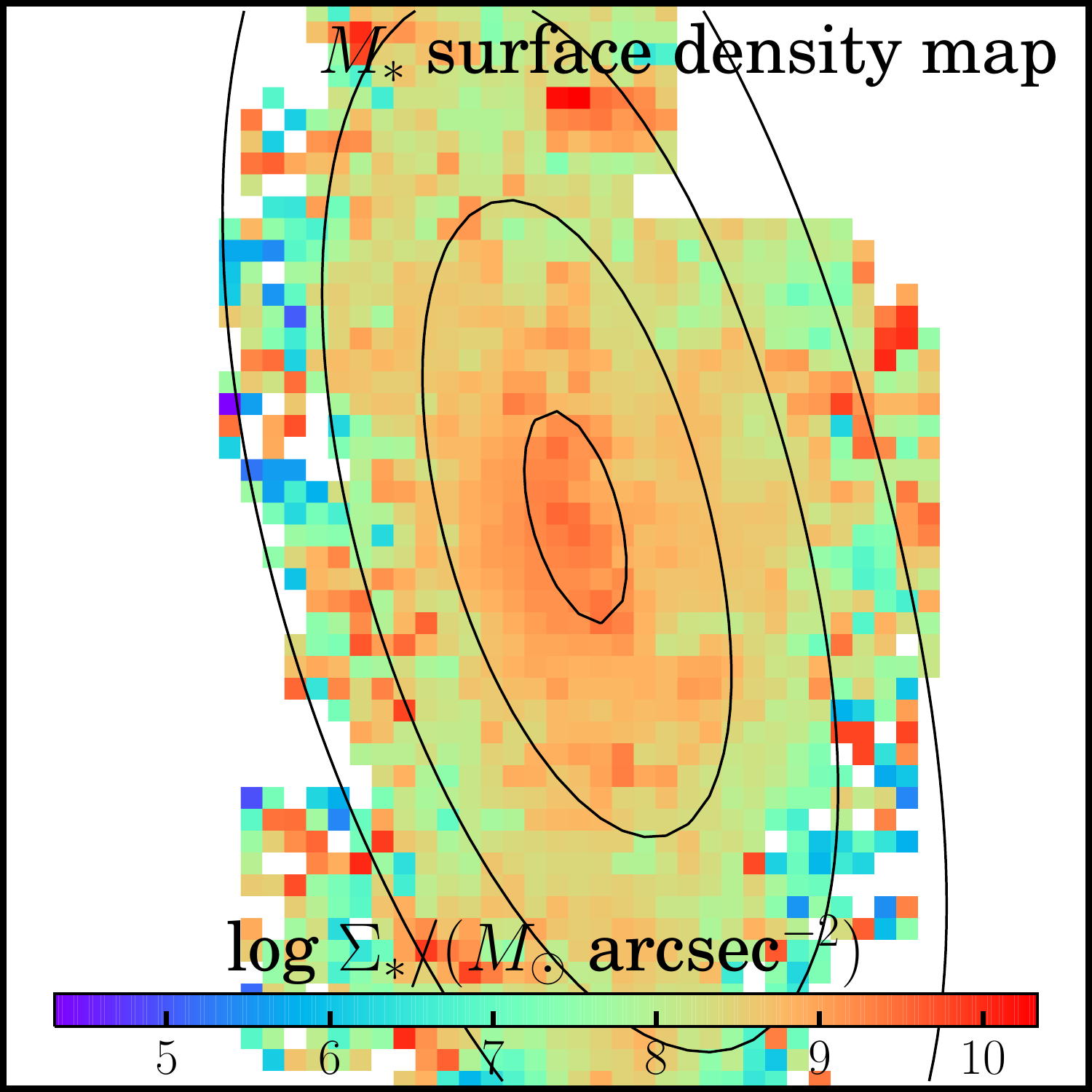}
    \includegraphics[width=.163\textwidth]{fig/baiban.png}
    \includegraphics[width=.163\textwidth]{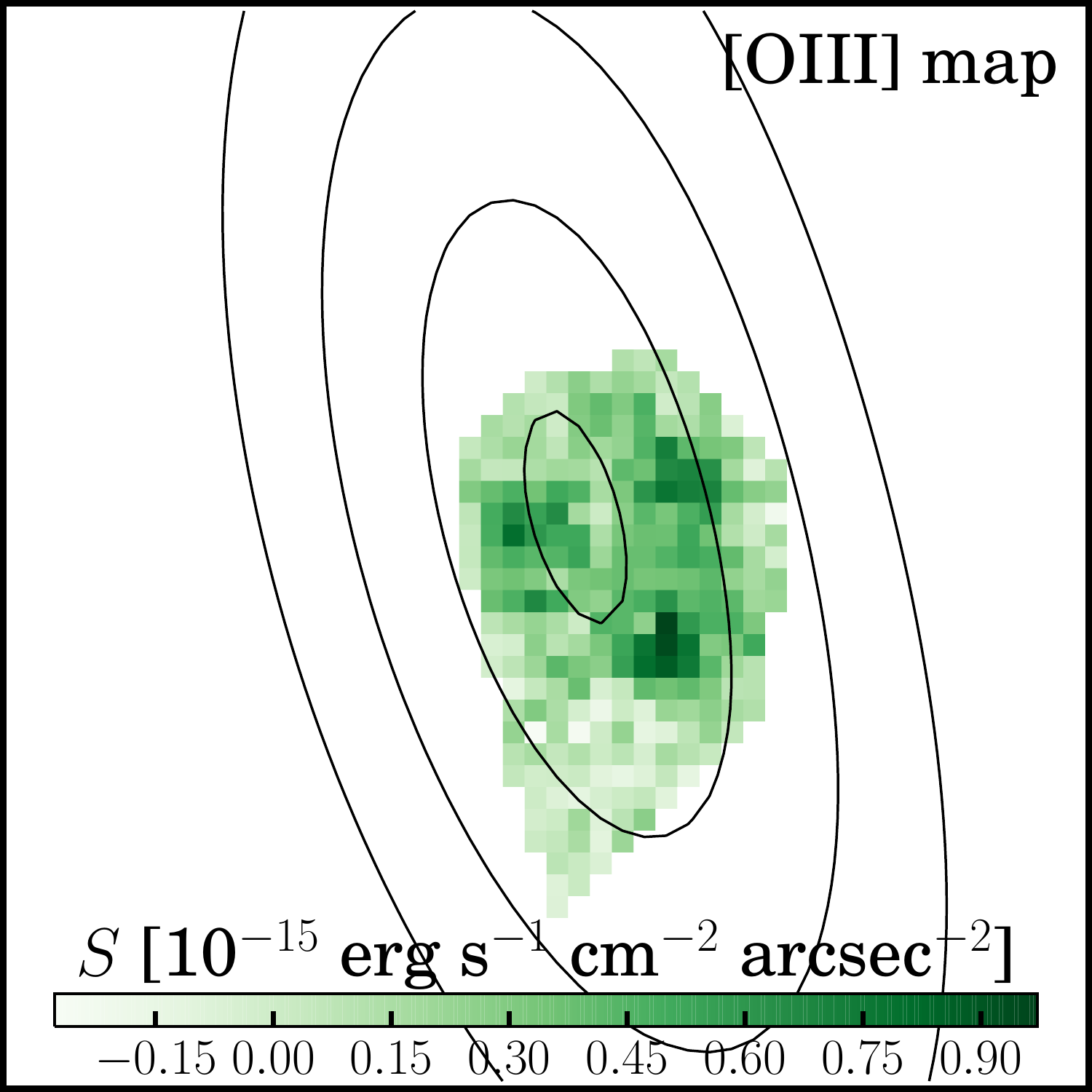}
    \includegraphics[width=.163\textwidth]{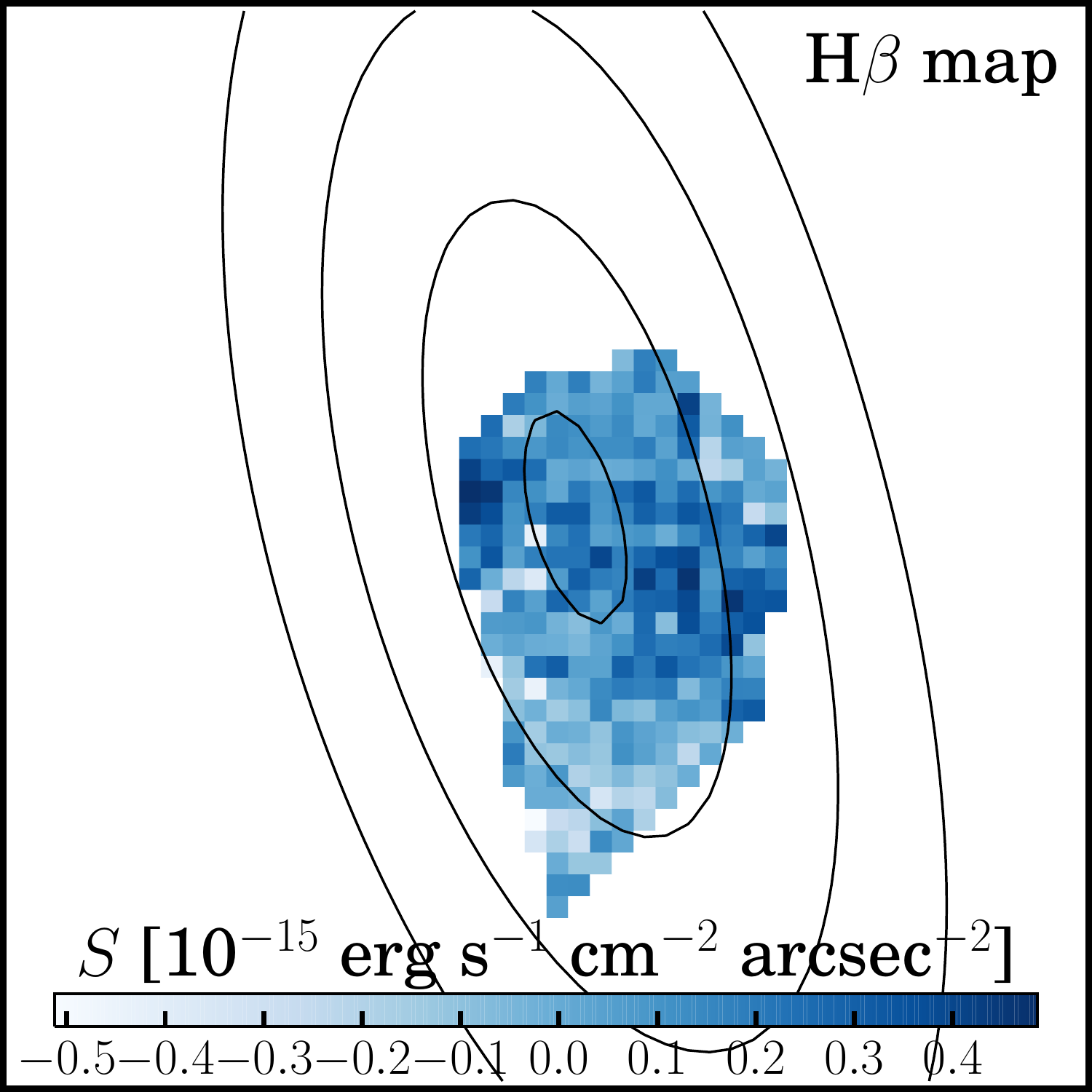}
    \includegraphics[width=.163\textwidth]{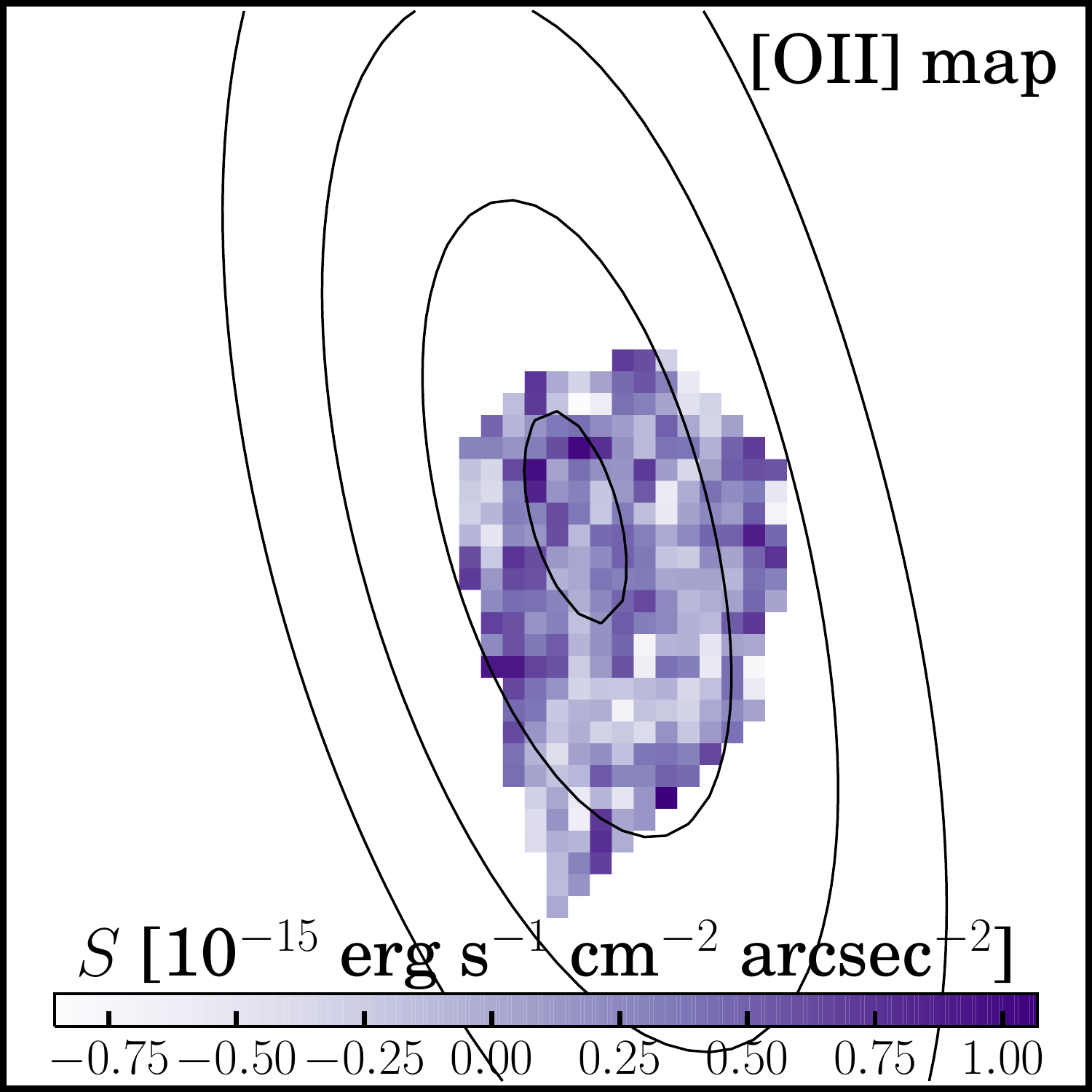}\\
    \includegraphics[width=.163\textwidth]{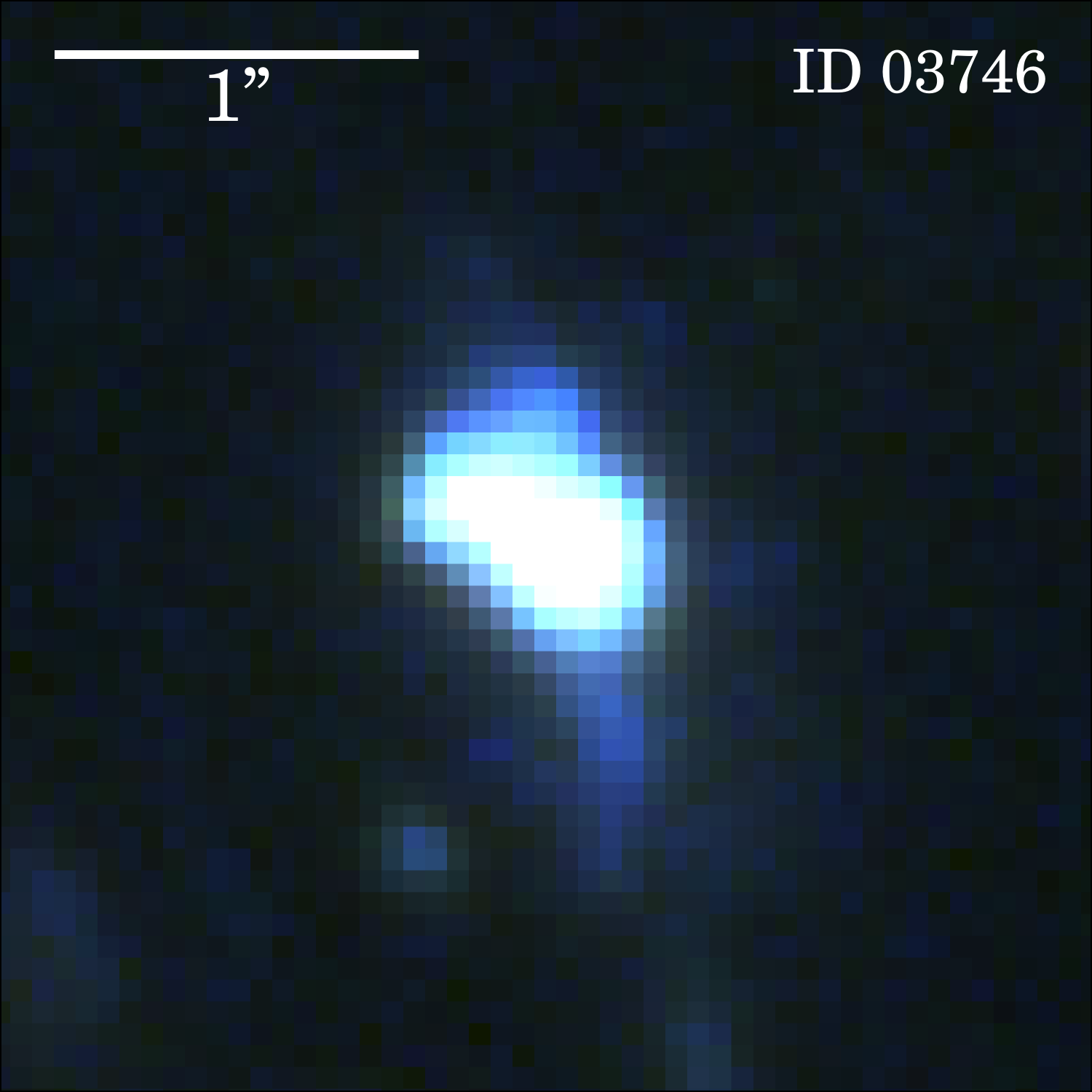}
    \includegraphics[width=.163\textwidth]{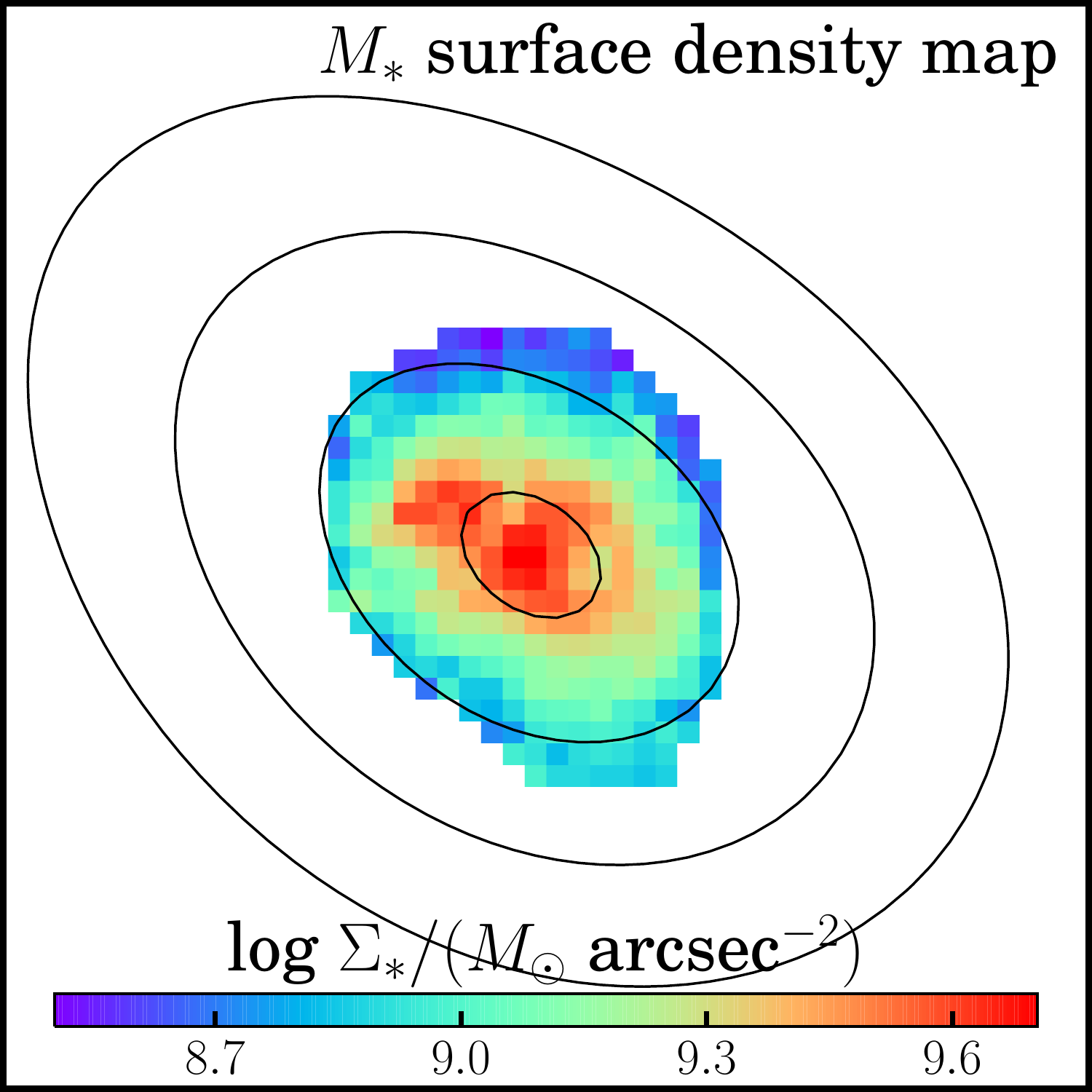}
    \includegraphics[width=.163\textwidth]{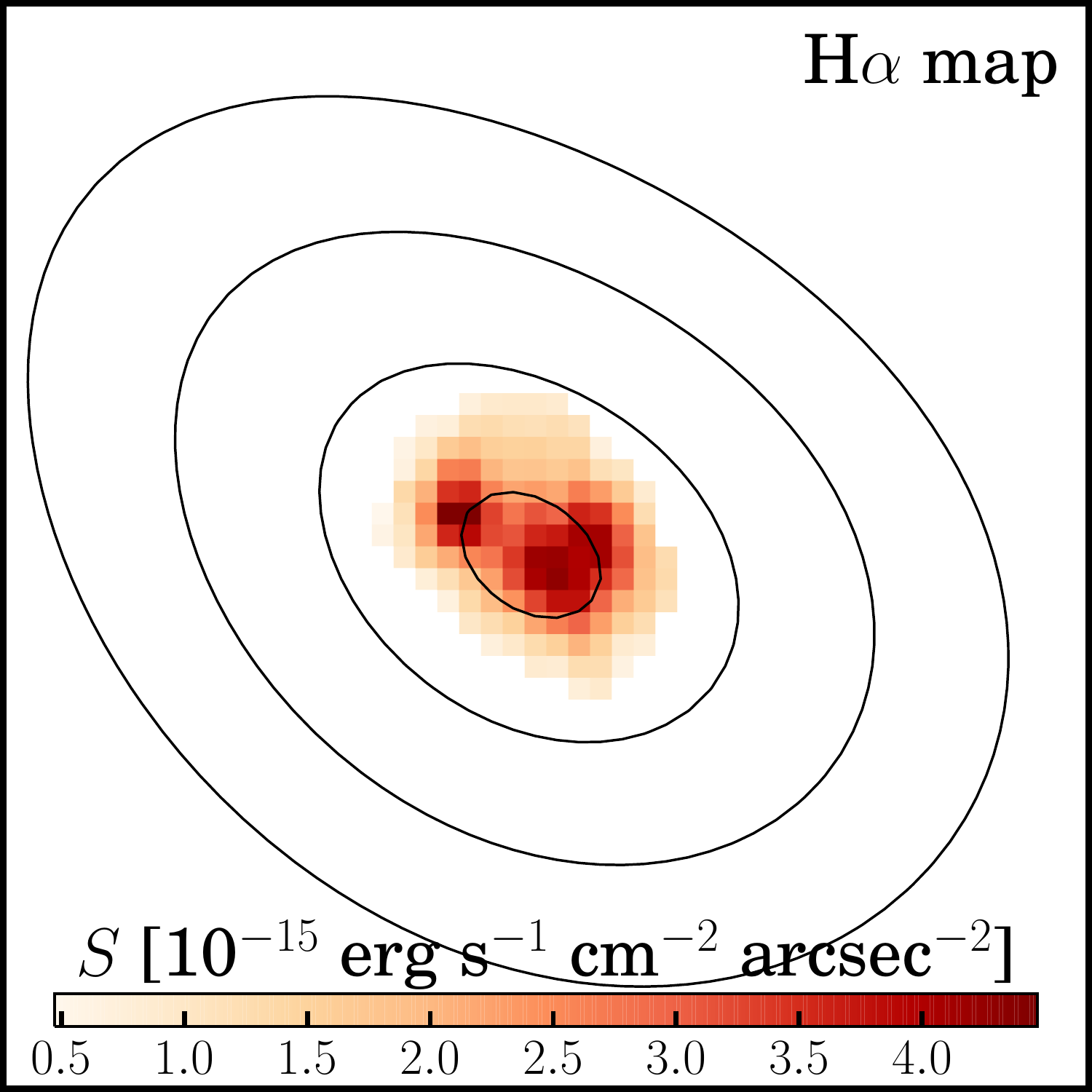}
    \includegraphics[width=.163\textwidth]{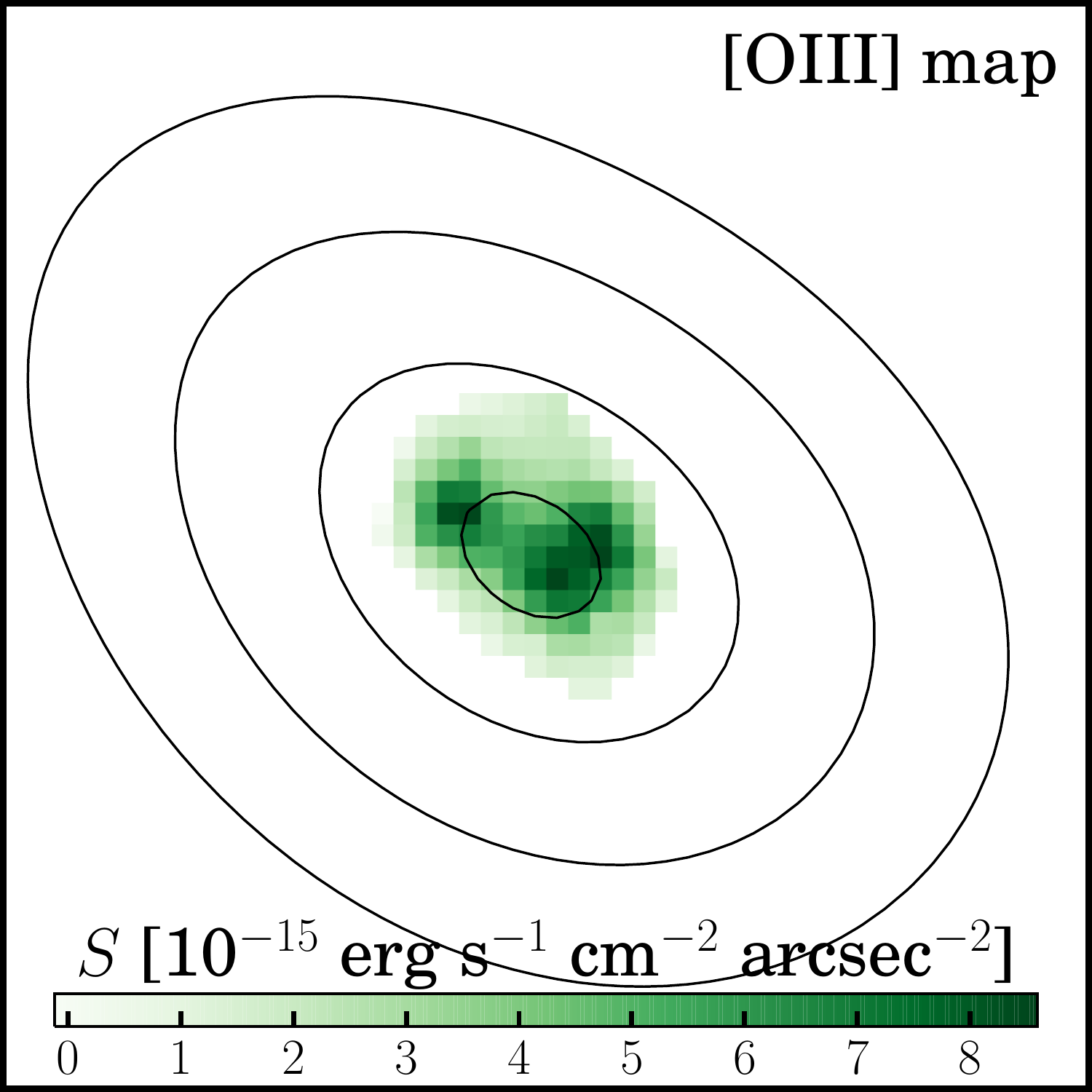}
    \includegraphics[width=.163\textwidth]{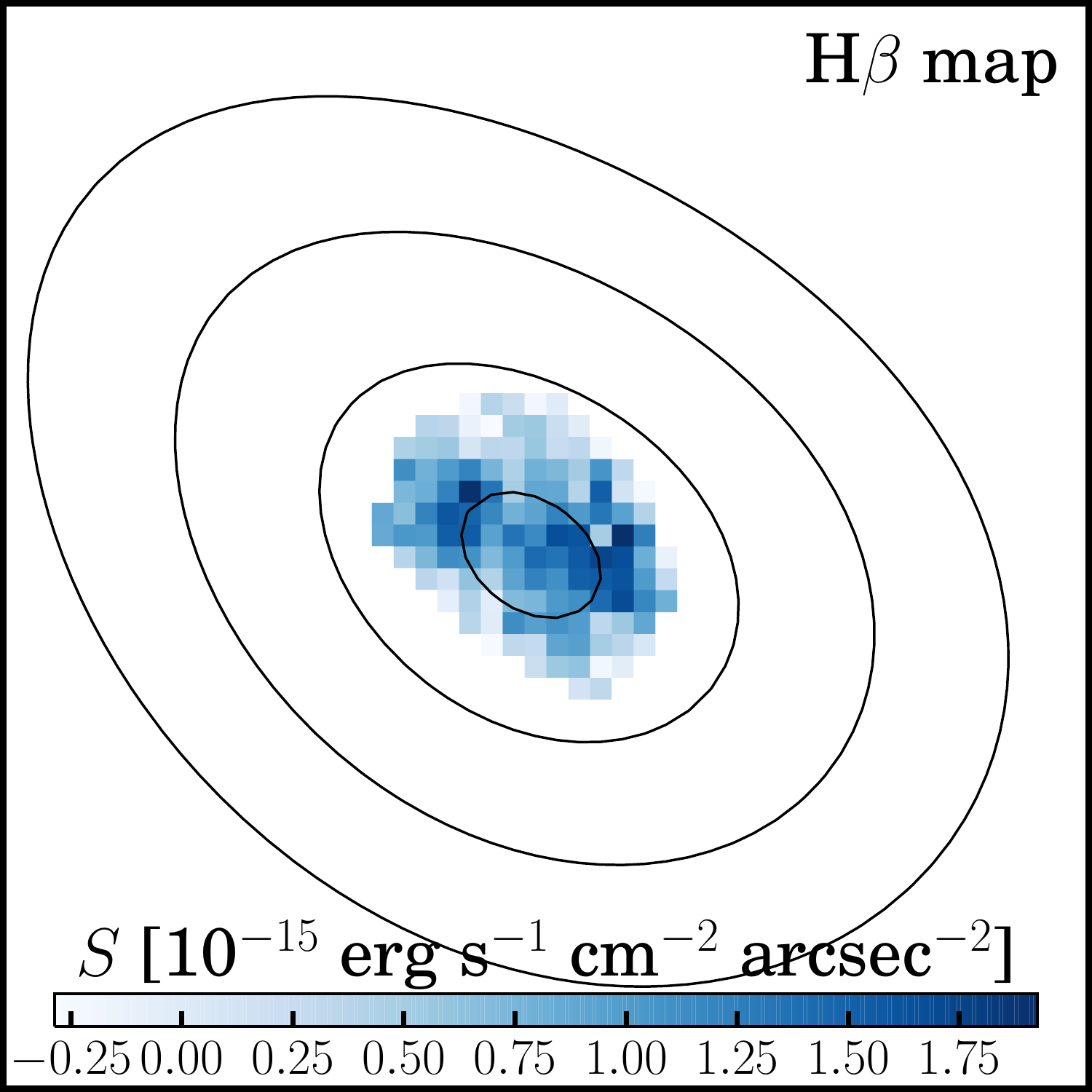}
    \includegraphics[width=.163\textwidth]{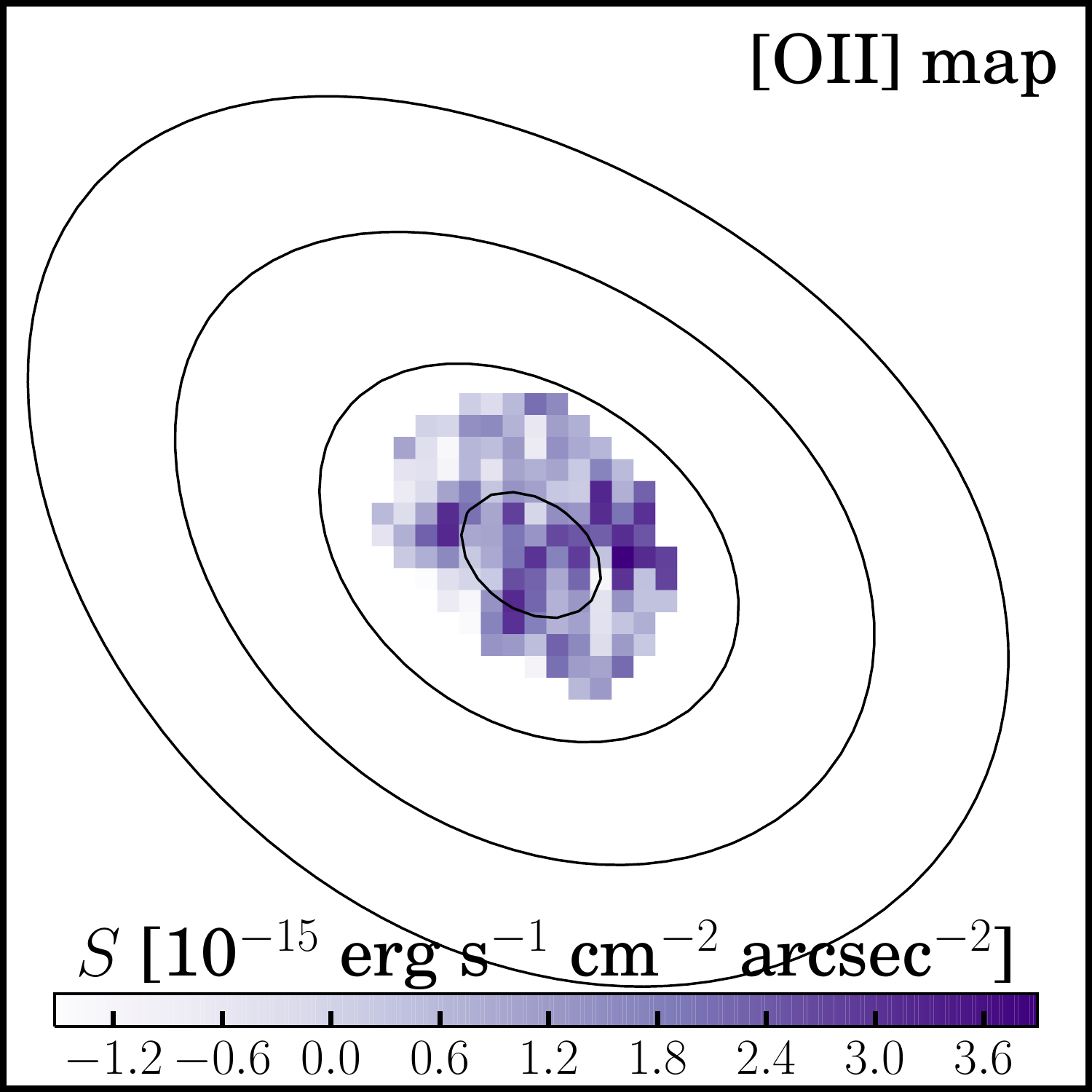}
    \caption{Multi-perspective view of our \mg sample except for the SN Refsdal host which is
    shown in Figure~\ref{fig:multiP_4054}.
    In each row, we show the zoom-in color composite stamp (the same as that in
    Figure~\ref{fig:RGBstamps}), the stellar mass surface density map, the combined surface
    brightness maps of \Ha (if available given object redshift), \OIII, \Hb, \OII for each
    object.
    The black contours overlaid represent the source plane de-projected galactocentric radii
    in 2 kpc interval, starting from 1 kpc. The spatial extent and orientation of the stamps
    for each object have been fixed.}
    \label{fig:multiP_rest}
\end{figure*}

\begin{figure*}
    \includegraphics[width=.163\textwidth]{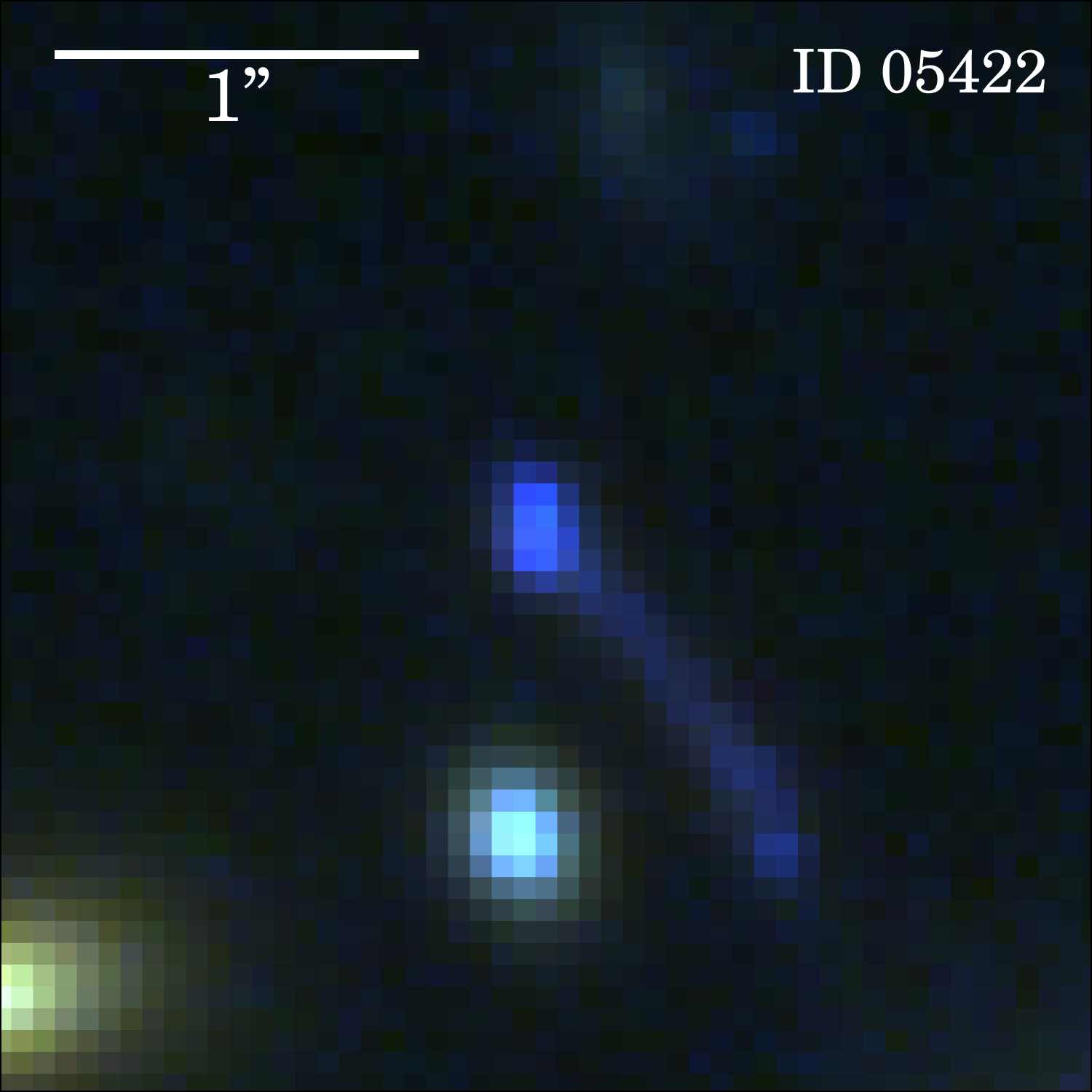}
    \includegraphics[width=.163\textwidth]{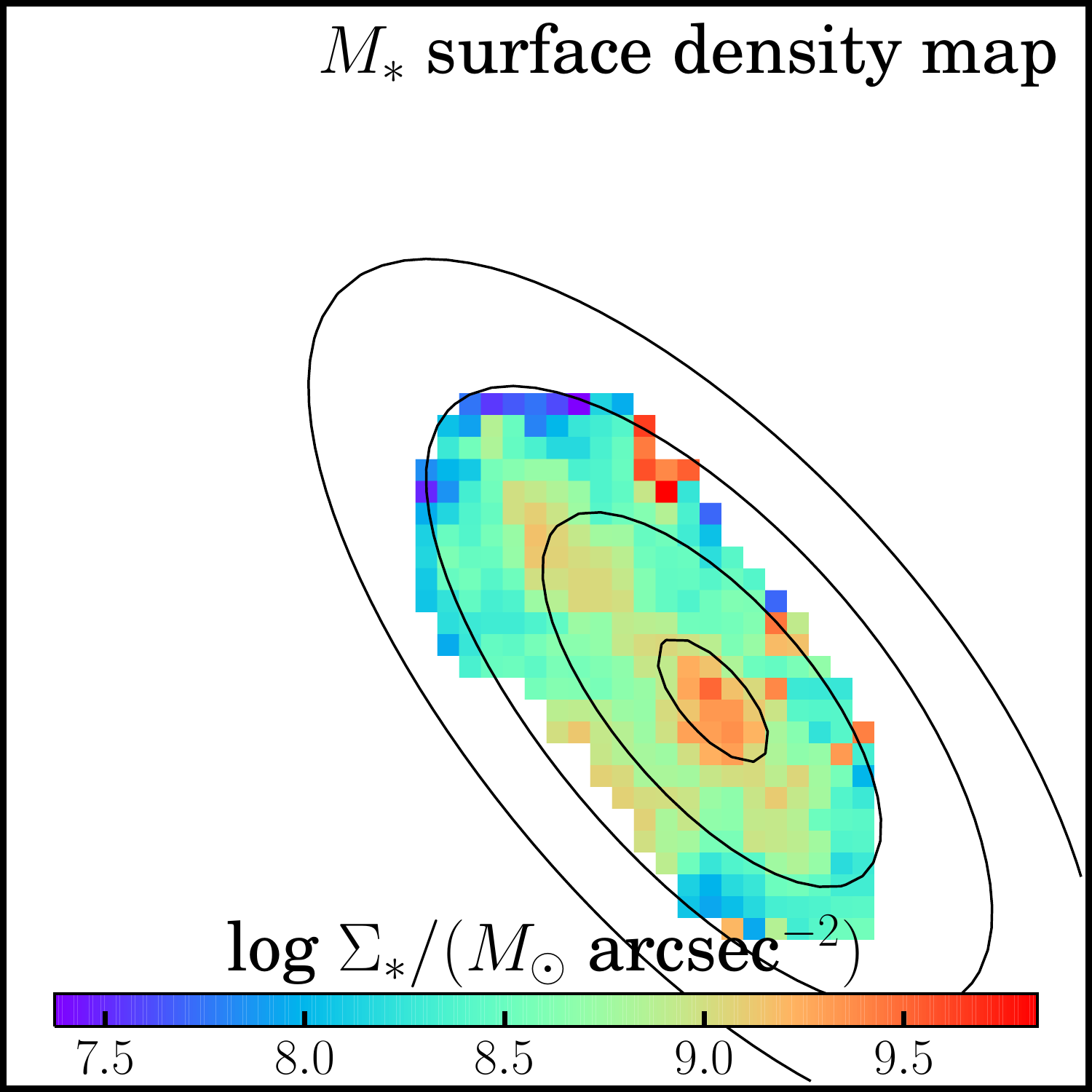}
    \includegraphics[width=.163\textwidth]{fig/baiban.png}
    \includegraphics[width=.163\textwidth]{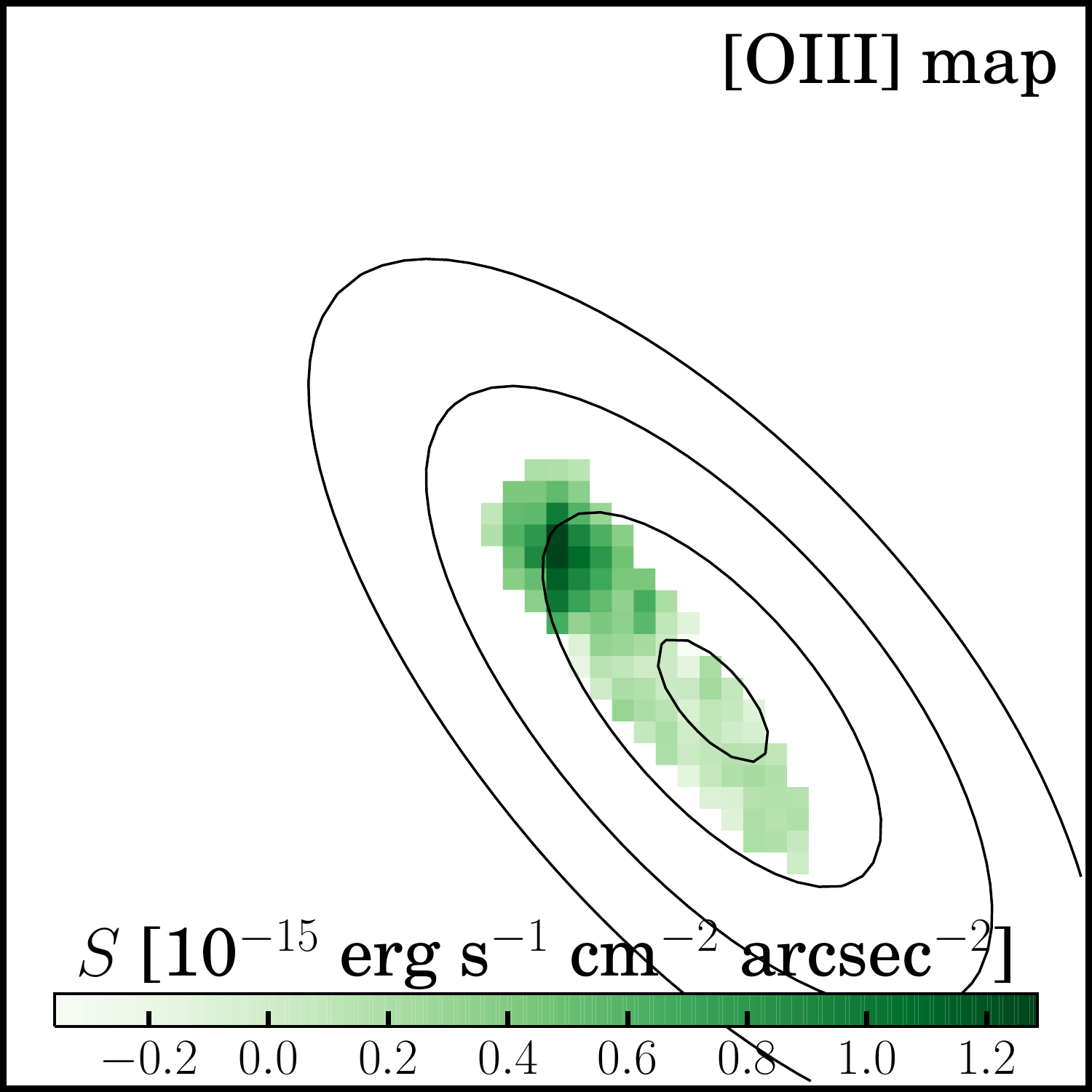}
    \includegraphics[width=.163\textwidth]{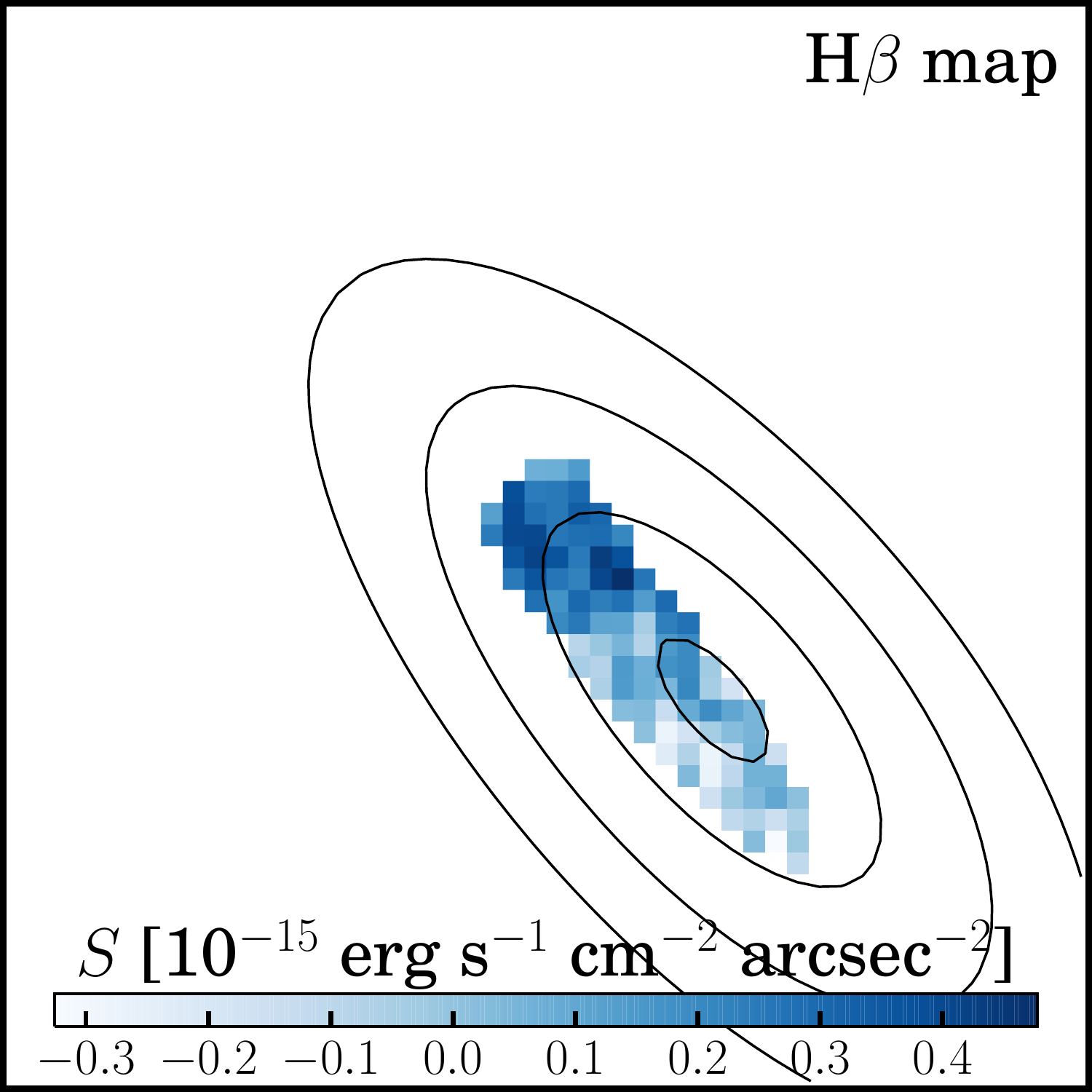}
    \includegraphics[width=.163\textwidth]{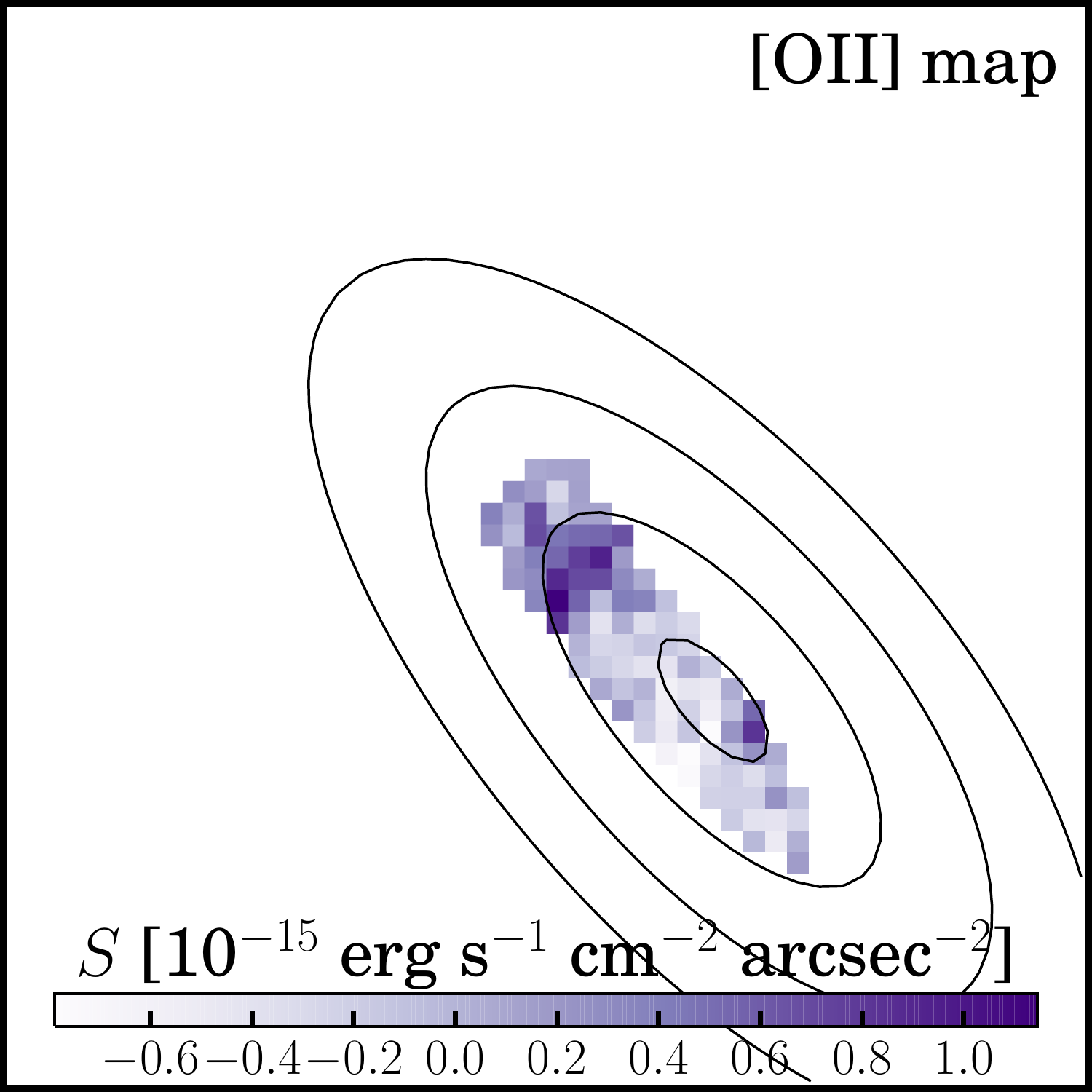}\\
    \includegraphics[width=.163\textwidth]{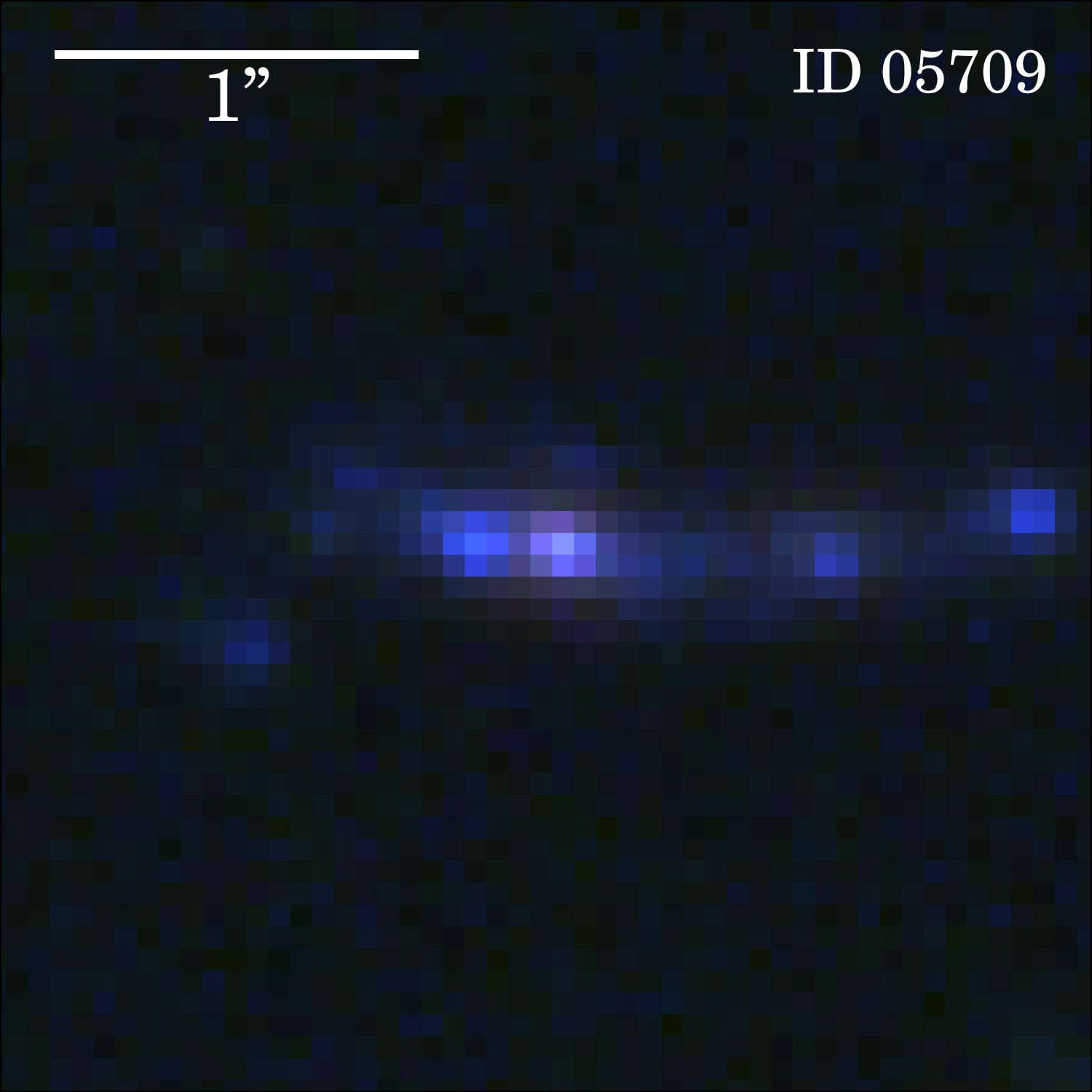}
    \includegraphics[width=.163\textwidth]{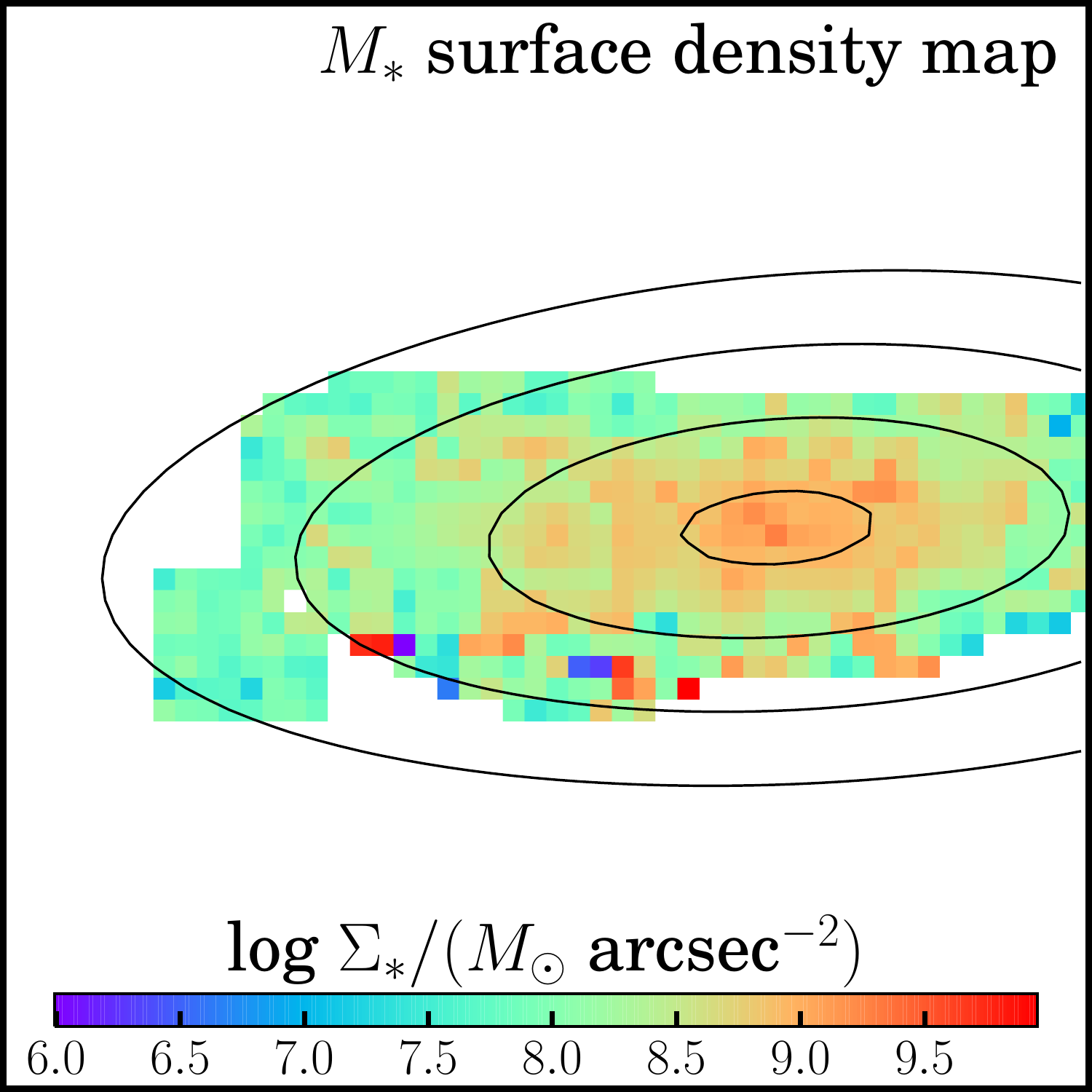}
    \includegraphics[width=.163\textwidth]{fig/baiban.png}
    \includegraphics[width=.163\textwidth]{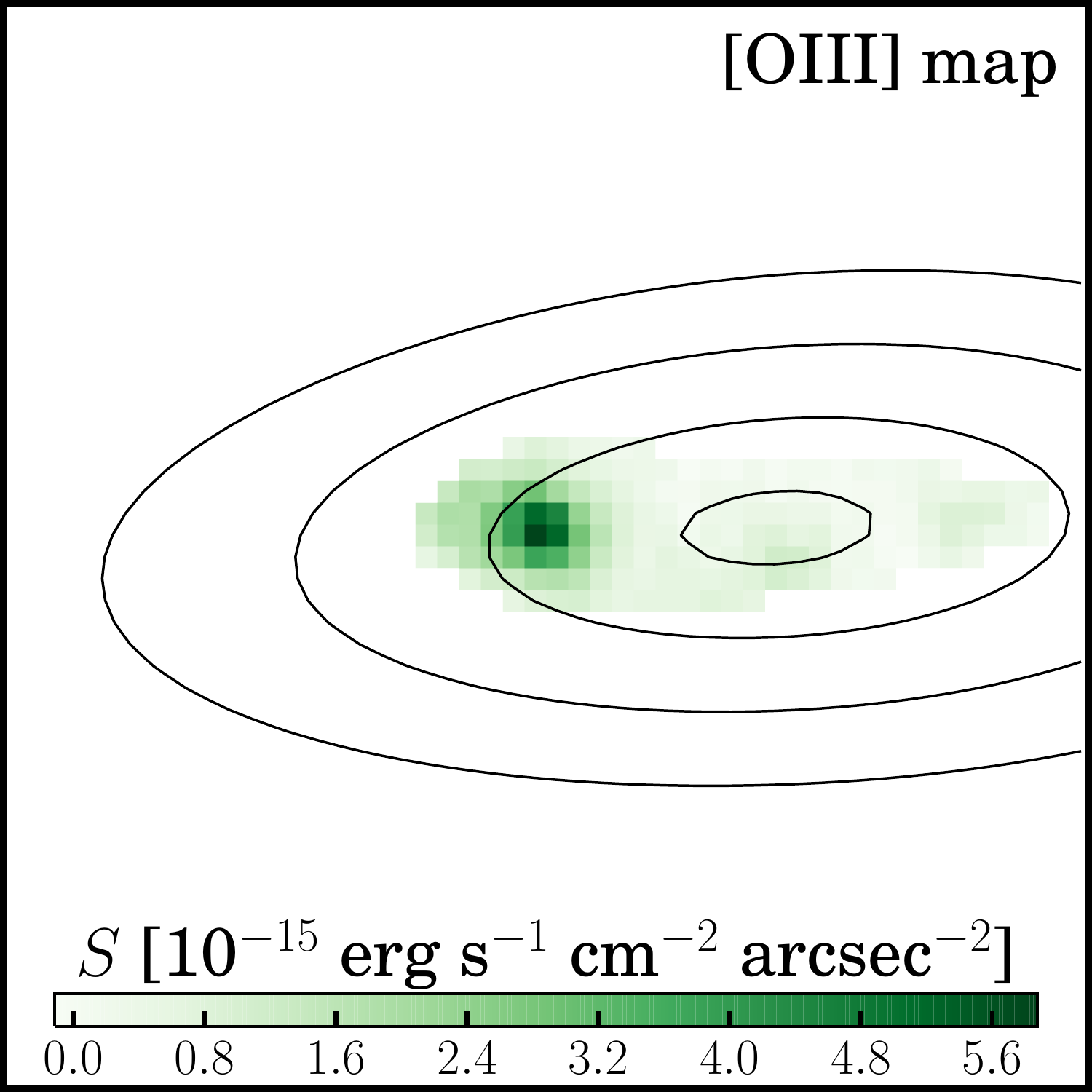}
    \includegraphics[width=.163\textwidth]{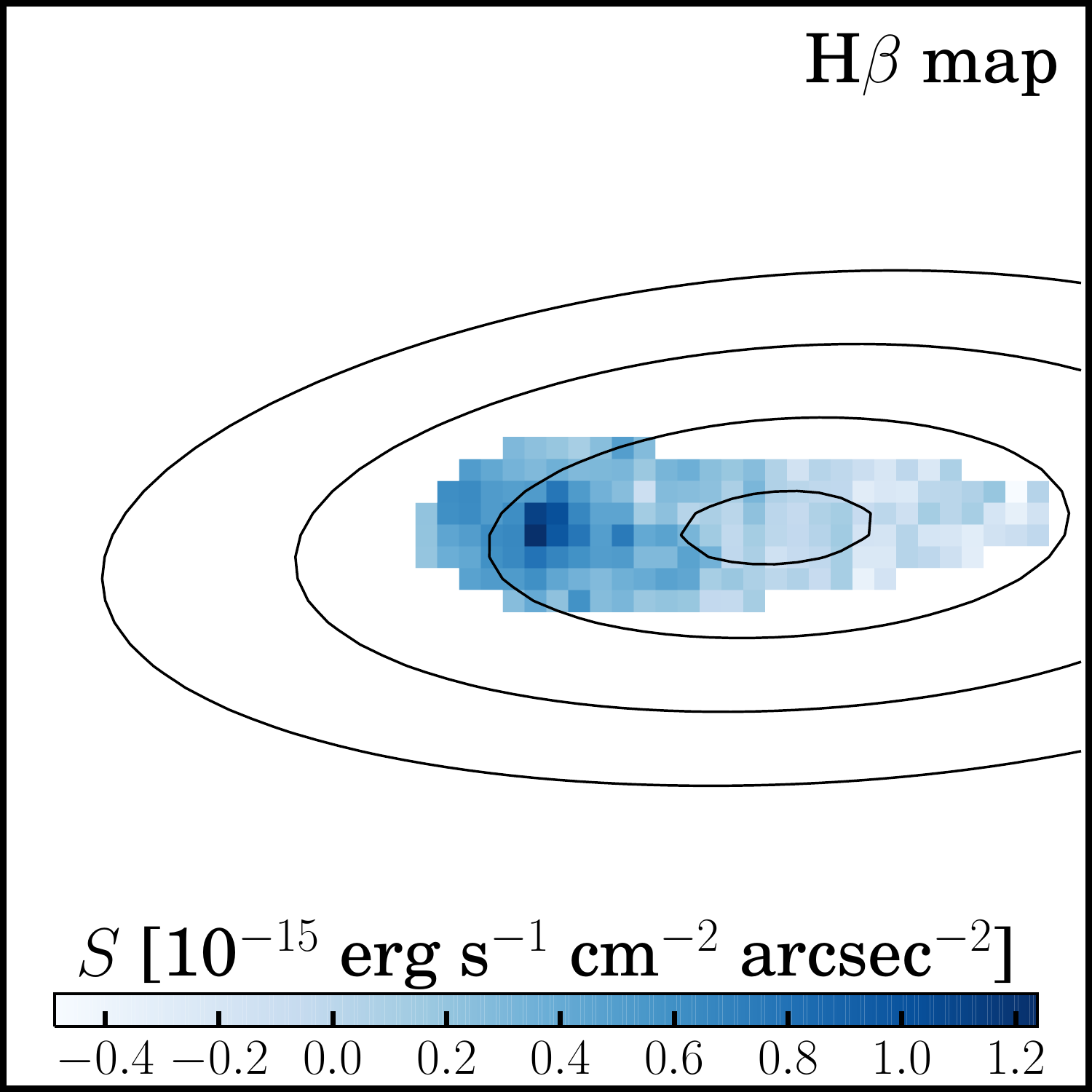}
    \includegraphics[width=.163\textwidth]{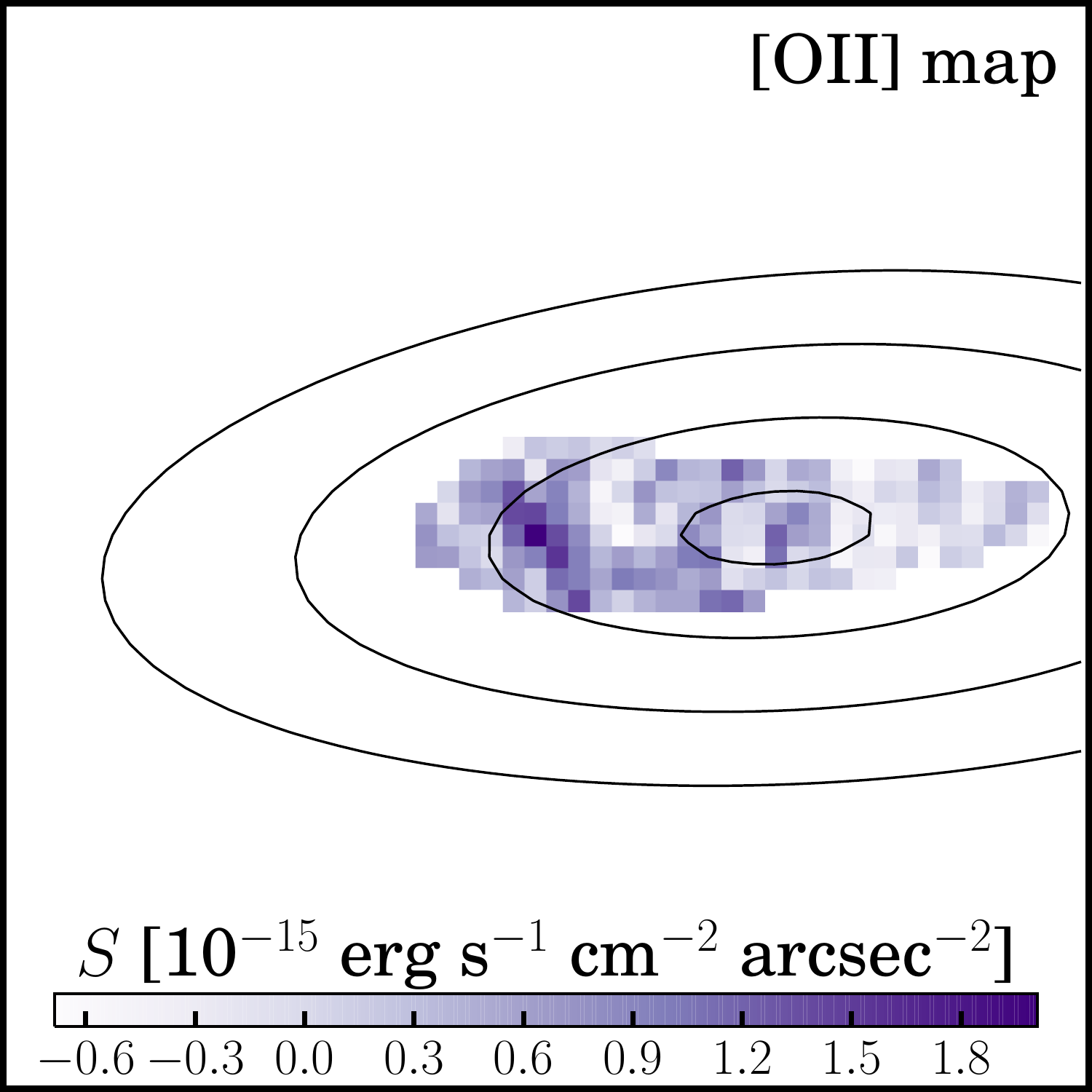}\\
    \includegraphics[width=.163\textwidth]{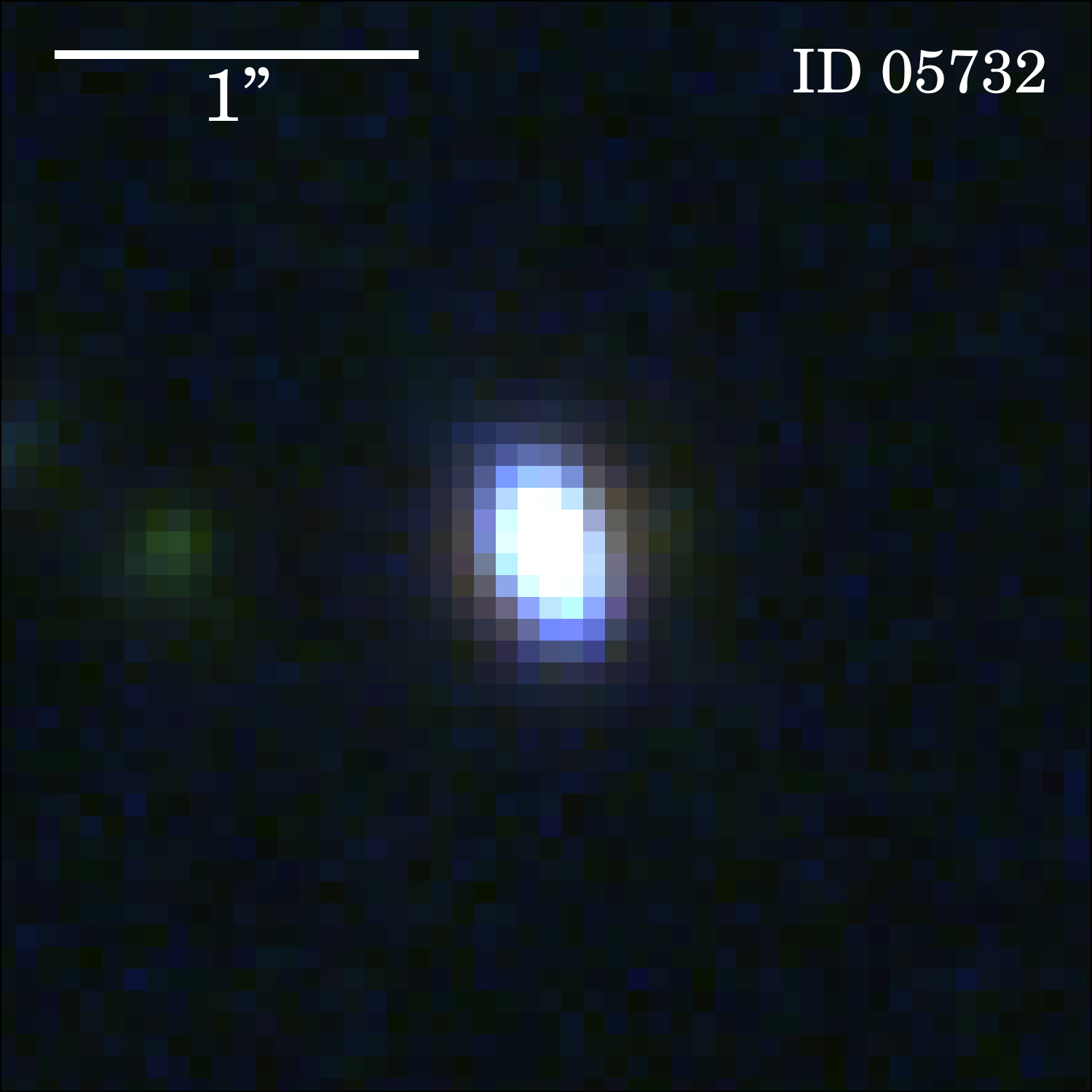}
    \includegraphics[width=.163\textwidth]{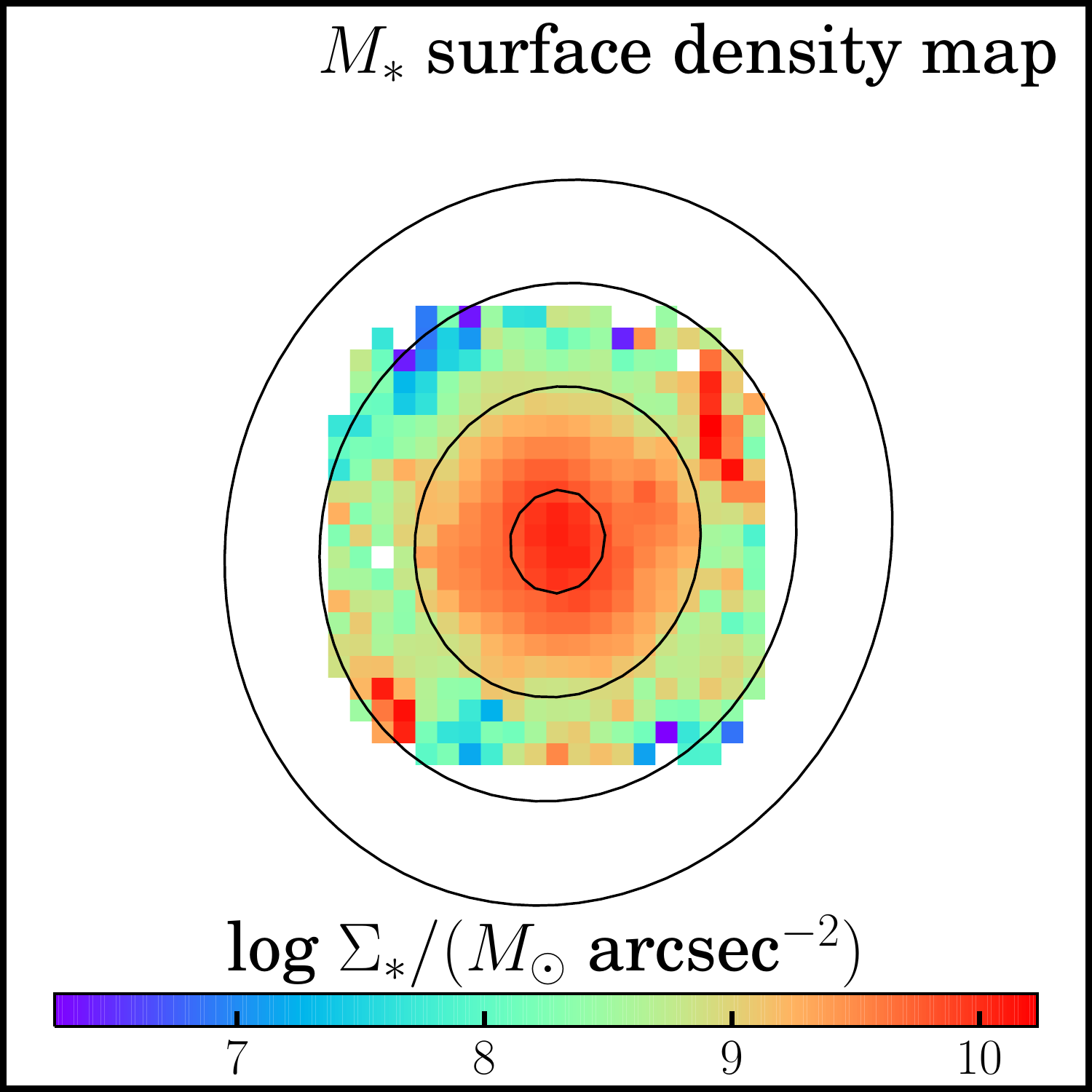}
    \includegraphics[width=.163\textwidth]{fig/baiban.png}
    \includegraphics[width=.163\textwidth]{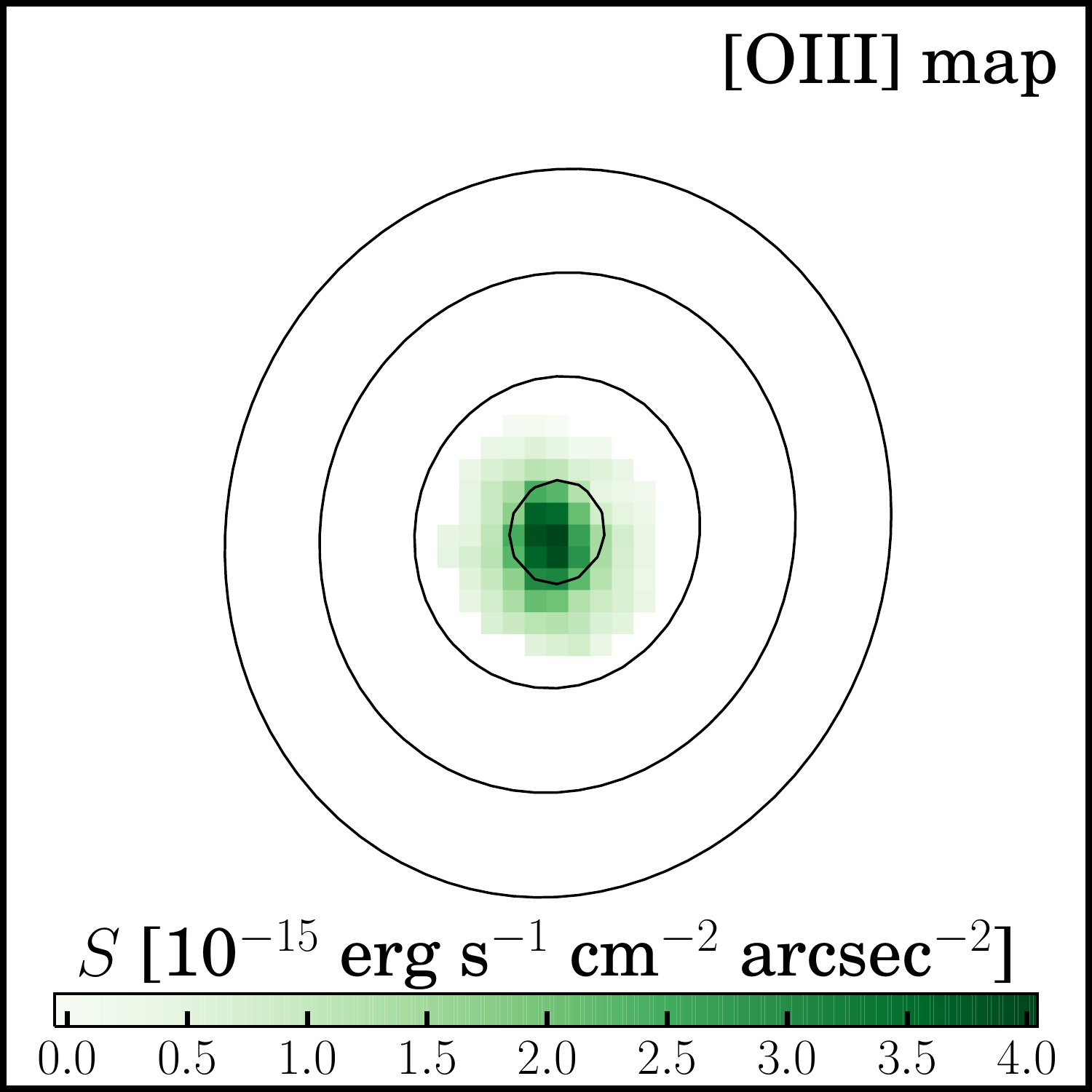}
    \includegraphics[width=.163\textwidth]{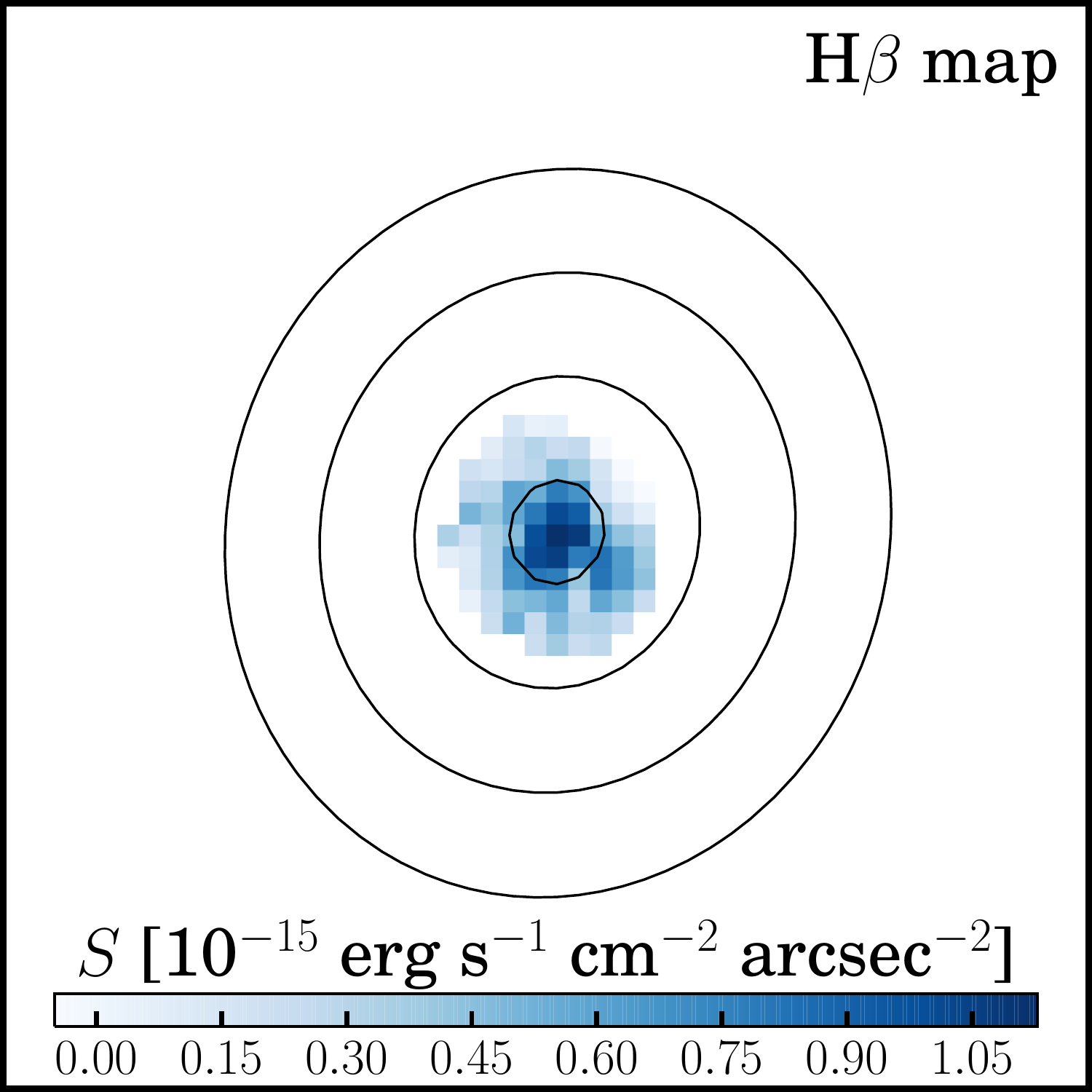}
    \includegraphics[width=.163\textwidth]{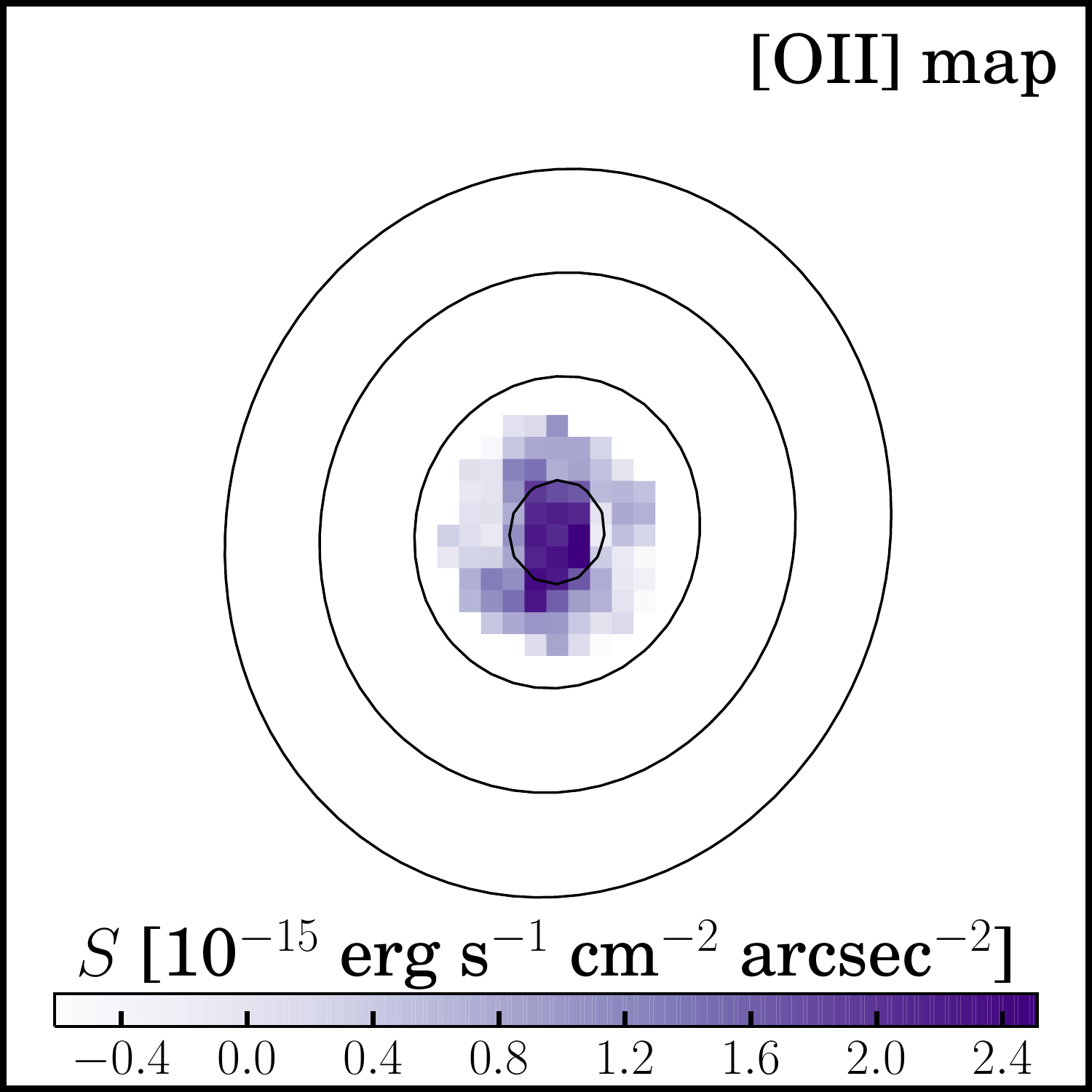}\\
    \includegraphics[width=.163\textwidth]{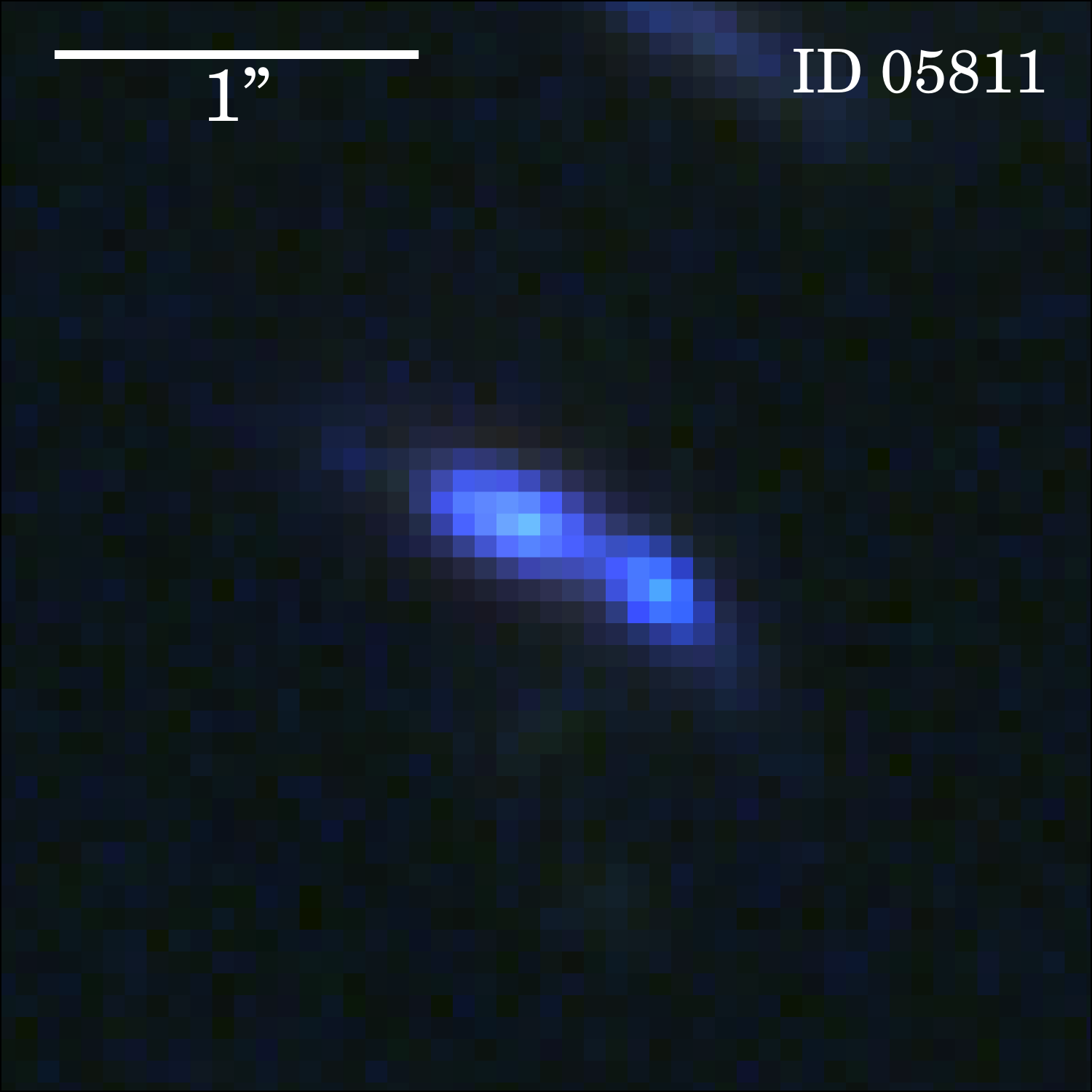}
    \includegraphics[width=.163\textwidth]{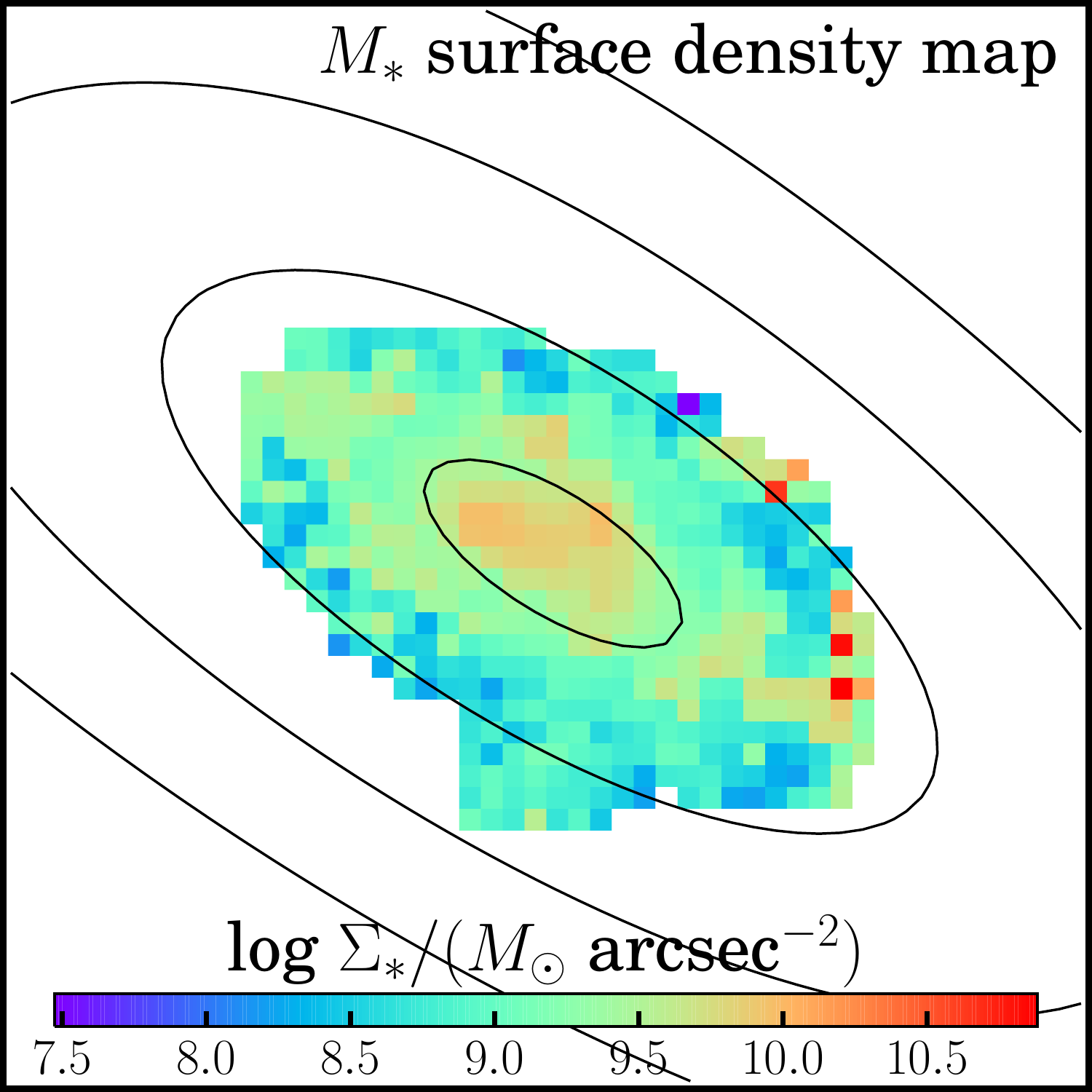}
    \includegraphics[width=.163\textwidth]{fig/baiban.png}
    \includegraphics[width=.163\textwidth]{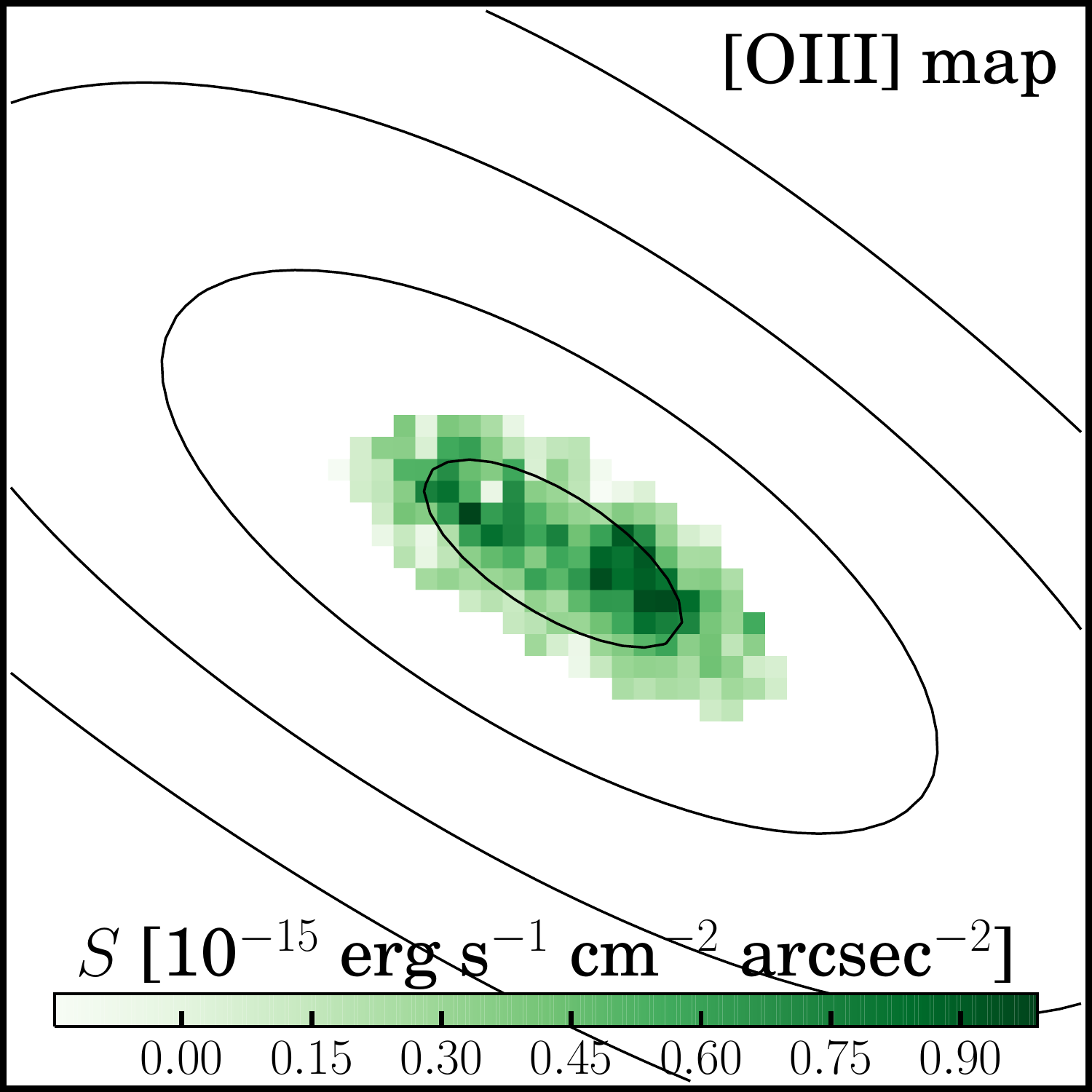}
    \includegraphics[width=.163\textwidth]{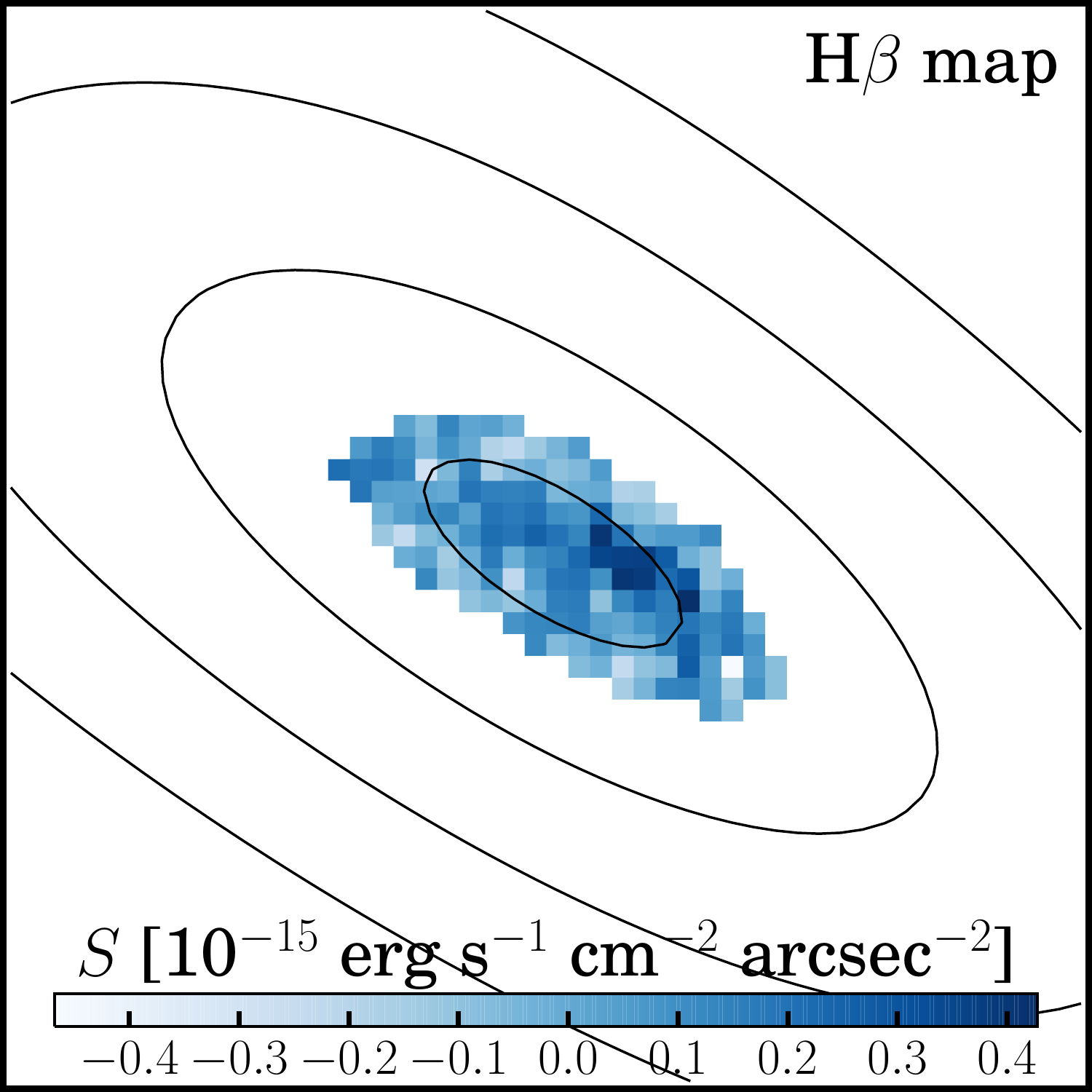}
    \includegraphics[width=.163\textwidth]{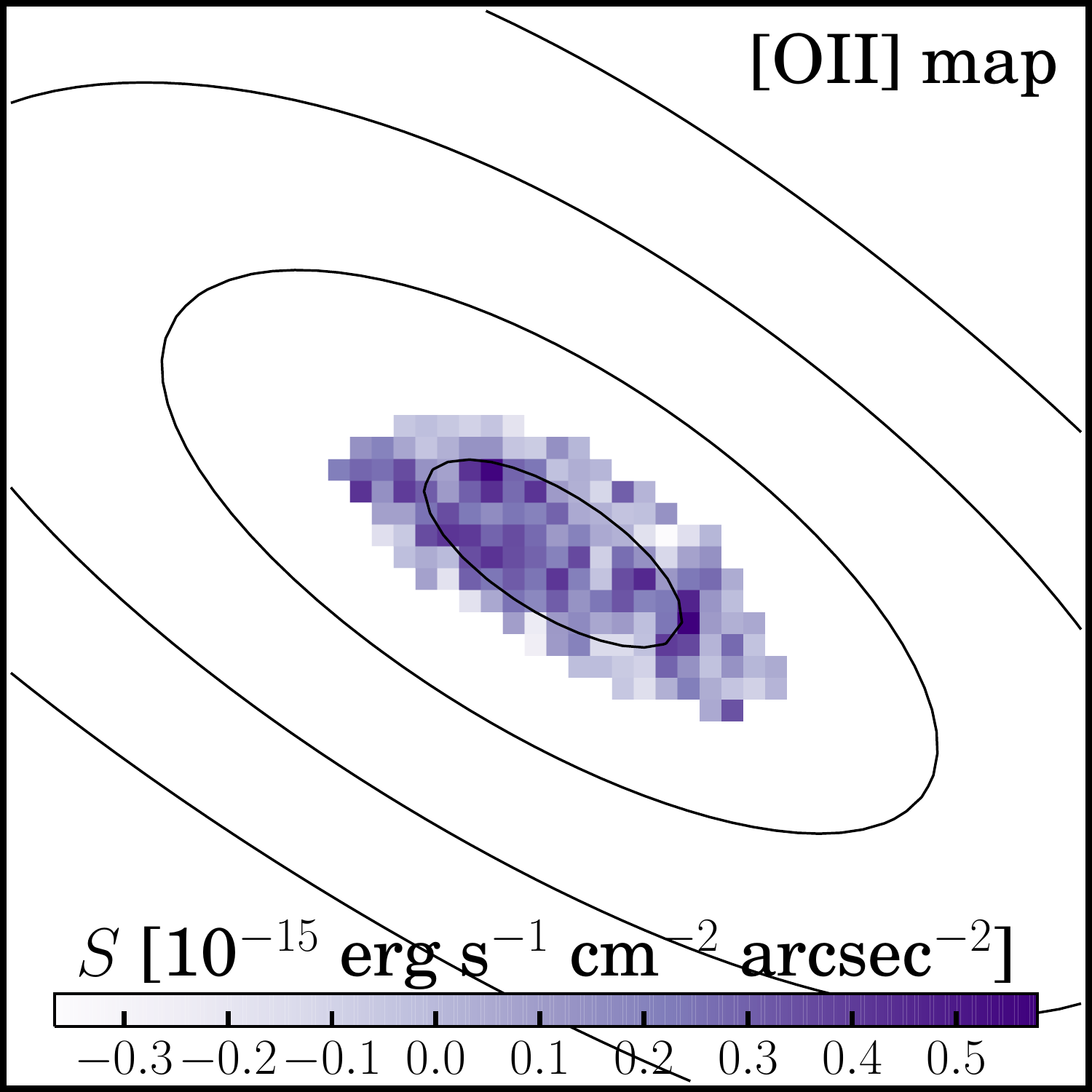}\\
    \includegraphics[width=.163\textwidth]{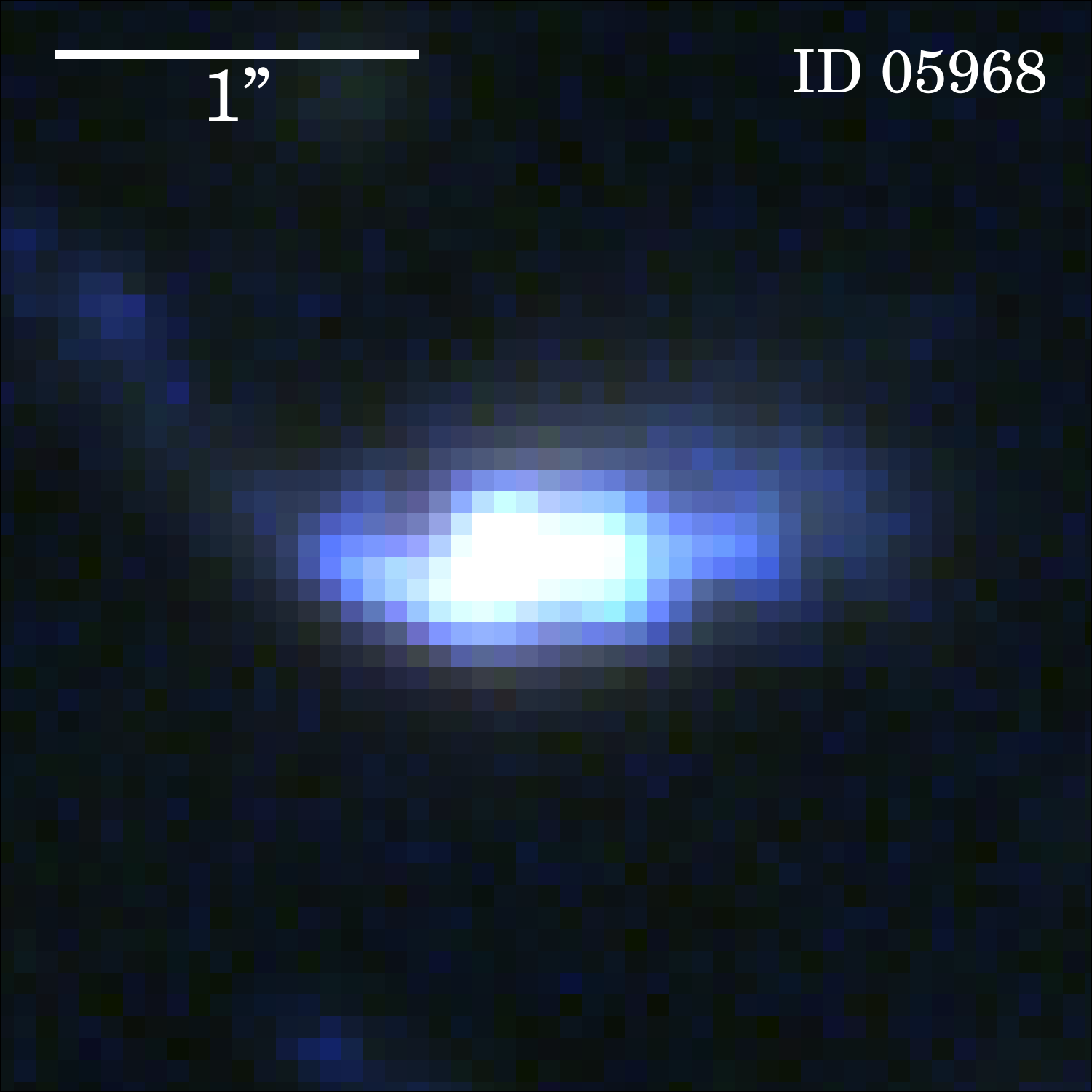}
    \includegraphics[width=.163\textwidth]{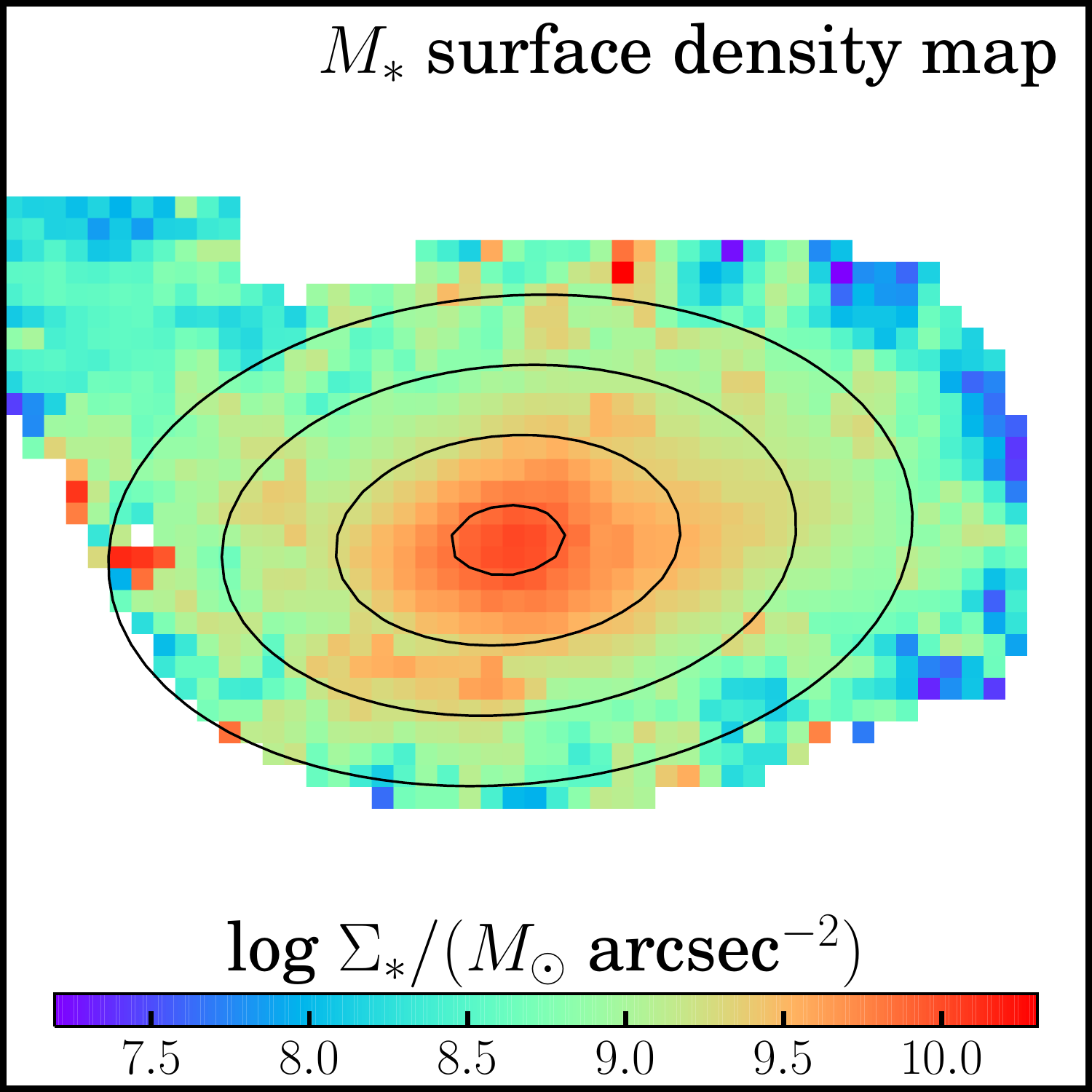}
    \includegraphics[width=.163\textwidth]{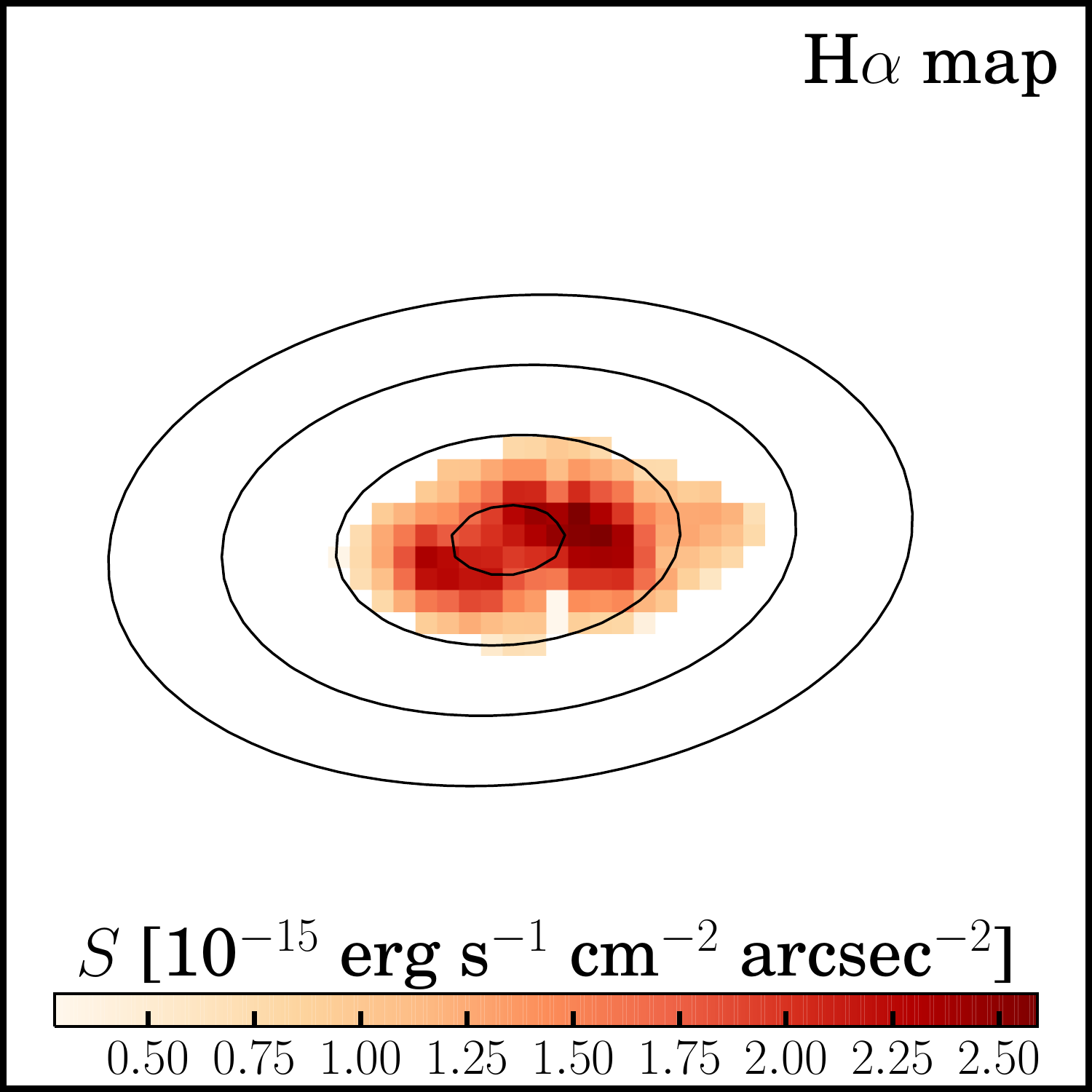}
    \includegraphics[width=.163\textwidth]{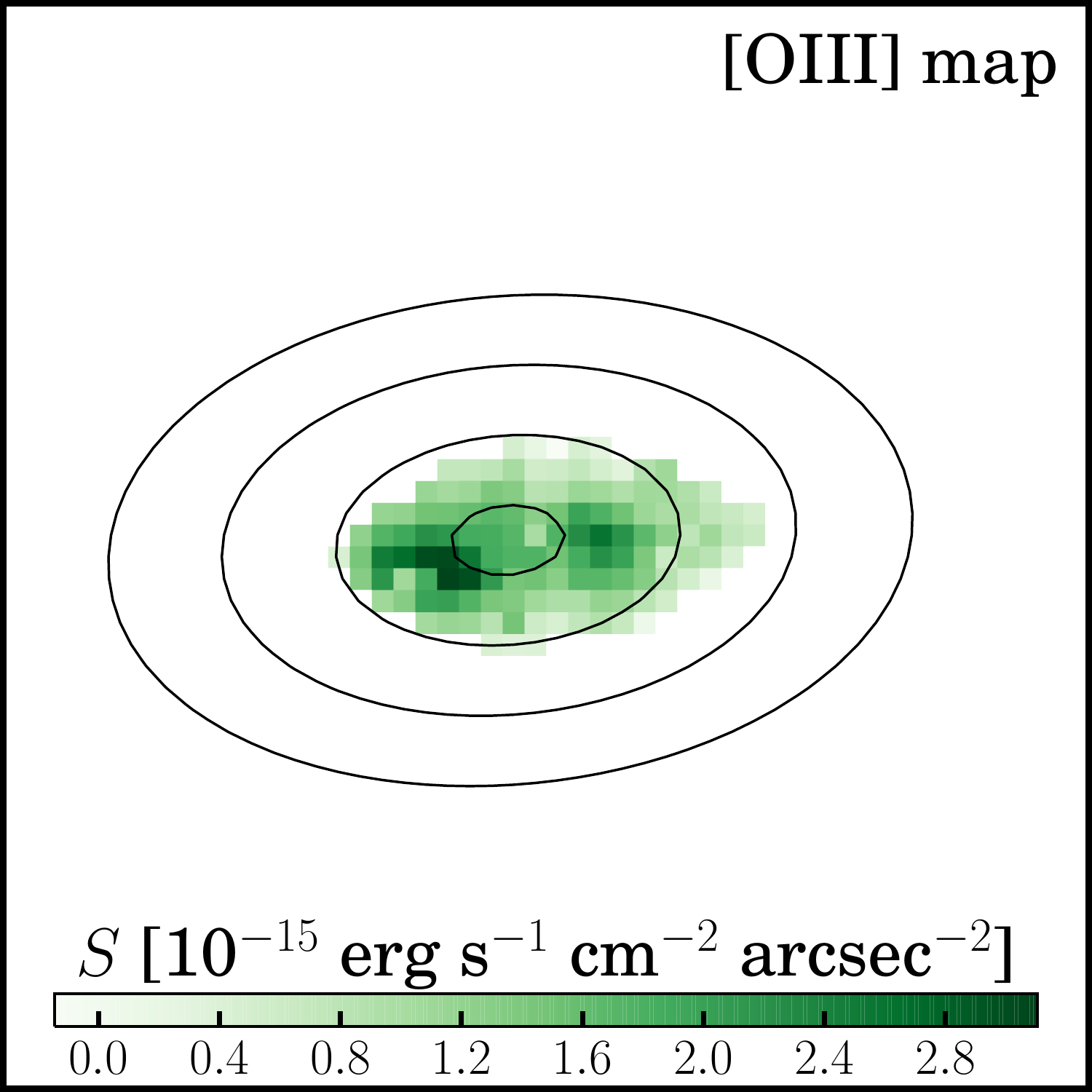}
    \includegraphics[width=.163\textwidth]{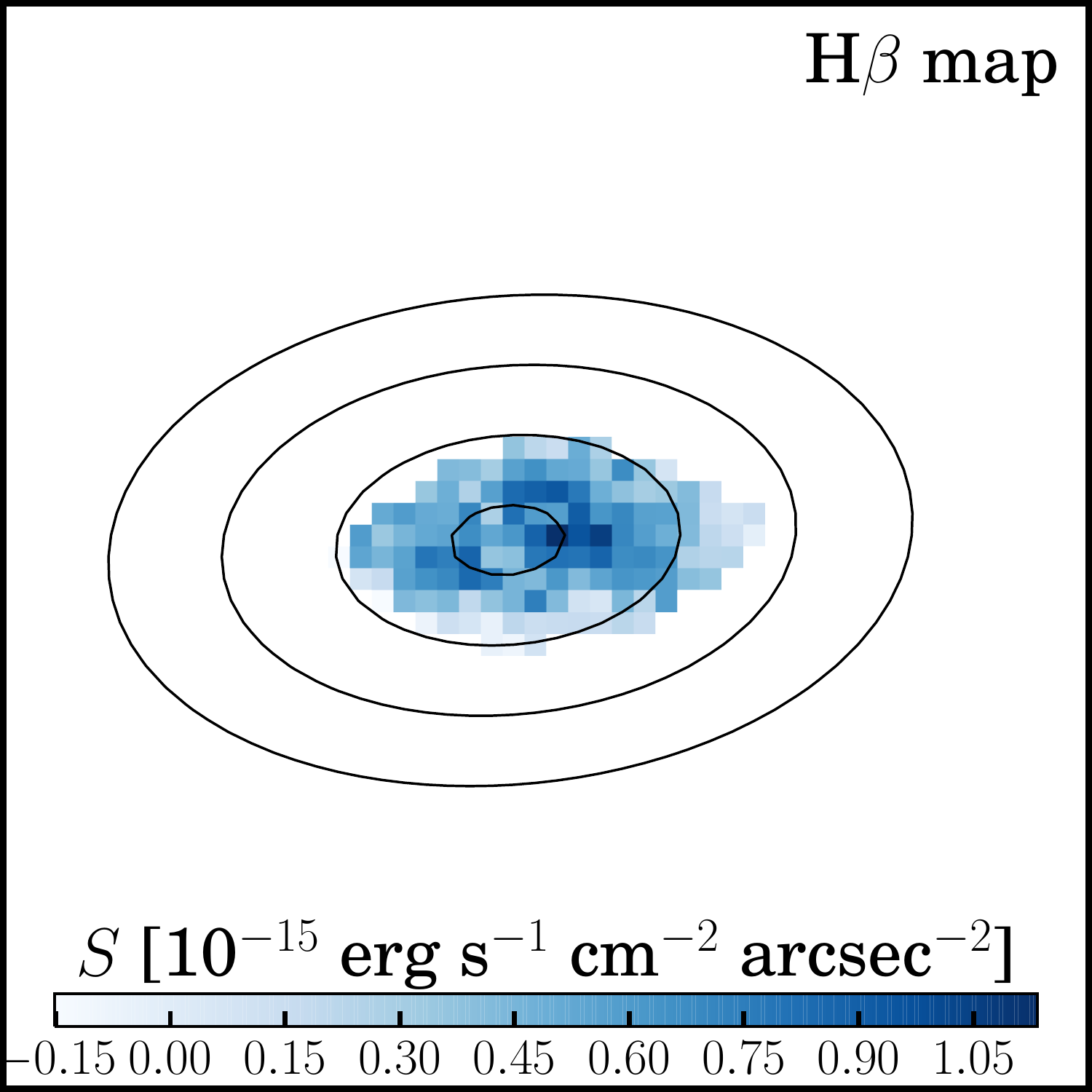}
    \includegraphics[width=.163\textwidth]{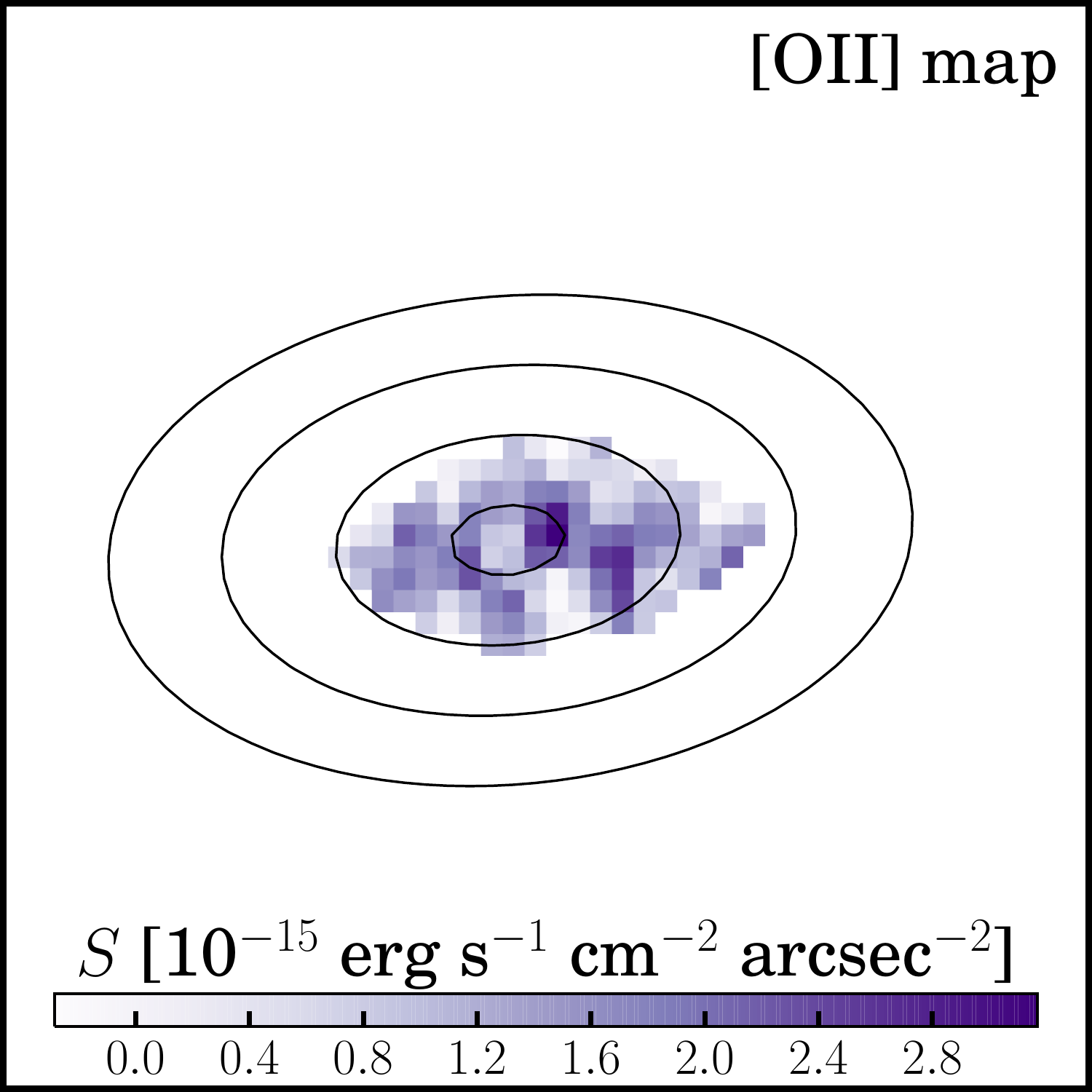}\\
    \includegraphics[width=.163\textwidth]{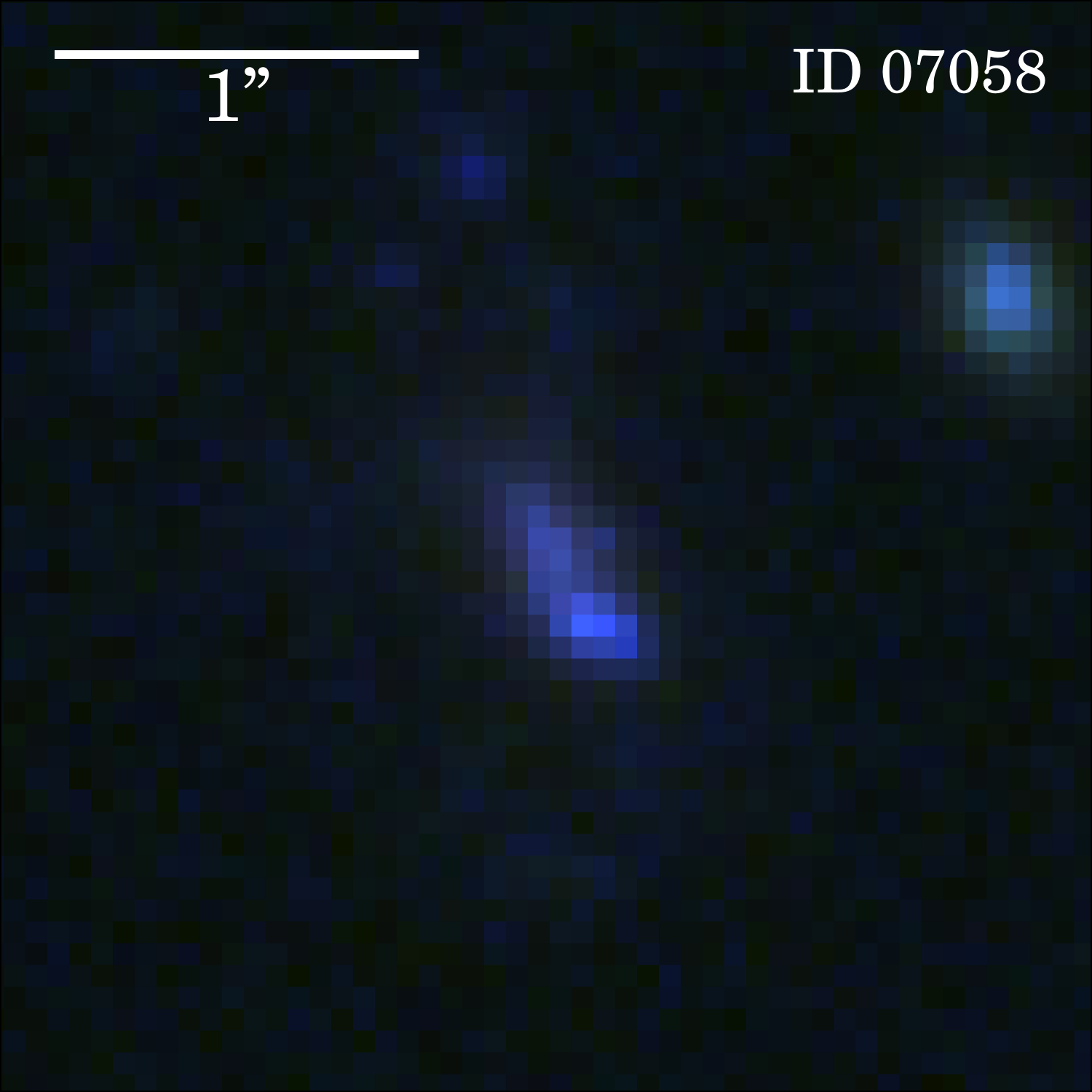}
    \includegraphics[width=.163\textwidth]{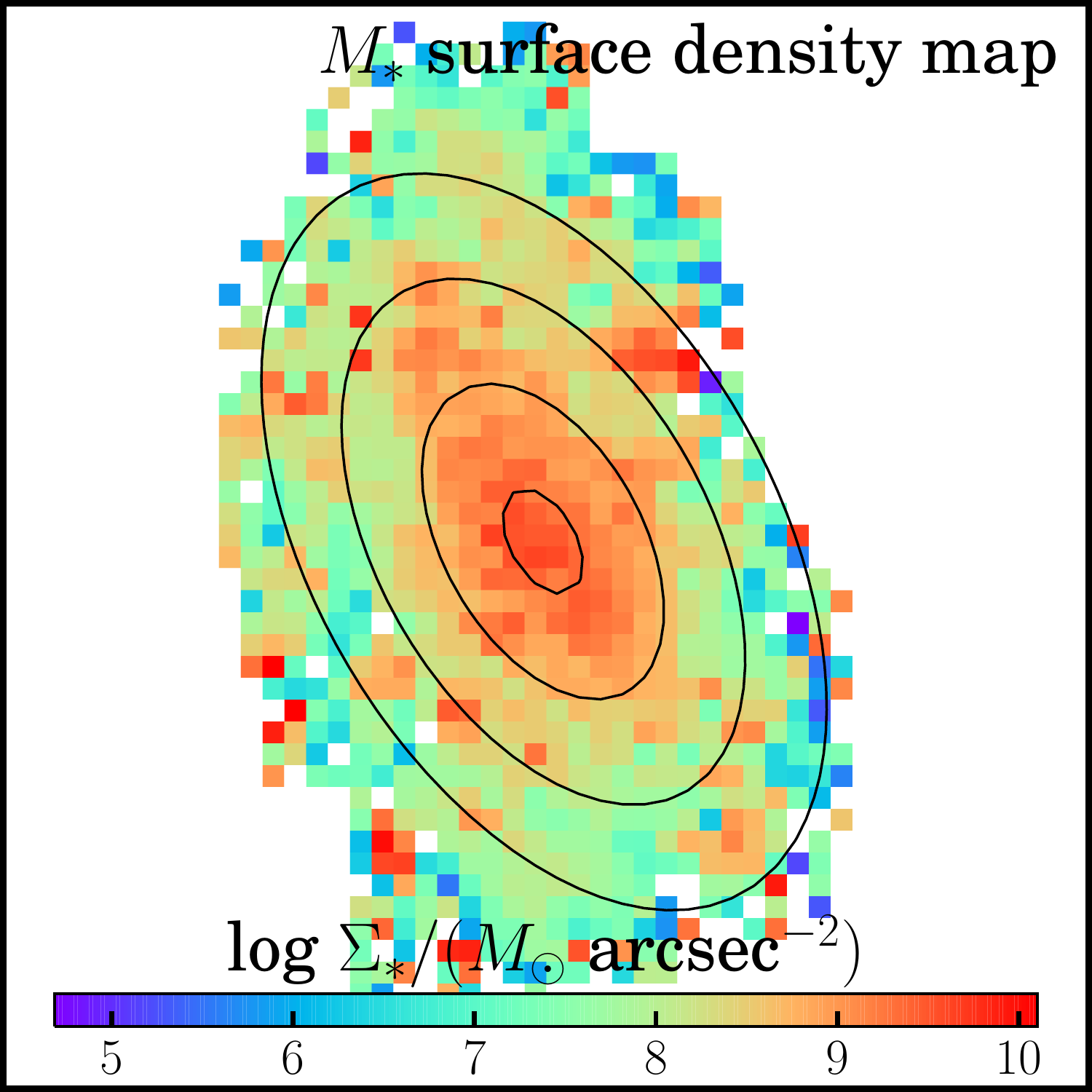}
    \includegraphics[width=.163\textwidth]{fig/baiban.png}
    \includegraphics[width=.163\textwidth]{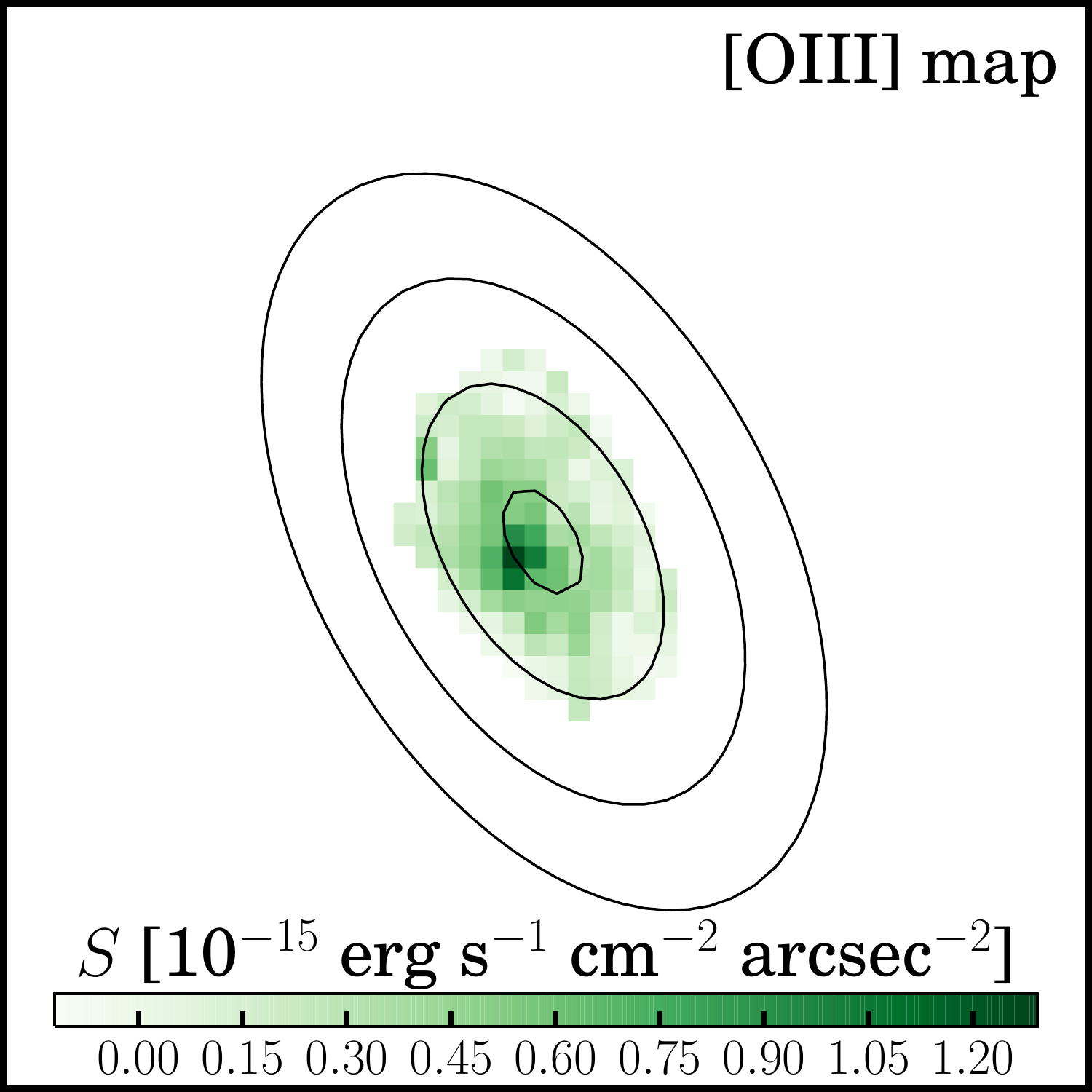}
    \includegraphics[width=.163\textwidth]{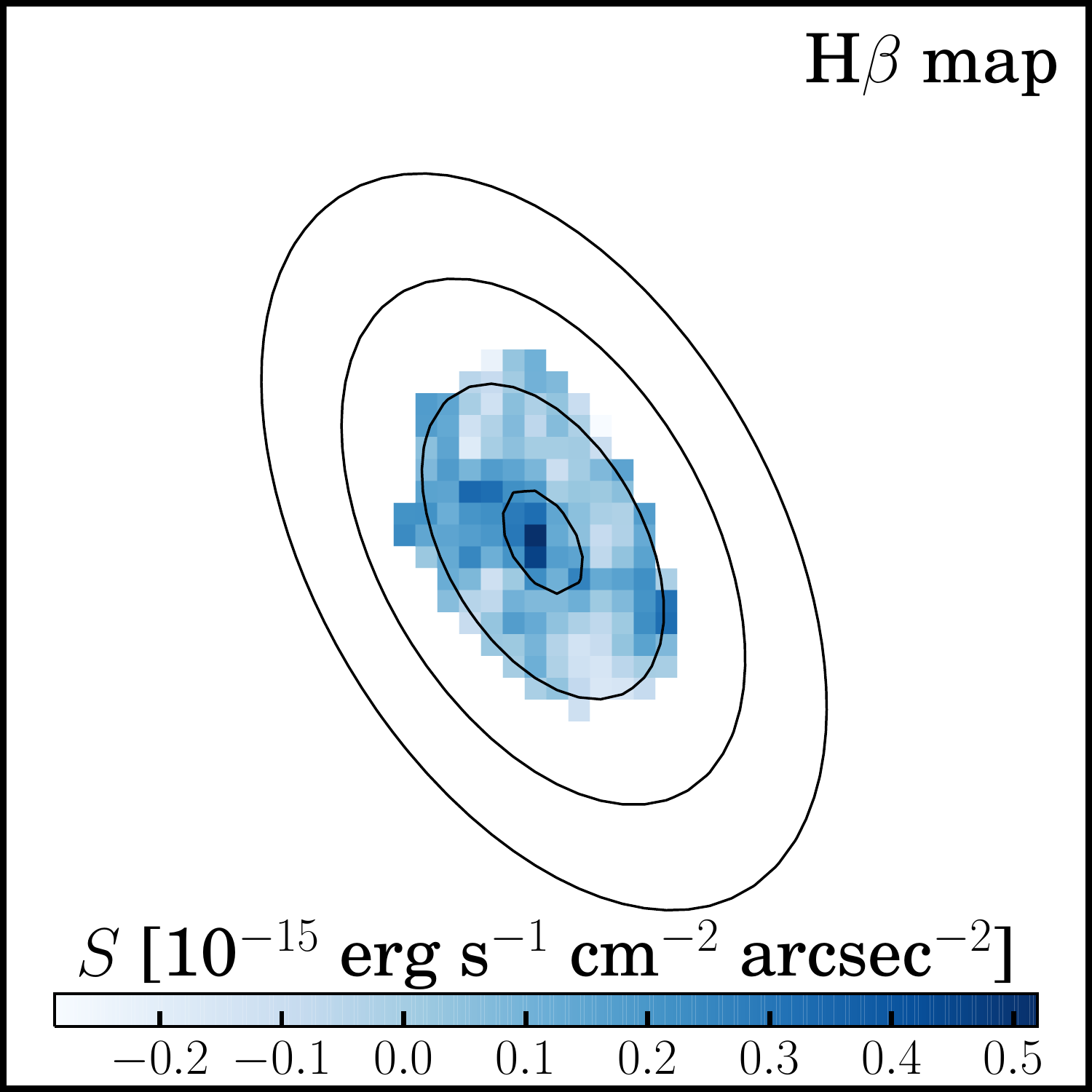}
    \includegraphics[width=.163\textwidth]{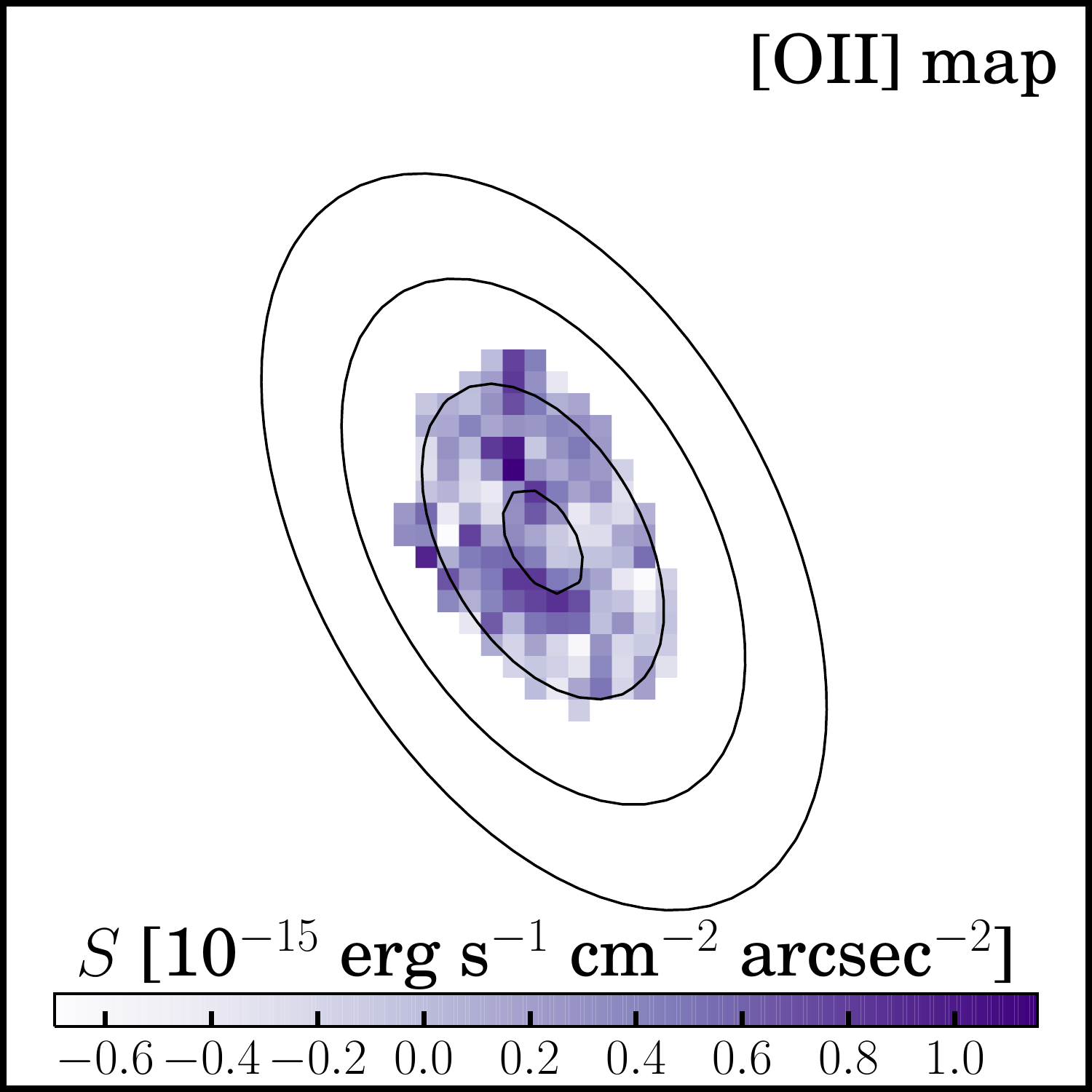}\\
    \includegraphics[width=.163\textwidth]{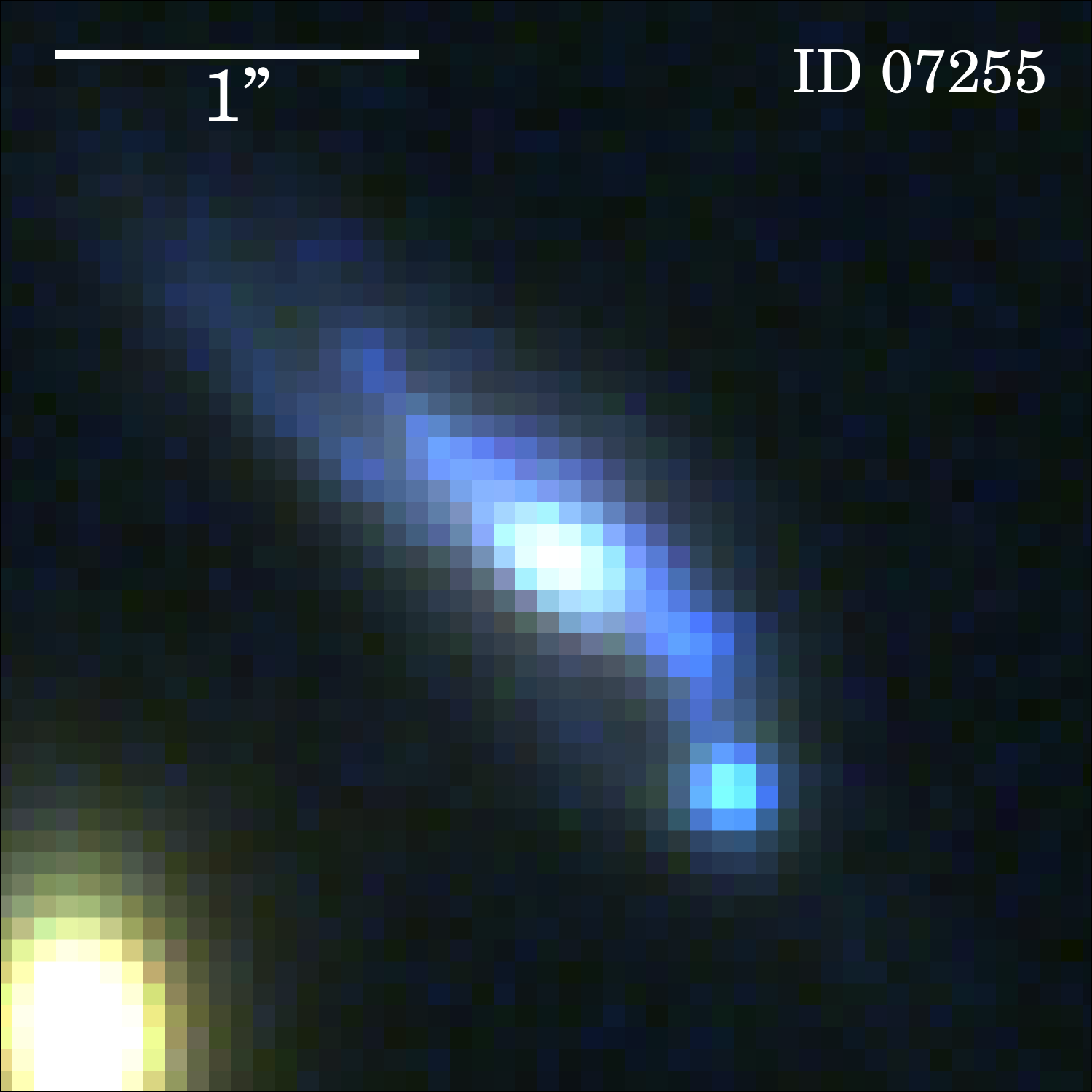}
    \includegraphics[width=.163\textwidth]{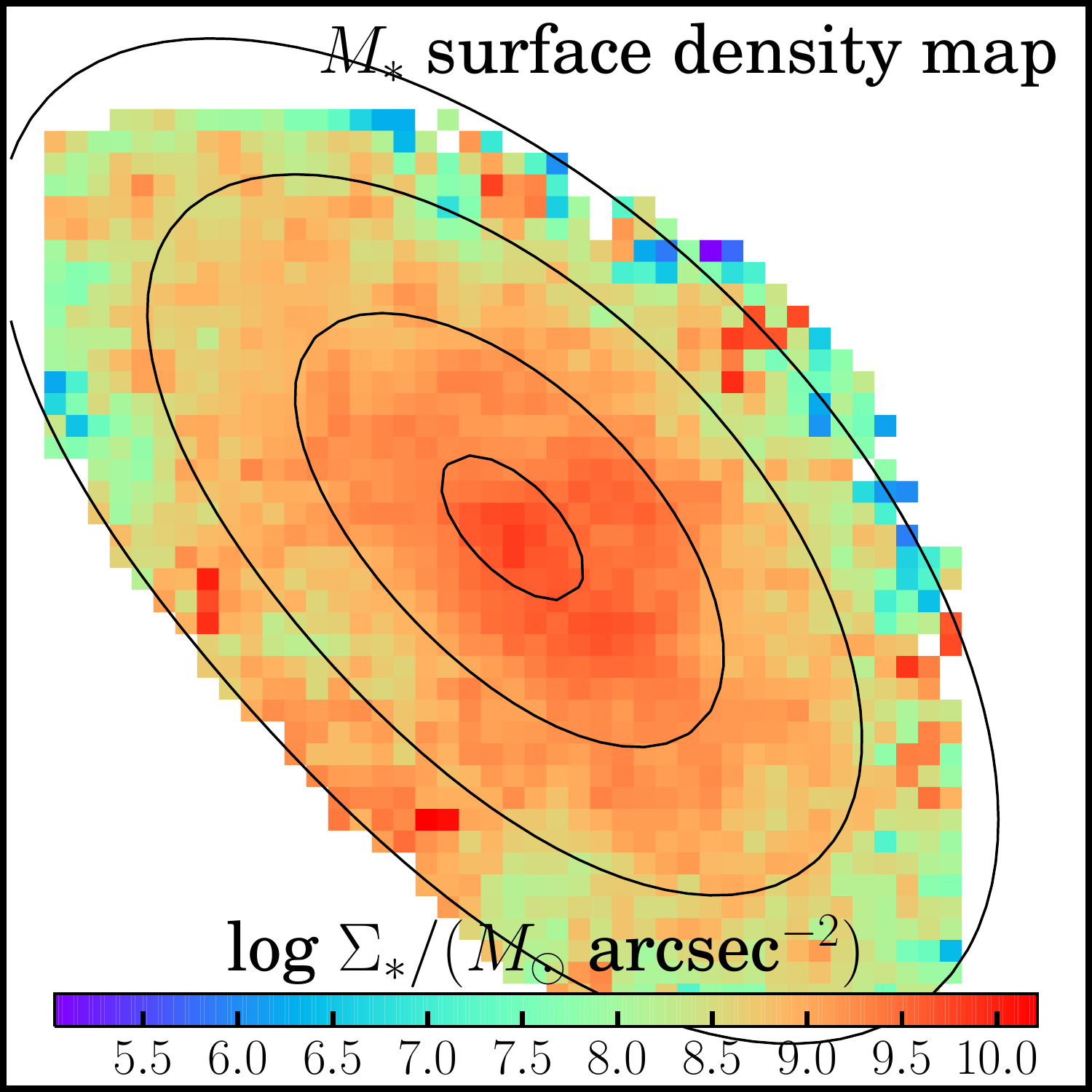}
    \includegraphics[width=.163\textwidth]{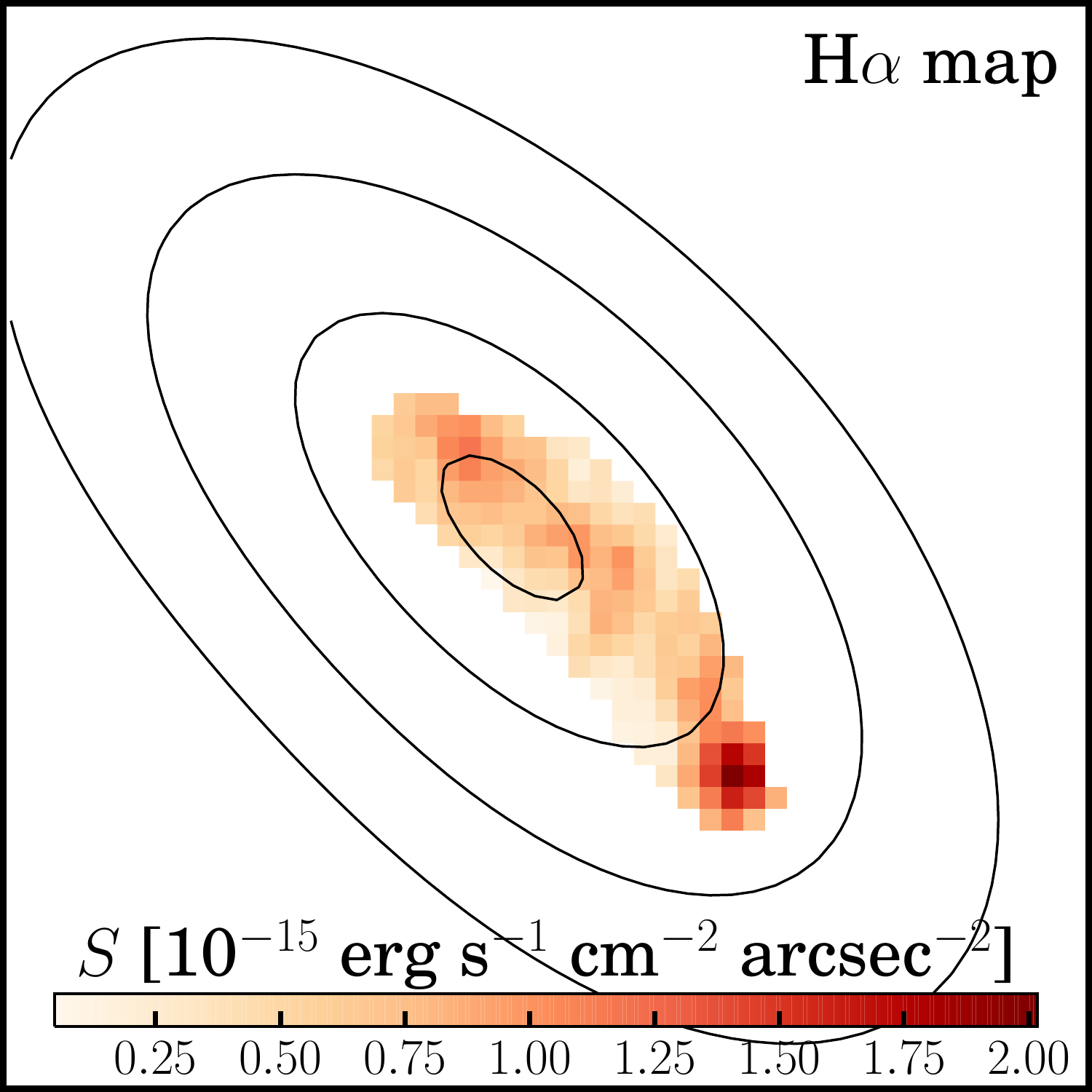}
    \includegraphics[width=.163\textwidth]{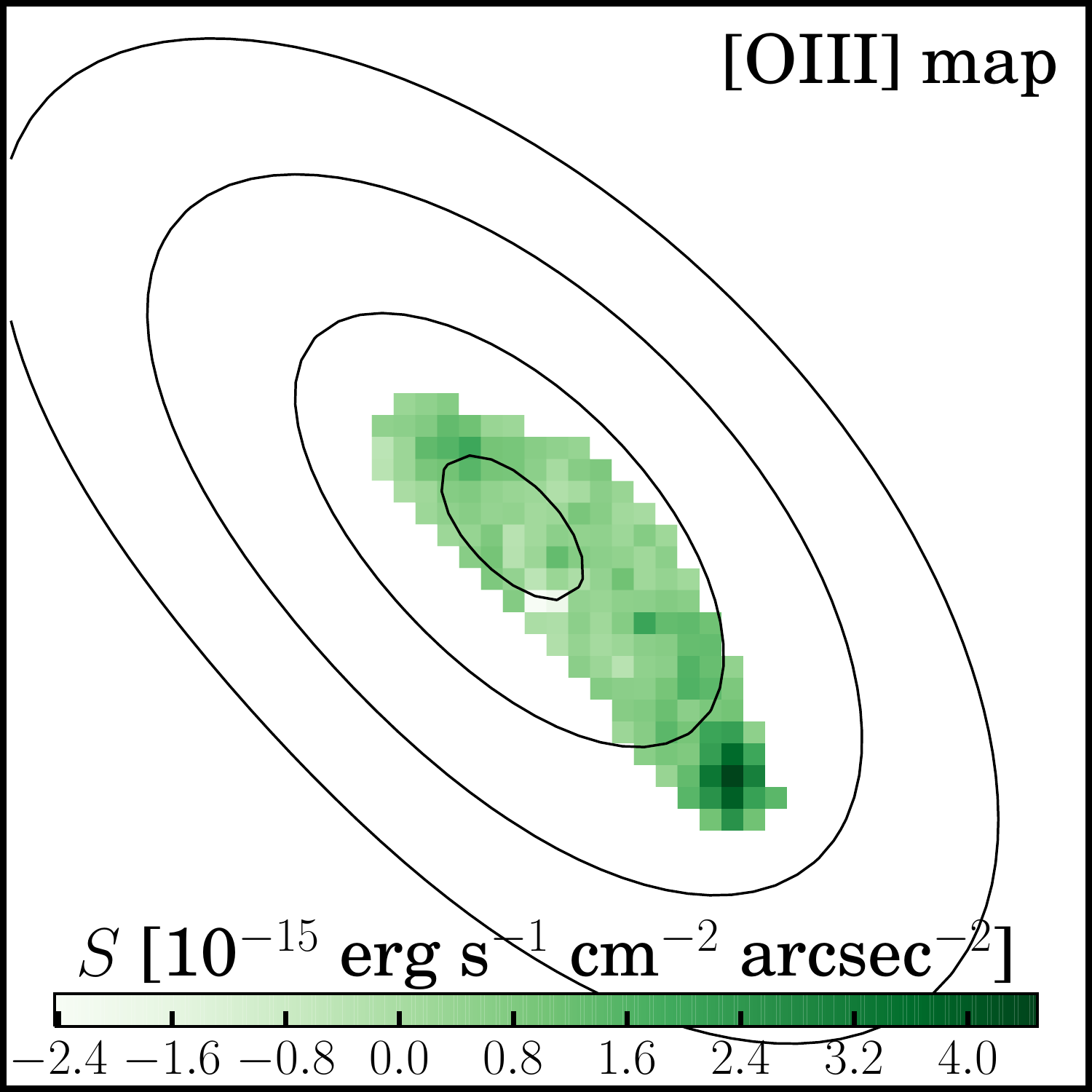}
    \includegraphics[width=.163\textwidth]{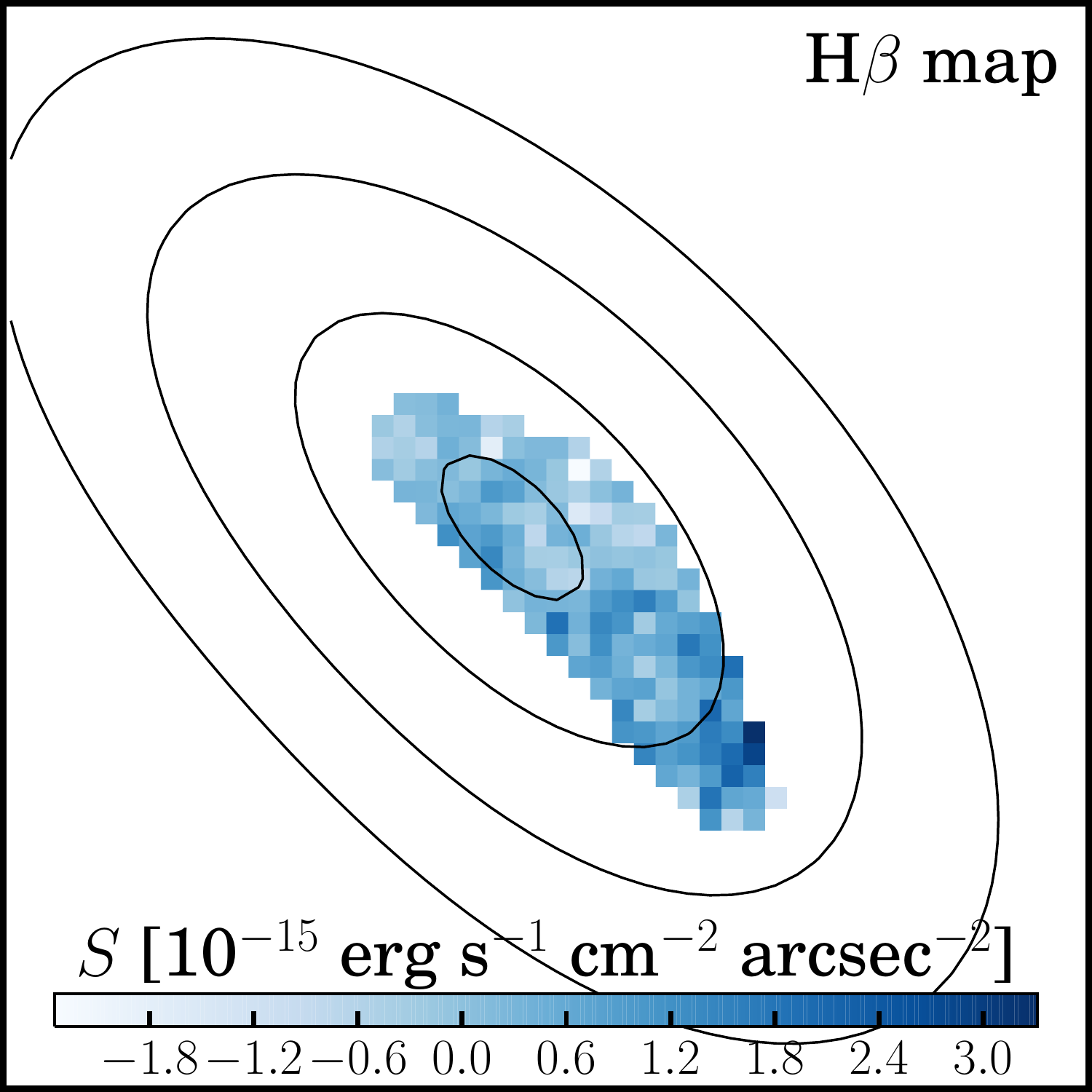}
    \includegraphics[width=.163\textwidth]{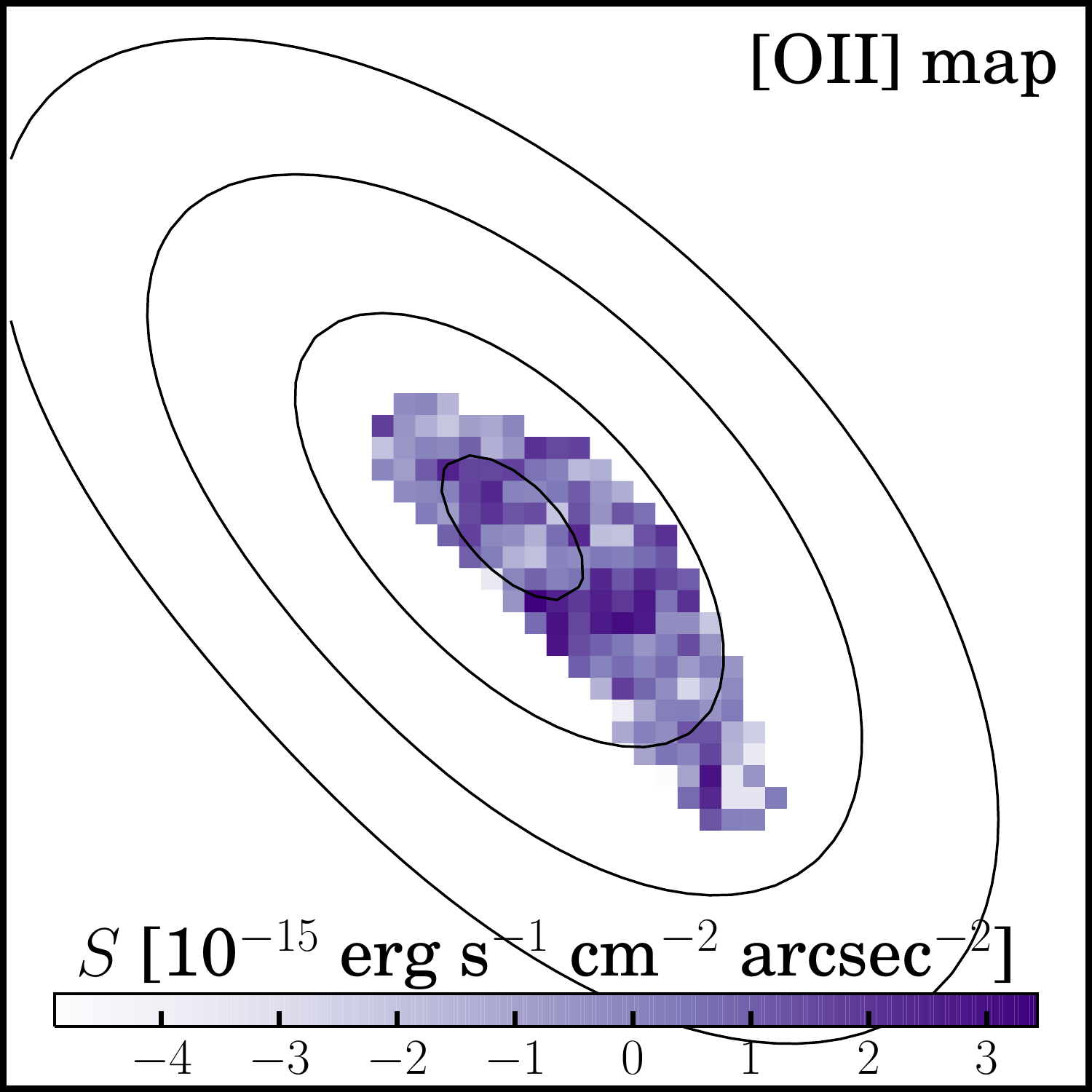}
    \contcaption{(cont.)}
\end{figure*}

\subsection{Emission line maps from grism spectra}\label{subsect:combELmaps}

The broad grism wavelength coverage provides spatially resolved maps of multiple ELs,
such as \OII, \Hg, \Hb, \OIII, \Ha$+$\NII, and \SII. To obtain pure EL maps, we first
use the \axe continuum models described in Section~\ref{sect:spec} to subtract contaminating
flux from neighboring sources. The continuum model for the target object is scaled to match
the locally observed levels, and then subtracted. We check that the flux residual in regions
near the ELs of interest follow a normal distribution with zero mean.

Due to the limited spectral resolution of \hst WFC3/IR grisms, \Hb and the \OIII$\lambda\lambda$4960,5008 doublets are
partially blended in the spatially extended sources. We adopt a direct de-blending technique
to separate these emission lines following the procedure used in \citet{2015AJ....149..107J}.
First, an isophotal contour is measured from the co-added \hst \H-band postage stamp,
typically at the level of $\sim$10 percent of the peak flux received from the source.
We then apply this contour to the 2D grism spectra at the observed wavelengths corresponding to
\OIII$\lambda\lambda$4960,5008 and \Hb, maximizing the flux given grism redshift uncertainty ($\sim$0.01).
We use the theoretical \OIII$\lambda\lambda$4960,5008 doublet flux ratio (\ie
\OIII$\lambda$5008/\OIII$\lambda$4960=2.98, calculated by \citet{Storey:2000jd}) to subtract
the flux contribution of \OIII$\lambda$4960 corresponding to the region where
\OIII$\lambda$5008 is unblended. We iterate this procedure to remove the \OIII$\lambda$4960
emission completely, resulting in pure maps of \Hb and \OIII$\lambda$5008.

Note that this direct de-blending will be compromised if the source is so extended that its
\OIII$\lambda$5008 and \Hb are blended. In practice, we fine-tune the isophotal contour level
to avoid this contamination. As a result, no cases in our sample have problem with our
direct de-blending method, and the resulting pure \OIII$\lambda$5008 and \Hb maps are
uncontaminated.

In the case of \Ha and \NII, the emission lines are separated by less than the grism spectral
resolution. Therefore we cannot use the direct technique to de-blend \Ha and \NII
emission, and instead treat the \NII/\Ha ratio as an unknown parameter in Section
\ref{subsect:bayes}.

After obtaining pure EL maps, we combine data taken at multiple PAs in
order to utilize the full depth of the grism exposures. There are in total six different
PA-grism combinations as given in Table~\ref{tab:obsdata}. At certain redshifts, ELs are
visible in the overlapping region between the two grisms at $\lambda\in[1.08,1.15]~\micron$
resulting in up to six individual maps of the same EL. We use the data taken at
PA=$119^{\circ}$ as a reference, as this minimizes errors arising from rotating the various
PAs into alignment. We first apply a small vertical offset to each EL map to account for the
slight tilt of the spectral trace (i.e. known offset between the dispersion direction and the
x-axis of extracted spectra; \citealt{Brammer:2012bu}). Next, we apply a small horizontal
shift to correct for uncertainties in the wavelength calibration and redshift. Finally we
rotate each EL map to align it with the reference orientation at PA=$119^{\circ}$.  The best
values of these three alignment parameters are found by maximizing the normalized
cross-correlation coefficient (NCCF) in order to achieve optimal alignment of multiple PAs:
\be\label{eq:ccf}
    \textrm{NCCF}=\frac{1}{n}\Sigma_{x,y}\frac{\(S_\refe(x,y)-\avg{S_\refe}\)\(S_\pa(x,y)-\avg{S_\pa}\)}{\sigma_{S_\refe}\sigma_{S_\pa}}.
\ee
Here $S$ is the \subr profile for a specific EL, $S_\refe$ corresponds to the reference
alignment stamp adopted as the PA=$119^{\circ}$ data, and $n$ is the number of spaxels in the
\subr profile. $\avg{S}$ and $\sigma_S$ represent the average and standard deviation of the
\subr, respectively.

Once the data from each PA are aligned, we vet the EL maps from each PA-grism combination and
reject those which show significant contamination-subtraction residuals or grism reduction defects.  We
combine the remaining maps of a given EL using an inverse variance weighted average
in order to properly account for the different exposure times and noise levels. The final
combined maps of each EL for each galaxy are shown in Figures~\ref{fig:multiP_4054} and
\ref{fig:multiP_rest}.

\subsection{Line-flux-based Bayesian inference of metallicity}\label{subsect:bayes}

%= = = = = = = = = = = = = = = = = = = = = = = = = = = = = = = = = = = = = = = =
% = = = = = = = = = = = = = = = = = = = = = = = = = = = = = = = = = = = = = = = = = =
% Include this table with \input{filename.tex}
% To rotate in emulateapj do: \begin{turnpage}\input{filename.tex}\end{turnpage}
% To display it on multiple pages do: \LongTables\input{filename.tex}
% - - - - - - - - - - - - - - - - - - - - - - - - - - - - - - - - - - - - - - - - - -
{\setlength\tabcolsep{2pt}
\begin{deluxetable}{llclccccccccc} \tablecolumns{13}
\tablewidth{0pt}
\tablecaption{Sampling parameters and their prior information.}
% - - - - - - - - - - - - - - - - - - - - - - - - - - - - - - - - - - - - - - - - - -
\tablehead{
    \colhead{Set} &
    \colhead{Parameter} &
    \colhead{Symbol (Unit)} &
    \colhead{Prior}
}
%---------------------------------------------------------------
\startdata
Vanilla & \gpm & \oh & flat: $[7.0,9.3]$ \\
        & nebular dust extinction & $\Av^{\rm N}$ & flat: $[0,4]$ \\
        & intrinsic \Hb flux & $f^\theo_{\Hb}$ ($10^{-17}$\Funit) & Jeffrey's: $[0,100]$ \\
\hline\smallskip
Extended & \NII to \Ha flux ratio & \NII/\Ha & Jeffrey's: [0, 0.5] \\
\hline
Derived & \sfr &  SFR (\Msun/yr) &  ---
\enddata
% - - - - - - - - - - - - - - - - - - - - - - - - - - - - - - - - - - - - - - - - - -
\tablecomments{Most constraints presented in this work are obtained under the ``Vanilla'' parameter set with the extended
parameter fixed as \NII/\Ha=0.05.}
\label{tab:param}
\end{deluxetable}

%= = = = = = = = = = = = = = = = = = = = = = = = = = = = = = = = = = = = = = = =

Generally speaking, two methods exist in deriving gas-phase oxygen abundance in star-forming galaxies at high redshifts.
``Direct'' determinations based on electron temperature measurements, which require high signal-to-noise ratio (SNR)
detections of auroral lines (\eg \OIII$\lambda$4364), are still extremely challenging beyond the local universe \citep[see][for a
rare example]{Sanders:2016uo}.
We therefore rely on the calibrations of strong collisionally excited EL flux ratios to estimate metallicities.

One of the most popular strong EL diagnostics is the flux ratio of \NII/\Ha calibrated by \citet{2004MNRAS.348L..59P}.  However, a
large offset (0.2-0.4 dex) between the loci of high-$z$ and present-day star-forming galaxies in the Baldwin-Phillips-Terlevich
\citep[BPT,][]{Baldwin:1981ev} diagram has recently been discovered, indicating that extending the locally calibrated \NII/\Ha to
high-$z$ can be problematic \citep{2015ApJ...801...88S,Sanders:2015gk}.  The interpretation of this offset is still the topic of
much debate. It has been interpreted as the existence of a fundamental relation between nitrogen-oxygen abundance ratio and \Mstar
\citep{Masters:2016vr}, a systematic dependence of strong EL properties with respect to Balmer line luminosity
\citep{Cowie:2016fv}, a combination of harder ionizing radiation and higher ionization states of the gas at high-$z$
\citep{2014ApJ...795..165S}, or an enhancement of nitrogen abundance in hot \HII regions \citep{Pilyugin:2010bx}. Fortunately, the
alpha-element BPT diagrams show no offset with $z$ \citep{2015ApJ...801...88S,Sanders:2015gk}, and therefore the diagnostics based
upon those ELs are more reliable.

In this work we adopt the calibrations of \citet[M08]{2008A&A...488..463M}, including the flux ratios of \OIII/\Hb, \OII/\Hb,
\NII/\Ha.
Given the potential systematics related to nitrogen ELs, we do not use the \NII/\Ha calibration of M08.
M08 fit the relation between these EL ratios and gas-phase oxygen abundance based upon ``direct'' measurements (from
\OIII$\lambda$4364) for \oh$<8.3$, and photoionization model results for higher metallicities. This provides a continuous
framework valid over the wide range \oh$\in(7.1,9.2)$.  We note that although there are alternative calibrations of metallicity
available in the literature and the applicability of these recipes is currently a hotly debated topic \citep[see,
\eg,][]{Blanc:2015hl,Dopita:2013bj}.  Here we are primarily interested in relative metallicity measurements.  Even though the
absolute measurements of metallicity may change if we used a different calibration, the gradients and morphological features in
the maps should be more robust to changes in the calibration. We will restrict our comparisons with other samples to only include
studies assuming the M08 calibrations.

Unlike previous work in which calibrated relations are applied to
various EL flux ratios \citep[see, \eg,][]{PerezMontero:2014jh,Bianco:2015hn}, we design a Bayesian statistical approach which
uses directly the individual EL fluxes as input, such that the information from one EL is only used once.
Our approach presents several advantages over those based on line flux ratios.
First, it properly accounts for flux uncertainties for lines that are weak
or undetected. Second, multiple lines can be taken into consideration
without the risk of double counting information.

Our approach simultaneously infers the gas-phase metallicity (\oh),
nebular dust extinction ($\Av^{\rm N}$), and intrinsic \Hb flux (which
is proportional to the star formation rate). For values of these three parameters we can compute the expected line
fluxes, and compare with measured values to compute the likelihood and
then the posterior probability.

Expected line fluxes are uniquely determined from the M08 calibrations
and the \citet{1989ApJ...345..245C} galactic extinction curve with \Rv=3.1,
assuming intrinsic $\Ha/\Hb=2.86$ and $\Hg/\Hb=0.47$ appropriate for case
B recombination and fiducial \HII region conditions \citep[see,
\eg,][]{Hummer:1987ed}. Together with measured EL fluxes and
uncertainties, the \chisq statistic in our inference procedure is
calculated as
\begin{equation}\label{eqn:chi2}
    \chisq = \sum_i \frac{\(f^\obs_{\el{i}} - R_i \cdot f^\theo_{\Hb}\)^2}
        {\(\sigma^\obs_{\el{i}}\)^2 + \(f^\theo_{\Hb}\)^2\cdot\(\sigma_{R_i}\)^2}.
\end{equation}
Here $\el{i}$ denotes the set of emission lines \OII, \Hg, \Hb, \OIII, \Ha, and \SII.
$R_i$ is the flux of \el{i} relative to \Hb, with $\sigma_{R_i}$ being intrinsic
scatter from the M08 calibrations. $f^\obs_{\el{i}}$ and
$\sigma^\obs_{\el{i}}$ represent the observed \el{i} flux and its
uncertainty. $f^\theo_{\Hb}$ is the intrinsic \Hb flux corrected for
dust extinction.

We apply this inference procedure to both galaxy-integrated line
fluxes (to be discussed in Section~\ref{sect:global}) as well as
individual spaxels in the EL maps (which will lead to \mgms described
in Section~\ref{sect:sra}).

Details of the prior assumptions applied to each parameter of the
inference procedure are explained in Table~\ref{tab:param}, and the coefficients for the EL
flux ratio diagnostics used in the statistical inference are given in
Table~\ref{tab:ELratioCoef}.  We additionally note the following:

\indent\textbullet~ The \NII/\Ha ratio is included as an additional parameter for galaxies at
$z\lesssim1.5$ where these lines are observed. We do not attempt to determine \NII/\Ha using
the M08 calibrations, as locally-calibrated diagnostics tend to underestimate the \NII/\Ha
ratio in high-$z$ galaxies whereas oxygen and other $\alpha$-element line ratios remain
constant with redshift \citep{2014ApJ...795..165S,2015ApJ...801...88S,2015ApJ...813..126J}.
Instead we either leave this parameter free (``extended'' priors), or fixed to $\NII/\Ha=0.05$
(``vanilla'' priors).  The vanilla value is typical of galaxies with
similar $z$ and oxygen EL ratios as our sample.  Fixing the value of $\NII/\Ha$ provides
faster convergence and does not significantly affect the inferred metallicity.  We therefore
report constraints for the ``vanilla'' parameter set. For galaxy ID 04054, the
SN Refsdal host galaxy, we fix its \NII/\Ha to be 0.11, as measured by \citet{Yuan:2011hj}.

\indent\textbullet~ For $\el{i} \in \{\Ha,~\Hg\}$, $R_i$ corresponds to the Balmer decrement and we set
$\sigma_{R_i}=0$ (\ie the intrinsic Balmer ratios are fixed with no assumed scatter). These
ratios are sensitive only to the nebular dust extinction $\Av^{\rm N}$.

\indent\textbullet~ For $\el{i} \in \{\OII,~\OIII\}$, $R_i$ and $\sigma_{R_i}$ are taken from the M08 calibrations.
These ratios are sensitive to the oxygen abundance \oh, whereas dust extinction correction is necessary for \OII/\Hb.

\indent\textbullet~ For $\el{i} = \SII$, $R_i$ and $\sigma_{R_i}$ come from our new
calibration derived following \citet{2015ApJ...813..126J}, using the same
data as the low-metallicity calibrations of M08. Although M08 do not provide any
calibrations for \SII, we expect our calibration to be self-consistent and reliable for
$\oh\lesssim8.4$. \SII/\Ha is primarily useful as a diagnostic to differentiate between the
upper and lower branches of the \OIII/\Hb calibration. This is valuable because \NII is not
directly measured and \OII is typically low signal-to-noise in the lower redshift galaxies
for which \SII is available, due to decreasing grism throughput at $\lambda<0.9$ \micron.

\indent\textbullet~ For $\el{i} =$ \Hb, $R_i = 1$ with $\sigma_{R_i}=0$.
\\

%The primary goal of the Bayesian inference is to determine metallicities. The posterior probability density distribution for
%metallicity is obtained after marginalizing over nuisance parameters, \ie,
%\begin{equation}
%    \cP(\oh) = \iint \exp(-\chisq/2) \cdot \cP_\textrm{prior}(f^\theo_{\Hb}) \d f^\theo_{\Hb} \cdot
%    \cP_\textrm{prior}(\Av^{\rm N}) \d \Av^{\rm N}.
%\end{equation}
We use the Markov Chain Monte Carlo sampler \emc \citep{ForemanMackey:2013io} to
explore the parameter space, setting the number of walkers to 100 with 5,000 iterations for each walker,
and a burnin of 2,000. The autocorrelation times are computed for
each parameter and are used to make sure chains have converged.
An example constraint result for the SN Refsdal host galaxy's global EL fluxes is shown in Figure~\ref{fig:corner}. While this
paper is primarily concerned with gas-phase metallicities, our method simultaneously provides constraints on $\Av^{\rm N}$ and
\sfr which can be compared to the results of SED fitting. We then derive the instantaneous \sfr from the intrinsic \Ha luminosity
calculated from $f^\theo_{\Hb}$ via the commonly-used calibration of \citet{Kennicutt:1998ki,Moustakas:2010ke}, \ie,
\be
    {\rm SFR} = 4.6\times10^{-42}~ \frac{L(\Ha)}{\rm erg/s} \quad [\Msun/\textrm{yr}],
\ee
valid for a \citet{Chabrier:2003ki} IMF. This SFR estimate does not rely on any assumptions about \sfh, and thus is a more direct measure of ongoing star formation than values given by SED fitting.

\begin{figure}
    \centering
    \includegraphics[width=.5\textwidth]{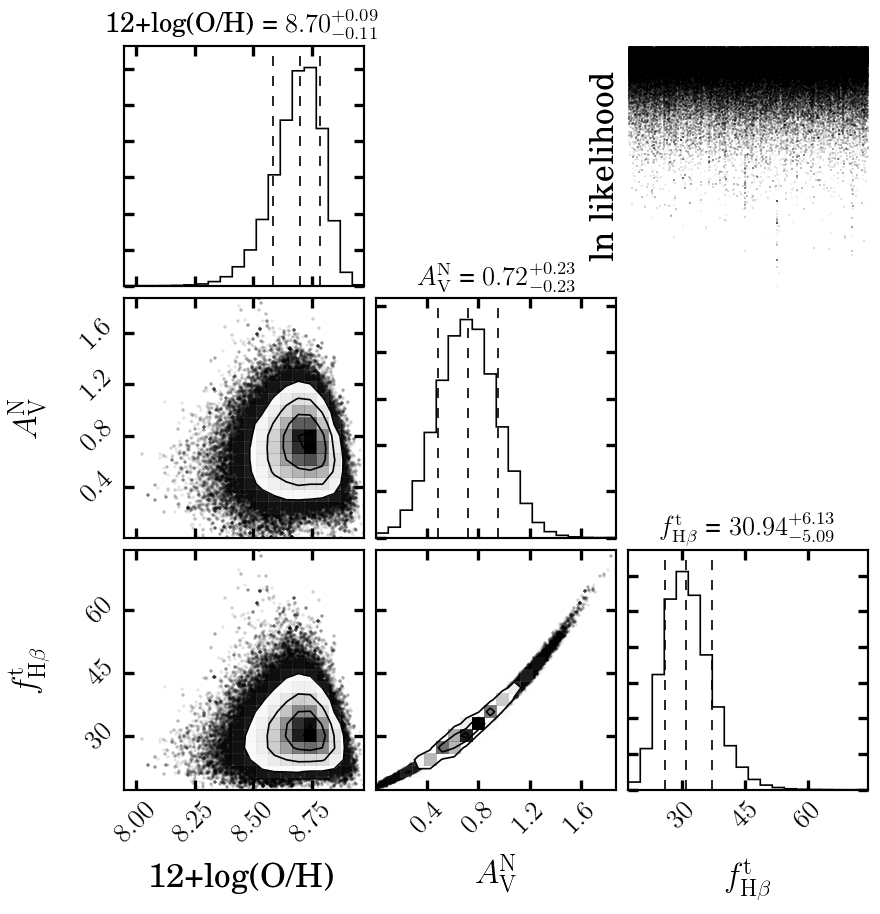}
    \caption{The marginalized 1D and 2D constraints on (\oh, $\Av^{\rm N}$, and
    $f^\theo_{\Hb}$), which are drawn from the integrated line fluxes of the Refsdal host
    galaxy (ID 04054), after all grism exposures are combined. The values on top of each
    column are the medians with 1-$\sigma$ uncertainties for each parameter. The plot in the
    upper right corner illustrates a good convergence of the sampling.}
    \label{fig:corner}
\end{figure}

%= = = = = = = = = = = = = = = = = = = = = = = = = = = = = = = = = = = = = = = =
% = = = = = = = = = = = = = = = = = = = = = = = = = = = = = = = = = = = = = = = = = =
% Include this table with \input{filename.tex}
% To rotate in emulateapj do: \begin{turnpage}\input{filename.tex}\end{turnpage}
% To display it on multiple pages do: \LongTables\input{filename.tex}
% - - - - - - - - - - - - - - - - - - - - - - - - - - - - - - - - - - - - - - - - - -
\begin{deluxetable}{ccccccccccccc} \tablecolumns{13}
\tablewidth{0pt}
\tablecaption{Coefficients for the EL flux ratio diagnostics used in our Bayesian inference.}
% - - - - - - - - - - - - - - - - - - - - - - - - - - - - - - - - - - - - - - - - - -
\tablehead{
    \colhead{$R$} & 
    \colhead{$c_0$} &
    \colhead{$c_1$} &
    \colhead{$c_2$} &
    \colhead{$c_3$} &
    \colhead{$c_4$} &
    \colhead{$c_5$}
}
%---------------------------------------------------------------
\startdata
\Ha/\Hb     &  0.4564 & \nodata & \nodata & \nodata & \nodata & \nodata \\
\Hg/\Hb     & -0.3279 & \nodata & \nodata & \nodata & \nodata & \nodata \\
\OIII/\Hb   &  0.1549 & -1.5031 & -0.9790 & -0.0297 & \nodata & \nodata \\
\OII/\Hb    &  0.5603 &  0.0450 & -1.8017 & -1.8434 & -0.6549 & \nodata \\
%\NeIII/\OII & -1.2608 & -1.0861 & -0.1470 & \nodata & \nodata & \nodata \\
\SII/\Ha    & -0.5457 &  0.4573 & -0.8227 & -0.0284 &  0.5940 &  0.3426
\enddata
% - - - - - - - - - - - - - - - - - - - - - - - - - - - - - - - - - - - - - - - - - -
\tablecomments{The value of EL flux ratio is calculated by the polynomial functional form,
\ie, $\log{R} = \sum_{i=0}^{5}c_i\cdot x^i$, where $x=\oh-8.69$. The ratio of \SII/\Ha is
calibrated by the work of \citet{2015ApJ...813..126J}, and the Balmer line ratios correspond
to the Balmer decrement, whereas the ratios between oxygen lines and \Hb are from M08.}
\label{tab:ELratioCoef}
\end{deluxetable}

%= = = = = = = = = = = = = = = = = = = = = = = = = = = = = = = = = = = = = = = =

\section{Global demographic properties}\label{sect:global}

Based upon the Bayesian analysis, we obtain the global measurements of \gpm, nebular dust extinction, and \sfr, for all sources in
our sample, as shown in Table~\ref{tab:srcprop}. First, we check for the presence of any active galactic nucleus (AGN)
contamination in our sample. Because not all galaxies in our sample reside in the redshift range where all BPT lines are
available, we resort to the mass-excitation diagram showing the \OIII/\Hb flux ratio as a function of \Mstar, first proposed by
\citet{Juneau:2011fz} to differentiate \sf galaxies and AGNs in local Sloan Digital Sky Survey (SDSS) observations.
\citet{Juneau:2014ca} further revised the demarcation scheme by applying line luminosity evolution and selection effects to a more
complete SDSS galaxy sample, and tested that this scheme agrees well with the bivariate distributions found in a number of
high-$z$ galaxy samples.
According to the \citet{Juneau:2014ca} classification scheme, all but one galaxy (ID 02607) have a negligible probability of being
AGN (see Figure~\ref{fig:MEx}).
Recently, \citet{Coil:2015dp} found that a +0.75 dex \Mstar shift in the \citet{Juneau:2014ca} scheme is required to avoid serious
AGN contamination in the \mosdef galaxy sample at $z\sim2.3$.
We verified that galaxy 02607 can still be safely classified as an AGN candidate, even considering the \citet{Coil:2015dp} shifted
classification scheme.
Because the M08 calibrations originate from \HII region observations and theoretical models, not designed for AGNs, our method
will not predict a reliable metallicity constraint for the AGN candidate ID 02607. We thereby exclude this source in the rest of
our population analysis. For the rest of our sample, comprising 13 star-forming galaxies, the constraint on \gpm is robust.

\begin{figure}
    \centering
    \includegraphics[width=.5\textwidth]{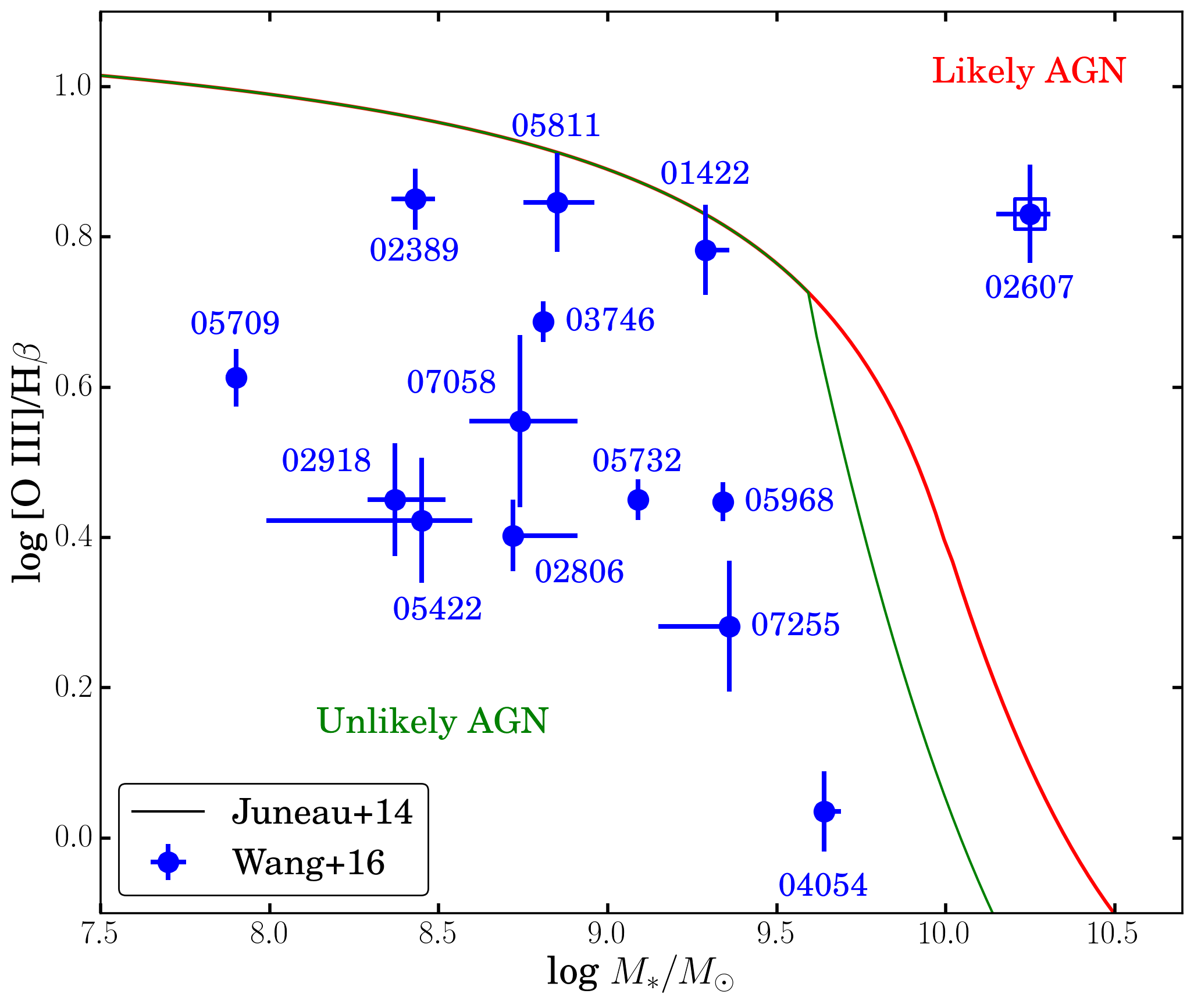}
    \caption{Mass excitation diagnostic diagram for our \mg sample. The curves are the
    demarcation lines updated by \citet{Juneau:2014ca}. The regions below the green curve
    have a low probability ($<10\%$) of being classified as AGNs, whereas galaxies located above the red curve are secure AGN
    candidates. In our sample, the only AGN candidate,  \ie, ID 02607, is marked by a square.}
    \label{fig:MEx}
\end{figure}

In Figure~\ref{fig:SFMS}, we plot the measured SFR as a function of \Mstar for our sample. In comparison, the loci of
mass-selected galaxies observed by the \kd survey and the ``\sfms'' \citep[SFMS,][]{Whitaker:2014ko,Speagle:2014dd} at similar
redshifts are also shown. We can see that selecting lensed galaxies based upon EL strength can probe deep into the low-mass regime
at high-$z$ that is currently inaccessible to mass complete surveys. As expected, our emission line selected targets probe the
upper envelope of the SFMS, so that a detail comparison is non-trivial and should take into account the selection functions of
each sample. The advantage of the emission line selection technique is that it is the most cost-effective way to obtain gas
metallicities for stellar masses down to 10$^7$ \Mstar at $z\sim2$. Using the same technique, we discovered an analog of local
group dwarf spheroidals \citep[\eg, Fornax,][]{Coleman:2008ca} at $z=1.85$ experiencing active star formation (with 1 dex offset
from the SFMS), in our previous work \citep{2015AJ....149..107J}.

\begin{figure}
    \centering
    \includegraphics[width=.5\textwidth]{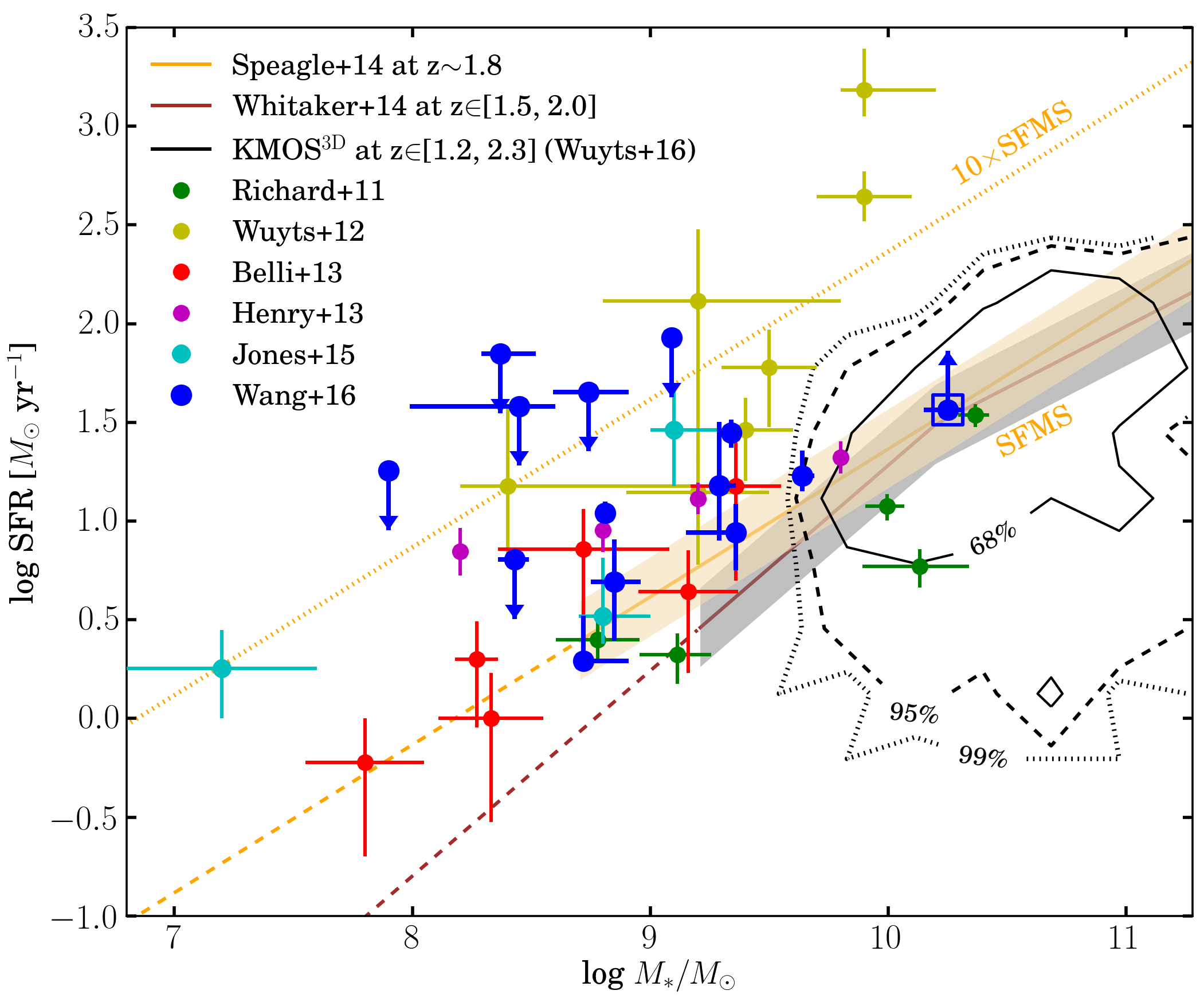}
    \caption{Star formation rate as a function of stellar mass for emission-line galaxies at
    $z\in[1.2,2.3]$. Our \mg sample is marked by blue points; the
    literature sample is color-coded by reference according to the legend. The star-forming
    main sequence compiled by
    \citet{Speagle:2014dd} and \citet{Whitaker:2014ko} are represented by brown and orange
    lines respectively, where the dotted part is extrapolated at masses below the mass
    completeness limit.  The shaded regions denote the typical scatter of star-forming main
    sequence, \ie 0.2-0.3
    dex. We also show the source density contours for the \kd survey at the same redshift
    range, where 68\%, 95\%, and 99\% of all \kd galaxies at $z\in[1.2,2.3]$ can be found
    within the black solid, dashed, and dotted contours respectively.
    It is found that our sample is highly complementary to the \kd sample in terms of stellar
    mass.}
    \label{fig:SFMS}
\end{figure}

We also collect all published \gpm measurements in emission-line galaxies in the redshift range of our sample
$1.2\lesssim z\lesssim2.3$, obtained exclusively from the strong EL calibrations
prescribed by M08. The reason we only select measurements based upon
the M08 calibrations is that adopting different strong EL calibrations
can give rise to different absolute metallicity scales offset by up to
$\sim$0.7 dex at the high-mass end, according to the comparative study conducted by
\citet{Kewley:2008be}.  In order to minimize the systematic uncertainty associated with EL
calibrations, we thus focus on this homogeneous M08 sample alone.

As shown in Figure~\ref{fig:MZR}, the M08-based sample includes five galaxies from
\citet{2011MNRAS.413..643R}, seven from
\citet{Wuyts:2012gb}, six from \citet{Belli:2013cn}, four
stack measurements by \citet{Henry:2013gx}, three interacting galaxies
at $z=1.85$ analyzed by \citet{2015AJ....149..107J}, and 13
star-forming galaxies presented in this work.  In total, this sample
consists of 38 independent measurements at $1.2\lesssim z\lesssim2.3$, with median
redshift $z_{\rm median}=1.84$. The mutual agreement
between our sample and that from the literature provides an
independent confirmation that our new Bayesian method leads to
constraints on \gpm consistent with previous studies also adopting
the same calibrations.  With the combined sample we
are able to derive the MZR at this redshift, spanning three orders of magnitude in stellar
mass.

We fit the following functional form to this sample of 38 galaxies in order to derive the
MZR, \ie,
\begin{equation}
    \oh = \alpha + \beta\log\(\Mstar/\Msun\) + N(\sigma),
\end{equation}
where $\alpha$, $\beta$, and $\sigma$ represent the intercept, the slope and the intrinsic
scatter, respectively.
The Python package \linmix\footnote{available at \url{https://github.com/jmeyers314/linmix}}
was employed to perform a Bayesian linear regression following the method proposed by \citet{Kelly:2007bv}. The median values and
1-$\sigma$ uncertainties for these three parameters drawn from the marginalized posteriors are
\begin{equation}
    \alpha=4.71^{+0.62}_{-0.64},~~\beta=0.40^{+0.07}_{-0.07},~~\sigma=0.03^{+0.02}_{-0.01}.
\end{equation}
The resulting MZR together with its 1-$\sigma$ confidence region is shown in
Figure~\ref{fig:MZR}.

\begin{figure}
    \centering
    \includegraphics[width=.5\textwidth]{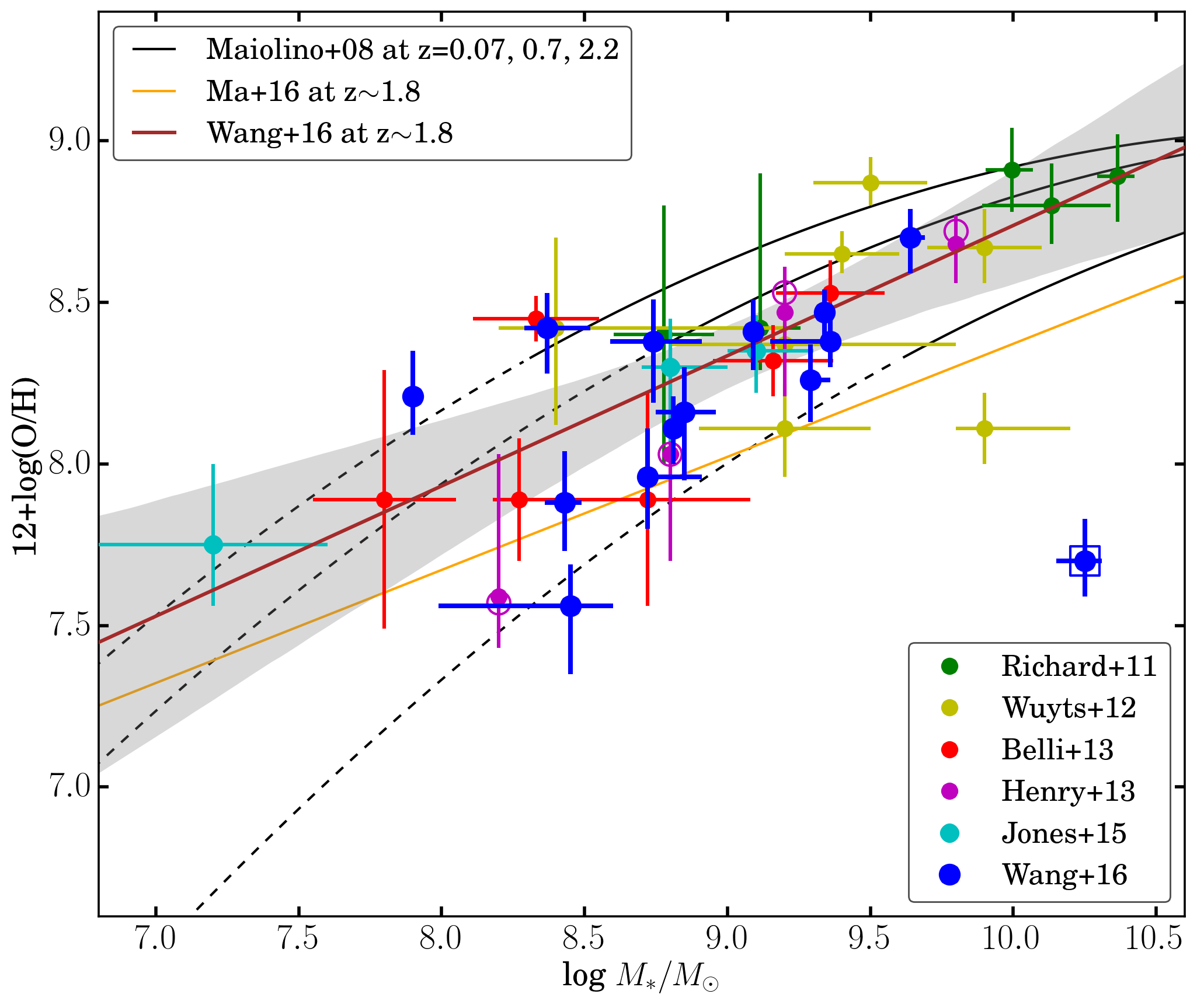}
    \caption{The relation between stellar mass and gas-phase oxygen abundance, from
    observations and simulations at $z\sim1.8$. The color-coding for all the points is the
    same as in Figure~\ref{fig:SFMS}. For the sake of consistency, all metallicity
    measurements are derived assuming the \citet{2008A&A...488..463M} strong EL calibrations.
    The black curves represent the 2nd order polynomials
    fitted by \citet{2008A&A...488..463M} to data sets at different redshifts, where dashed
    parts correspond to extensions beyond mass completeness limits of those data sets.  The
    thick brown line denotes our best-fit linear relation based upon all data
    points except the AGN candidate (\ie ID 02607) marked by a square.
    The shaded band marks the 1-$\sigma$ confidence region.  For the \citet{Henry:2013gx}
    stack data points, we also show the measurements without dust correction as open circles,
    which are not included in the fit.
    The thin orange line is the prediction from the cosmological zoom-in simulations
    conducted by \citet{Ma:2016gw} at the same redshift range, which is consistent with our
    measurement at 1-$\sigma$.}
    \label{fig:MZR}
\end{figure}

Recently, \citet{Guo:2016wk} presented a comprehensive study of the MZR and its scatter at
$z$=0.5-0.7 from data acquired by large spectroscopic surveys in the \candels fields
\citep{Grogin:2011hx,Koekemoer:2011br}. \citet{Guo:2016wk} also summarized a variety of
theoretical predictions of the MZR slope predicted by diverse approaches, including those
predicted by the energy-driven wind ($\sim$0.33) and the momentum-driven wind ($\sim$0.17)
equilibrium models proposed by \citet{2008MNRAS.385.2181F,2012MNRAS.421...98D}.
Our inferred slope is only marginally compatible with the prediction given by the
energy-driven wind model at 1-$\sigma$ confidence level, whereas strongly disfavors the
momentum-driven wind model. Considering the majority of the galaxies in our sample have
relatively lower masses, our result confirms the finding by \citet{Henry:2013gx} that
momentum-driven winds cannot be the primary workhorse that shapes the MZR in the low-mass
regime (below 10$^{9.5}$ \Msun) at $z\gtrsim1$.

Interestingly, our result is consistent within 1-$\sigma$ with the formula of the MZR
evaluated at $z=z_{\rm median}$, \ie, $\oh = 0.35 \log\(\Mstar/\Msun\) + 4.87$, prescribed by
\citet{Ma:2016gw}, using the FIRE (Feedback in Realistic Environment) cosmological zoom-in
simulations \citep{GalaxiesonFIREFe:2014dn}.  This suite of simulations reaches a spatial
resolution as high as 1-10 pc, 1-2 orders of magnitude better than that in conventional
large-volume cosmological hydrodynamic simulations \citep[\eg][]{2013MNRAS.434.2645D}, and
enable a more realistic and self-consistent treatment of multiphase ISM, star formation,
galactic winds and stellar feedback, than those in the semi-analytic models \citep[see][for a
recent review]{2015ARA&A..53...51S}.  As a result, these zoom-in simulations are capable of
reproducing many observed relations, \eg, the galaxy stellar mass functions, the evolution of
cosmic SFR density and specific SFR (sSFR). The consistency between the shapes of our MZR and
that by \citet{Ma:2016gw} indicates that small-scale physical processes are important in
producing the cumulative effects of galactic feedback, and hence high resolution simulations
are needed for accurate results.

Aiming at testing the validity of the FMR using the M08-based sample at $z\sim1.84$ compiled
in this work, we calculate the predicted values for metallicity from the measurements of
\Mstar and SFR, according to Eq.~(2) in \citet{Mannucci:2011be}, who extended the FMR to
masses down to $10^{8.3}\Msun$.
Figure~\ref{fig:FMR} shows the difference between the FMR-predicted
values and the actual measurements of metallicity.  We find that the entire
M08-based sample is consistent with the FMR given the measurement
uncertainties and intrinsic scatter. The median offset for the entire
sample is 0.07 dex, with median uncertainty 0.14 dex.
The median offset and uncertainty for our galaxy sample analyzed in this work are $-0.07$ dex and 0.12 dex, respectively. Among
other individual datasets, the only one that shows a
marginally significant deviation from the FMR is that by
\citet{Wuyts:2012gb}, which has median offset 0.22 dex and median
uncertainty 0.11 dex. We speculate that it is due to the fact that the
\citet{Wuyts:2012gb} dataset is the only one in our compiled M08-based
sample which relies solely upon the \NII/\Ha flux ratio, which can be
a biased tracer of metallicity at high-$z$ \citep{2015ApJ...801...88S,Sanders:2015gk}.
This result reiterates the necessity of combining multiple EL flux ratio diagnostics
simultaneously in order to suppress the systematics associated with local calibrations in the
accurate measurement of high-$z$ metallicity.

\begin{figure}
    \centering
    \includegraphics[width=.5\textwidth]{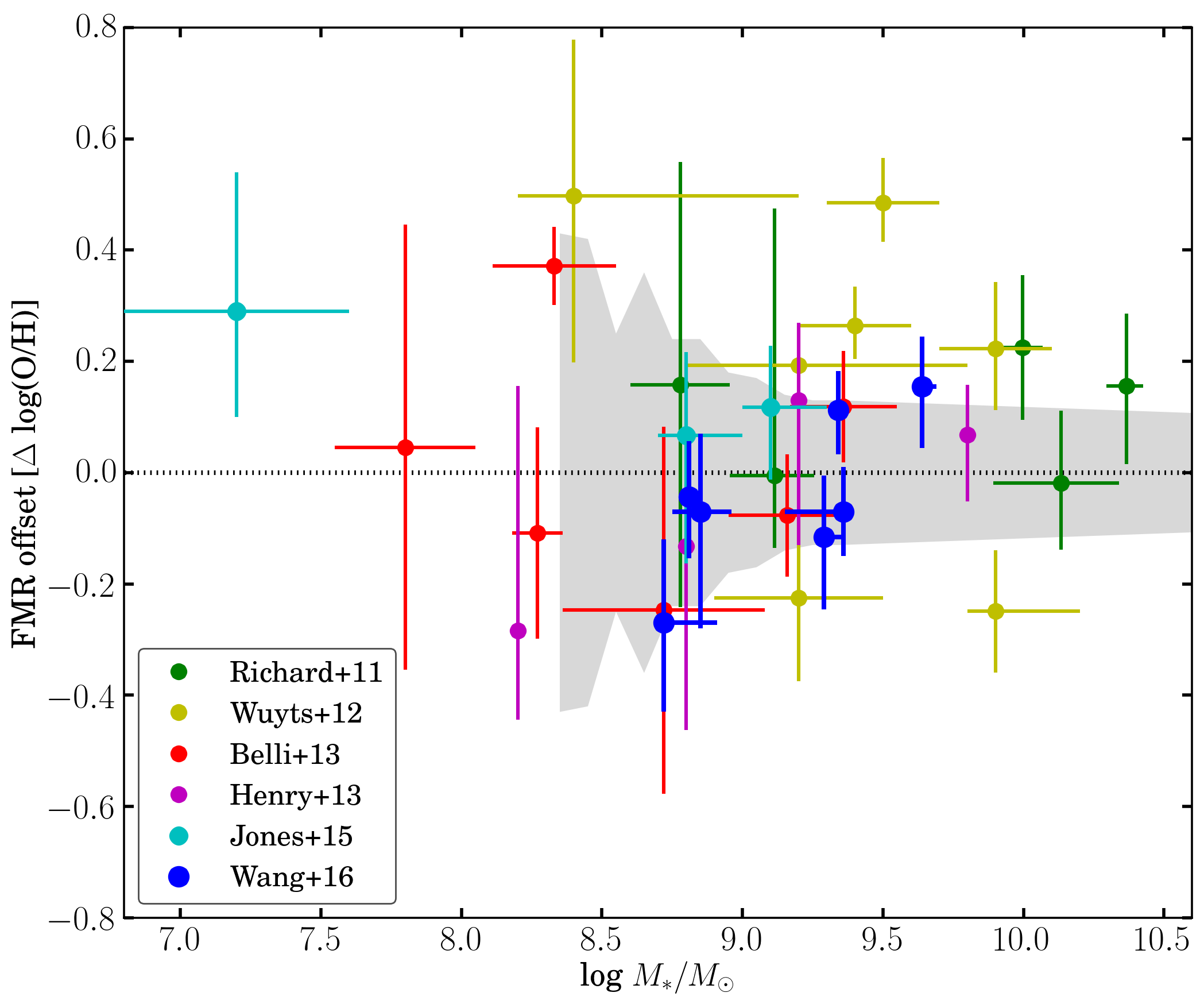}
    \caption{The offset from the fundamental metallicity relation first proposed by
    \citet{2010MNRAS.408.2115M} for the $z\sim1.8$ metallicity measurements derived assuming
    the \citet{2008A&A...488..463M} calibrations.  The color-coding for the points is the
    same as in Figure~\ref{fig:SFMS}.  The shaded region shows the intrinsic scatter of the
    FMR extended to low-mass regime given by
    \citet{Mannucci:2011be}.}
    \label{fig:FMR}
\end{figure}

\section{Spatially resolved analysis}\label{sect:sra}

In this section, we present and discuss in depth the new and unique results obtained from the spatially resolved analysis of our
sample, beyond the reach of the conventional integrated measurements in Section~\ref{sect:global}.
The high spatial resolution maps of \gpm from \hst grism
spectroscopy are described in Sect~\ref{subsect:oh12grad}, and the maps of EL kinematics
undertaken with ground-based IFU spectrographs are presented in Section~\ref{subsect:kinem}.
We make notes on individual galaxies in Section~\ref{subsect:indvd}.

\subsection{Gas-phase metallicity maps at sub-kpc resolution}\label{subsect:oh12grad}

%= = = = = = = = = = = = = = = = = = = = = = = = = = = = = = = = = = = = = = = =
% = = = = = = = = = = = = = = = = = = = = = = = = = = = = = = = = = = = = = = = = = =
% Include this table with \input{filename.tex}
% To rotate in emulateapj do: \begin{turnpage}\input{filename.tex}\end{turnpage}
% To display it on multiple pages do: \LongTables\input{filename.tex}
% - - - - - - - - - - - - - - - - - - - - - - - - - - - - - - - - - - - - - - - - - -
\begin{deluxetable}{ccccccccccccc} \tablecolumns{13}
\tablewidth{0pt}
\tablecaption{Gas-phase metallicity gradients measured by two different methods.}
% - - - - - - - - - - - - - - - - - - - - - - - - - - - - - - - - - - - - - - - - - -
\tablehead{
    \colhead{ID} &
    \multicolumn{2}{c}{Metallicity gradient [dex/kpc]} \\
    &
    \colhead{Individual spaxel} &
    \colhead{Radial annulus}
}
%---------------------------------------------------------------
\startdata
 01422  &  0.02  $\pm$ 0.08   &    0.06  $\pm$  0.04  \\
 02389  & -0.07  $\pm$ 0.04   &   -0.13  $\pm$  0.03  \\
 02806  & -0.01  $\pm$ 0.02   &    0.04  $\pm$  0.02  \\
 03746  & -0.03  $\pm$ 0.03   &    0.04  $\pm$  0.02  \\
 04054  & -0.04  $\pm$ 0.02   &   -0.07  $\pm$  0.02  \\
 05709  & -0.22  $\pm$ 0.05   &   -0.19  $\pm$  0.06  \\
 05732  &  0.06  $\pm$ 0.05   &    0.08  $\pm$  0.02  \\
 05811  & -0.18  $\pm$ 0.08   &   -0.40  $\pm$  0.07  \\
 05968  & -0.01  $\pm$ 0.02   &    0.02  $\pm$  0.01  \\
 07255  & -0.16  $\pm$ 0.03   &   -0.21  $\pm$  0.03
\enddata
% - - - - - - - - - - - - - - - - - - - - - - - - - - - - - - - - - - - - - - - - - -
\label{tab:oh12grad}
\end{deluxetable}

%01422  &  0.03 $\pm$  0.08   &   0.11  $\pm$   0.05    \\
%02389  & -0.02 $\pm$  0.05   &  -0.14  $\pm$   0.07    \\
%02806  & -0.01 $\pm$  0.01   &  -0.02  $\pm$   0.02    \\
%02918  & -0.15 $\pm$  0.06   &  -0.35  $\pm$   0.07    \\
%03746  & -0.03 $\pm$  0.03   &   0.04  $\pm$   0.03    \\
%04054  & -0.05 $\pm$  0.02   &  -0.08  $\pm$   0.03    \\
%05709  & -0.12 $\pm$  0.06   &  -0.36  $\pm$   0.09    \\
%05732  & -0.01 $\pm$  0.09   &   0.08  $\pm$   0.06    \\
%05811  & -0.26 $\pm$  0.07   &  -0.24  $\pm$   0.05    \\
%05968  & -0.01 $\pm$  0.03   &   0.01  $\pm$   0.01    \\
%07255  & -0.16 $\pm$  0.03   &  -0.20  $\pm$   0.03

%= = = = = = = = = = = = = = = = = = = = = = = = = = = = = = = = = = = = = = = =

To estimate the spatial distribution of the constrained parameters we apply the analysis
described in Section~\ref{sect:global} to each individual spaxel in the EL maps.
In Figure~\ref{fig:oh12grad}, we show the maps of best-fit
metallicity and the corresponding conservative uncertainty (the larger side of asymmetric
error bars given by the Bayesian inference) measured for galaxies in our \mg sample.  We bin
spaxels 2 by 2 to regain the native WFC3/IR pixel scale.
In the maps, we only include spaxels with at least one EL
(primarily \Ha or \OIII) detected with $\geq5\sigma$ significance. The \mgs measured from
these individual spaxels are shown in the right column of Figure~\ref{fig:oh12grad}. We also
measure \mgs using another independent method, \ie, via radial binning.  The range of each
radial annulus is determined by requiring its SNR $\gtrsim15$.  In
order to avoid biasing toward low metallicities, we put the SNR threshold on \Ha whenever
available (\ie $z\lesssim1.5$) instead of on \OIII, since the latter correlates tightly with
\oh, in the lower branch of \OIII/\Hb, where the majority of our sample resides.  Based upon
our SNR threshold criterion, three sources in our sample, IDs 02918, 05422 and 07058, do not
have enough spaxels to constitute a trustworthy metallicity map, and therefore were
removed from the \sra.  So in total, we present 10 star-forming galaxies whose \mgs (in unit of dex/kpc) are measured. Linear fits
to the radial gradients are given in Table~\ref{tab:oh12grad}. We note that the metallicity maps are not always symmetric around
the center of stellar mass. Therefore, while gradients are a useful statistic to describe the overall behavior and to compare with
numerical simulations, they do not contain all the available information about the apparent diversity of morphologies.  Thus, it
is helpful to keep the 2D maps in mind while interpreting the gradients.

\begin{figure*}
    \centering
    \includegraphics[width=\textwidth]{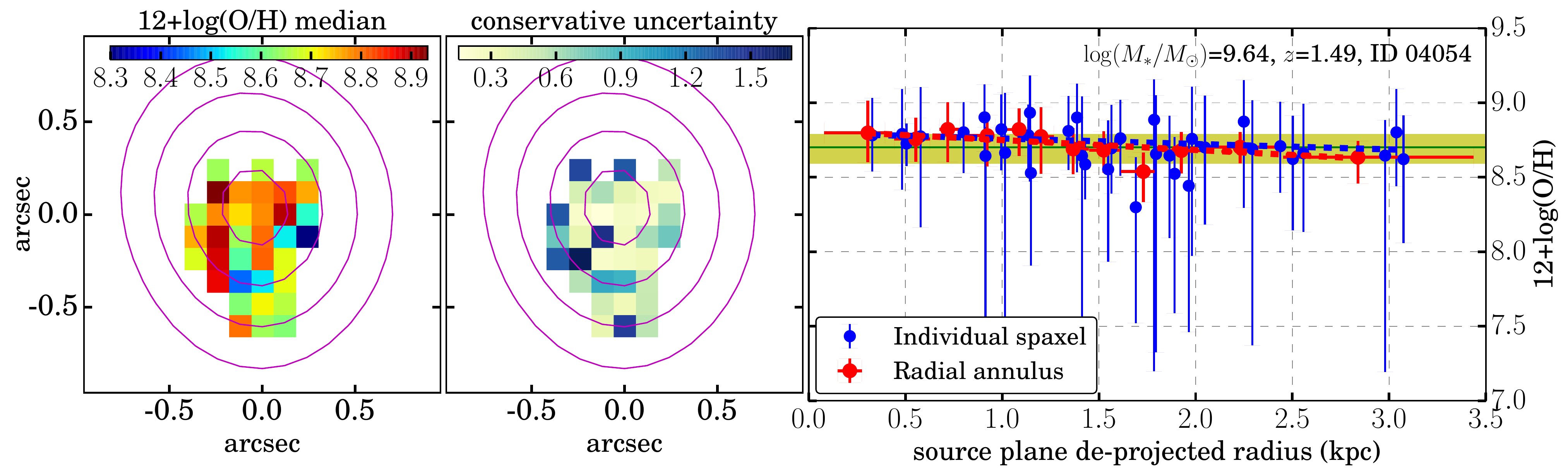}
    \includegraphics[width=\textwidth]{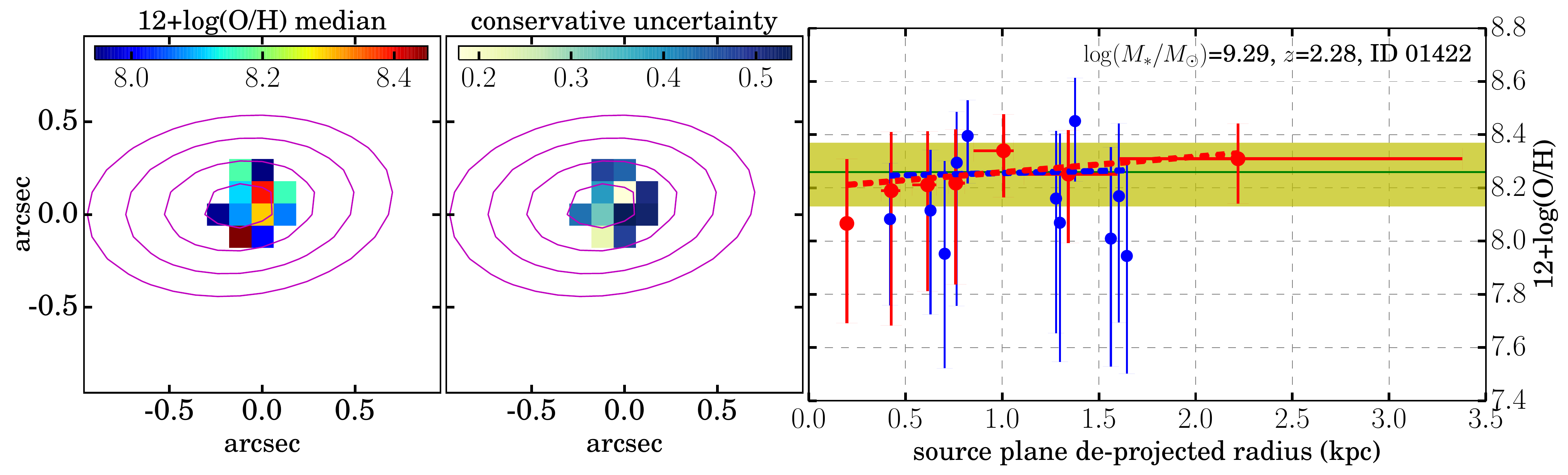}
    \includegraphics[width=\textwidth]{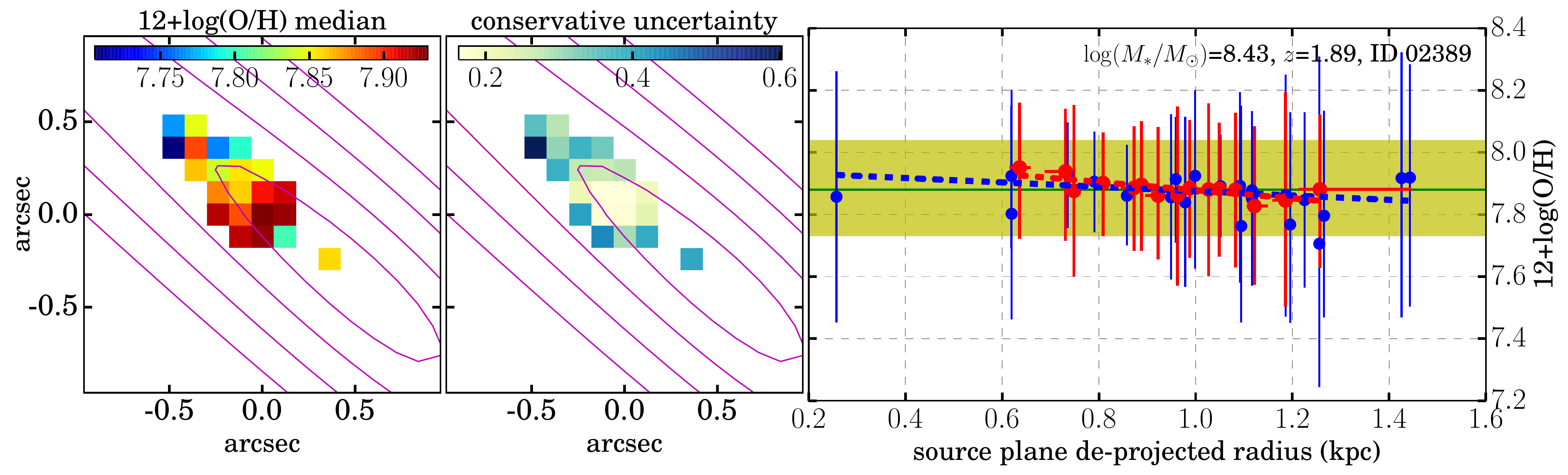}
    \caption{Maps of metallicity constraints (median value in the left column and
    conservative uncertainty --- the larger side of asymmetric 1-$\sigma$ error bars --- in
    the center column) and plots of radial metallicity gradients in the right column. Note
    that unlike what we show in the combined EL maps (Figures~\ref{fig:multiP_4054} and
    \ref{fig:multiP_rest}), here the metallicity maps are rebinned to a scale of 0$\farcs$13,
    corresponding to the native resolution of WFC3/IR. In the maps, only spaxels with the
    signal-to-noise ratio of \Ha or \OIII larger than 5 are shown.  The same source plane
    de-projected galactocentric radii that are denoted by black contours in
    Figures~\ref{fig:multiP_4054} and \ref{fig:multiP_rest} are represented by magenta
    contours, with the only difference being that contours are in 1 kpc intervals now. In the
    panels in the right column, blue and red points correspond to metallicity measurements
    for individual spaxels and radial annuli, respectively.  The radial gradients derived
    based upon these measurements are shown by the dashed lines in corresponding colors.  The
    yellow band and green horizontal line mark the global constraint on \oh, from integrated
    line fluxes.}
    \label{fig:oh12grad}
\end{figure*}

\begin{figure*}
    \centering
    \includegraphics[width=\textwidth]{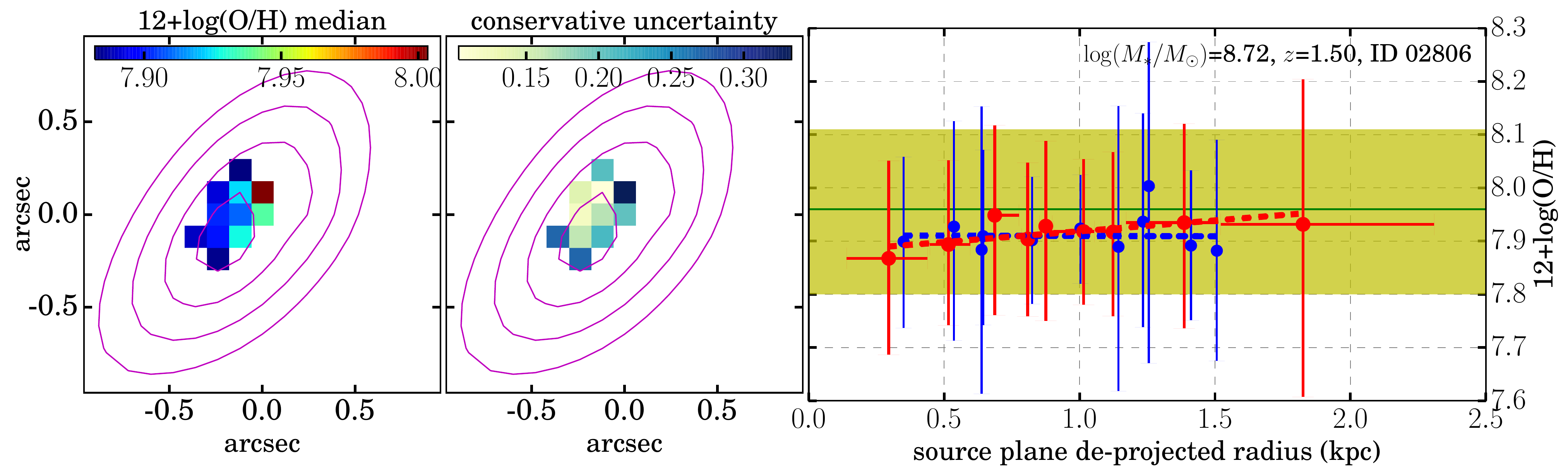}
    \includegraphics[width=\textwidth]{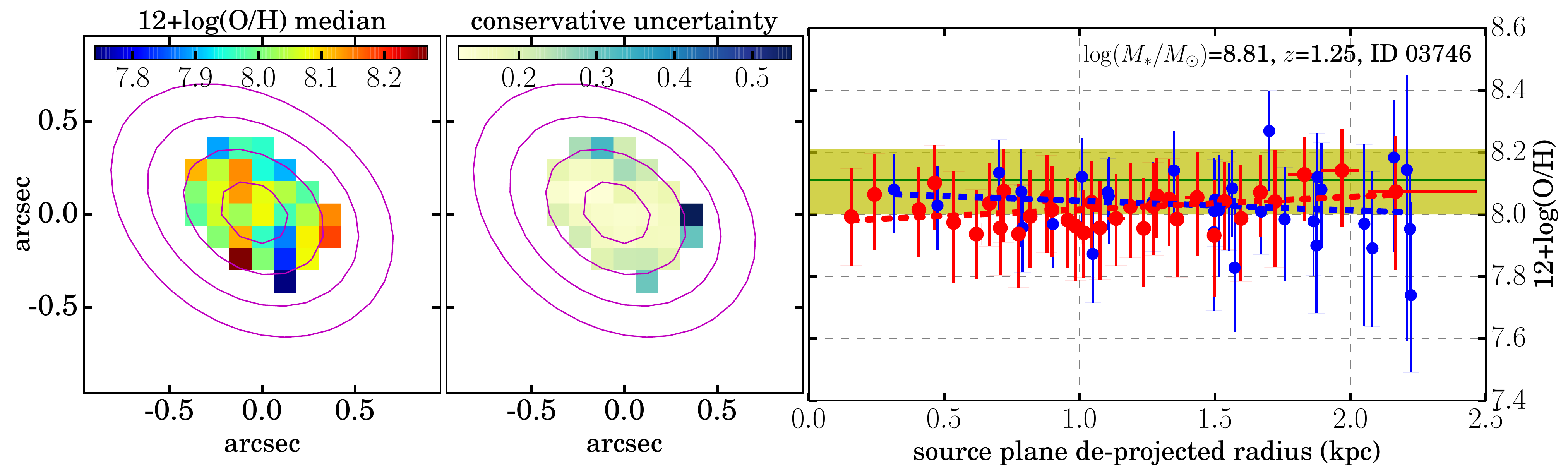}
    \includegraphics[width=\textwidth]{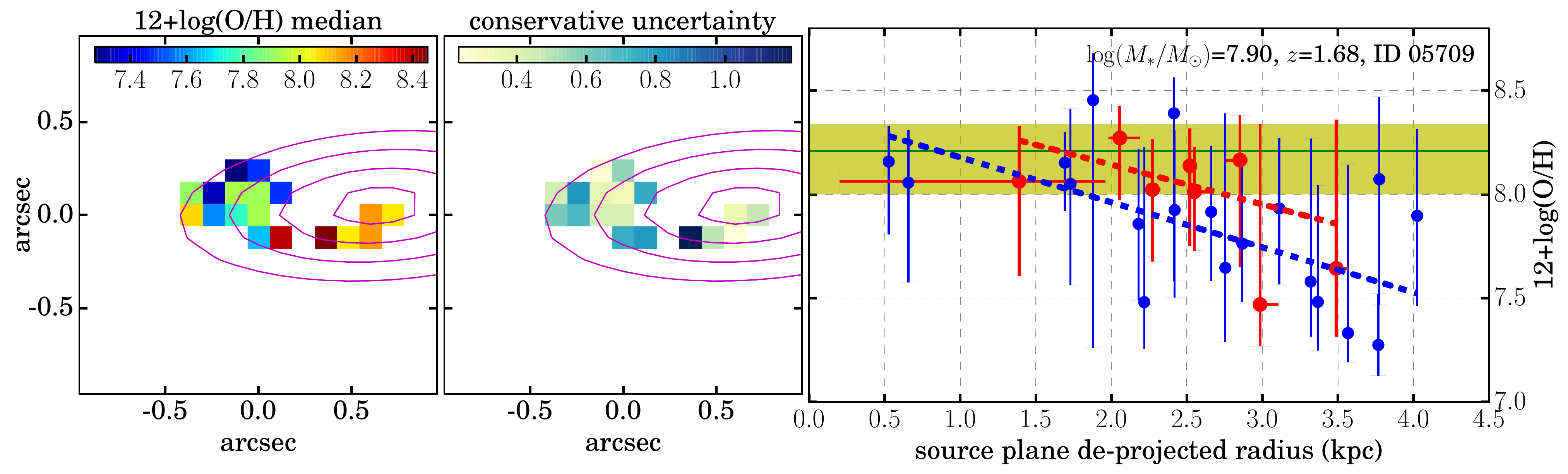}
    \includegraphics[width=\textwidth]{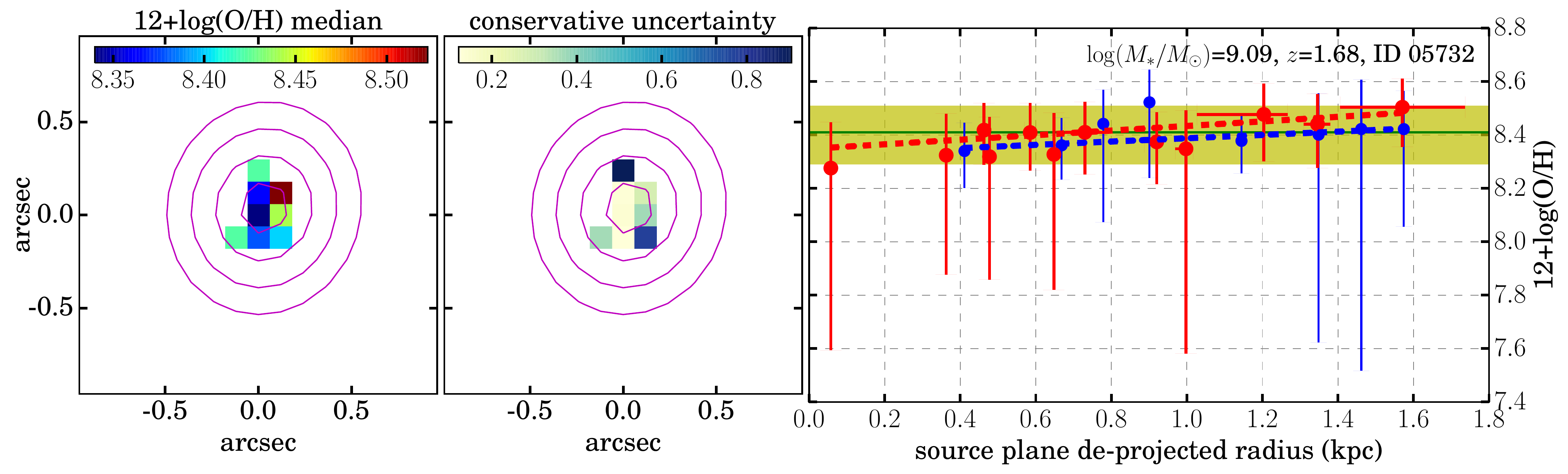}
    \contcaption{(cont.)}
\end{figure*}

\begin{figure*}
    \centering
    \includegraphics[width=\textwidth]{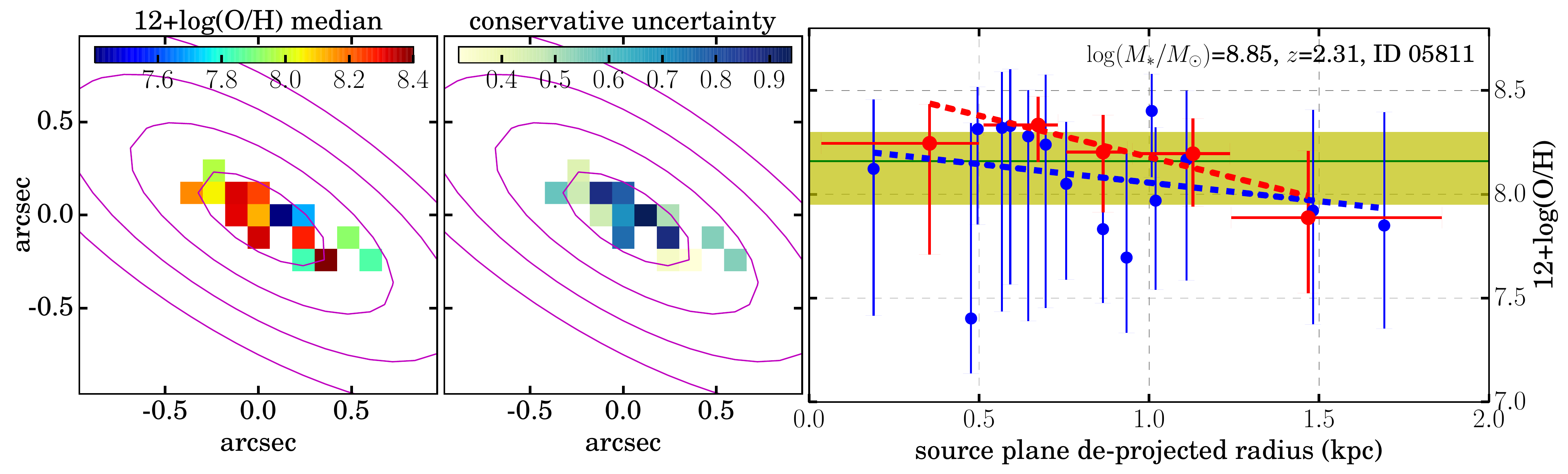}
    \includegraphics[width=\textwidth]{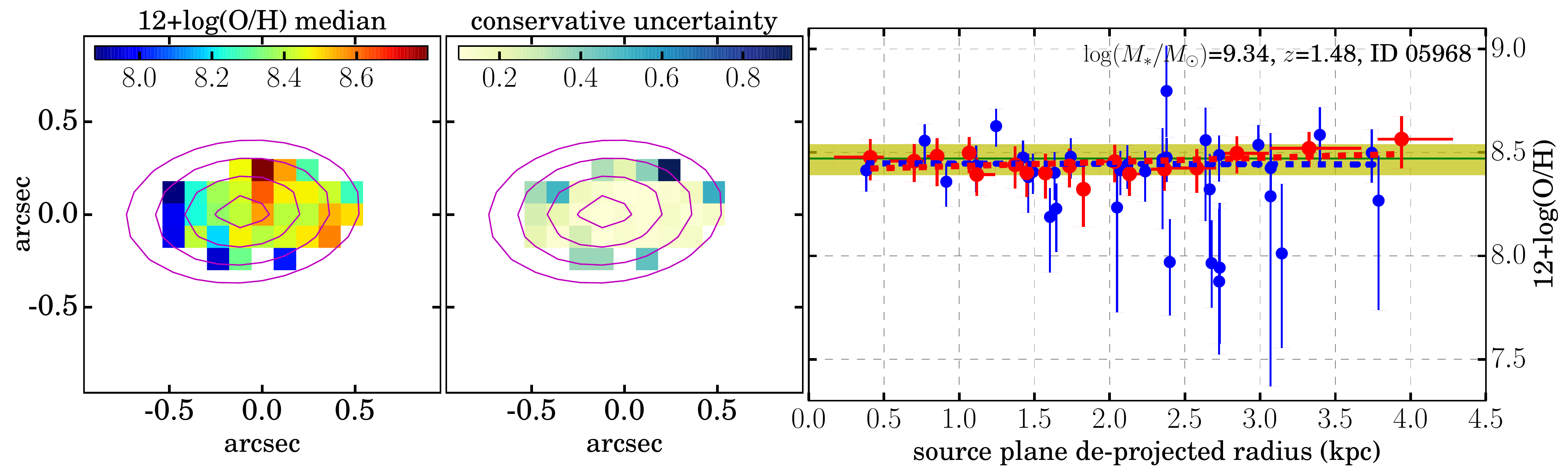}
    \includegraphics[width=\textwidth]{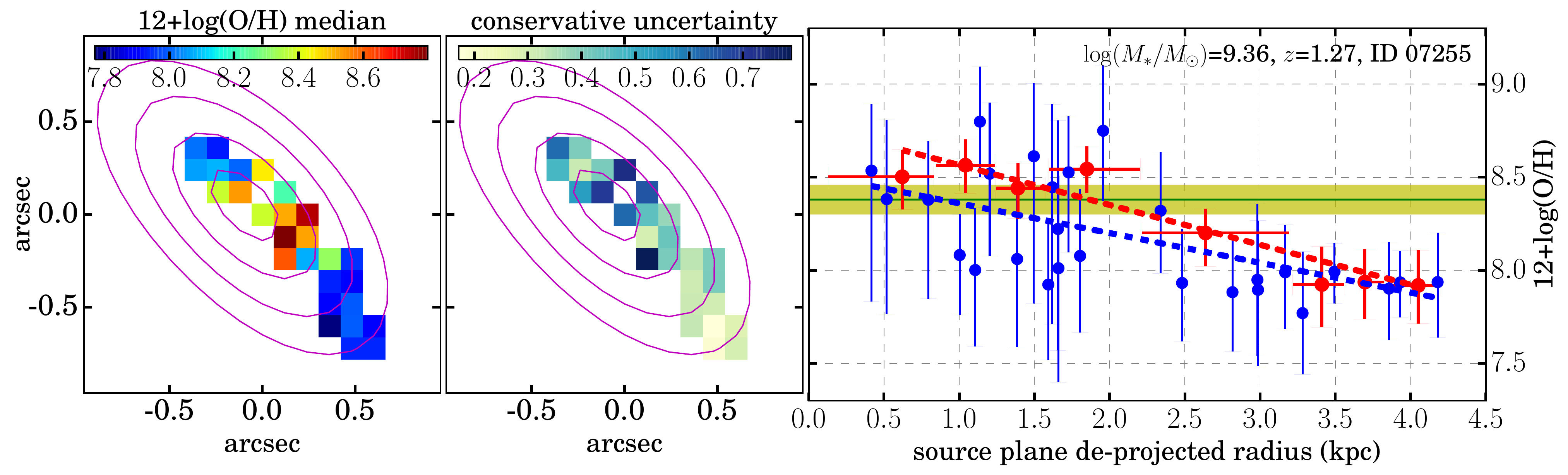}
    \contcaption{(cont.)}
\end{figure*}

As a result, seven galaxies have gradients consistent with being flat at 2-$\sigma$, among which three show almost uniformly
distributed metals (IDs 02806, 03746 and 05968), two have marginally positive gradients (IDs 01422 and 05732), and the other two
display mildly negative gradients (IDs 02389 and 04054). Apart from these, the other three galaxies in our sample have very steep
negative gradients: IDs 07255, 05709 and 05811. In particular, after the lensing magnification correction, the stellar mass of
galaxy ID 05709 is estimated to be $\sim10^{7.9}$ \Msun. This is for the first time sub-kpc scale spatially resolved analysis has
been done on such low-mass systems at high-$z$.

In order to verify that our results are not contaminated by ionizing radiation from AGN, we
examine spatially resolved BPT diagrams. This approach is only possible for sources at
$z\lesssim1.5$, where \Ha is detectable. Furthermore, due to the low spectral resolution of
\hst grisms, we slightly modify the BPT diagram to plot flux ratio of \OIII/\Hb as a function
of \SII/\Ha+\NII, using the ``vanilla'' value for \NII/\Ha. As Figure~\ref{fig:BPT} shows, we
do not identify any strong correlation between source plane galactocentric radius and
deviation from the \HII region loci, suggesting that there is no AGNs hidden in the center of
these galaxies.
Moreover, the integrated EL fluxes of all three galaxies lie in close proximity to the
extreme starburst model prescribed by \citet{Kewley:2006ib}, which confirms our working assumption that AGN contributions are
negligible for our sample, except in one case (\ie ID 02607, see Figure~\ref{fig:MEx}).

\begin{figure*}
    \centering
    \includegraphics[width=.33\textwidth]{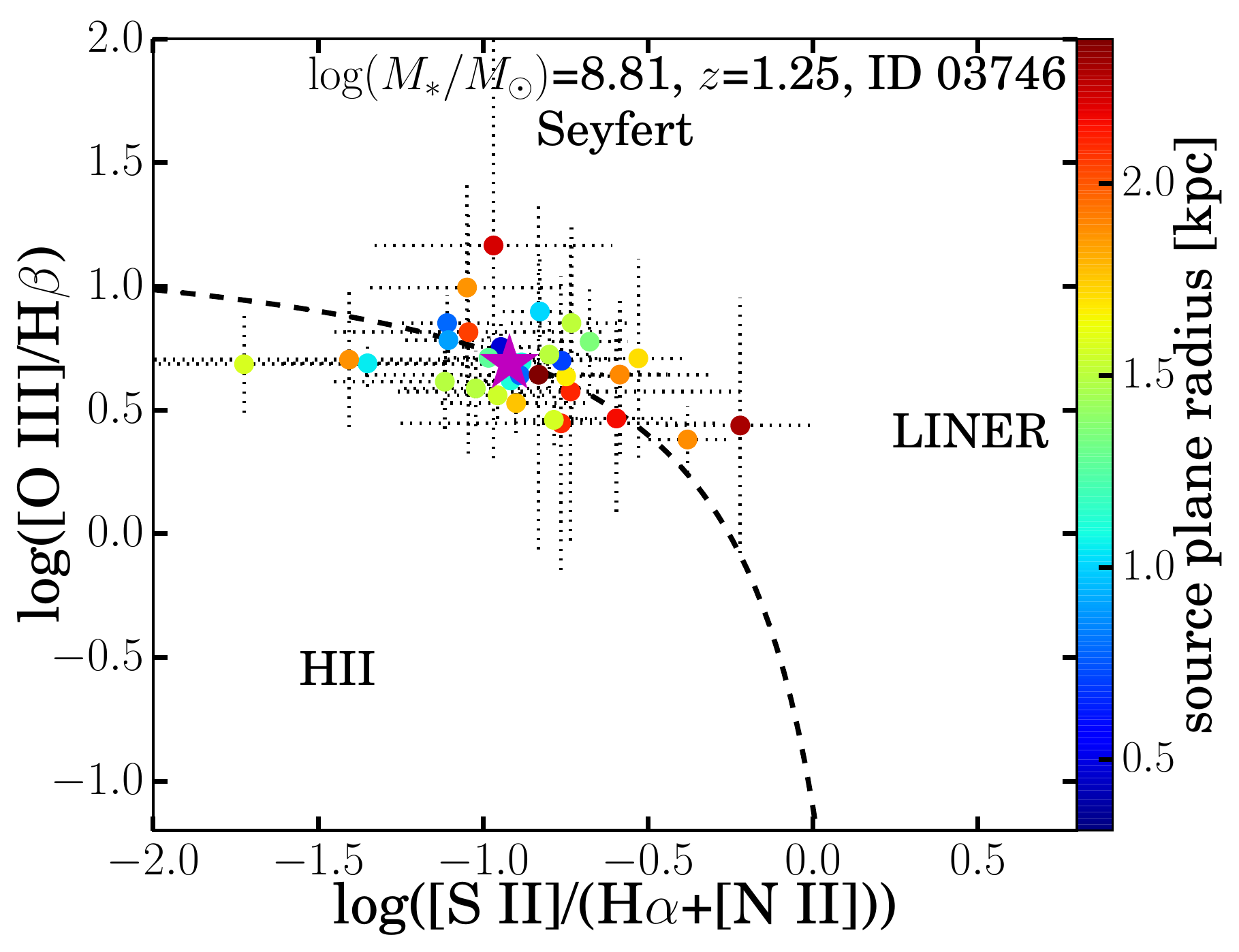}
    \includegraphics[width=.33\textwidth]{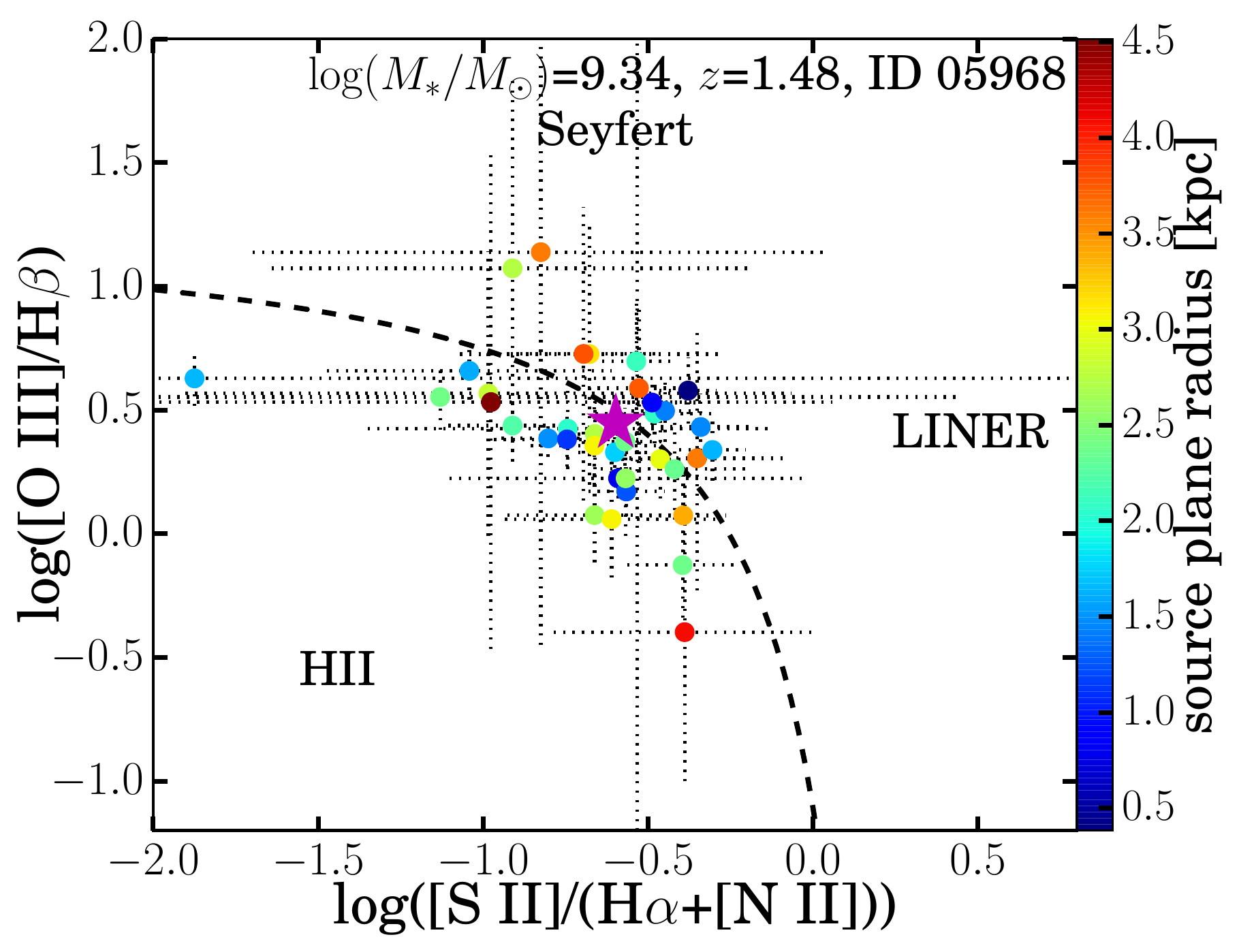}
    \includegraphics[width=.33\textwidth]{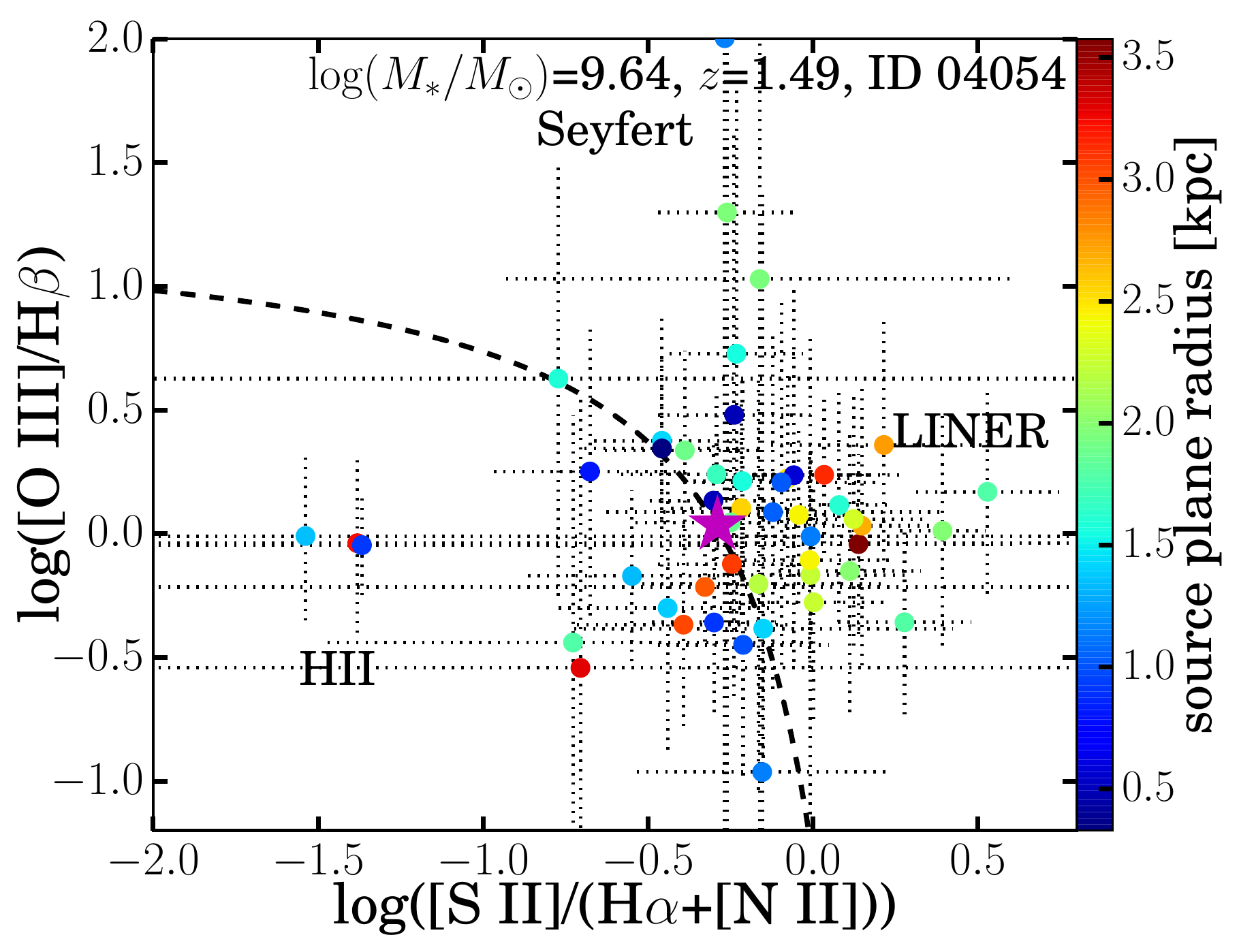}
    \caption{Spatially resolved BPT diagrams for three representative sources in our \mg
    sample.  The points in each panel correspond to spaxels in the combined EL maps for each
    source, color coded by the source plane de-projected galactocentric radius.  Note that
    similar to what we show in Figure~\ref{fig:oh12grad}, all spaxels represented by colored
    points here are rebinned to recover the native WFC3/IR pixel scale (0$\farcs$13).
    The magenta star denotes the position where the entire galaxy would lie, calculated from
    integrated EL fluxes, with the length of the errorbar comparable to the size of the
    symbol. The dashed curve is adapted from the extreme starburst scenario given by
    \citet{Kewley:2006ib}, assuming the ``vanilla'' value for \NII/\Ha.
    Given the 1-$\sigma$ uncertainties, all the spaxels are broadly consistent with being
    \HII regions.}
    \label{fig:BPT}
\end{figure*}

\subsubsection{Cosmic evolution of metallicity gradients}

\begin{figure*}
    \centering
    \includegraphics[width=\textwidth]{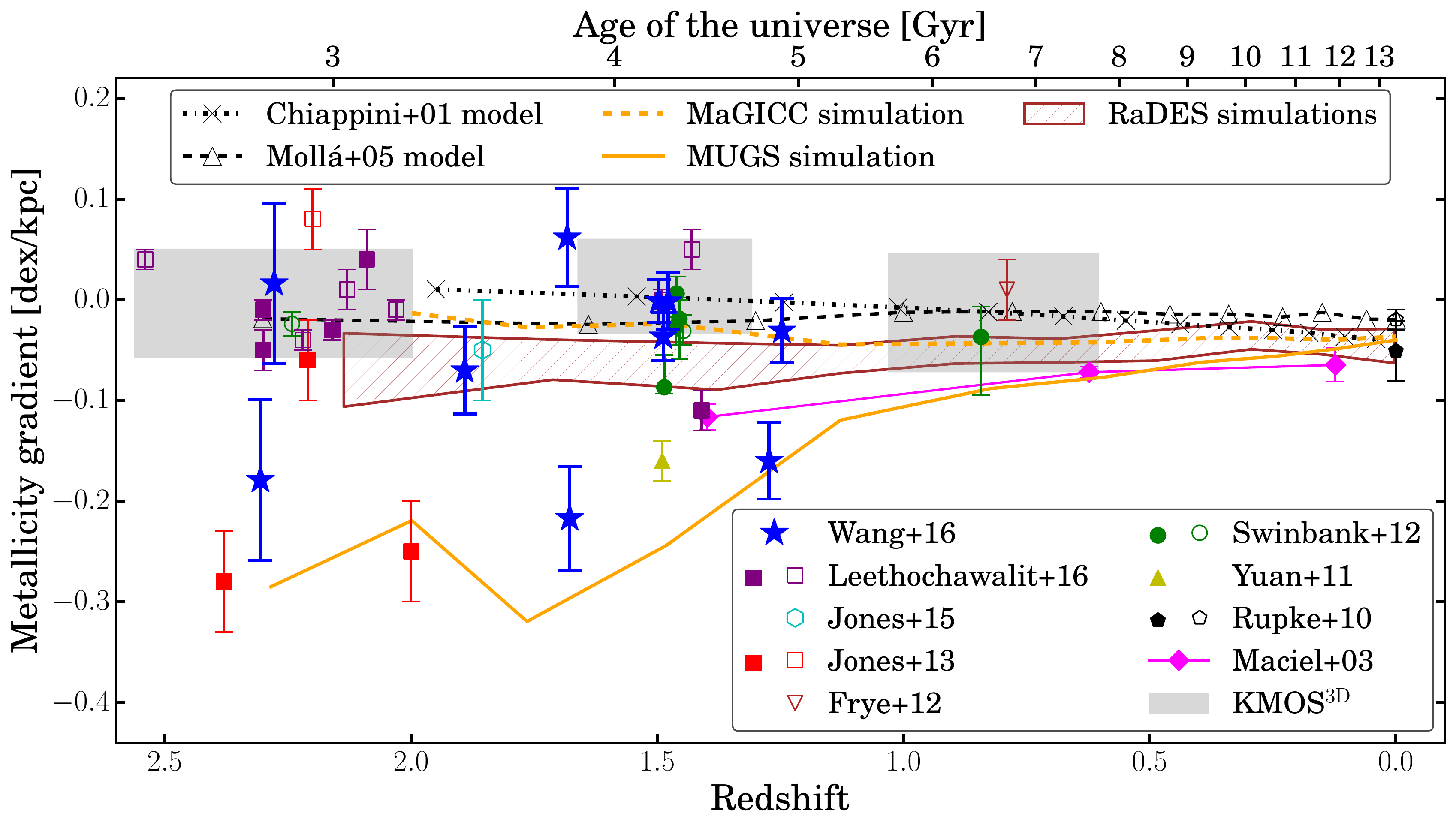}
    \caption{Evolution of metallicity gradients with redshift. The blue stars represent
    the \mgs measured in this work, and the cyan hexagon is from our previous work
    \citep{2015AJ....149..107J}. We include all the high-$z$ \mgms
    with sub-kpc resolution in the current literature: lensed galaxies analyzed by
    \citet{Yuan:2011hj}, \citet{Frye:2012dr},
    \citet{2013ApJ...765...48J}, and \citet{2015arXiv150901279L}, non-lensed
    galaxies observed with AO \citep{2012MNRAS.426..935S}.
    We also show the spread of the \kd measurements obtained with seeing-limited conditions
    \citep{Wuyts:2016th},
    the averages of local \mgs \citep{Rupke:2010cg}, and a trend of
    the Milky Way gradient evolution based upon planetary nebula estimates
    \citep{Maciel:2003kv}.
    Except for the blue stars, hollow and filled symbols correspond to interacting/merging
    and isolated systems, respectively.
    In addition, the predictions from different analytical chemical evolution models and
    numerical simulations are also shown as comparisons to the observational results (see
    Section~\ref{subsect:oh12grad} for more details).}
    \label{fig:oh12gradVSz}
\end{figure*}

We plot the evolution of metallicity gradient across cosmic time in
Figure~\ref{fig:oh12gradVSz}. Here we include all existing high-$z$ gradient measurements
obtained with sufficiently high spatial sampling, \ie finer than \kpc resolution in the
source plane. Together with the 10 new \mgms from the individual spaxel method presented
here, in Figure~\ref{fig:oh12gradVSz}, we show one highly magnified galaxy of an interacting
triple at $z$=1.85 analyzed in our previous work \citep{2015AJ....149..107J}.  We also
include 11 measurements by \citet{2015arXiv150901279L} (5 isolated and 6 merging), 4 by
\citet{2013ApJ...765...48J} (3 isolated and 1 merging), 1 sub-\Lstar post-merger at $z\sim1$
by \citet{Frye:2012dr}, 1 isolated grand-design spiral at $z$=1.49 by \citet{Yuan:2011hj},
and 7 data points from \citet{2012MNRAS.426..935S} (5 isolated and 2 interacting). The
scatter of these high-resolution gradient measurements is $\sim$0.12 dex and $\sim$0.22 dex
at $z\sim1.5$ and $z\sim2.3$ respectively.  The spread of recent gradient measurements from
the seeing-limited \kd survey is also overlaid in Figure~\ref{fig:oh12gradVSz}.  Note that
the PSF FWHM given by the median seeing of their observations is 0\farcs6, which results in a
5 \kpc resolution at $z\sim2$, twice the size of the half-light radius of an \Lstar galaxy at
this redshift.  We thus only focus on interpreting the sub-kpc resolution gradient
measurements, in order to avoid possible biases toward null values associated poor spatial
sampling \citep[see, \eg,][]{2013ApJ...767..106Y}.  Although corrected for beam smearing, the
\kd gradient measurements still display small scatter than that of the high resolution
results.  When possible, we also divide each data set in terms of sources being isolated or
kinematically disturbed (\ie undergoing merging).  In part of our sample, signatures of
post-merger tidal remnants can be clearly identified, which strongly indicates that
gravitational interaction plays a key role in shaping the spatial distribution of metals in
these galaxies (IDs 02806, 03746, and 05968).

In order to compare theoretical predictions with observations, we incorporate into Figure~\ref{fig:oh12gradVSz} the predictions
for the \mg cosmic evolution from canonical chemical evolution models based upon the ``inside-out'' disk growth paradigm given by
\citet[][C01]{Chiappini:2001ds} and \citet[][M05]{Molla:2005eq}.  C01 constructed a two-phase accretion model for galaxy
formation.  The first infall period corresponds to the formation of the dark matter halo and the thick disk, with an exponentially
declining infall rate at a fixed $e$-folding time scale ($\sim$1 Gyr).  In this phase, the infall of pristine gas is rapid and
isotropic, giving birth to a relatively extended profile of star formation.  The thin disk is formed in the second phase, where
gas is preferentially accreted to the periphery of the disk, with the $e$-folding time scale proportional to galactocentric radius
(\vsv the ``inside-out'' growth).  M05 proposed their prescription by treating the ISM as a multiphase mixture of hot diffuse gas
and cold condensing molecular clouds, and calibrating against different star-formation efficiencies.  Here we show the high
star-formation efficiency solution.  The common Kennicutt-Schmidt law \citep{Kennicutt:1998id,Schmidt:1959bp} is implemented to
estimate \sfr.  Popular nucleosynthesis prescriptions for type Ia/II SNe and AGB stars are employed by both to construct the
yields of dominant isotopes (\eg H, He, C, N, O, Fe).  As shown in Figure~\ref{fig:oh12gradVSz}, the majority of the high-$z$
measurements are generally compatible with both model predictions, with the M05 model showing better consistency, largely ascribed
to its more accurate assumptions of ISM structure and more realistic treatment of star formation.  Note that these chemical
evolution models do not take galactic feedback or radial gas flows into account.  So without the consideration of strong physical
mechanisms that stir and mix up the enriched gas with relatively unenriched ambient ISM, conventional analytical models are still
able to match the \mgs observed at high redshifts.

In addition to the predictions from these analytical models, we also compare our measurements
with results from numerical simulations.  First, we show the fiducial representative (\ie the
realization g15784) of the McMaster Unbiased Galaxy Simulations
\citep[MUGS,][]{Stinson:2010ix} using the gravitational N-body and SPH code \gasoline.  This
simulation employed the ``conventional'' feedback scheme, \ie depositing $\sim$10-40\%
kinetic energy from SN explosions into heating ambient ISM. Furthermore, it adopted a high
star-formation threshold (gas particle density $>$1 cm$^{-3}$), which makes star forming
activities highly concentrated in the galaxy center at early stages of disk formation.  As a
result, a steep gradient (\ie $\gtrsim-0.2$ dex/kpc) is persisting at $z>1$ and then
significantly flattens with time.  Although greatly offset from most of the measurements (and
other theoretical predictions) as seen in Figure~\ref{fig:oh12gradVSz}, this particular
evolutionary track is actually quite consistent with some of the observed steep gradients
(galaxies 07255, 05709, 05811 in our analysis).  In comparison, the same galaxy (g15784) is
re-simulated with an enhanced feedback prescription of sub-grid physics as part of the Making
Galaxies In a Cosmological Context \citep[MaGICC,][]{Stinson:2013ex} program.  A factor of
2.5 difference in the energy output rate from SN feedback diverges the evolutionary track of
\mgs given by the MUGS and MaGICC simulations, making the evolution of the MaGICC gradient
less dramatic and more similar to the M05 model solution.  This demonstrates that amplifying
the feedback strength can significantly flatten the \mgs in early star-forming galaxies
\citep{Gibson:2013jw}.
This is actually adopted by some recent studies \citep{2015arXiv150901279L,Wuyts:2016th} to
explain their \mg measurements, which are also shown in Figure~\ref{fig:oh12gradVSz}.
However, as we have already mentioned, this is not the only explanation possible.

Lastly, in Figure~\ref{fig:oh12gradVSz}, we also show another set of numerical results, from
an ensemble of 19 galaxies in different environments (9 in isolated field and 10 in loose
groups), simulated by the \ramses Disk Environment Study \citep[RaDES,][]{Few:2012jl} using
the adaptive mesh refinement code \ramses.  Unlike what is shown for the MUGS and MaGICC
simulations (\ie just one single realization), the full range of the whole suite of RaDES
simulations is marked in Figure~\ref{fig:oh12gradVSz}, which provides a statistical
comparison sample, although the scatter of the RaDES simulations cannot capture the scatter
seen in actual measurements.
Generally speaking, the evolutionary trends of the abundance gradients of RaDES galaxies lie
somewhat in-between the two tracks of g15784 in MUGS and MaGICC simulations, but still quite
consistent with the majority of the high-$z$ gradient measurements.
Because no mass loading is assumed in the RaDES simulations, outflows do not play any role in
this observed consistency.
Moreover, since a lower star-formation threshold is used ($>$ 0.1 cm$^{-3}$), which is
similar to what the analytical chemical evolution models assumed, star formation in RaDES
galaxies occurs more uniformly in the early disk formation phase.  This strongly suggests
that having more extended star formation can also result in galaxies possessing shallower
gradients, confirming the tight link between star formation profile and metallicity gradient
\citep{Pilkington:2012ib}.

\subsubsection{Mass dependence of metallicity gradients}

In Figure~\ref{fig:oh12gradVSmstar}, we plot metallicity gradient as a function of stellar mass for a subset of all
gradient measurements shown in Figure~\ref{fig:oh12gradVSz}, where reliable \Mstar estimates are available.
Notably, in the low-mass regime of $\Mstar\lesssim10^9\Msun$, all the current sub-kpc
resolution \mgms at $z\gtrsim1$ are obtained from our work, and our measurements even extend
to below $10^8$ \Msun.
This illustrates the power of combining the space-based grism spectroscopy with
gravitational lensing as a means to recover reliable metal abundance distributions in
early star-forming disks in the low-mass regime.
From these measurements, we can tentatively observe an intriguing correlation between \Mstar and \mg, consistent with the galaxy
formation ``downsizing'' picture \citep[see, \eg,][]{Brinchmann:2004hy,Fontanot:2009fp} that more massive galaxies are more
evolved and their metal distributions flatter since they are in a later phase of disk mass assembly where star formation occurs in
a more coherent mode throughout the disk.
This is also seen (although in a much smaller \Mstar range) in the RaDES simulations that more massive galaxies tend to have
shallower (less negative) gradients \citep{Few:2012jl}.

As a comparison, we include the linear fit to the low spatial resolution \mg measurements
from the \kd survey in the similar $z$ range \citep[see Figure 6 in][]{Wuyts:2016th}.
Because the mass completeness limit for the \kd data set is $\sim$10$^{9.7}$\Msun, beyond
that limit we show the extrapolation of the linear fit.
As we have already seen in Figure~\ref{fig:SFMS}, our analysis is highly complementary to the \kd result in terms of \Mstar
associated with \mgms, albeit at very different spatial resolution.
However, the low-mass extension of the \kd fitting relation systematically deviates from
our gradient measurements (especially the ones below $10^9\Msun$), and is also in contrast with the ``downsizing'' picture.
This can be attributed to possible biases associated with the low spatial resolution of the \kd observations and the small mass
range of their sample.

We also show the 2-$\sigma$ spread of the galaxy \mgs from the FIRE cosmological zoom-in simulations \citep{Ma:2016ww}.
This suite of simulations has been used to study the MZR as we discussed in Section~\ref{sect:global} \citep{Ma:2016gw}.
The simulations show a diversity of gradient behaviors at four redshift epochs ($z$=0, 0.8, 1.4, 2.0). At $z=2$, the scatter of
their simulation results can reach 0.2 dex when radial gradients are measured in the central 2 kpc region, quite consistent with
the scatter ($\sim0.22$ dex) seen in high-resolution gradient measurements at $z\sim2.3$.
However, their simulations still cannot reproduce some of the steep gradient measurements, especially some of our results in the
low-mass regime.
According to \citet{Ma:2016ww}, galactic feedback from intense starburst episodes on 0.1-1 Gyr time scale can effectively flatten
\mgs.
Hence for the galaxies measured with steep \mgs, galactic feedback must be of limited influence, at least in affecting metal
distribution in \HII regions on such short time scale.

In short, we observe a tentative mass dependence of \mgs that low-mass galaxies have steeper gradients compared with high-mass
galaxies at high redshifts. A larger set of accurately measured and uniformly analyzed \mgs well sampled in \Mstar (in the regime
of $\Mstar\lesssim10^9\Msun$ in particular) is needed in order to fully explore the validity of the \mg ``downsizing'' picture and
whether galactic feedback plays a crucial role in altering this picture.

\begin{figure*}
    \centering
    \includegraphics[width=\textwidth]{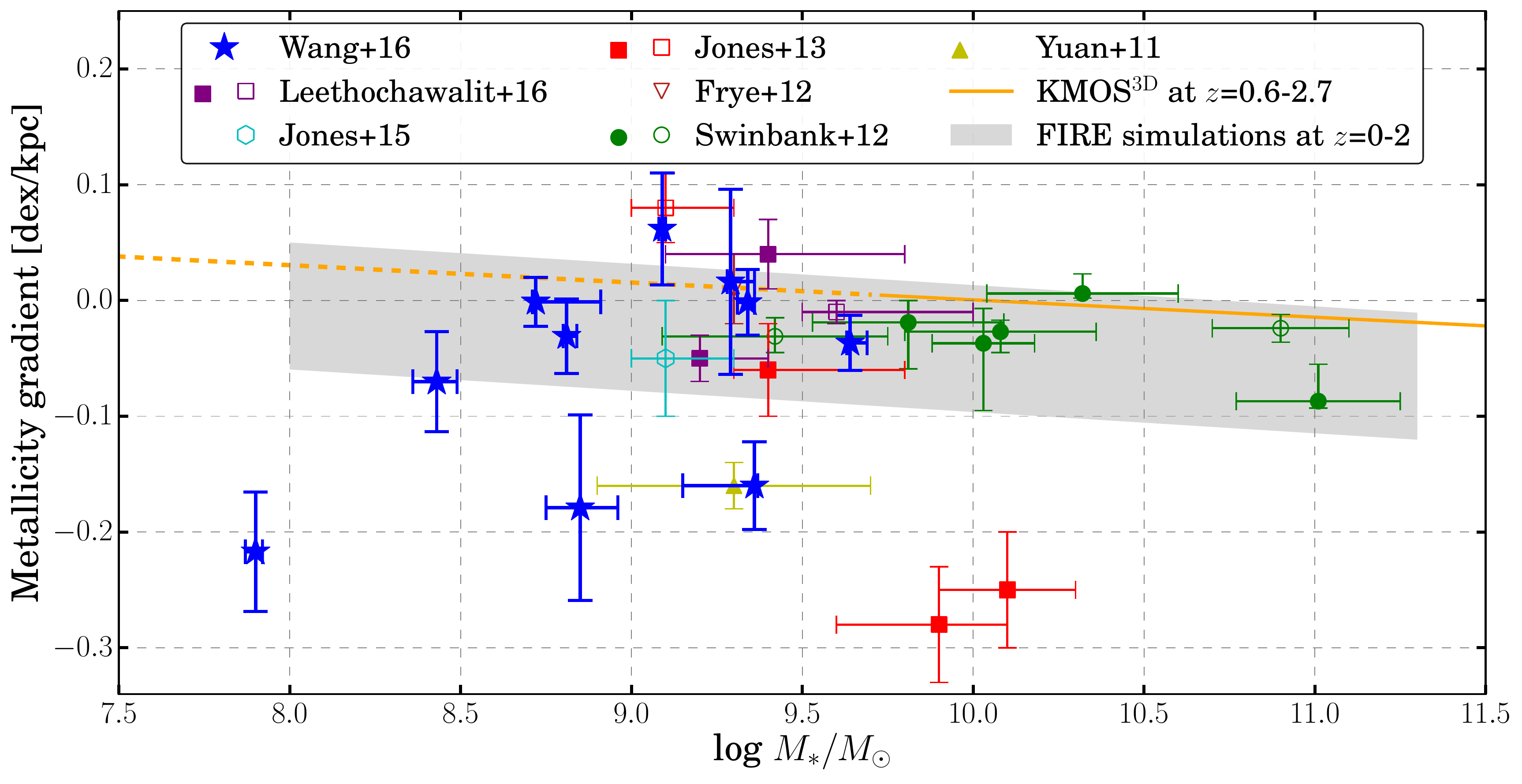}
    \caption{Observed metallicity gradients as a function of \Mstar. The color-coding of symbols is the same as in
    Figure~\ref{fig:oh12gradVSz}. Except for the blue stars, hollow and filled symbols correspond to interacting/merging and
    isolated systems, respectively.
    Most of the \citet{2015arXiv150901279L} gradient measurements do not have reliable \Mstar estimates due to the lack of near IR
    photometry. The orange line shows a simple linear fit to the \mg results at z=0.6-2.7 from the \kd observations
    \citep{Wuyts:2016th}.
    The mass complete range is marked by the solid line, and the dashed line is only an
    extension from the linear fit.
    The 2-$\sigma$ interval of the FIRE simulation results presented by \citet{Ma:2016ww} is
    shown as the shaded region.
    The only work to date that presents \mgms in the $\Mstar\lesssim10^9\Msun$ regime is this work, which even extends to below
    $10^8$\Msun.}
    \label{fig:oh12gradVSmstar}
\end{figure*}

\subsection{Emission line kinematics}\label{subsect:kinem}

%= = = = = = = = = = = = = = = = = = = = = = = = = = = = = = = = = = = = = = = =
% = = = = = = = = = = = = = = = = = = = = = = = = = = = = = = = = = = = = = = = = = =
% Include this table with \input{filename.tex}
% To rotate in emulateapj do: \begin{turnpage}\input{filename.tex}\end{turnpage}
% To display it on multiple pages do: \LongTables\input{filename.tex}
% - - - - - - - - - - - - - - - - - - - - - - - - - - - - - - - - - - - - - - - - - -
\begin{deluxetable}{ccccccccccccc} \tablecolumns{13}
\tablewidth{0pt}
\tablecaption{Velocity dispersion for objects with ground-based IFU data available.}
% - - - - - - - - - - - - - - - - - - - - - - - - - - - - - - - - - - - - - - - - - -
\tablehead{
    \colhead{ID} & 
    \colhead{Instrument} &
    \colhead{Emission feature} & 
    \colhead{$\sigma_{\rm obs}$ (\kms)\tablenotemark{a}} & 
    \colhead{$\sigma_{\rm int}$ (\kms)\tablenotemark{b}}
}
%---------------------------------------------------------------
\startdata
02389  & \muse  & \CIII  & $63\pm4$  & $19^{+11}_{-19}$  \\
02806  & \muse  & \OII   & $42\pm3$  & $24\pm5$  \\
03746  & \muse  & \OII   & $74\pm9$  & $63^{+10}_{-11}$  \\
04054  & \muse  & \OII   & $68\pm6$  & $59\pm7$  \\
05709  & \kmos  & \OIII  & $41\pm7$  & $17^{+13}_{-17}$  \\
05968  & \kmos  & \OIII  & $85\pm6$  & $76\pm7$
\enddata
% - - - - - - - - - - - - - - - - - - - - - - - - - - - - - - - - - - - - - - - - - -
\tablenotetext{a}{Not corrected for instrument resolution}
\tablenotetext{b}{Corrected intrinsic velocity dispersion}
\label{tab:kinem}
\end{deluxetable}

%= = = = = = = = = = = = = = = = = = = = = = = = = = = = = = = = = = = = = = = =

Kinematic information is valuable for interpreting \mgs and their evolution. Where available,
we derive gas-phase kinematics for galaxies observed with \muse or \kmos, by fitting the
strongest available emission feature. This is typically \CIII$\lambda\lambda$1907,1909,
\OII, or \OIII$\lambda\lambda$4960,5008. We fit each doublet as a
sum of two Gaussian components with equal velocity dispersion ($\sigma$) and redshift, plus a
constant continuum level. The fits are weighted by the inverse-variance spectrum estimated
from sky regions in the data cubes. Best-fit values for the spatially integrated spectra are
given in Table~\ref{tab:kinem}. The intrinsic velocity dispersion $\sigma_{\rm int}$ is calculated by subtracting the instrument
resolution (determined from sky ELs) in quadrature from the observed best-fit velocity dispersion $\sigma_{\rm obs}$.
For \CIII and \OII we determine the best-fit doublet ratio of
the integrated spectra, while the \OIII doublet ratio is fixed to its intrinsic atomic value.
For individual spaxels, we fix the doublet ratios to their integrated values in order to
avoid spurious fits. We also spatially smooth each data cube with a 3-pixel (0\farcs6) kernel
to improve SNR in the diffuse extended regions. We attempt to fit every spaxel in the
smoothed data cube surrounding each object. Bad fits are rejected on the basis of low SNR and
unrealistic values (typically requiring $20~\kms<\sigma_{\rm obs}<200~\kms$, and
velocities within 200 \kms of the integrated systemic value).  Figure~\ref{fig:kinem} shows
velocity fields of each source from all spaxels with acceptable EL fits.
We discuss the kinematics of individual galaxies in the context of their metallicity
gradients and other properties in Section~\ref{subsect:indvd}.

\begin{figure*}
    \centerline{
    \includegraphics[width=0.7\columnwidth]{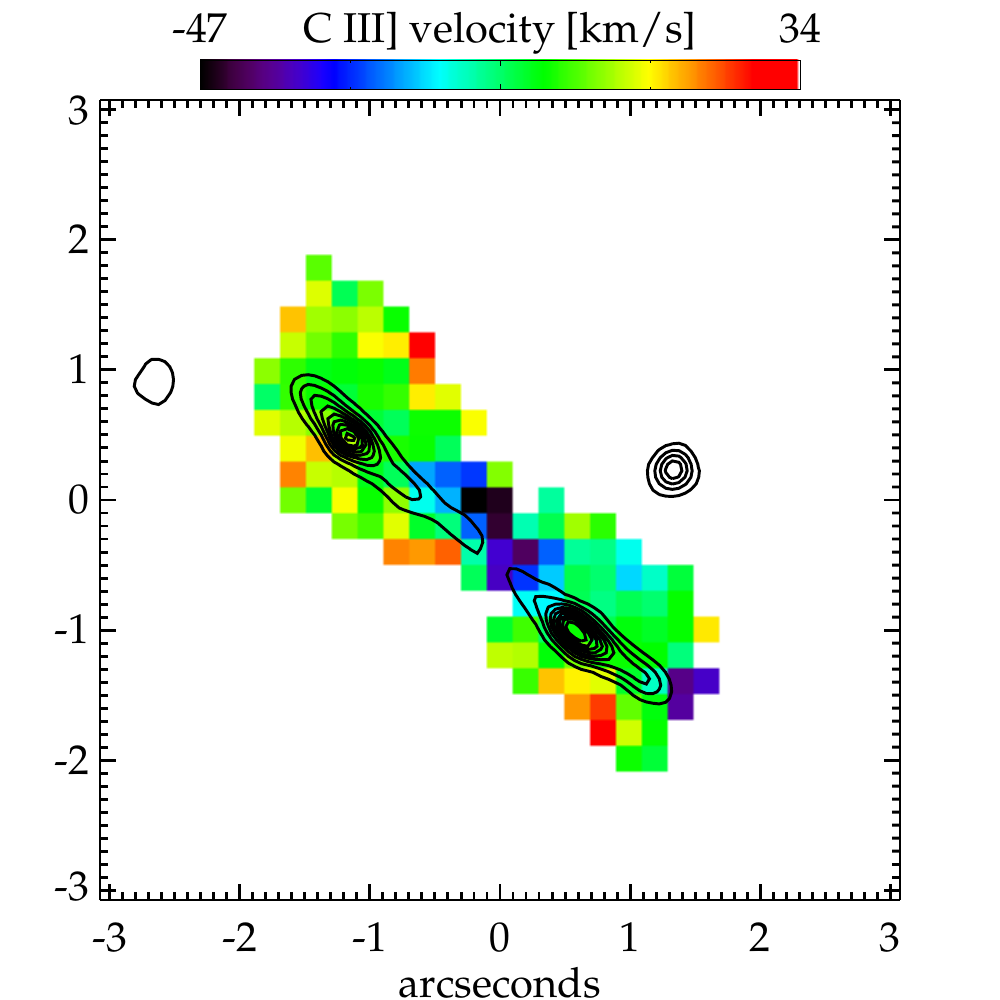}
    \includegraphics[width=0.7\columnwidth]{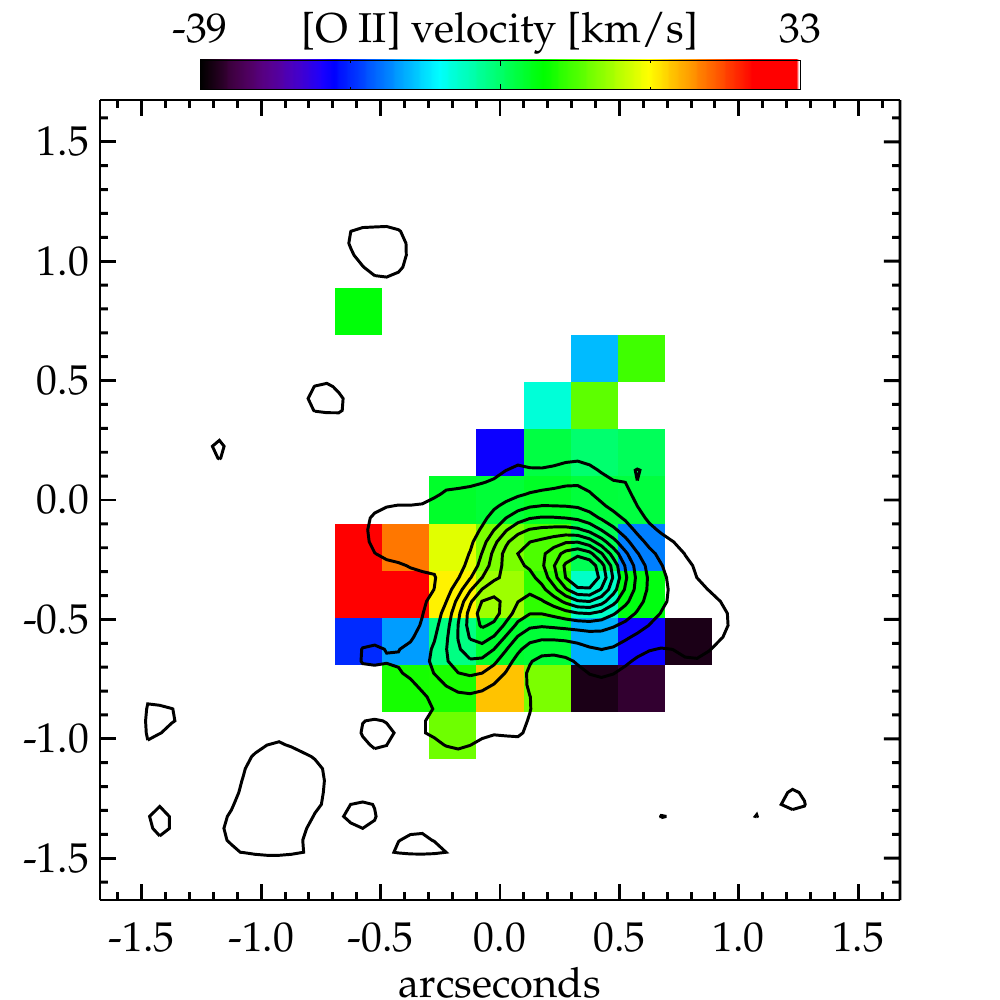}
    \includegraphics[width=0.7\columnwidth]{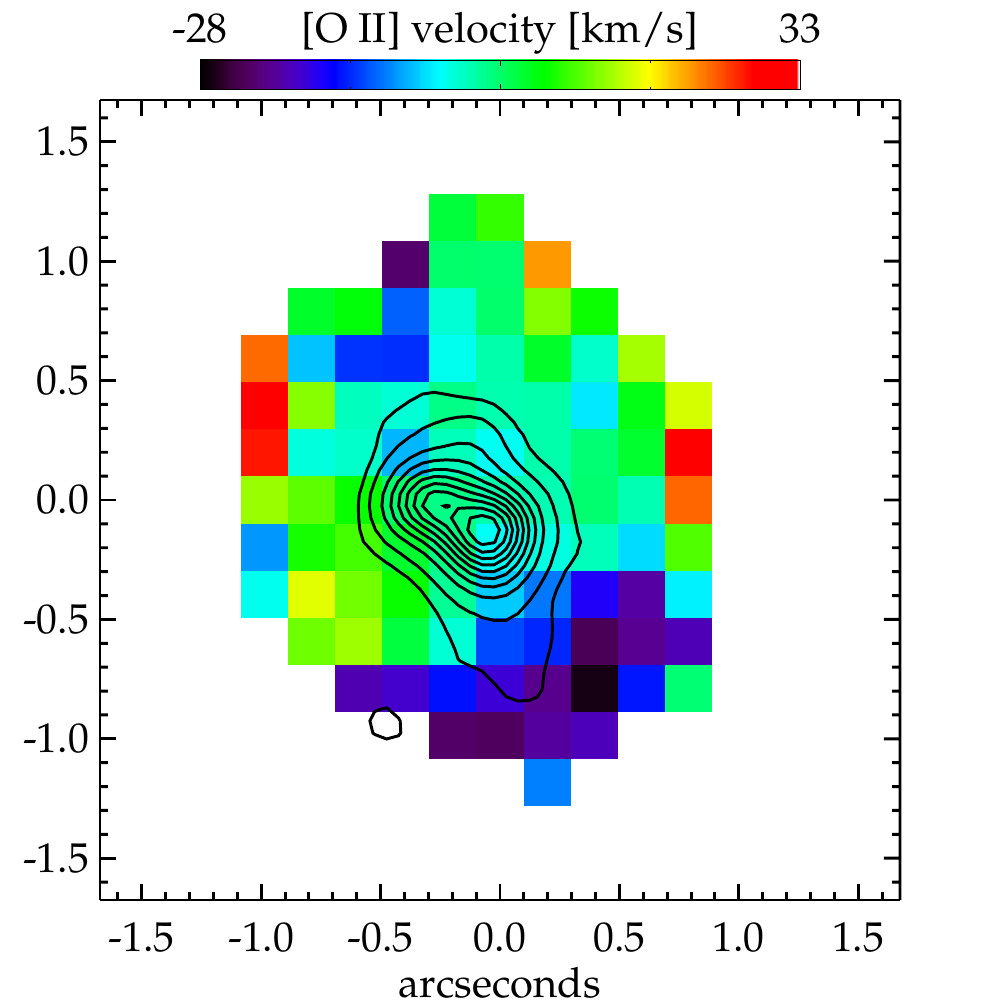}}
    \centerline{
    \includegraphics[width=0.7\columnwidth]{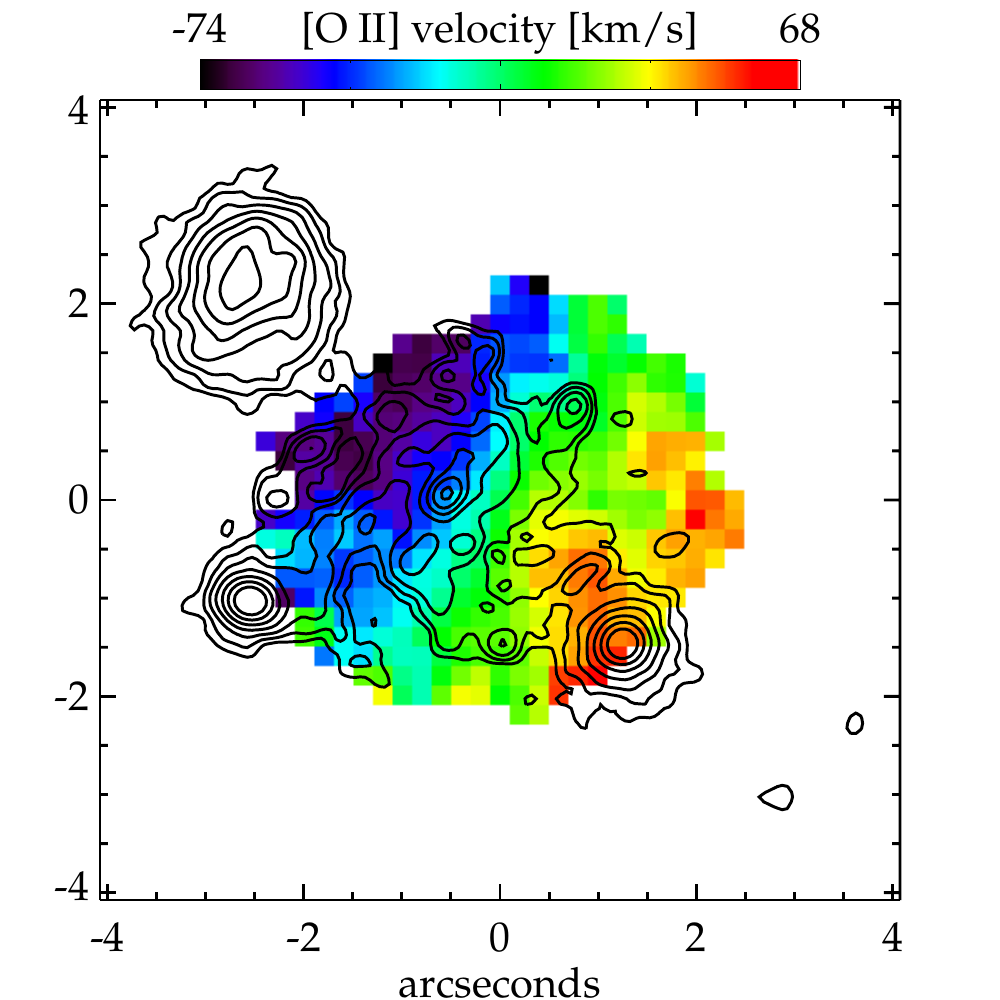}
    \includegraphics[width=0.7\columnwidth]{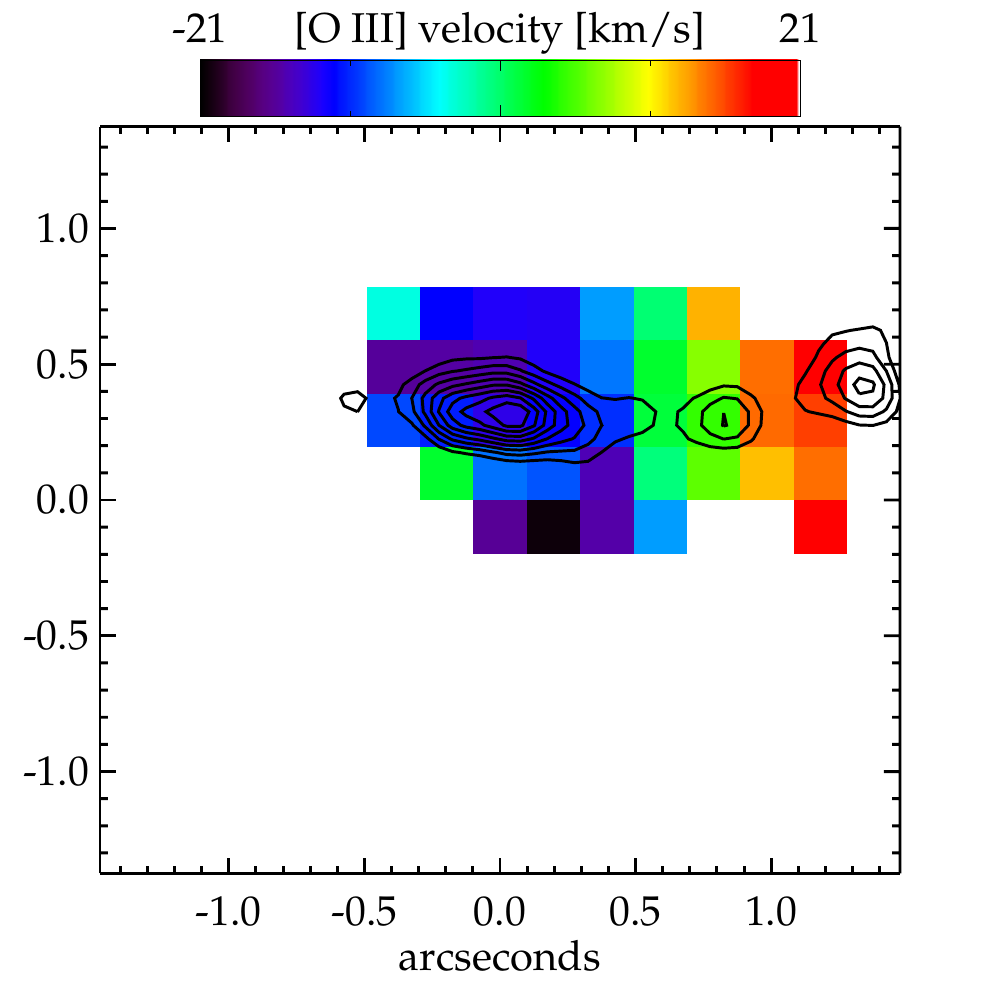}
    \includegraphics[width=0.7\columnwidth]{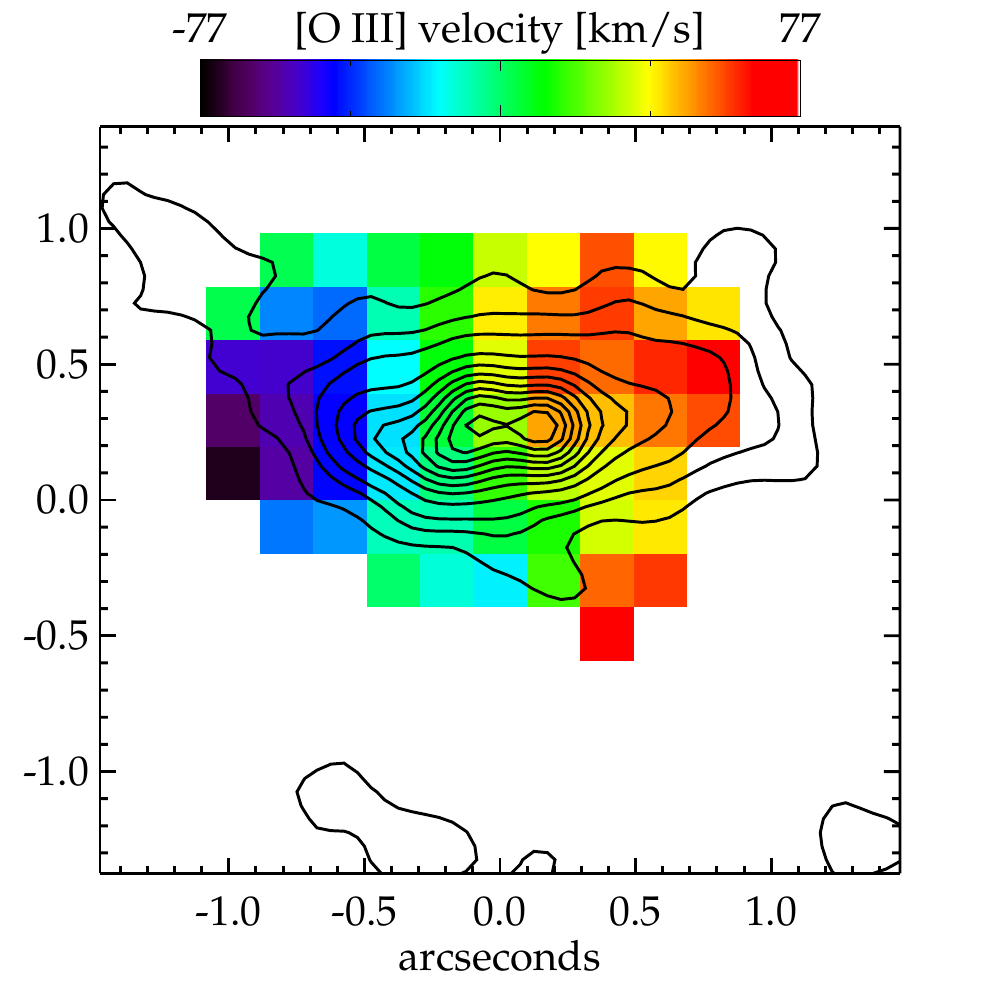}}
    \caption{Velocity fields derived from integral field spectra. \textbf{Top row}: IDs
    02389, 02806, 03746 (left to right) observed with \muse. \textbf{Bottom row}: ID 04054
    observed with \muse (see also Figure~8 in \citet{2016ApJ...822...78G}) and IDs 05709,
    05968 (left to right) observed with \kmos.  In all cases, the color scale shows EL
    velocity centroids in each spaxel, with contours showing \hst/ACS \I band continuum.
    The kinematics map for galaxy 05709 (shown in the lower middle panel) is the first one to
    date presented for star-forming galaxies with $\Mstar\lesssim10^8\Msun$ at $z\gtrsim1$.}
    \label{fig:kinem}
\end{figure*}

\subsection{Notes on individual galaxies}\label{subsect:indvd}

In this section, we highlight some noteworthy sources in our \mg sample. For
reference, all the measurements of global properties, metallicity gradients, and kinematic
decompositions in these sources are presented in Tables \ref{tab:srcprop},
\ref{tab:oh12grad}, and \ref{tab:kinem}, respectively.

{\bf 02389}: This is the most highly magnified source in our sample, with $\mu=43.25_{-5.78}^{+7.03}$, given by the \SJ version 3
model.  It is one of two connected double images that straddle the lensing critical curve at source redshift $z=1.89$, forming a
highly sheared arc more than 5 arcsec.  This is confirmed by the velocity shear of $\sim50$ \kms detected in both components of
the fold arc.  After lensing correction, the stellar mass of this source is $\log(\Mstar/\Msun)=8.43_{-0.07}^{+0.06}$, consistent
with the very low dynamical mass indicated by both the velocity shear and integrated velocity dispersion.  The relatively high
$V/\sigma \gtrsim 2$ of this source shows that it is a disk galaxy. The \mg is mildly negative, \ie, $-0.07\pm0.04$, consistent
with the prediction by analytical chemical evolution models for disk galaxies at similar redshifts.
%(likely higher after accounting for beam smearing and inclination).

{\bf 02806}: This is a star-forming galaxy at $z=1.50$ with very flat \mg of $-0.01\pm0.02$.  We removed the contamination from a
marginally overlapping cluster member (with greenish color in the color composite stamp shown in Figure~\ref{fig:multiP_rest}).
Note that such a process would not be possible with seeing-limited data quality.  The \muse observation does not spatially resolve
this source well, and its \OII emission line is detected with low SNR near the red edge of \muse spectral coverage. A tentative
velocity shear is seen at low significance. The faint fuzzy elongation to the south-east of this galaxy suggests a likely recent
merging event, which accounts for its gradient being flat. (See sources 03746 and 05968 for more outstanding cases where
post-merger features can be identified.) The global metallicity of this source ($\oh=7.91^{+0.16}_{-0.16}$) is slightly lower than
the average given by the MZR at similar \Mstar, in agreement with the study of \citet{MichelDansac:2008gp} suggesting that the
infall of low-metallicity gas during mergers reduces the average metallicity.

{\bf 03746}: This is an intense starburst galaxy with $\textrm{SFR}=10.95_{-0.58}^{+1.61}$
\Msun/yr at $z=1.25$, as revealed by the BPT diagram in Figure~\ref{fig:BPT}. It also
displays a flat \mg of $-0.03\pm0.03$.  Moreover the EL maps of this source show highly
clumpy star-forming regions, which do not have direct correspondence in the broad-band
photometry.  The \muse observation gives a velocity shear of $\sim$20 \kms, and an integrated
dispersion of $\sim$60 \kms, which suggests that this system is dispersion dominated
($V/\sigma<1$).  There is an extended feature southward to the source center, which is
probably a tidal remnant, consistent with this source being in a recent post-merger phase.
Besides, given the irregularities seen in the central region of the \Mstar map, this system
is still long before dynamical relaxation, and we likely capture this merger system almost
right after the coalescence.  Therefore we conclude that the flat \mg recovered in this
system is from tidal interactions. See galaxy 05968 below for an even more outstanding
post-merger case.

{\bf 04054}: This is one of the multiple images of the SN Refsdal host galaxy, a highly magnified almost face-on spiral at
$z=1.49$. It is labeled as image 1.3 by \citet{2015arXiv151005750T}\footnote{It is located at the minimum of its lensing time
delay surface, any intrinsic variabilities in the source are observed first in this image.  Unfortunately in the case of SN
Refsdal, its image is estimated to appear $\sim$ 12 years ago, long before the deployment of WFC3.}.  Our global demographic
analysis yields the following measurements: $\log(\Mstar/\Msun)=9.64_{-0.01}^{+0.05}$, $\textrm{SFR}=16.99_{-2.80}^{+5.71}$
\Msun/yr, $\oh8.70_{-0.11}^{+0.09}$.
Compared with the previous analysis of this galaxy \citep[][not necessarily the same image
though]{2015MNRAS.450.1812L,Rawle:2016dg}, our work benefits from a more complete wavelength coverage of imaging data (from \B to
\H bands) at a greater depth.  In particular, we used the most updated lens model, optimized specifically for this spiral galaxy
\citep{2015arXiv151005750T}, giving a reliable estimate of the magnification.
\citet{Rawle:2016dg} measured SFR$\sim$5 \Msun/yr on image 1.1 of the SN Refsdal host from \herschel IR photometry.
However their value is derived assuming a magnification factor of $\mu=23.0$, given by an older lens model for \clsan
\citep{Smith:2009cu}.
If the magnification value of $\mu=8.14$ given by the latest \glafic model is adopted instead, their measurement yields
SFR$\sim$14, comparable to our SFR measurement.
As a result, our measurements are fairly consistent with the \sfms, the MZR and the FMR at similar redshifts. Given the high
\Mstar and \oh, this galaxy is significantly evolved by $z\sim1.5$ and with the still relatively high SFR, it will likely turn
into a galaxy more massive than our Milky Way at $z=0$.  According to the kinematic map derived from the \muse spectroscopy of the
\OII EL doublet, we obtain the velocity shear to be $\sim$110 \kms (see also Figure~8 in \citet{2016ApJ...822...78G}) and the
integrated dispersion $\sim$60 \kms, in good agreement with \citet{2015MNRAS.450.1812L} who found $2v_{2.2}=118\pm6$ \kms, and
local (spatially resolved) velocity dispersion of 50$\pm$10 \kms, based upon adaptive optics data.

The measured \mg for the SN Refsdal host based upon our analysis of image 1.3 is $-0.04\pm0.02$ from the individual spaxel method,
and $-0.07\pm0.02$ from the radial annulus method. This is not as steep as the previous result reported by \citet{Yuan:2011hj},
\ie $-0.16\pm0.02$, measured from the AO assisted \osiris IFU observation of image 1.1 of the SN Refsdal host, assuming the
\citet{2004MNRAS.348L..59P} \NII/\Ha empirical calibration.  After the discovery of SN Refsdal,
\citet{2015arXiv151209093K} obtained a 1-hr \mosfire spectra on the nuclear region and the SN explosion site in image 1.1, and
measured $\oh=8.6\pm0.1$ and $\oh=8.3\pm0.1$, respectively, using the same \NII/\Ha calibration. The spatial offset between these
two measurements in terms of galactocentric radius is $8.2\pm0.5$ \kpc, given by the \glafic model. As a result, the recent
\mosfire observations indicate a gradient being $-0.04\pm0.02$, highly consistent with our measurements.

Moreover, we obtain a robust constraint on the nebular dust extinction, \ie, $\Av^{\rm N}=0.72_{-0.23}^{+0.23}$. Compared with the
stellar dust extinction retrieved from  SED fitting, \ie, $\Av^{\rm S}=1.10_{0.01}^{0.01}$, we see that in this galaxy, the dust
reddening of the ionized gas in \HII regions is less severe than that of the stellar continuum (see another source, \ie ID 05968,
from which the same conclusion is drawn).  This is different from what has been assumed previously: the color excess of nebular
ELs is larger than that of the stellar continuum, \ie, $E_{\rm S}(B-V)=(0.44\pm0.03)E_{\rm N}(B-V)$ \citep{Calzetti:2000iy}.
Recently, \citet{Reddy:2015ho} conducted a systematic study of dust reddening using the early observations from the \mosdef survey
and found that only 7\% of their sample is consistent with $E_{\rm N}(B-V)\gtrsim E_{\rm S}(B-V)/0.44$. They also discovered that
as sSFR (derived from Balmer EL luminosities) increases, $E_{\rm N}(B-V)$ diverges more significantly from $E_{\rm S}(B-V)$, with
a scatter of $\sim$0.3-0.4 dex at sSFR$\sim3\times10^{-9}$ yr$^{-1}$, which can account for the difference we see in the total
dust attenuation (\Av) of the line and continuum emissions of this galaxy.

{\bf 05709}: This galaxy has the lowest stellar mass in our \mg sample ($\log(\Mstar/\Msun)=7.90_{-0.03}^{+0.02}$) after
lensing correction\footnote{The magnification value we used in correcting for lensing is $7.10_{-0.36}^{+0.43}$, given by the
\glafic model.  The magnification value given by the \SJ model is similar, \ie, $8.43_{-0.60}^{+0.73}$.}. This is for the first
time any diffraction-limited \mg analysis and seeing-limited gas kinematics measurement have been achieved in star-forming
galaxies with such small \Mstar at the cosmic noon. This galaxy shows a steep \mg, \ie $-0.22\pm0.05$.
The velocity shear is measured to be $\sim$25 \kms, with integrated dispersion being $\sim$20 \kms. The inferred low dynamical
mass is consistent with the result from SED fitting mentioned above. In light of the kinematic result $V/\sigma \gtrsim 1$, this
galaxy is probably a clumpy thick disk.  More interestingly, observed from a multi-perspective point of view (see
Figure~\ref{fig:multiP_rest}), this galaxy shows a clumpy star-forming region at the periphery of its disk (having a distance of
3-4 kpc to the mass center), which has a tremendous amount of line emission, yet very little stellar mass, and has low \gpm. Also
see galaxy ID 07255 for a more dramatic case.

{\bf 05968}: With a tidal tail more appreciably shown to the north-east, and almost perpendicular to the galactic plane, this
galaxy (at $z=1.48$) is safely classified as a post-merger. However note that this identification wound not be possible and this
source would have been surely classified as a thick disk due to its symmetric morphology and kinematic properties, without the
ultra deep \hff imaging data.
Compared with the \Mstar and EL maps of ID 03746, the \Mstar map of this galaxy is more smooth and EL maps less clumpy, which
indicates that merging has taken place longer before than in source ID 03746.  Still the star formation is high, \ie
$\textrm{SFR}=27.95_{-4.41}^{+4.70}$ \Msun/yr.  The global metallicity ($\oh=8.47_{-0.08}^{+0.07}$) and stellar mass
($\log(\Mstar/\Msun)=9.34_{-0.03}^{+0.02}$) are quite comparable to the MZR at similar redshifts.  The \mg, as expected for
systems which have experienced recent mergers \citep[see \eg][]{Kewley:2010eg}, is constrained to be robustly flat
($-0.01\pm0.02$).  The EL kinematic analysis yields a velocity shear of $\sim$120 \kms, and an integrated dispersion being
$\sim$75 \kms. The velocity field appears typical of intermediate-mass disk galaxies at this redshift, which infers that this
post-merger system has been significantly relaxed and regained at least some of the rotation support.  This is also one of the few
cases where robust constraints on nebular dust extinction can be derived, \ie $\Av^{\rm N}=0.23^{+0.15}_{-0.13}$, which is
less than the dust extinction of stellar continuum obtained from SED fitting $\Av^{\rm S}=0.60_{-0.01}^{+0.20}$, as already seen
in the case of ID 04054.

{\bf 07255}: This galaxy is highly consistent with the SFMS at $z=1.27$ formulated by
\citet{Speagle:2014dd}, with $\log(\Mstar/\Msun)=9.36_{-0.21}^{+0.01}$ and
$\textrm{SFR}=8.73_{-3.10}^{+3.53}$ \Msun/yr. By $z=0$, it will turn into a sub-Milky
Way-sized galaxy according to \citet{Behroozi:2013fg}.
Its slightly lower than average metallicity of $\oh=8.38_{-0.08}^{+0.08}$ can be a result of
low-metallicity gas inflow, following the popular interpretation of the FMR at high $z$.
This hypothesis would also explain its slightly enhanced SFR.  In the metallicity map the
inflow can be identified with a bright star-forming clump in the lower right corner, which
consists of very little \Mstar, but is dominating in the EL maps, similar to the case of
galaxy 05709 discussed above.  This indicates that in this system, the metal enrichment time
scale is much larger than the star-forming time scale, which is in turn much larger than the
gas infall time scale. Unfortunately, at the moment we do not have EL kinematic observation
on this system.

The metallicity gradient for this galaxy is constrained to be very steep, \ie $-0.16\pm0.03$ using the individual spaxel method,
and $-0.21\pm0.03$ from the radial annulus method.  We also dissected this galaxy to separate the star-forming clump from the bulk
part and re-do the analysis. It is found that the \HII regions closely related to the clump (with at least 2.5 kpc distance to the
galaxy mass center) have significantly lower metallicity and provide more than one third of the entire star formation, \ie,
$\oh=8.04^{+0.14}_{-0.18}$, $\Av^{\rm N}<0.78$, $\textrm{SFR}=3.05^{+1.21}_{-1.10}$, compared with the quantities measured in the
bulk of the galaxy (within 2 kpc to the mass center), \ie, $\oh=8.51^{+0.08}_{-0.10}$, $\Av^{\rm N}=0.63^{+0.59}_{-0.41}$,
$\textrm{SFR}=5.78^{+3.41}_{-2.47}$.  In both regions, EL \Ha is detected at a significance of $\gtrsim 30\sigma$.  This again
demonstrates that measurements of global quantities can be to some extent misleading, and the importance of detailed spatially
resolved information.  Cases like this galaxy should be exceedingly interesting, since their steep metallicity gradients at such
high $z$ are not predicted by current mainstream numerical simulations (equipped with strong galactic feedback) or conventional
analytical chemical evolution models.  Through studying systems like this, we can also learn how gas is accreted into the inner
disk, and more importantly whether the accretion happens alongside star formation or long
before the gas clouds can collapse.

\section{Summary and discussion}\label{sect:conclu}

\subsection{Summary}

In this paper, we took advantage of the deep \hst grism spectroscopic data, acquired by the
\glass and SN Refsdal follow-up programs, and obtain \gpm maps in high-$z$ star-forming
galaxies. The \hst grisms provide an un-interrupted window for observing the cosmic evolution
of metallicity gradients through $z$ from 2.3 to 1.2 (corresponding to the age of the
universe from roughly 2.5 to 5.5 Gyr), including the redshift range $1.7<z<2.0$, unattainable
from ground-based observations due to low atmospheric transmission. Another advantage of the
HST grisms is that multiple ELs are available within a single setup and all the potential
issues related to combining data with different setups and atmospheric conditions are
eliminated. Our main conclusions can be summarized as follows:
\begin{enumerate}

    \item We presented 10 maps of \gpm (as shown in Figure~\ref{fig:oh12grad}), measured at sub-kpc resolution in star-forming
    galaxies in the redshift range of $1.2\lesssim z\lesssim2.3$, in the prime field of \clsan.  This field is so far one of the
    deepest fields exposed by \hst spectroscopy, at a depth of 34(12) orbits of G141(G102),
    reaching a 1-$\sigma$ flux limit of 1.2(3.5)$\times$$10^{-18}$ \Funit, without lensing correction.

    \item We developed a novel Bayesian statistical approach to
    jointly constrain \gpm, nebular dust extinction, and \Ha-based
    dust-corrected star formation rate. In determining these
    quantities, our method does not rely on any assumptions about
    star-formation history, nor the conversion of dust reddening from
    stellar phase to nebular phase.  Unlike the majority of
    previous work on deriving metallicity from strong EL calibrations,
    our method does not compare directly the observed EL flux ratios
    with the calibrated ones. Instead we work directly with EL fluxes, which
    effectively avoids using redundant information and circumvents
    possible biases in the low SNR regime.

    \item The metallicity maps we obtained show a large diversity
    of morphologies. Some maps display disk-like shapes and have \mgs
    consistent with those predicted by analytical chemical evolution
    models (\ie IDs 04054, 01422, 02389, 05732), whereas other
    disk-shaped galaxies have exceedingly steep gradients (\ie IDs
    07255, 05709, 05811) and bright star-forming clumps at the
    periphery of their disks. These large-offset star-forming clumps, containing very little stellar mass, are estimated to have
    lower metallicity than the corresponding global values, which can be interpreted as an evidence for the infall of low
    metallicity gas enhancing star formation.
    We also recovered three systems with nearly uniform spatial distribution of metals (\ie IDs 02806, 03746, 05968). In these
    systems, we can identify possible signatures of post-merger tidal remnants, which suggests that the observed flat gradients
    are likely caused by gravitational interactions.

    \item Collecting all existing sub-kpc resolution \mgms at high-$z$, we study the cosmic evolution and mass dependence of \mgs.
        \begin{itemize}
            \item We found that predictions given by analytical chemical evolution models assuming a relatively extended
            star-formation profile in the early phase of disk growth can reproduce the majority of observed \mgs at high-$z$,
            without involving strong galactic feedback or radial outflows. This confirms the tight link between star formation and
            metal enrichment.
            \item We tentatively observed an correlation between stellar mass and \mg, which is in accord with the ``downsizing''
            galaxy formation scenario that more massive galaxies are more evolved into a later phase of disk growth, where they
            experience more coherent mass assembly and metal enrichment, than lower mass galaxies.
        \end{itemize}
    A larger sample of uniformly analyzed metallicity maps well sampled in the full \Mstar range (especially in the low mass
    regime, \ie, $\Mstar\lesssim10^{9}\Msun$) at high redshifts is needed to further investigate the cosmic evolution and mass
    dependence of \mgs.

    \item We presented 6 maps of EL kinematics obtained from ground-based IFU instruments. Thanks to the powerful synergy of
    lensing magnification and diffraction-limited data, we push for the first time the mass limit of these spatially resolved
    analysis (the measurements of \mgs and EL kinematics) to below $10^8\Msun$.

    \item We also compiled a sample of 38 global metallicity measurements all derived from the \citet{2008A&A...488..463M} EL
    calibrations in the current literature (including 13 galaxies presented in this work). This sample, at $z\sim1.8$, spans three
    orders of magnitude in \Mstar. We measured the MZR and tested the FMR using this sample.
        \begin{itemize}
            \item The slope of the MZR constrained by this sample rules out the momentum-driven wind model by
            \citet{2012MNRAS.421...98D} at 3-$\sigma$ confidence level. Our MZR is consistent with the theoretical prediction
            given by recent cosmological zoom-in simulations, suggesting that high spatial resolution simulations are favorable in
            reproducing the metal enrichment history in star-forming galaxies.
            \item Given the intrinsic scatter and measurement uncertainties, we do not see any significant offset from the FMR
            except the subsample of \citet{Wuyts:2012gb}. This discrepancy is likely due to the fact that the \citet{Wuyts:2012gb}
            dataset is the only subsample which relies exclusively on \NII/\Ha to estimate metallicity. We therefore advocate to
            avoid nitrogen lines when estimating \gpm in high-$z$ \HII regions.
        \end{itemize}

\end{enumerate}

\subsection{Interpretation of flat/steep \mgs at high-$z$}

We now turn to the interpretation of our observed metallicity gradients and maps. In general, there are four physical mechanisms
that could flatten a galaxy's \emph{pre-existing} negative metallicity gradient (not exclusively in high-$z$ scenarios):\\
\indent\textbullet~Efficient radial mixing by tidal effects from gravitational
interactions, as seen in galaxy IDs 02806, 03746, and 05968 in our sample and the triply
imaged arc 4 in our previous work \citep{2015AJ....149..107J}.\\
\indent\textbullet~Rapid recycling of metal-enriched material by enhanced galactic
feedback, as seen in the results of the MaGICC numerical simulations for example.\\
\indent\textbullet~Turbulence mixing driven by thermal instability as elucidated by
\citet{Yang:2012ia}.\\
\indent\textbullet~Funnelling of cold streams of low metallicity gas directly into the galaxy
center, as suggested by \citet{Dekel:2009fz}\footnote{This mechanism is invoked to
explain the inverted gradients (lower abundances in the inner disk regions) seen in some of
the AMAZE/LSD observations \citep{Cresci:2010hr}. But we note in their observations,
the near-IR integral field spectrometer \sinf was used in seeing-limited mode.}

Besides \emph{pre-existing} negative gradients being flattened by these processes, galaxies
can also inherit a broadly flat \mg from relatively extended star-formation profile in the
early disk-formation phase and coherent mass assembly for the disk growth. Because there is
strong evidence that the mass assembly for Milky Way progenitors takes place uniformly at all
radii, maintaining almost the same \Mstar profile from $z\sim2.5$ to $z\sim1$
\citep{vanDokkum:2013fg,Morishita:2015cz}. However the uncertainties associated with current
measurements are still insufficient to distinguish numerical simulations with enhanced
feedback prescriptions from conventional analytical chemical evolution models assuming more
uniform star formation (see Figure~\ref{fig:oh12gradVSz}).

One perhaps obvious but nevertheless important caveat common to most galaxy evolution work is
that we only have access to \emph{cross-sectional} data, \ie, observations of different
objects at one specific epoch. Therefore any interpretations of plots like
Figure~\ref{fig:oh12gradVSz}, which try to tie these data with \emph{longitudinal} theories
(describing the evolution of one specific object), should be made with caution.

Perhaps an even more intriguing interpretation of Figure~\ref{fig:oh12gradVSz}, which is
robust against the aforementioned caveat, is that there exist some really steep \mgs at high
$z$, such as our galaxy IDs 07255, 05709, and 05811.
Based upon current measurements, the occurrence rate of high-$z$ \mgs more negative than $-0.1$ dex/kpc is 6/35.
We can derive the following constraints on the physical processes that can contribute to
these steep \mgs observed beyond the local universe:\\
\indent\textbullet~galactic feedback (including supernova explosions and stellar winds) must
be of limited influence (via being confined for instance), \\
\indent\textbullet~star-formation efficiency is low, or at least lower than their local
analogs (in simulations, this can be achieved by having a higher star-formation threshold or
less concentrated molecular gas, as shown by the MUGS runs), \\
\indent\textbullet~star formation at early times is highly centrally
concentrated, and the mass assembly of the inner regions is much faster than that of the outer regions, \\
\indent\textbullet~alternatively, steep gradients are the result of low-metallicity
gas falling onto the outskirts of their disks, implying that the dynamic time scale is much
shorter than the star-forming time scale, which is much shorter than the metal enrichment
time scale, as seen in our galaxy 07255.

We need larger samples with robust measurements in order to more accurately quantify the
occurrence rate of these steep \mgs in high-$z$ star-forming disks.
The existence of a distinct population of galaxies with steep \mgs, if confirmed by future
observations constraining their abundance, is a powerful test of galaxy formation models,
because this is not predicted by most current theoretical models.
Hence, these galaxies with steep \mgs should be the prime targets for new numerical
predictions: future \jwst observations will constrain their properties in great detail,
allowing discriminating tests of current theories.

Furthermore, most current numerical simulations are aimed at reproducing present-day Milky
Way analogs, surrounded by a dark matter halo with Virial mass $\sim10^{12}\Msun$. It would
be useful to have large samples of simulated sub-\Lstar galaxies (\Mstar below
$5\times10^9\Msun$ at $z\sim2$) in order to investigate how metals are recycled in these less
massive systems and provide a more suitable comparison to our \mgms at high $z$.  With
improvements on both the observational and theoretical sides, we will be able to answer why
in these high-$z$ star-forming disk galaxies, the \mgs can be established so early and why
they are so steep compared with those found in local analogs.

\acknowledgements

We acknowledge support by NASA through grant HST-GO-13459 (PI: Treu).
This paper is based on observations made with the NASA/ESA Hubble Space Telescope, and
utilizes gravitational lensing models produced by PIs Brada{\v c}, Natarajan \& Kneib (CATS),
Merten \& Zitrin, Sharon, and Williams, and the GLAFIC and Diego groups. This lens modeling
was partially funded by the HST Frontier Fields program conducted by STScI.  STScI is
operated by the Association of Universities for Research in Astronomy, Inc. under NASA
contract NAS 5-26555. The lens models were obtained from the Mikulski Archive for Space
Telescopes (MAST).
Ground based data have been obtained with the MUSE and KMOS spectrographs at the ESO VLT
Telescopes.
TJ acknowledges support provided by NASA through Program \# HST-HF2-51359 through a grant from the Space Telescope Science
Institute, which is operated by the Association of Universities for Research in Astronomy, Inc., under NASA contract NAS 5-26555.
XW is greatly indebted to Ryan Sanders for a careful reading of an early version of this manuscript, and to Sirio Belli, Sandy 
Faber, Cheng Li,
Hui Li, Zhiyuan Li, Shude Mao, Houjun Mo, Alice Shapley, Yiping Shu, Alessandro Sonnenfeld, Lixin Wang, Tao Wang, Anita Zanella, 
and Zheng Zheng for helpful discussions.
XW also acknowledges the warm hospitality at the Tsinghua Center for Astrophysics where a large portion of this paper was written.
Last but not the least, XW thanks Dr. Xiao-Lei Meng for her everlasting love and support.

%========================================================================
\bibliographystyle{apj}
\bibliography{bibtexlib}
% NOTE: do NOT put emulateapj at the current path, otherwise will mess up bibtex
%========================================================================

\end{document}